\documentclass[journal]{IEEEtran}

\usepackage{hyperref}
\usepackage[flushleft]{threeparttable}
\usepackage{textcomp}
\usepackage{gensymb}
\usepackage[euler]{textgreek}
\usepackage{amsfonts}
\usepackage{bm}

\usepackage[table,xcdraw]{xcolor}
\usepackage{multirow}
\usepackage{hhline}

\usepackage{algorithm,algorithmic}

\usepackage[binary-units]{siunitx}
\DeclareSIUnit{\count}{count}
\DeclareSIUnit{\year}{yr}
\DeclareSIUnit{\north}{N}
\DeclareSIUnit{\east}{E}

\ifCLASSINFOpdf
	\usepackage[pdftex]{graphicx}
\else
	\usepackage[dvips]{graphicx}
\fi
\usepackage[cmex10]{amsmath}
\usepackage{array}

\ifCLASSOPTIONcompsoc
 \usepackage[caption=false,font=normalsize,labelfont=sf,textfont=sf]{subfig}
\else
 \usepackage[caption=false,font=footnotesize]{subfig}
\fi
\usepackage{url}

\begin{document}

\title{Field Evaluation of Four Low-cost PM Sensors and\\Design, Development and Field Evaluation of\\A Wearable PM Exposure Monitoring System}

\author{Wei-Ying~Yi,
		Yu~Zhou,
		Ya-Fen~Chan,
		Yee~Leung,
		Kam-Sang~Woo,
		Wen-Wei~Che,
		Kai-Hon~Lau,
		Jia-Min~Chen,
		and~Kwong-Sak~Leung
\thanks{W.Y. Yi, J.M. Chen and K.S. Leung are with the Department
of Computer Science and Engineering, Y. Zhou, Y.F. Chan, and Y. Leung are with the Institute of Future Cities and the Department of Geography and Resource Management, and K.S. Woo is with the Department of Medicines and Therapeutics, The Chinese University of Hong Kong, Sha Tin, N.T., Hong Kong, China; W.W. Che and K.H. Lau are with the Department of Civil and Environmental Engineering, The Hong Kong University of Science and Technology, Clear Water Bay, Kowloon, Hong Kong, China (e-mail: wyyi1991@gmail.com, 1155092190@link.cuhk.edu.hk, ksleung@cse.cuhk.edu.hk, yuzhou@cuhk.edu.hk, yfchan@cuhk.edu.hk, yeeleung@cuhk.edu.hk, kamsangwoo@cuhk.edu.hk, wenweiche@ust.hk, and alau@ust.hk).}
}

\markboth{Journal of Latex Class Files,~Vol.~1, No.~1, June~2022}%
{Yi \MakeLowercase{\textit{et al.}}: Evaluation of Low-cost Particulate Matter Sensors and Development of Wearable Particulate Matter Exposure Monitoring System}

\maketitle

\begin{abstract}
In order to mitigate the significant biases or errors in epidemiological research studying specific associations between particulate matter (PM) and health, which are introduced by the coarse and/or inadequate assessments of PM exposure from conventional PM monitoring paradigm, a personalized monitoring system consisting of a low-cost wearable PM device is proposed. However, owing to the absence of a unifying evaluation protocol for low-cost PM sensors, the evaluation results from the existing studies and the performance specifications from their datasheets are of limited reference values when attempting to determine the best candidate sensor for the proposed monitoring system. In this regard, the authors appeal to the research community to develop a standardized evaluation protocol for low-cost PM sensors and devices, and a unifying attempt is established in this manuscript by adopting the definitive terminology provided in international documents and the evaluation metrics regarded as best practices. Collocated on the rooftop of the Hong Kong University of Science and Technology Air Quality Research Supersite, four empirically selected PM sensors were first compared against each other and calibrated against two reference monitors. The calibrated sensors were then evaluated against the reference following the protocol. The PlanTower PMS-A003 sensor was eventually selected for the wearable device as it typically outperformed the others in terms of affordability, portability, detection capability, data quality, as well as humidity and condensation insusceptibility. An automated approach was proposed to identify and remove the condensation associated unusual sensor measurements. The wearable device in the proposed system have better affordability and portability as well as similar usability and data accessibility compared to those existing devices recognized. The 10 initially implemented devices were also evaluated and calibrated at the Supersite and showed comparable results as the PMS-A003. Additional 120 units were manufactured of which 51 have been delivered to the subjects to acquire their daily PM\textsubscript{2.5} exposure assessments for investigating the association with subclinical atherosclerosis.
\end{abstract}


\section{Introduction}
\label{Sect:Introduction}

As one of the most serious air pollutants, particulate matter (PM), especially the particles with aerodynamic diameters \cite{hinds1999aerosol3p6} below \SI{2.5}{\micro\meter}, known as PM\textsubscript{2.5}\footnote{According to the ISO 23210:2009, the PM\textsubscript{2.5} is referred to as particulate matter that passes through a size-selective inlet with a \SI{50}{\percent} efficiency cut-off at \SI{2.5}{\micro\meter} aerodynamic diameter.}, increasingly and adversely affects the global environment and human health. From the environmental perspective, PM produced by the anthropogenic activities, such as industrial processes, automobile emissions, and household fuel combustions, makes a significant amount of contribution to the total concentration and leads to haze, acid rain, climate change, and so forth \cite{guo2014elucidating,elser2016new,wang2005ion,tai2010correlations}. From the human health perspective, exposure to PM, especially to particles with smaller sizes, causes numerous negative effects including respiratory system diseases \cite{guaita2011short,xing2016impact}, cardiovascular diseases \cite{pope2004cardiovascular,miller2007long}, and premature mortality \cite{samoli2008acute,apte2015addressing}. In addition, previous studies provided suggestive evidence of that long-term exposure to high concentration level of PM\textsubscript{2.5} may cause reproductive and developmental outcomes, and cancers \cite{radwan2016exposure,ross2009integrated}. In the current pandemic of coronavirus infection disease (COVID-19), recent analyses also suggested preliminary evidence of a positive relationship between air pollution, the concentration of PM\textsubscript{2.5} in particular, and COVID-19 cases and deaths \cite{wu2020exposure,cole2020air,yongjian2020association}.

Given the board health risks associated with PM, public awareness of their personal exposures to PM are hindered due to the lack of PM information with high spatio-temporal resolution provided by governmental agencies, while the PM concentrations in urban areas may present significant spatio-temporal inhomogeneity \cite{bell2011community,steinle2013quantifying,yi2015survey,de2017comparison}. Conventionally, the governmental PM information is acquired by a few fixed stations composed of approved instruments. These instruments are expensive, and require skillful staff, regular maintenance, and stringent operation conditions \cite{chow1995measurement}. Densely deploying a large number of these stations is impractical. PM information in Hong Kong for example, is provided by the Environmental Protection Department (EPD) with in total 18 fixed stations comprising instruments certified by the Hong Kong Laboratory Accreditation Scheme (HOKLAS) \cite{epd2019air}. Although spatially and temporally fine-grained PM information can be estimated via modeling and/or interpolation \cite{krige1966two,zheng2013u,qi2018deep,he2018satellite,leung2019integration}, few estimation techniques have spatial resolution higher than \SI{1 x 1}{\kilo\meter}. Besides, the PM information measured by or estimated according to the outdoor monitoring stations may not reflect the actual exposures of individuals when they are indoors \cite{lee2000indoor,bo2017assessment} where people often spend as much as \SI{90}{\percent} of their time \cite{klepeis2001national}.

The deficiencies of conventional PM monitoring paradigm along with the public acute concerns about the adverse PM-related health effects, particularly in the developing countries and regions, foster the development of personal PM exposure monitoring systems \cite{snyder2013changing,cheng2014aircloud,zhuang2015airsense,steinle2015personal,tian2016mypart,chen2017open,morawska2018applications,genikomsakis2018development,tryner2019design,chatzidiakou2019characterising,brilliant2019pico,plumelabs2019flow,atmotube2019atmotube,habitatmap2019air,airviz2019speck,kaiterra2019air}, which are complementary to the conventional ones, with portable or wearable monitoring devices utilizing sensors that are compact, low-cost, power-efficient, and yet instantaneously providing data with quality at a certain degree. These emerging monitoring systems are typically referred to as practical applications of participatory sensing \cite{burke2006participatory} or community sensing \cite{krause2008toward}. Public participants taking the portable monitoring devices alongside are aware of the real-time PM concentrations in their vicinity while contributing sensing data to the community. Identifying micro-environmental hot spots, determining personal exposure levels, and exploring less exposed commuting strategies are achievable with and for single participant. Additionally, data aggregated from a large number of active participants unlock the opportunities for constructing a dense PM concentration map and providing PM exposure forecasts or early warnings that further encourage public participation. Ultimately, public awareness of and engagement towards environmental protection may likely be increased.

Regarding the many aspects that could be beneficial from the paradigm shift of the PM exposure monitoring, this study focuses on addressing the significant biases or errors, which are introduced by the coarse and/or inadequate PM exposure assessments from conventional approaches, in epidemiological studies researching specific pollution-health associations \cite{zeger2000exposure}. This is a pilot study for a research project investigating the association between the daily PM\textsubscript{2.5} exposure levels and the subclinical atherosclerosis of the adolescents and the adults in Hong Kong and Chongqing. In order to assess an individual's everyday PM exposure level, a dedicated system embodying a wearable PM\textsubscript{2.5} monitoring device carried by each subject was proposed. Four different sensors that are capable of detecting the presence of PM\textsubscript{2.5} were evaluated to determine the most suitable one for constructing the wearable monitoring device. Using the selected PM sensor model, ten monitoring devices were then assembled and evaluated. A mobile application and a back-end server were also developed for data visualization, relay, management, and analytics. Initial results suggested that the implemented system met its intended purposes. Based on the users' feedbacks, 120 monitoring devices with a few slight modifications were manufactured, and 51 of them have been delivered to the subjects to date (31 January 2021).

Among the many techniques for detecting the presence of PM \cite{yi2015survey,kulkarni2011aerosol,giechaskiel2014review,amaral2015overview}, the optical method, light-scattering in particular, is widely applicable in commercial instruments because of its simple yet effective detection mechanism that can measure PM continuously and safely, and provide high sensitivity even at low concentrations. According to section \ref{SubSect:LightScattering}, light-scattering based instruments can be categorized as optical particle counter (OPC) or photometer/nephelometer (PNM) depending on their detection principles. Thanks to the technology advancements, great improvements in miniaturization and cost-reduction of the OPCs and the PNMs have been achieved, even in millimeter scale with few US dollars \cite{li2014miniaturized,dong2016silicon}, since they were first introduced \cite{kerker1997light}. These compact and low-cost instruments (hereinafter referred to as PM sensors) are extensively adopted in personal PM exposure monitoring systems. In the remainder of this section, the necessity for evaluating the PM sensors in environment close to the final deployment one is presented in section \ref{SubSect:WhyEvaluation}. Then, the benefits of developing a wearable PM exposure monitoring system in-house are illustrated in section \ref{SubSect:WhyDevelopment}. Finally, the contributions of this manuscript are summarized in section \ref{SubSect:Contributions}.

\subsection{Why evaluate the performances of the PM sensors again?}
\label{SubSect:WhyEvaluation}

While the PM sensors are becoming a mainstay of research that investigates the pollution-health associations in personal level, only a few of them have been rigorously and scientifically evaluated in laboratory or in the field. The literature on their performances is either thin \cite{lewis2016validate} or mixed \cite{crilley2018evaluation}. Further, multiple PM sensors utilizing the state-of-the-art technologies have been launched recently on the emerging market. Many questions remain regarding the reliability and data quality of these sensors and their suitability for the target applications \cite{kumar2015rise}, which are the major hindrances for deploying them at a large scale. Publishing uncertain information from PM sensors that have not been thoroughly evaluated may result in various negative outcomes. For example, it may cause either unnecessary public concern or complacency about the pollution levels and the associated health risks \cite{judge2014regulation}. Misled people might question the authenticity of the pollution information provided by governmental agencies, which further leads to undermining government credibility.

Although the PM sensors' datasheets, which typically list the sensor performances including operation ranges, detection limits, and accuracies, are available from their manufactures, they are of limited value because there exist ambiguities in the assessment processes and evaluation metrics accomplished by the manufactures. Stringently relying on the performances listed in a sensor's datasheet or directly comparing different sensors' performances based on their datasheets may end up with mistaken decisions. These datasheets, at best, provide a rough qualitative baseline for screening out those PM sensors that are inapplicable for the intended purposes.

As detailed in section \ref{SubSect:LightScattering}, a PM sensor retrieves the mass concentration of sampled aerosol by analyzing the intensity of the scattered light from illuminated particles with multiple assumptions on the aerosol's properties. Despite most of PM sensors have been well calibrated by the manufactures before shipment utilizing aerosols with known properties, noticeable uncertainties are introduced to the PM sensors' measurements when the actual aerosol's properties, which vary significantly over time and across locations, do not meet these assumptions. Likewise, a PM sensor's measurement responses are heavily dependent on the meteorological conditions, such as relative humidity (RH) and temperature (TEMP), causing further uncertainties in measurements. On top of that, the responses to meteorological condition changes from a PM sensor vary as well depending on the aerosol's properties, e.g., the aerosol's chemical composition. Directly adopting the evaluation results from existing studies conducted on various occasions may not diminish, but in turn, raise the uncertainties in concentration measurements \cite{zusman2020calibration}.

Owing to the absence of a unifying evaluation protocol in the literature \cite{castell2017can}, which causes inconsistencies in reference methods (gravimetric, beta attenuation, etc.), ambiguities in evaluation metrics (e.g., definition of accuracy), diversities in assessment criteria (uncertainty \cite{union2008directive}, coefficient of variation \cite{ecfr2020part53subpartC}, etc.) and analytical techniques (e.g., sampling/averaging interval and regression models), etc., it is extremely difficult to make inter-comparison on a sensor's performances among different studies and draw generalized conclusions regarding their evaluation results. Existing studies tend to be of limited value when attempting to determine the most suitable sensor for the intended application.

Consequently, there is an apparent need for evaluating the PM sensors in environment close to the final deployment one instead of simply adopting the sensor performances available in the datasheets and/or the literature. For a compact and low-cost PM sensor, it is neither critical nor possible to meet the same performances as the research-grade instruments and the conventional monitors. Instead, the purpose of evaluating the PM sensors is to achieve sufficient degree of knowledge about their performances under deployed environment with different circumstances and ultimately improve the confidence in their measurements.

In this manuscript, three models of OPCs and one model of PNM, namely the Alphasense OPC-N2\footnote{The OPC-N2 sensor model was evaluated in this study. This sensor model was upgraded to OPC-N3 by the manufacturer while the authors were drafting this manuscript.} \cite{alphasense2019opcn3}, the Honeywell HPMA115S0 \cite{honeywell2019hpma115s0}, the PlanTower PMS-A003 \cite{plantower2019pms}, and the Sharp DN7C3CA007 \cite{sharp2019dn7c}, were selected empirically for performance evaluation by comparing the time synchronized data pairs acquired by these PM sensors and two collocated reference instruments, i.e., a TSI DustTrak II Aerosol Monitor 8532 \cite{tsi2019dusttrakii} research-grade instrument and a Thermo Scientific Model 5030i SHARP Monitor \cite{thermo2019model5030i} federal equivalent method (FEM) instrument. After a comprehensive consideration, the PMS-A003 sensor was selected for constructing the wearable device utilized in the proposed personal PM exposure monitoring system. While drafting this manuscript, three recently launched sensor models, namely the Honeywell HPMA115C0 \cite{honeywell2019hpma115c0}, the Sensirion SPS30 \cite{sensirion2019sps30}, and the Alphasense OPC-R1 \cite{alphasense2019opcr1}, which are considered to be the potential candidates for the wearable device, have been identified. Their performance evaluations are left for future work.

\subsection{Why develop the wearable PM monitoring device in-house?}
\label{SubSect:WhyDevelopment}

\begin{table*}[ht]
\centering
\caption{Comparisons of commercial PM monitoring devices.}
\label{tab:CompareCommercialPMDevices}
\renewcommand{\arraystretch}{1.06}
\begin{threeparttable}
\begin{tabular}{|c|r|r|r|r|r|r|}

\hline
\textbf{Name}
& \multicolumn{1}{c|}{\textbf{PiCO Home} \cite{brilliant2019pico}}
& \multicolumn{1}{c|}{\textbf{Flow2} \cite{plumelabs2019flow}}
& \multicolumn{1}{c|}{\textbf{Atmotube PRO} \cite{atmotube2019atmotube}}
& \multicolumn{1}{c|}{\textbf{AirBeam2} \cite{habitatmap2019air}}
& \multicolumn{1}{c|}{\textbf{Speck} \cite{airviz2019speck}}
& \multicolumn{1}{c|}{\textbf{Laser Egg} \cite{kaiterra2019air}}
\\ \hline 

\begin{tabular}[c]{@{}c@{}}\textbf{Price} \\ \textbf{(\SI[detect-weight]{}{US\$})} \end{tabular}
& $99$ & $159$ & $179$ & $249$ & $149$ & $149$
\\ \hline 

\begin{tabular}[c]{@{}c@{}}\textbf{Dimensions} \\ \textbf{(\SI[detect-weight]{}{\milli\meter})}\end{tabular}
& $48 \times 48 \times 21$
& $125 \times 40 \times 35$
& $86 \times 50 \times 22$
& $132 \times 98 \times 28$
& $114 \times 89 \times 94$
& $106 \times 88$\tnote{f}
\\ \hline

\begin{tabular}[c]{@{}c@{}}\textbf{Weight} \\ \textbf{(\SI[detect-weight]{}{\gram})}\end{tabular}
& $48$ & $70$ & $104$ & $142$ & $165$ & $285$
\\ \hline

\begin{tabular}[c]{@{}c@{}}\textbf{Battery} \\ \textbf{(\SI[detect-weight]{}{\hour})}\end{tabular}
& external power
& $24$ to $72$\tnote{b}
& up to $168$\tnote{d}
& $10$
& external power
& $8$
\\ \hline

\textbf{Accessibility}
& \begin{tabular}[c]{@{}r@{}}mobile app,\\multi-color LEDs\end{tabular}
& \begin{tabular}[c]{@{}r@{}}mobile app,\\animated LEDs\end{tabular}
& mobile app
& \begin{tabular}[c]{@{}r@{}}mobile app,\\web app\end{tabular}
& \begin{tabular}[c]{@{}r@{}}mobile app,\\web app, screen\end{tabular}
& \begin{tabular}[c]{@{}r@{}}mobile app,\\screen\end{tabular}
\\ \hline

\textbf{Connectivity}
& \begin{tabular}[c]{@{}r@{}}Bluetooth LE,\\Wi-Fi\end{tabular}
& Bluetooth LE
& Bluetooth 5.0
& \begin{tabular}[c]{@{}r@{}}Bluetooth 2.0,\\Wi-Fi, GSM\end{tabular}
& Wi-Fi
& Wi-Fi
\\ \hline

\begin{tabular}[c]{@{}c@{}}\textbf{Sampling}\\\textbf{Interval}\end{tabular}
& unknown
& \SI{1}{\minute}
& \SI{10}{\minute}
& \begin{tabular}[c]{@{}r@{}}\SI{1}{\second}\\ or \SI{1}{\minute}\tnote{e}\end{tabular}
& \SI{1}{\minute}
& unknown
\\ \hline

\textbf{PM Sensor}
& \begin{tabular}[c]{@{}r@{}}PlanTower\\PMS-A003\tnote{a}\end{tabular}
& dedicated\tnote{c}
& \begin{tabular}[c]{@{}r@{}}Sensirion\\SPS30\end{tabular}
& \begin{tabular}[c]{@{}r@{}}PlanTower\\PMS-7003\end{tabular}
& \begin{tabular}[c]{@{}r@{}}SYhitech\\DSM501A\end{tabular}
& \begin{tabular}[c]{@{}r@{}}PlanTower\\PMS-3003\end{tabular}
\\ \hline

\begin{tabular}[c]{@{}c@{}}\textbf{Detection}\\\textbf{Principle}\end{tabular}
& OPC & OPC & OPC & OPC & PNM & OPC
\\ \hline

\begin{tabular}[c]{@{}c@{}}\textbf{Pollution}\\\textbf{Categories}\end{tabular}
& \begin{tabular}[c]{@{}r@{}}PM\textsubscript{2.5}, PM\textsubscript{10},\\VOCs, CO\textsubscript{2}\end{tabular}
& \begin{tabular}[c]{@{}r@{}}PM\textsubscript{1.0}, PM\textsubscript{2.5},\\PM\textsubscript{10}, VOCs, NO\textsubscript{2}\end{tabular}
& \begin{tabular}[c]{@{}r@{}}PM\textsubscript{1.0}, PM\textsubscript{2.5},\\PM\textsubscript{10}, VOCs\end{tabular}
& \begin{tabular}[c]{@{}r@{}}PM\textsubscript{1.0}, PM\textsubscript{2.5},\\PM\textsubscript{10}\end{tabular}
& PM\textsubscript{2.5}
& \begin{tabular}[c]{@{}r@{}}PM\textsubscript{1.0}, PM\textsubscript{2.5},\\PM\textsubscript{10}\end{tabular}
\\ \hline

\end{tabular}
	\begin{tablenotes}
		\item [a] PiCO Home is speculated by the authors that it utilizes the PMS-A003 sensor for PM measurements according to its 3D assembly drawing.
		\item [b] Flow2's battery life is dependent on the frequency of its idle mode usage.
		\item [c] Flow2 utilizes a dedicated OPC instead of an off-the-shelf one for PM measurements.
		\item [d] Atmotube PRO can operate for 168 hours on a single charge if the PM sensor is used 10 times a day for about 10 minutes each time.
		\item [e] AirBeam2 uploads the data to the mobile application every second via Bluetooth, or to the web application every minute via Wi-Fi or GSM.
		\item [f] Laser Egg is in a cone structure.
	\end{tablenotes}
\end{threeparttable}
\end{table*}

\begin{table*}[ht]
\centering
\caption{Comparisons of dedicated PM monitoring devices developed in existing studies.}
\label{tab:CompareDedicatedPMDevices}
\renewcommand{\arraystretch}{1.06}
\begin{threeparttable}
\begin{tabular}{|c|r|r|r|r|r|r|}
\hline

\textbf{Name}
& \multicolumn{1}{c|}{\textbf{AirSense} \cite{zhuang2015airsense}}
& \multicolumn{1}{c|}{\textbf{Dylos Backpack} \cite{steinle2015personal}}
& \multicolumn{1}{c|}{\textbf{MyPart} \cite{tian2016mypart}}
& \multicolumn{1}{c|}{\textbf{LASS4U} \cite{chen2017open}}
& \multicolumn{1}{c|}{\textbf{MARS} \cite{tryner2019design}}
& \multicolumn{1}{c|}{\textbf{PAM} \cite{chatzidiakou2019characterising}}
\\ \hline
\begin{tabular}[c]{@{}c@{}}\textbf{Price} \\ \textbf{(\SI[detect-weight]{}{US\$})}\end{tabular}
& $< 50$\tnote{a}
& $> 425$\tnote{d}
& $< 50$
& $170$
& $> 350$\tnote{i}
& $2500$
\\ \hline

\begin{tabular}[c]{@{}c@{}}\textbf{Dimensions} \\ \textbf{(\SI[detect-weight]{}{\milli\meter})}\end{tabular}
& $100 \times 100 \times 50$
& largest\tnote{e}
& $40 \times 55 \times 23$
& $147 \times 54 \times 88$
& $140 \times 70 \times 40$
& $130 \times 90 \times 100$
\\ \hline

\begin{tabular}[c]{@{}c@{}}\textbf{Weight} \\ \textbf{(\SI[detect-weight]{}{\gram})}\end{tabular}
& unknown
& heaviest\tnote{e}
& unknown
& unknown
& $350$
& $400$
\\ \hline

\begin{tabular}[c]{@{}c@{}}\textbf{Battery} \\ \textbf{(\SI[detect-weight]{}{\hour})}\end{tabular}
& unknown
& $6$
& $24$\tnote{f}
& external power
& $23$ to $32$\tnote{j}
& $10$ or $20$\tnote{l}
\\ \hline

\textbf{Accessibility}
& microSD card
& built-in memory
& \begin{tabular}[c]{@{}r@{}}mobile app,\\animated LEDs\end{tabular}
& \begin{tabular}[c]{@{}r@{}}mobile app, web app,\\screen, microSD card\end{tabular}
& \begin{tabular}[c]{@{}r@{}}mobile app,\\microSD card\end{tabular}
& \begin{tabular}[c]{@{}r@{}}web app,\\microSD card\end{tabular}
\\ \hline

\textbf{Connectivity}
& off-line
& off-line
& Bluetooth LE
& LoRa
& \begin{tabular}[c]{@{}r@{}}Bluetooth 4.2,\\Wi-Fi\end{tabular}
& GPRS
\\ \hline

\begin{tabular}[c]{@{}c@{}}\textbf{Sampling}\\\textbf{Interval}\end{tabular}
& \SIrange{2}{4}{\second}\tnote{b}
& \SI{1}{\minute}
& \SI{6}{\minute}\tnote{g}
& \SI{5}{\second}
& \SI{5}{\second}
& \begin{tabular}[c]{@{}r@{}}\SI{20}{\second}\\or \SI{1}{\minute}\tnote{k}\end{tabular}
\\ \hline

\textbf{PM Sensor}
& \begin{tabular}[c]{@{}r@{}}Sharp\\GP2Y1010AU0F\end{tabular}
& \begin{tabular}[c]{@{}r@{}}Dylos\\DC1700\end{tabular}
& dedicated
& \begin{tabular}[c]{@{}r@{}}PlanTower\\PMS-3003\end{tabular}
& \begin{tabular}[c]{@{}r@{}}PlanTower\\PMS-5003\tnote{k}\end{tabular}
& \begin{tabular}[c]{@{}r@{}}Alphasense\\OPC-N2\end{tabular}
\\ \hline

\begin{tabular}[c]{@{}c@{}}\textbf{Detection}\\\textbf{Principle}\end{tabular}
& PNM
& OPC
& OPC
& OPC
& OPC
& OPC
\\ \hline

\begin{tabular}[c]{@{}c@{}}\textbf{Pollution}\\\textbf{Categories}\end{tabular}
& dust\tnote{c}
& PM\textsubscript{2.5}
& PM\textsubscript{2.5}, PM\textsubscript{10}\tnote{h}
& PM\textsubscript{2.5}, CO\textsubscript{2}
& PM\textsubscript{2.5}
& \begin{tabular}[c]{@{}r@{}}PM\textsubscript{1.0}, PM\textsubscript{2.5},\\
PM\textsubscript{10}, CO, O\textsubscript{3},\\
NO, NO\textsubscript{2}, noise\end{tabular}
\\ \hline

\end{tabular}
	\begin{tablenotes}
		\item [a] AirSense's unit cost was not listed in the original study; it is estimated by the authors according to the components used.
		\item [b] AirSense's sampling interval is estimated by the authors based to the time stamps of the air quality profile illustrated in the original study.
		\item [c] AirSense measures the dust instead of PM\textsubscript{2.5} as claimed in the original study because no size-selective filter or impactor is utilized.
		\item [d] Dylos Backpack's unit cost was not listed in the original study; it is estimated by the authors according to the components used.
		\item [e] Dylos Backpack's size and weight were not listed in the original study; it has visually the largest size and weight among the other five devices.
		\item [f] MyPart's battery life depends on the sampling interval; it can operate for around 24 hours on a single charge if it takes 1 reading every 6 minutes. 
		\item [g] MyPart can take a PM measurement by pressing a button or periodically at a user configurable sampling interval.
		\item [h] MyPart is not reporting the mass concentrations but the counts of the small and the large sized particles. The exact size ranges of the small and large sized particles require further studies, and they were approximating the PM\textsubscript{2.5} and the PM\textsubscript{10}, respectively. 
		\item [i] MARS's unit cost was not listed in the original study; it is estimated by the authors according to the components used.
		\item [j] MARS can operate for either 23 hours with \SI[per-mode=symbol]{1}{\liter\per\minute} flow rate or 32 hours with \SI[per-mode=symbol]{0.25}{\liter\per\minute} flow rate on a single charge.
		\item [k] MARS utilizes a modified PlanTower PMS-5003 PM sensor with custom-designed housing.
		\item [l] PAM can operate for either 10 hours at 20-second sampling interval or 20 hour at 1-minute sampling interval on a single charge.
	\end{tablenotes}
\end{threeparttable}
\end{table*}

While a significantly large portion of studies \cite{tsujita2005gas,moltchanov2015feasibility,johnston2019city} were dedicated to measuring the personal pollution exposure levels by densely deploying a large quantity of low-cost and fixed monitoring stations in the field, systems embracing the concept of participatory or community sensing, which consist of a vast population of monitoring devices privately held by participants recording and contributing their nearby pollution information, are more appreciated by the authors because the acquired information from these systems reveals the actual personal exposure levels. The application scenarios of these systems require the monitoring devices to be enabled with the following capabilities:
\begin{itemize}
	\item affordability for a large-scale deployment to improve the spatio-temporal resolution of acquired PM information;
	\item portability to avoid burdening participants and discouraging their initiatives;
	\item usability to prevent participants being overwhelmed by tedious interactions or interruptions;
	\item data accessibility with user-friendly interfaces and real-time visualizations; and
	\item data reliability with mitigated uncertainty sufficient for the intended applications.
\end{itemize}
Nevertheless, the monitoring devices in existing systems may either not fulfill one or the other capabilities listed above or present one or more aspects that can be enhanced.

For the commercially available devices summarized in Tab. \ref{tab:CompareCommercialPMDevices}, namely the PiCO Home \cite{brilliant2019pico}, the Flow2 \cite{plumelabs2019flow}, the Atmotube PRO \cite{atmotube2019atmotube}, the AirBeam2 \cite{habitatmap2019air}, the Speck \cite{airviz2019speck}, and the Laser Egg \cite{kaiterra2019air}, their astonishingly low prices in comparison to the research-grade instruments ($\approx\SI{2000}[US\$]{}$) or the conventional monitors ($\approx\SI[group-separator={,}]{30000}[US\$]{}$) make massive deployment of these devices in real world economically practical. Enabled by the on-board screens or LEDs, mobile applications, and/or web applications, these devices provide excellent usability and data accessibility such as automatic and continuous recording and uploading of PM concentration, visualization of historical and instantaneous PM information, and indication of the air quality index (AQI) using distinctive icons or colors. The AirBeam2 and the Speck further provide easy-to-use web interfaces for downloading the PM concentration data for in-depth analysis while the Flow2 is capable of generating air quality forecasts that estimate the health impacts in the near future.

However, excluding the AirBeam2, the Speck, and the Laser Egg that have been rigorously evaluated in the literature \cite{mukherjee2017assessing,feinberg2018long,manikonda2016laboratory,zikova2017estimating,scaqmd2017laseregg,zuo2018using}, few studies have focused on the performances of the three remaining devices in the field except the technical specifications available from their manufacturers. Extensive performance evaluations are needed before drawing any conclusions on whether the uncertain quality of data from these three remaining devices can be of value for the intended applications or not. In terms of portability, the excessive size and significant weight of the Speck and the Laser Egg make them more suitable for stationary scenario, such as monitoring indoor PM concentrations at fixed location, instead of taking them along. Moreover, for portable scenario, the Flow2 and the Atmotube PRO surpass the AirBeam2 substantially with much more compact structures. Although the PiCO Home has the minimum size and weight among the commercial devices listed in Tab. \ref{tab:CompareCommercialPMDevices}, its portability is limited and heavily dependent on the external power source that powers it up using wired connection. Overall, the authors generally recognize that the Flow2 and the Atmotube PRO are good examples of wearable devices for personal PM exposure monitoring.

For the dedicated monitoring devices developed in existing scientific studies as illustrated in Tab. \ref{tab:CompareDedicatedPMDevices}, namely the AirSense \cite{zhuang2015airsense}, the Dylos Backpack \cite{steinle2015personal}, the MyPart \cite{tian2016mypart}, the LASS4U \cite{chen2017open}, the MARS \cite{tryner2019design}, and the PAM \cite{chatzidiakou2019characterising}, except the AirSense, all of them have been rigorously and scientifically evaluated in laboratory and/or in the field. The evaluation results demonstrated the suitabilities for their intended applications. Apart from that, their evaluation and development procedures were well documented, which provide key reference values to the authors when evaluating the PM sensors and constructing the in-house wearable PM exposure monitoring system.

However, the AirSense and the Dylos Backpack, which were cursorily assembled from off-the-shelf components, are in their prototyping stages and not yet ready for real-world deployment. Besides, they have no wireless connectivity and the acquired data are processed in an off-line manner which requires professional skills and causes significant time lags in PM exposure awareness. Further considering the AirSense is measuring the concentrations of dust instead of PM\textsubscript{2.5} and the Dylos Backpack has visually the largest size and heaviest weight among the devices introduced in Tab. \ref{tab:CompareDedicatedPMDevices}, both of them are omitted from the remaining discussion. For the other four devices, the MyPart offers the most exceptional affodability. In contrast, the PAM integrating with multiple miniaturized sensors has the highest unit price similar to the research-grade instrument \cite{tsi2019dusttrakii} used as reference monitor in this manuscript. In terms of portability, the MyPart also outperforms the others considering the wrist-worn capability enabled by its compact structure; the MARS and the PAM have moderate portability while the LASS4U with external power supply requirement is more suitable for the stationary scenario. In regard to data reliability, the MARS is distinguished from others because of the on-board active filter sampler that facilitates gravimetric correction on its PM sensor's real-time data and compositional analyses of the collected PM. The open hardware, open software, and open data concepts adopted in the LASS4U, which encourage public participation as well as improve the overall system cost-efficiency and sustainability, are considered to be good practices and worth learning.

In this study, a device named Wearable Particle Information Node (WePIN) monitoring the personal PM exposure levels was implemented and evaluated. The WePIN is compact and lightweight, and suitable for wearable scenario. It also equips with a microSD card for storing data locally and a Bluetooth low energy (Bluetooth LE or BLE) link to communicate with the mobile application. The WePIN for data acquisition, the mobile application for data visualization and relay, and the back-end server for data management and analytics compose the proposed personal PM exposure monitoring system. Building the system in-house provides opportunities for addressing the imperfections in the existing systems and further adopting their proven good practices into the proposed one. It as well enables the authors with more controls over the system such as deploying dedicated techniques on the wearable device to enhance its power efficiency, employing advanced algorithms in the mobile application to achieve higher data quality, and applying specified analytic approaches in the back-end server to meet different data consumers' needs.

\subsection{Contributions}
\label{SubSect:Contributions}

To the best of the author's knowledge, this is the first study attempts to develop a standardized evaluation protocol for the low-cost PM sensors or devices to characterize the quality of their measurements, and involves a large number of low-cost wearable devices to collect the daily PM\textsubscript{2.5} exposure levels of each participant. The major contributions of this work are:
\begin{itemize}
	\item conducting a comprehensive review on the evaluations of low-cost PM sensors and on the development of personal PM exposure monitoring systems;
	
	\item introducing the detection principles of the light-scattering based instruments and the sources of uncertainty in their measurements;
	
	\item summarizing the specifications and literature of the four nominated PM sensors and the two reference monitors;
	
	\item attempting to establish a standardized evaluation protocol with definitive terminology and unified evaluation metrics for low-cost PM sensors/devices in collocated conditions;

	\item performing collocated field evaluations of the four nominated PM sensors of which three have not been extensively evaluated in the literature;

	\item proposing a real-time automated approach for identifying and eliminating the PM sensors' unusual measurements with condensed tiny droplets detected as particles;

	\item developing a personal PM exposure monitoring system consisting of a wearable PM device for data acquisition, a mobile application for data visualization and relay, and a back-end server for data management and analytics;

	\item performing collocated field evaluation of the first 10 PM devices and manufacturing additional 120 units to collect the daily PM\textsubscript{2.5} exposures of the adolescents and adults in Hong Kong and Chongqing.
\end{itemize}

This manuscript is organized as follows. Section \ref{Sect:RelatedWorks} presents the literature review of sensor evaluation and existing systems. Then, section \ref{Sect:Method} introduces the detection principles of light-scattering based instruments, summarizes the specifications as well as literature of the nominated PM sensors and reference monitors, and describes the experimental setups and principle assumptions for collocated field evaluation and calibration of the PM sensors and WePINs. Standardized terminology given in international documents and evaluation metrics recognized as best practices are also provided in section \ref{Sect:Method} for approaching a unifying evaluation protocol for low-cost PM sensors or devices. The development of the proposed system, consisting of a wearable PM device (i.e., WePIN), a mobile application, and a back-end server, for monitoring individuals' daily PM\textsubscript{2.5} exposure levels is detailed in section \ref{Sect:SystemImplementation}. The collocated field evaluation and calibration results of the four PM sensors and the ten PM devices are discussed in section \ref{Sect:ResultsAndDiscussion}; an automated approach able to identify and remove the PM sensors' unusual measurements, caused by detecting the condensed droplets as particles, in real time is also proposed in this section. Finally, the conclusions of this manuscript are drawn in section \ref{Sect:Conclusion}.

\section{Related Works}
\label{Sect:RelatedWorks}

This section presents a comprehensive literature review on the performance evaluations of PM sensors and the existing personal PM exposure monitoring systems. The experimental setups and evaluation metrics assessing the field and/or laboratory performances of various PM sensors and the valuable insights drawn from their experimental results are illustrated in section \ref{SubSect:SensorEvaluation} while an overview, including pros and cons, of existing personal PM exposure monitoring systems is given in section \ref{SubSect:PersonalPMSystem}. Note that, owing to the absence of a unifying protocol for performance evaluation, the identical symbols and terminology among the existing studies might have different definitions; the original studies may be consulted for details.

\subsection{Performance Evaluations of PM Sensors}
\label{SubSect:SensorEvaluation}

Study \cite{wang2015laboratory} evaluated and calibrated three models of PNMs (i.e., Shinyei PPD42NS, Sharp GP2Y1010AU0F, and Samyoung DSM501A) in the laboratory. Six performance aspects, namely linearity of response, precision of measurement, limit of detection, dependence on particle composition, dependence on particle size, and influences from RH and TEMP, were examined. A TSI AM510 monitor or a TSI scanning mobility particle sizer (SMPS) was selected accordingly as reference instrument. The ordinary least-squares (OLS) and the reduced major axis (RMA) regressions were conducted and yielded similar results, which indicated that the data uncertainty of the AM510 had minimal influence. The results demonstrated that the measurements from these sensors and the AM510 agreed well ($R^2=0.95$, $0.98$, and $0.89$, respectively, on 30-second averaged PM\textsubscript{2.5}) at the range below \SI[per-mode=symbol]{1000}{\micro\gram\per\cubic\meter}, but there existed significant uncertainties in their measurements at low concentrations ($<\SI[per-mode=symbol]{200}{\micro\gram\per\cubic\meter}$). Proper modifications such as regulating the sampling air flow of a sensor might improve its data quality. The experimental results also indicated that the particle composition and the size distribution indeed affected the responses of these three models of sensors. Besides, while the TEMP (\SIrange{5}{32}{\celsius}) had negligible effects on the sensors' outputs, the water vapor at high RH conditions absorbed the sensors' infrared radiation and caused overestimation.

A model of OPC (PlanTower PMS-1003/3003) used by the PurpleAir \cite{purpleair2019real} community air quality network was evaluated in wind-tunnel and ambient environments in study \cite{kelly2017ambient}. In the wind-tunnel environment, two research-grade instruments (TSI 3321 and GRIMM 1.109) were selected as references. In the ambient environment, the PMS sensors were evaluated against the federal reference method (FRM), two federal equivalent methods (FEMs; TEOM-FDMS and BAM), and one research-grade instrument (GRIMM 1.109). Underestimation and overestimation of the particle counts were spotted in wind-tunnel and ambient environments, respectively. Although the results in ambient environment demonstrated that the PMS sensors correlated well with each other ($R^2>0.99$ on 1-hour averaged PM\textsubscript{2.5} under identical $CF$ settings) and the PMS sensors had better correlations ($R^2>0.88$ on 24-hour averaged PM\textsubscript{2.5}) with the FRM than a Shinyei PPD42NS ($R^2=0.72$ \cite{holstius2014field} and $R^2=0.53$ \cite{gao2015distributed}), the PMS sensors overestimated the ambient PM and began to exhibit nonlinear responses, which could be corrected by setting $CF=atmos$ instead of $CF=1$, when concentration reached approximately \SI[per-mode=symbol]{40}{\micro\gram\per\cubic\meter}. Additionally, the PMS sensors' responses varied depending on the particle composition and particle size distribution. The experimental results indicated that the particle size distribution provided by a PMS sensor was based on a theoretical model rather than a measurement; different calibration models should be selected accordingly across locations with various meteorological conditions.











In study \cite{crilley2018evaluation}, the precision and the accuracy of a model of OPC (Alphasense OPC-N2) were evaluated at typical urban background sites. The precision was assessed by putting 14 OPC-N2 sensors at a single site and investigating the variation in their outputs ($CV=\SI{25}{\percent}$ on 5-minute averaged PM\textsubscript{2.5}). The accuracy was determined by comparing the outputs of the sensors against that of two research-grade light-scattering based instruments (TSI 3330 and GRIMM 1.108; $R^2=0.63$ and $0.76$, respectively, on 5-minute averaged PM\textsubscript{2.5}) and a FEM (TEOM-FDMS; $R^2=0.72$). Significant positive artefacts were observed in the OPC-N2 sensors' outputs during extreme humid conditions ($>\SI{85}{\percent}$). A RH correction factor was developed based upon the \textkappa-K\"{o}hler theory using average bulk particle aerosol hygroscopicity; it significantly improved the correlations between the OPC-N2 sensors' outputs and the reference measurements. However, each OPC-N2 responded differently to the RH artefact, and the \textkappa-value was related to the particle composition which may vary over the monitoring period. Hence, the study recommended calibrating the OPC-N2 sensors individually and frequently. No RH dependence was observed for the two research-grade light-scattering based instruments because their built-in pumps warmed the sheath flow and lowered the RH. Overall, the precision of OPC-N2 would be suitable for applications requiring a vast number of instruments, and a properly calibrated and RH corrected OPC-N2 can accurately measure the PM concentrations.









A 26-week comparison campaign involving four models of OPCs (namely the Nova Fitness SDS011, the Winsen ZH03A, the PlanTower PMS-7003, and the Alphasense OPC-N2) and a TEOM-FDMS was conducted in study \cite{badura2018evaluation}. Three copies of each sensor model were placed in a common box collocated with the TEOM reference instrument. All these sensors were exposed to the ambient air (the TEMP and RH ranged from \SIrange{-8}{36}{\celsius} and from \SIrange{27}{94}{\percent}, respectively); they operated stably throughout the entire campaign. When acceptable reproducibilities were reported from the PMS-7003, the SDS011, and the ZH03A ($CV=\SI{5.9}{\percent}$, $\SI{6.5}{\percent}$, and $\SI{10.8}{\percent}$, respectively, on 1-minute averaged PM\textsubscript{2.5}), the OPC-N2 had moderate reproducibility ($CV=\SI{20.0}{\percent}$). The PMS-7003 and the SDS011 were agreed well with the TEOM ($R^2=0.84$ to $0.92$ and $R^2=0.81$ to $0.88$, respectively, on 1-minute to 24-hour averaged PM\textsubscript{2.5}), but moderate and comparatively poor correlations against the TEOM were observed for the ZH03A ($R^2=0.72$ to $0.84$) and the OPC-N2 ($R^2=0.49$ to $0.60$), respectively. The study stated that a sensor model with high reproducibility could relax the calibration effort by weakening the needs for individual calibration, and a strong correlation between sensor and reference could simplify the calibration procedure utilizing simple linear regression. When comparing the dispersions of outputs between the OPCs and the TEOM, along with increasing RH, significant, moderate, minor, and minimal changes were spotted for the OPC-N2, the SDS011, the ZH03A, and the PMS-7003, respectively. Such analysis highlighted the necessity of employing RH corrections on the OPC-N2's and the SDS011's outputs. Due to the properties of aerosols used by the manufacturers for calibrating the sensors may differ from that of aerosols in the monitored air, the study suggested that calibration or recalibration should be made in the final deployment environment.









Study \cite{zheng2018field} characterized the performance capabilities of a model of OPC (PlanTower PMS-3003) in both the suburban (low-concentration, 1-hour averaged PM\textsubscript{2.5} $\leq\SI[per-mode=symbol]{10}{\micro\gram\per\cubic\meter}$) and the urban regions (high-concentration, 1-hour averaged PM\textsubscript{2.5} $\geq\SI[per-mode=symbol]{36}{\micro\gram\per\cubic\meter}$). A Met One Instrument E-BAM-9800 research-grade monitor, one Thermo Scientific 5030 SHARP (FEM), and/or two Teledyne T640s (FEM) were selected as reference instruments accordingly. Although the PMS sensors ($CF=1$) exhibited a nonlinear response when the PM\textsubscript{2.5} concentrations exceeded $\SI[per-mode=symbol]{125}{\micro\gram\per\cubic\meter}$, excellent sensor precision ($R^2>0.97$ among sensor units on 1-hour averaged PM\textsubscript{2.5}) were observed. The experimental results demonstrated that the type (detection mechanism) and the precision (at low-concentration level) of a reference instrument were critical in sensor evaluations, and the sensors' performances generally improved when the concentration increased ($R^2=0.40$ and $R^2=0.66$ in low- and high-concentration regions, respectively, on 1-hour averaged PM\textsubscript{2.5}) and the average interval prolonged ($R^2=0.66$ to $0.94$ on 1-minute to 24-hour averaged PM\textsubscript{2.5}). Serious RH impacts on the PMS sensors' outputs were identified under frequently high RH conditions. Within the original study, the empirical equation for RH correction introduced in \cite{zhang1994mie} was employed across the entire RH range due to various deliquescence RH values were reported in the literature. In suburban regions, the RH corrected sensors correlated much better ($R^2=0.93$ on 1-minute averaged PM\textsubscript{2.5}) with the reference, compared with the uncorrected ones ($R^2=0.66$). However, no improvement was found when applying the identical factors to the urban data sets, implying that the RH correction factors were highly specific to study sites with different particle chemical, micro-physical, and optical properties \cite{laulainen1993summary}. In regard to the TEMP impact on sensors' outputs, the Akaike's information criterion (AIC) was used to determine its significance, and the TEMP adjustment had marginal improvements. The study concluded that the PM\textsubscript{2.5} measurement errors can confined within around \SI{10}{\percent} of the reference concentrations if the PMS-3003 sensors were properly calibrated and RH corrected.

Study \cite{jayaratne2018influence} investigated the particle number concentration (PNC) and the particle mass concentration (PMC) responses of a model of OPC (PlanTower PMS-1003) to increasing RH as well as presence of fog in both laboratory and field tests. In the laboratory test, the sensor was placed inside a chamber with constant PM\textsubscript{2.5} PMC (\SI[separate-uncertainty=true,per-mode=symbol]{10\pm1}{\micro\gram\per\cubic\meter}) and adjustable RH. The sensor's PM\textsubscript{2.5} PMC values began to increase when the RH exceed \SI{78}{\percent} and achieved a maximum deviation of \SI{80}{\percent} while RH reached \SI{89}{\percent}. A hysteresis effect with no marked reduction in sensor's PM\textsubscript{2.5} PMC values until RH had decreased to \SI{50}{\percent} was observed. In the field test, the sensor was evaluated against two collocated TEOMs (FEMs; PM\textsubscript{2.5} and PM\textsubscript{10}) having heaters; a high-end nephelometer detected the presence of fog. The TEOMs exhibited little variation in the PM\textsubscript{2.5} PMC ($\approx\SI[per-mode=symbol]{10}{\micro\gram\per\cubic\meter}$), but the sensor overestimated it substantially (deviation $\approx\SI{90}{\percent}$ at $RH=\SI{94}{\percent}$) during periods of fog. Both the sensor's PNC values in all size bins and the PNC ratio between the large ($>\SI{1.0}{\micro\meter}$) and small (\SIrange{0.3}{0.5}{\micro\meter}) sized particles grew when fog occurred. The experimental results indicated that deliquescent growth of particles significantly affected the sensor's PNC and PMC values and the sensor detected the fog droplets in the air as particles. The study concluded that it was unlikely to derive any appropriate RH corrections due to no direct relationship between the RH and the sensor's PNC and PMC values, and the PMS-1003 sensor cannot be used to ascertain whether or not the air quality standards were being met.

Three copies of a model of OPC (PlanTower PMS-A003) were evaluated in study \cite{zamora2018field} by exposing them to laboratory air, indoor air, and outdoor air. The Thermo Scientific pDR-1200 research-grade nephelometer (gravimetrically corrected 1-minute averaged PM\textsubscript{2.5}) or the Met One Instrument BAM-1020 monitor (FEM; 1-hour averaged PM\textsubscript{2.5}) was selected as reference monitor accordingly. In all scenarios, these sensors were highly correlated with each other ($R^2\geq0.86$) and with the reference monitors ($R^2\geq0.89$). In laboratory experiments, the sensors and pDR were placed inside a common chamber and exposed to different polydispersed particle sources that their particle size distribution information was determined by the TSI 3082 SMPS and the TSI 3321 APS. These sensors exhibited a wide range of accuracy (quantified as complement of absolute relative error; \SIrange{13}{87}{\percent}) depending on the particle sources, which was partially due to misclassification of the particle sizes resulting from differences in the particle optical properties. Experiments showed that the density and size distribution of particles significantly affected the sensor accuracy, and the PMS sensor was not capable of handling monodispersed particles correctly. In indoor experiments, the sensors and the pDR were exposed to cooking emissions and residential air. These sensors demonstrated excellent precision ($CV\leq\SI{9}{\percent}$) and accuracy ($\geq\SI{92}{\percent}$) that were appreciated for personal monitoring applications. In outdoor experiments, the sensors were collocated with the BAM. While the TEMP had negligible influences on their performances, the overall sensor accuracy was low ($\SI{37}{\percent}$) and strongly dependent on the RH. The sensor accuracy was high ($>\SI{85}{\percent}$) at low RH ($<\SI{40}{\percent}$), but significant reductions were observed when RH was above $\SI{50}{\percent}$. Acceptable sensor precision ($CV=\SI{10}{\percent}$) was reported throughout the outdoor measurement period.


\subsection{Personal PM Exposure Monitoring Systems}
\label{SubSect:PersonalPMSystem}

A context-sensing device named AirSense for personal air quality monitoring was designed, implemented, and evaluated in \cite{zhuang2015airsense}. This device was a portable and cost-effective platform equipped with multiple sensors acquiring the air quality and the contextual information. This kind of information was used to estimate the association between air quality levels and user activities, and minimize the adverse health effects. It had been tested across different contextual scenes and considered to be feasible for daily continuous air quality monitoring. However, the on-board PNM (Sharp GP2Y1010AU0F) had not yet been evaluated and the uncertain quality of the collected data may affect the air quality level estimations. Besides, the PNM was measuring the mass concentration of dust instead of PM\textsubscript{2.5} as claimed because no size-selective filter nor impactor had been used to remove the irrelevant particles.

To comprehensively assess the individual's PM\textsubscript{2.5} exposure risks on a daily basis, study \cite{steinle2015personal} measured the near-complete exposure pathways across different micro-environments that a person encountered everyday. A diary (used to keep the time-activity notes) and a backpack, that is, the Dylos Backpack (embedded with an OPC and a GPS receiver), were assigned to each individual. Their exposure pathways were estimated based on the contextual and time-activity information and the pollution information acquired from the diary and the backpack, respectively. The OPC (Dylos DC1700) was rigorously validated against multiple high-end instruments. The excessive size and weight of the Dylos Backpack may severely burden the participants. Additionally, participants were aware of their exposures with considerable time lags because all information acquired was processed in an off-line manner.

In response to the health risks and mortality associated with the airborne PM, a personal and crowd-sourced environmental monitoring device called MyPart was developed in study \cite{tian2016mypart}. It was claimed that the MyPart had substantial enhancements over the existing systems in terms of accessibility, flexibility, portability, and accuracy. A dedicated OPC, which was capable of distinguishing and counting particles in different sizes, was integrated into the MyPart. This OPC had been validated in both the laboratory and the real world against a research-grade hand-held particle counter. Multiple LEDs were embedded on the MyPart and a novel mobile application was developed for real-time air quality indication and visualization, respectively. One major issue of the MyPart was that the dedicated OPC only counted the small and the large sized particles; it was not able to derive any mass concentrations from them. Moreover, the exact size ranges of the small and the large sized particles should require further investigations.

A expandable and sustainable participatory urban sensing framework for PM\textsubscript{2.5} monitoring was proposed, developed, and deployed in study \cite{chen2017open}. Such framework, featured with open system architecture including open hardware, open software, and open data, was capable of providing data services to improve environmental awareness, trigger on-demand responses, and assist governmental policy-making. In order to identify the suitable PM sensors, laboratory and field validations on various PM sensors used in existing monitoring systems were performed by comparing them with professional instruments. Four types of monitoring devices using the selected sensors were developed but only the LASS4U equipped with an OPC (PlanTower PMS-3003) was recognized regarding its usability and portability. In addition to the devoted web application for data visualization, a number of open data APIs were designed to facilitate more innovations and applications based on the acquired PM\textsubscript{2.5} information.

A portable PM\textsubscript{2.5} monitor named Mobile Aerosol Reference Sampler (MARS) was developed in study \cite{tryner2019design}. It utilized an active filter sampler, which was in line with the adopted OPC (PlanTower PMS-5003), to facilitate gravimetric correction of the OPC's real-time data and chemical characterization of the collected PM\textsubscript{2.5}. A dedicated cyclone inlet having \SI{2.5}{\micro\meter} cut-point was developed as well to remove the irrelevant particles. Laboratory tests were conducted to evaluate the performances of this filter sampler in comparison to a sophisticated reference instrument; correction factors were derived for the OPC when exposed to different test aerosols. Featured by the filter sampler, the MARS improved the data quality of the low-cost OPC's real-time measurements and provided insight into the PM sources and potential toxicity. Field experiment was also conducted and it demonstrated that the OPC employed in the MARS was able to identify intermittent PM\textsubscript{2.5} pollution events in home environment.

Incorrect and biased health estimation caused by inaccurate quantification of personal air pollution exposure severely limits the causal inference in epidemiological research. Although the emerging affordable and miniaturized air pollution sensors are facilitating a potential paradigm shift in capturing the personal pollution exposure during daily life, concerns remain in regard to their suitability for the scientific, health, and policy-making purposes. In study \cite{chatzidiakou2019characterising}, a portable and personal air pollution monitor (i.e., the PAM), which was able to measure multiple chemical and physical parameters simultaneously, was evaluated in different environments. The PAM manifested excellent reproducibility and good agreement with standard instruments. With proper calibration and post-processing, the PAM yielded more accurate personal exposure estimations compared to the conventional fixed stations. However, since multiple advanced sensors including an OPC (Alphasense OPC-N2) were used in a PAM unit, the relatively poor affordability, portability, and usability might hinder its applicability in large-scale personal PM exposure monitoring systems.

\section{Materials and Methods}
\label{Sect:Method}

The detection principles of the light-scattering based instruments measuring the aerosol's mass concentration along with the potential sources of uncertainty in their measurements are first introduced in section \ref{SubSect:LightScattering}. Then, section \ref{SubSect:Instrumentation} describes the details of the four models of PM sensors evaluated and the two reference instruments utilized in this study, including their technical specifications and literature reviews. Following that are the experimental setups and the principle assumptions of the collocated field evaluation and calibration for the four nominated PM sensors as well as the ten implemented wearable monitoring devices (i.e., the WePIN detailed in section \ref{Sect:SystemImplementation}) as presented in section \ref{SubSect:ExperimentalSetup}. Lastly, in section \ref{SubSect:TerminologyAndEvaluationMetrics}, the authors are appealing for a standardized evaluation protocol for low-cost sensors or devices to characterize the quality of their measurements; definitions of internationally standardized terminology related to the measurement quality and evaluation metrics of the best practices in quantifying the performances of a sensor or device are given as the basis for approaching a unifying evaluation protocol.

\subsection{Light Scattering}
\label{SubSect:LightScattering}

The light-particle interactions such as scattering, absorption, and extinction have been well studied in the literature \cite{hulst1957light,bohren1998absorption}. Whenever an individual particle is illuminated by an incident light, it re-radiates the incident light in all directions (i.e., scattering) and simultaneously transforms part of it into other forms of energy (i.e., absorption). Extinction is referred to as the sum of scattering and absorption. In this manuscript, instead of covering every single aspect of the light-particle interactions, only the basic concepts are introduced for general understanding. Moreover, each individual particle is assumed to have spherical shape and the incident light is approximated to be electromagnetic plane wave. Note that the aerodynamic diameter of a particle is most likely different from its actual diameter and a proper conversion may require \cite{decarlo2004particle}.

The intensity and the pattern of the light scattered by an individual particle is highly dependent on the particle size and the wavelength $\lambda$ of the incident light. For a spherical particle, its size is represented as $\pi d$ where $d$ is the diameter. When the particle size is far smaller than the wavelength of incident light ($\pi d < 0.1\lambda$), the scattered light is simply modeled by the Rayleigh scattering. When the particle size increases, the Rayleigh scattering regime breaks down due to interferences developed through phase variations over the particle's surface, and the scattered light is then modeled by the more complex Mie scattering. Even though the Mie scattering regime has no upper limitation on the particle size, it converges to the limit of simple geometric optics if the particle size is much larger than the wavelength of incident light ($\pi d>100\lambda$) \cite{hahn2009light}.

Given that the typical light-scattering based instruments for PM measurement are operating with visible or near infrared light ($\lambda\approx$ \SIrange{0.4}{1.1}{\micro\meter}), the light scattered by most of the particles of interest (e.g., $d\leq\SI{2.5}{\micro\meter}$) falls into the Mie scattering regime. The intensity of the light scattered by an individual particle at observation angle $\theta$ is given as follows \cite{mie1908beitrage,hodkinson1965response,matzler2002matlab,chen2018measurements}:
\begin{equation}
\label{Eq:ScatteringLightIntensity}
I_s=\frac{I_0\lambda^2}{8\pi^2R^2}\cdot[i_1(d,\lambda,n,\theta)+i_2(d,\lambda,n,\theta)]
\end{equation}
where $I_0$ is the intensity of incident light, $R$ is the distance between the particle and the observation point, $n$ is the refractive index of the particle, and $i_1(d,\lambda,n,\theta)$ and $i_2(d,\lambda,n,\theta)$, which are formulated as the summation of infinity series, are the vertically and the horizontally polarized (with respect to the scattering plane defined by the directions of the incident and scattered light) intensity functions of the scattered light, respectively. For a settled instrument structure with monochromatic incident light, the $I_0$, $\lambda$, $R$, and $\theta$ are constants. For the sake of clarity, only the light scattered at observation angle $\theta$ is discussed; Eq. \ref{Eq:ScatteringLightIntensity} is simplified as follows:
\begin{equation}
\label{Eq:SimplifyScatteringLightIntensity}
I_s=K\cdot[i_1(d,n)+i_2(d,n)]
\end{equation}
where, in practice, $K$ is a constant value achieved by calibration processes using particles with known properties instead of theoretical calculation.

\begin{figure}[!t]
\centering
	\subfloat[]{
		\includegraphics[width=0.8\columnwidth]{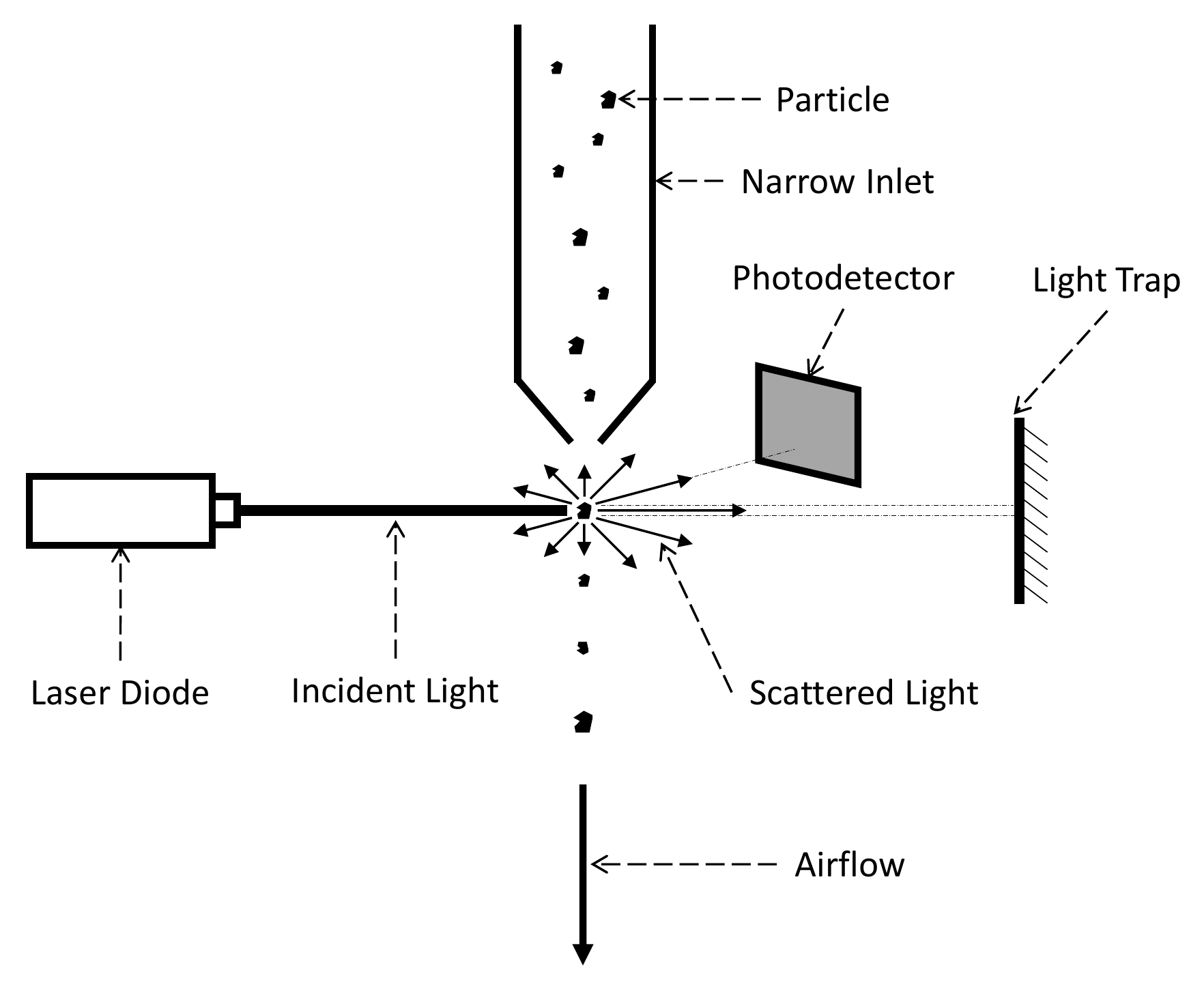}
		\label{subfig:GenDiaOPC}}
\hfil
	\subfloat[]{
		\includegraphics[width=0.9\columnwidth]{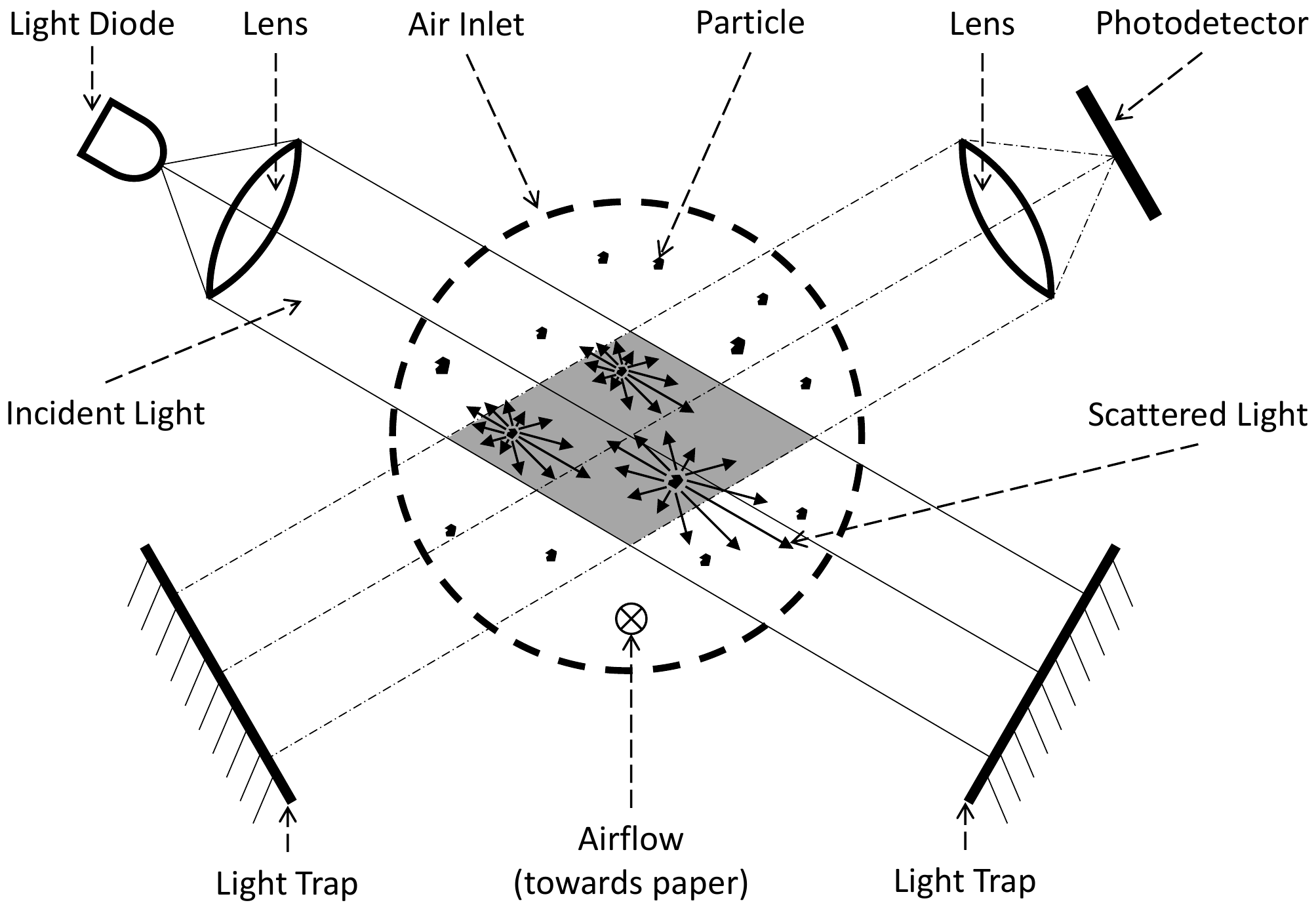}
		\label{subfig:GenDiaPhotometer}}
\caption{Schematic diagrams of (a) an optical particle counter (OPC) and (b) a photometer/nephelometer (PNM).}
\label{fig:GenDiaLightScatterIns}
\end{figure}

Depending on whether the instrument is detecting the light scattered by a single particle individually or an ensemble of particles simultaneously, these instruments can be categorized as optical particle counter (OPC) or photometer/nephelometer (PNM) as shown in Fig. \ref{fig:GenDiaLightScatterIns}. The calculations of mass concentration for the OPC and the PNM described in the following are general approaches; the exact ones are stated as proprietary and thus unknown \cite{sousan2016evaluation}.

\subsubsection{Optical Particle Counter (OPC)}
\label{SubSubSect:OPC}

An OPC detects the light scattered by an individual particle at a time, and deduces the particle size and count simultaneously. The aerosol's mass concentration is then determined based on the particle size and count. Because the particle size is derived from the scattered light, a size-selective filter or impactor is not essential for an OPC. In order to sample one single particle at a time, a narrow sampling inlet is required to physically constrain the particles to pass through the illuminating zone. A sufficiently powerful air-pump is required to draw the sampled particles through and avoid blockage at the narrow inlet. For miniaturized OPC, the narrow inlet and air-pump are removed due to the power and size constraints. Instead, a miniature OPC detects the scattered light at the virtual sensing zone formed by the intersection of a focused light beam (e.g., laser) and a narrow particle beam (constructed by a tiny inlet and a steady airflow generated by a micro-fan) inside the chamber. The volume of the virtual sensing zone is so small that only one particle is illuminated at a time in most cases until the particle coincidence errors (multiple particles are illuminated at a time) occur owing to extreme high concentrations. A miniature OPC can typically detect particles with diameters ranging from \SIrange{0.3}{10}{\micro\meter} because the small sized ones ($<\SI{0.3}{\micro\meter}$) can not scatter light sufficiently while the large sized ones ($>\SI{10}{\micro\meter}$) are not easy to be drawn into the virtual sensing zone \cite{koehler2015new}.

In a certain time interval $T$, the number of particles $N$ is determined by the number of electronic pulses detected by the photodetector. Because it is unlikely to get the refractive index $n$ of each illuminated particle in the field, it is assumed to be homogeneous for all particles in practice and predefined by the manufacturers with typical value $n=1.5+0i$. The diameter $d$ is then approximated from the intensity of the corresponding pulse using Eq. \ref{Eq:SimplifyScatteringLightIntensity}, and the size distribution $f_N(d)$ of the $N$ particles sampled within interval $T$ can be achieved. Given the density $\rho$ and the diameter $d$ of an illuminated particle, its mass is calculated as follows:
\begin{equation}
\label{Eq:OPCPatcileMass}
m=\frac{\rho \pi d^3}{6}
\end{equation}
where the particle has spherical shape. Same as the refractive index $n$, the density $\rho$ of each particle is most likely inaccessible in the field and it is assumed to be homogeneous for all particles and predefined by the manufacturers with typical value $\rho=\SI[per-mode=symbol]{1.65}{\gram\per\cubic\centi\meter}$. The mass concentration of the particles with diameters below $D$ is then expressed as follows: 
\begin{equation}
\label{Eq:OPCMassConcentration}
C_m=\frac{N}{Q \cdot T} \cdot \sum_{d=0}^{d=D}[f_N(d) \cdot \frac{\rho \pi d^3}{6}]
\end{equation}
where $Q$ is the sample flow rate generated by the air-pump or the micro-fan.

\subsubsection{Photometer/Nephelometer (PNM)}
\label{SubSubSect:Photometer}

A PNM detects the combined light scattered simultaneously by an ensemble of particles at a time and directly deduces the aerosol's mass concentration. Because this value is derived from the light scattered by all particles presented inside the optical sensing volume, a size-selective filter or impactor is needed to remove irrelevant particles (e.g., particles with aerodynamic diameter over \SI{2.5}{\micro\meter} when measuring PM\textsubscript{2.5}). Compared to an OPC, a PNM utilizes the fastest technique for measuring the mass concentration of aerosol and it performs well even in extreme high concentration.

At time instant $t$, $N$ particles are illuminated simultaneously by the incident light beam. Assuming these illuminated spherical particles have identical diameter $d$ and density $\rho$, the mass concentration of the aerosol sampled at time instant $t$ is calculated as:
\begin{equation}
\label{Eq:PhotometerPatcileMass}
C_m=\frac{N}{V} \cdot \frac{\rho \pi d^3}{6}
\end{equation}
where $V$ is the optical sensing volume formed by the intersection of the focused incident light beam and the observation coverage region. For simplicity reason, only the scattered light observed at angle $\theta$ is taken into account. According to Eq. \ref{Eq:ScatteringLightIntensity}, the intensity of the combined light scatted by $N$ particles at time instant $t$ is calculated as:
\begin{equation}
\label{Eq:PhotometerScatteringLightIntensity}
I_t = N \cdot \frac{I_0\lambda^2}{8\pi^2R^2}\cdot[i_1(d,\lambda,n,\theta)+i_2(d,\lambda,n,\theta)]
\end{equation}
and the mass concentration sensitivity \cite{smith1987response} of the instrument is then achieved from Eq. \ref{Eq:PhotometerPatcileMass} and Eq. \ref{Eq:PhotometerScatteringLightIntensity} as:
\begin{equation}
\label{Eq:PhotometerMassConcentrationSensitivity}
\frac{I_t}{C_m} = \frac{3I_0\lambda^2}{4\pi^3R^2} \cdot \frac{1}{\rho d^3} \cdot V \cdot [i_1(d,\lambda,n,\theta)+i_2(d,\lambda,n,\theta)]
\end{equation}
where the $I_0$, $\lambda$, $R$, $\theta$, and $V$ are constants given a settled instrument structure with monochromatic incident light. It can be further simplified as:
\begin{equation}
\label{Eq:PhotometerSimplifyMassConcentrationSensitivity}
\frac{I_t}{C_m} = K' \cdot \frac{1}{\rho d^3} \cdot [i_1(d,n)+i_2(d,n)]
\end{equation}
where $K'$ is achieved by calibration processes using particles with known properties in practice.

For most aerosols, the particle sizes are not uniform, and therefore, a better representation of the diameter $d$ in Eq. \ref{Eq:PhotometerSimplifyMassConcentrationSensitivity} is needed. The particle size distribution $f(d)$ of the aerosol is frequently adopted. The modified representation of the mass concentration sensitivity is as follows:
\begin{equation}
\label{Eq:PhotometerSimplifyMassConcentrationSensitivityMeanDiameter}
\frac{I_t}{C_m} = K' \cdot \sum_{d=0}^{d=D}[\frac{f(d)}{\rho d^3} \cdot [i_1(d,n)+i_2(d,n)]
\end{equation}
where $D$ is the largest diameter of the particles limited by the size-selective filter or impactor.

Similar to an OPC, the refractive index $n$ and the density $\rho$ of each particle are most likely inaccessible in the field for a PNM. These parameters are predefined by the manufacturers with typical values. Although more advanced PNMs, such as polar nephelometer \cite{castagner2007particle,espinosa2017retrievals}, are capable of determining the particle size distribution $f(d)$ of aerosol in real time, for miniature PNMs, $f(d)$ is predefined by the manufacturers with typical value achieved experimentally that may vary over time and across locations.

\subsubsection{Uncertainty of Instruments}
\label{SubSubSect:UncertaintyOfInstruments}

According to the detection principles of an OPC and a PNM described in section \ref{SubSubSect:OPC} and section \ref{SubSubSect:Photometer}, especially the commercially available ones having size, cost, and power constrains that are referred to as PM sensors in this manuscript, their reported mass concentrations of certain PM are derived based upon the following assumptions on the sampled aerosol:
\begin{itemize}
	\item particles with spherical shape;
	\item particles with homogeneous density $\rho$;
	\item particles with homogeneous refractive index $n$; and
	\item aerosol with known particle size distribution $f(d)$ (for PNM only)
\end{itemize}
where the $\rho$, $n$, and $f(d)$ are the aerosol properties that are usually pre-defined by the manufacturers with constant values achieved experimentally.

These strong assumptions are generally not held for most aerosols in the field \cite{gorner1995photometer,vincent2007aerosolD,jimenez2009evolution}, but enable the PM sensors to approximate the mass concentration of the sampled aerosol based on the intensity of the scattered light. In fact, the properties of aerosols vary over time and across locations. Although studies \cite{mishchenko1996t,sorensen2001light} have been conducted to inspect the light scattered by nonspherical particles, and techniques \cite{kelly1992measurement,zhao1999determination} have been proposed to determine the properties of the sampled aerosol in the field, to the best of the authors' knowledge, they are not available to the PM sensors on the market. Whenever the aerosol being sampled has properties different from the predefined ones, the PM sensors are most likely reporting biased measurements even though they were well calibrated when manufactured. Biases may occur as well when the aerosols comprise a significant portion of ultra small particles ($\pi d<0.1\lambda$) because the scattered light from these particles, if detectable, should be accurately modeled using Rayleigh scattering regime instead of Mie scattering regime. Experimental results from existing studies \cite{wang2015laboratory,zamora2018field,northcross2013low,sousan2016inter}, in which PM sensors were exposed to different types of aerosols, revealed that the particle shape, size distribution, and refractive index affected the scattered light and played a major role in response variation.

Additionally, the atmospheric RH affects the properties of aerosols and further causes noticeable artefacts in the reported mass concentration measurements from a PM sensor \cite{yoon2006influences,zieger2013effects,liu2018influence}. According to the previous studies, atmospheric aerosols may contain hygroscopic particles, such as NaCl and NaNO\textsubscript{3}, that absorb water vapor and grow in size when the RH exceeds their deliquescent points (normally resulting in higher intensity of scattered light and therefore overestimated measurements) \cite{hu2010hygroscopicity,gupta2015hygroscopic,molnar2020aerosol}. Research also indicated that the density $\rho$ and the refractive index $n$ of a particle are dependent on the RH levels \cite{stein1994measurements,hanel1972computation}.

Besides, the water vapor in the atmosphere may condense into tiny liquid droplets when the RH approaches \SI{100}{\percent}. The liquid droplets accumulated on the surfaces of the electronic components inside a PM sensor may bias the reported mass concentration measurements by altering the dielectric properties of these components, reducing the scattered light received by the photodetector, etc. Moreover, the condensed tiny liquid droplets that remain suspended in the atmosphere forming as mist or fog are possibly detected as particles by a PM sensor and lead to extensively large positive artefacts in the reported concentration measurements. 

While there is a causal relationship between aerosols and adverse health effects, the presence of water in the aerosols has nothing to do with it, and therefore, current air quality standards only consider the dry and nonvolatile portion of the aerosols. When measuring the mass concentration of aerosols for regulatory purposes, the water portion must be eliminated. The conventional PM monitoring systems utilize instruments with heaters or dryers employed at their inlets to condition the sampled aerosols. For a PM sensor without any heater or dryer, the reported mass concentrations of aerosols may be significantly different from that measured by a conventional instrument especially under high RH conditions.

Although the uncertainties in measuring the mass concentrations of aerosols with PM sensors are inevitable, they could be substantially mitigated with proper efforts. The biases due to variations in the aerosols' properties over time and across locations could be considerably reduced by calibrating the PM sensors initially and frequently in environment close to the final deployment one \cite{badura2018evaluation,rai2017end}. Additionally, the artefacts resulting from differences of meteorological conditions, RH in particular, could be appropriately compensated by invoking the corresponding meteorological information \cite{crilley2018evaluation,zheng2018field}.

\subsection{Instrumentation}
\label{SubSect:Instrumentation}

\begin{figure*}[!t]
\centering
	\subfloat[]{
		\includegraphics[width=0.92\linewidth]{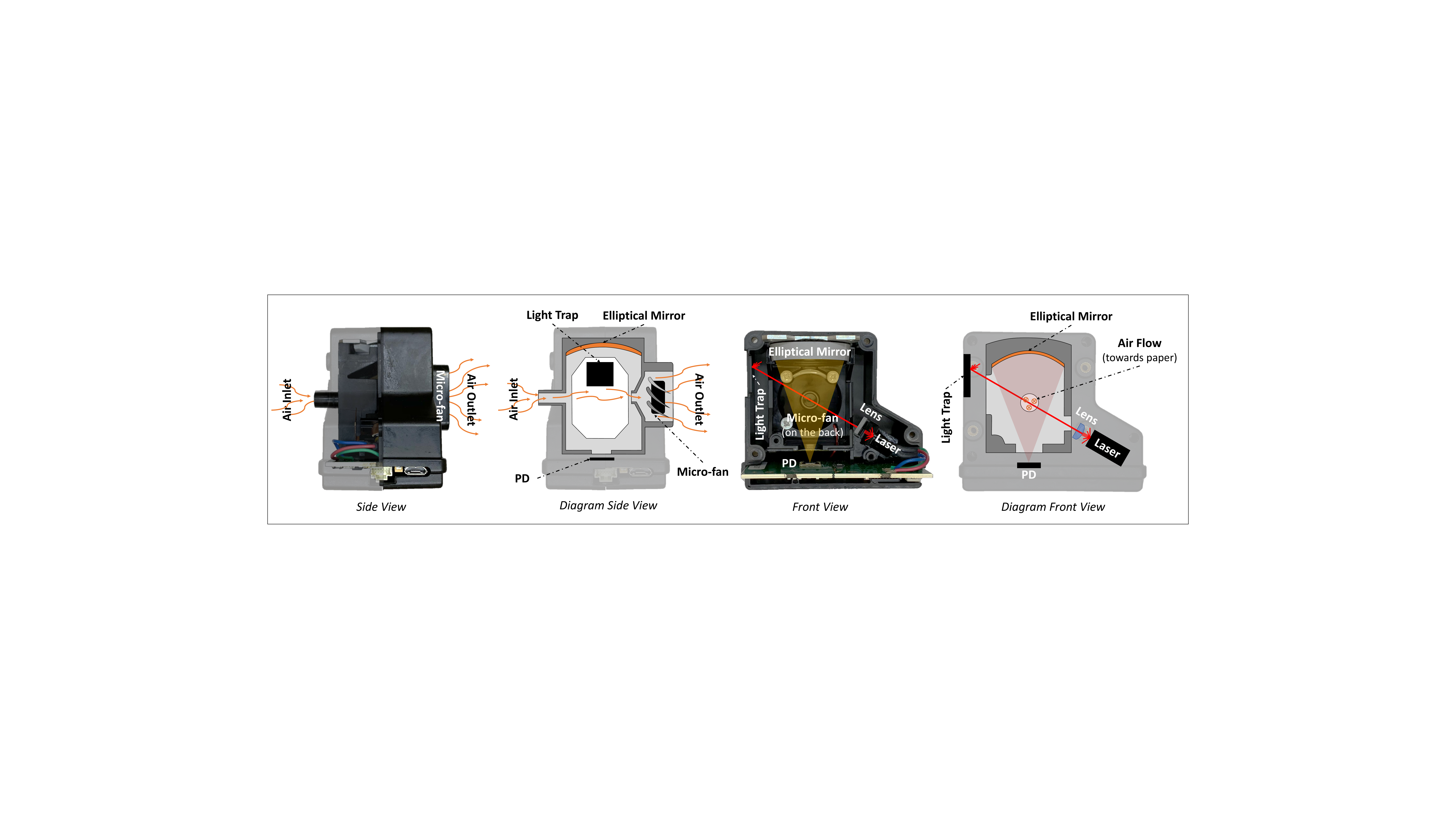}
		\label{subfig:DiagramOPCN}}
\hfil
	\subfloat[]{
		\includegraphics[width=0.92\linewidth]{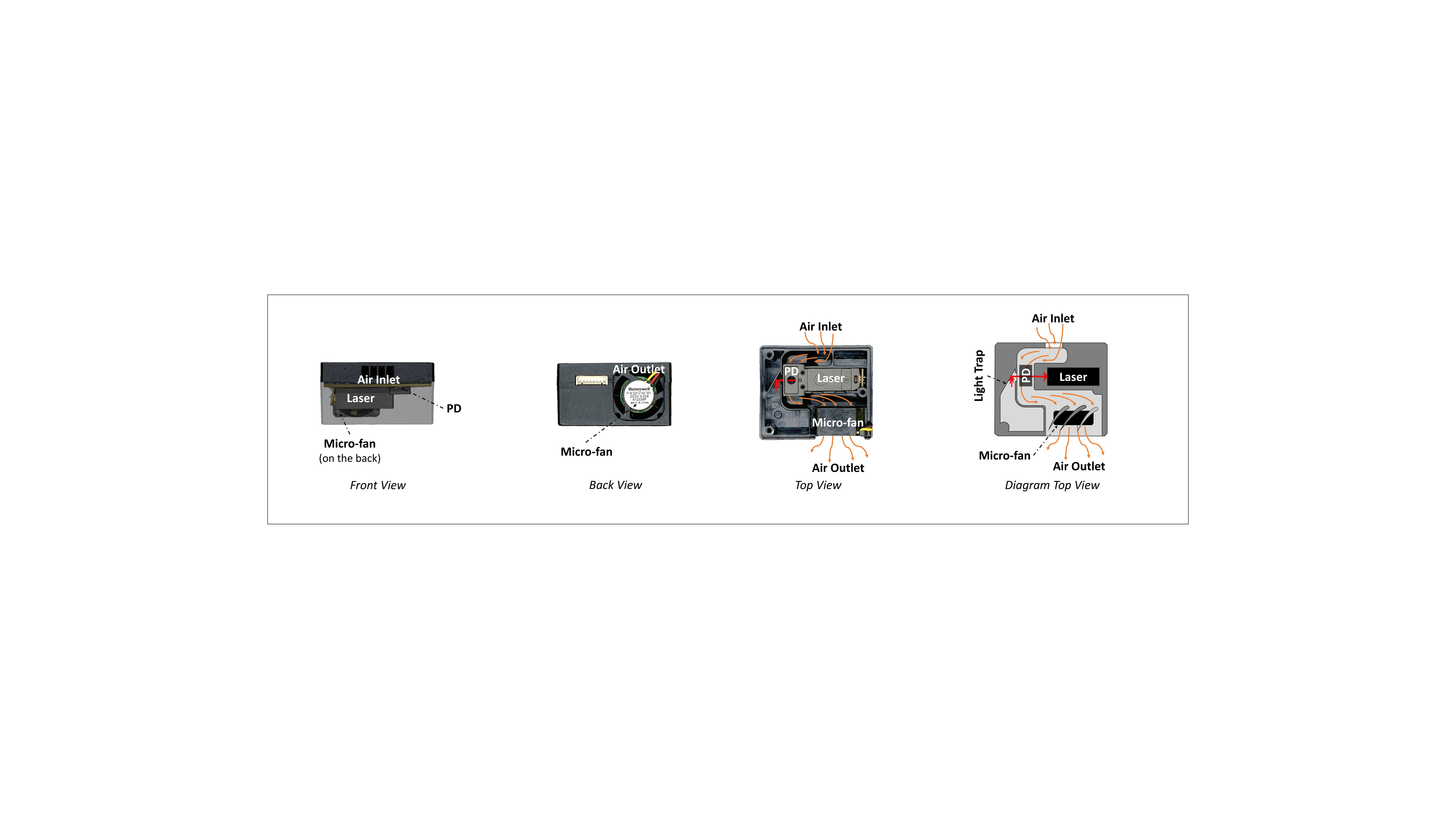}
		\label{subfig:DiagramHPMA}}
\hfil
	\subfloat[]{
		\includegraphics[width=0.92\linewidth]{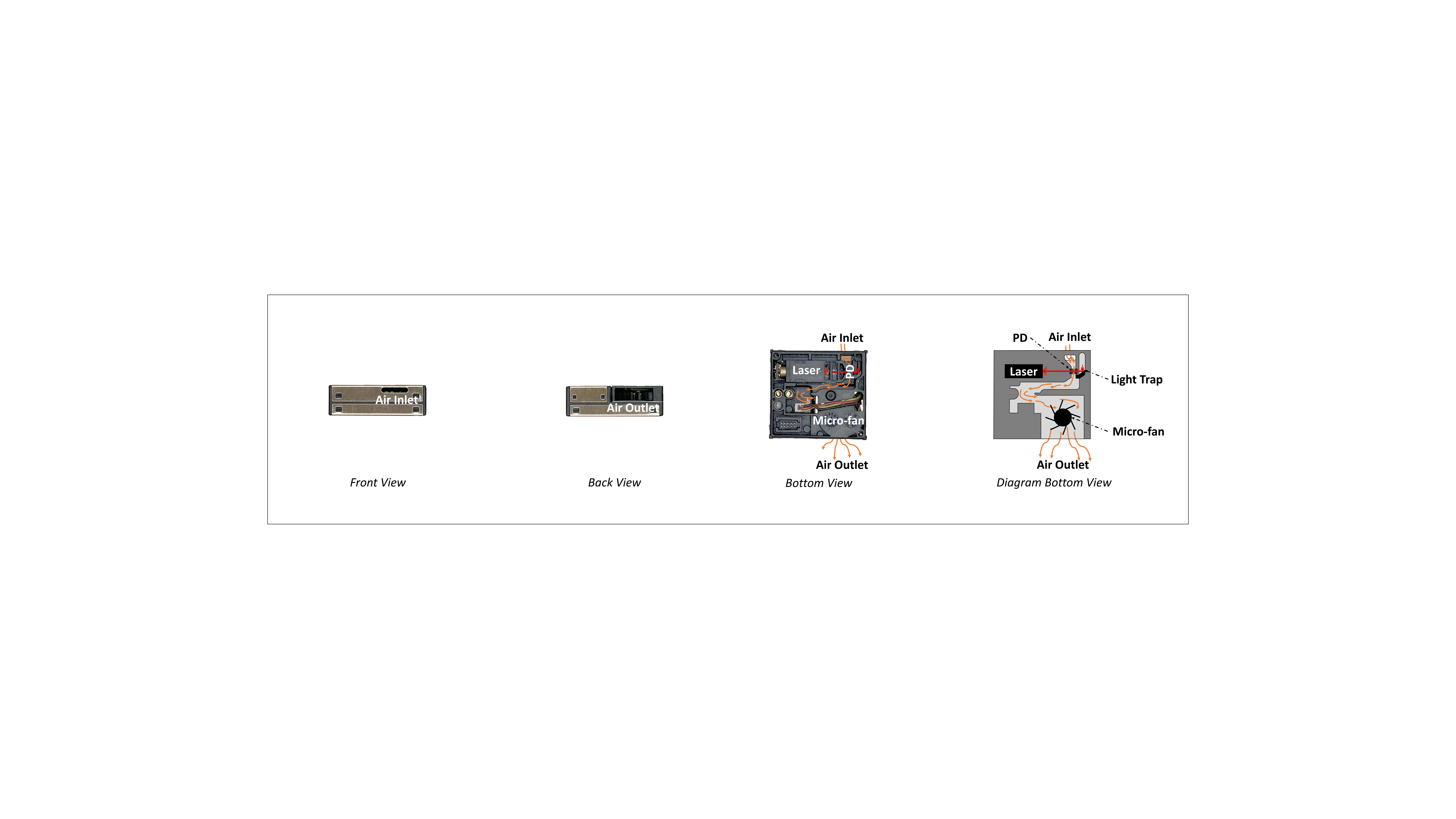}
		\label{subfig:DiagramPMSA}}
\hfil
	\subfloat[]{
		\includegraphics[width=0.92\linewidth]{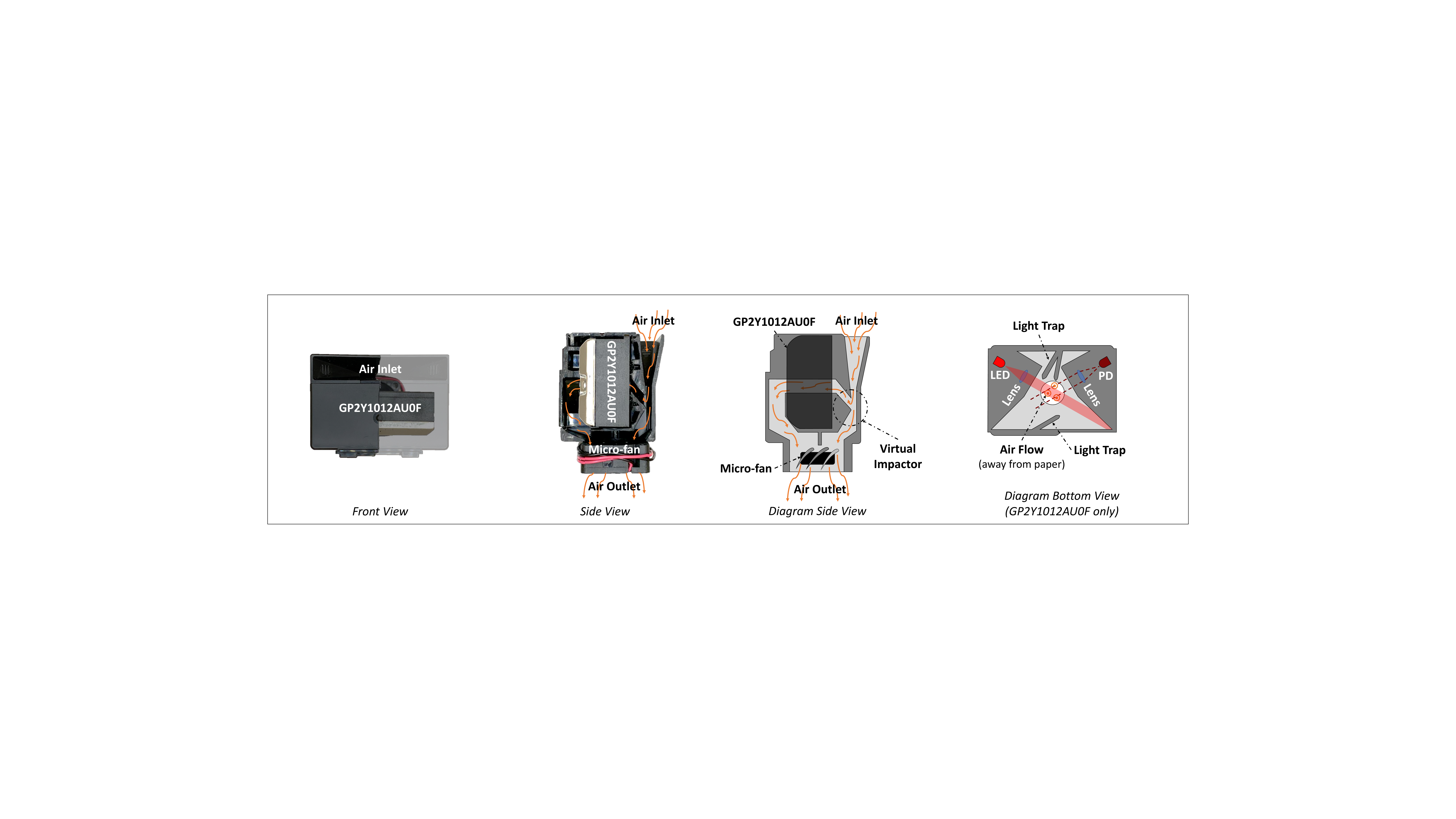}
		\label{subfig:DiagramDN7C}}
\caption{Images and schematic diagrams of (a) Alphasense OPC-N2 (OPCN); (b) Honeywell HPMA115S0 (HPMA); (c) PlanTower PMS-A003 (PMSA); and (d) Sharp DN7C3CA007 (DN7C).}
\label{fig:SchematicDiagramOfSensorEvaluated}
\end{figure*}

\begin{table*}[ht]
\centering
\caption{Technical details of the PM sensors evaluated in this manuscript.}
\label{tab:SensorSpec}
\renewcommand{\arraystretch}{1.2}
\begin{threeparttable}
\begin{tabular}{|c|r|r|r|r|}
\hline
\textbf{Name}
& \multicolumn{1}{c|}{\textbf{Alphasense OPC-N2}\tnote{a}}
& \multicolumn{1}{c|}{\textbf{Honeywell HPMA115S0}}
& \multicolumn{1}{c|}{\textbf{PlanTower PMS-A003}}
& \multicolumn{1}{c|}{\textbf{Sharp DN7C3CA007}}
\\ \hline

\textbf{Abbreviation}
& OPCN & HPMA & PMSA & DN7C
\\ \hline

\textbf{Detection Principle}
& OPC & OPC & OPC & PNM
\\ \hline

\textbf{Price (\SI[detect-weight]{}{US\$})}
& $400$ & $35$ & $15$ & $30$
\\ \hline

\textbf{Dimensions (\SI[detect-weight]{}{\milli\meter})}
& $72 \times 63.5 \times 60$
& $43 \times 36 \times 23.7$
& $38 \times 35 \times 12$
& $53 \times 51 \times 40$
\\ \hline

\textbf{Weight (\SI[detect-weight]{}{\gram})}
& $110$ & $35$ & $20$ & $50$
\\ \hline

\textbf{Supply Voltage (\SI[detect-weight]{}{\volt})}
& $5$ & $5$ & $5$ & $5$
\\ \hline

\textbf{Supply Current (\SI[detect-weight]{}{\milli\ampere})}
& $\approx175$ & $\leq80$ & $\leq100$ & $\leq100$
\\ \hline

\textbf{Light Source}
& red LD (\SI{658}{\nano\meter})
& red LD ($\approx$ \SI{650}{\nano\meter})\tnote{b}
& red LD ($\approx$ \SI{650}{\nano\meter})\tnote{b,c}~
& infrared LED ($\approx$ \SI{950}{\nano\meter})\tnote{d}
\\ \hline

\textbf{Scattering Angle (\SI[detect-weight]{}{\degree})}
& \SIrange{30}{90}{}\tnote{e}
& $90$
& $90$
& $60$
\\ \hline

\textbf{Particle Size Range (\SI[detect-weight]{}{\micro\meter})}\tnote{f}
& \SIrange{0.38}{17}{} & \SIrange{0.3}{5}{} & \SIrange{0.3}{10}{} & \SIrange{0.5}{2.5}{}
\\ \hline

\textbf{Particle Size Bins}
& $16$ & not accessible\tnote{g} & $6$ & not detectable
\\ \hline

\textbf{Reported PM}
& PM\textsubscript{1.0}, PM\textsubscript{2.5}, PM\textsubscript{10}
& PM\textsubscript{2.5}, PM\textsubscript{10}\tnote{h}
& PM\textsubscript{1.0}, PM\textsubscript{2.5}, PM\textsubscript{10}
& PM\textsubscript{2.5} analog\tnote{i}
\\ \hline

\textbf{Concentration Range (\SI[detect-weight,per-mode=symbol]{}{\micro\gram\per\cubic\meter})}
& \SIrange{0.01}{2000}{} (PM\textsubscript{2.5})\tnote{j}
& \SIrange{0}{1000}{} (PM\textsubscript{2.5})
& \SIrange{0}{500}{} (PM\textsubscript{2.5})
& \SIrange{25}{500}{} (PM\textsubscript{2.5})
\\ \hline

\textbf{Resolution (\SI[detect-weight,per-mode=symbol]{}{\micro\gram\per\cubic\meter})}
& $0.01$
& $1$
& $1$
& ADC dependent\tnote{i}
\\ \hline

\textbf{Response Time (\SI[detect-weight]{}{\second})}
& \SIrange{1}{10}{} & $<6$ & $<1$ & $<1$
\\ \hline

\textbf{Operating TEMP (\SI[detect-weight]{}{\celsius})}
& \SIrange{-20}{50}{}
& \SIrange{-10}{50}{}
& \SIrange{-10}{60}{}
& \SIrange{-10}{60}{}
\\ \hline

\textbf{Operating RH (\SI[detect-weight]{}{\percent})}
& \SIrange{0}{95}{} (non-condensing)
& \SIrange{0}{95}{} (non-condensing)
& \SIrange{0}{99}{} (non-condensing)
& \SIrange{10}{90}{} (non-condensing)
\\ \hline

\end{tabular}
\begin{tablenotes}
		\item [a] The firmware version 18 is installed in the OPC-N2. The manufacturer upgraded the OPC-N2 to OPC-N3 while the authors were drafting this manuscript.
		\item [b] The wavelength of light source is unspecified in the datasheet; a red LD is utilized in the sensor and its typical wavelength is at $\approx\SI{650}{\nano\meter}$.
		\item [c] The wavelength of the LDs utilized in PMS-1003 and PMS-3003 from PlanTower was estimated at \SI[separate-uncertainty=true]{650\pm10}{\nano\meter} with a spectrophotometer in \cite{kelly2017ambient}.
		\item [d] The core of a DN7C is a Sharp GP2Y1012AU0F \cite{sharp2019gp2y} PNM. The light source's wavelength is unspecified in the datasheet; the authors presume that the Sharp GL480E00000F \cite{sharp2019irled} infrared LED is utilized in the PNM and its typical wavelength is at $\approx\SI{950}{\nano\meter}$.
		\item [e] An elliptical mirror is utilized to focus all scattered light that strikes it onto the PD. The scattering angles are estimated by the authors.
		\item [f] Spherical diameter deduced from Mie scattering theory for OPCN, HPMA, and PMSA; aerodynamic diameter restricted by the virtual impactor for DN7C.
		\item [g] The HPMA detects and counts the particles in diameter from \SIrange{0.3}{5}{\micro\meter}, but neither particle size bin nor particle count information is accessible.
		\item [h] The HPMA uses the particle count data (\SIrange{0.3}{5}{\micro\meter}) to derive PM\textsubscript{2.5} mass concentration readings; PM\textsubscript{10} readings are calculated from PM\textsubscript{2.5} readings.
		\item [i] The DN7C does not output the PM\textsubscript{2.5} mass concentration directly but the analog voltage proportional to it. Proper conversion is needed in order to get the PM\textsubscript{2.5} mass concentrations, and the resolution of PM\textsubscript{2.5} mass concentration is dependent on that of the ADC (analog-to-digital converter) utilized.
		\item [j] The maximum concentration value \SI[per-mode=symbol]{1.5e6}{\micro\gram\per\cubic\meter} (PM\textsubscript{10}) stated in the datasheet is a theoretical value before the electronics fall over. The OPCN can measure PM\textsubscript{2.5} with mass concentration up to \SI[per-mode=symbol]{2000}{\micro\gram\per\cubic\meter} in real world (confirmed with its manufacturer).
	\end{tablenotes}
\end{threeparttable}
\end{table*}

Four models of PM sensors as illustrated in Fig. \ref{fig:SchematicDiagramOfSensorEvaluated}, namely Alphasense OPC-N2 (OPCN) \cite{alphasense2019opcn3}, Honeywell HPMA115S0 (HPMA) \cite{honeywell2019hpma115s0}, PlanTower PMS-A003 (PMSA) \cite{plantower2019pms}, and Sharp DN7C3CA007 (DN7C) \cite{sharp2019dn7c}, were evaluated against two collocated reference instruments, i.e., a TSI DustTrak II Aerosol Monitor 8532 (DUST) \cite{tsi2019dusttrakii} research-grade instrument and a Thermo Scientific Model 5030i SHARP Monitor (REFN) \cite{thermo2019model5030i} FEM instrument. The technical details of these nominated PM sensors are provided in Tab. \ref{tab:SensorSpec}. All four sensor models have been evaluated in existing studies and it is expected that the performance assessments on each sensor model from different studies are mixed, if not thin, due to the significant variations in the experimental setup (target PM, reference instruments, averaging intervals, etc.) and the environment (meteorological conditions, aerosol properties, concentration levels, etc.). The technical specification and the literature review of each sensor model and reference instrument are briefly described below.

\subsubsection{Alphasense OPC-N2 (OPCN)}
\label{SubSubSect:OPCNIntro}

This OPC applies a red laser diode (LD) with wavelength at \SI{658}{\nano\meter} to generate the incident light sheet interacting with individual particle drawn by a micro-fan. An elliptical mirror is adopted in its scattering chamber to focus the scattered light towards the dual-element photodetector (PD). The angle between the laser sheet and the normal to the PD face was estimated at \SI{60}{\degree} as shown in Fig. \ref{subfig:DiagramOPCN}\footnote{A 3D image showing the detailed internal structure of an OPC-N2 sensor is available in study \cite{blanchard2018uncertainty}.}. The intensity of the scattered light is used to derive the particle size (spherical) based on the Mie scattering formula that has been laboratory-calibrated by the manufacturer with polystyrene latex (PSL) spheres having known diameter and refractive index (RI). Taking an average RI of value $1.5+0i$, the OPCN (in firmware version 18) is able to detect particles in spherical diameters from \SIrange{0.38}{17}{\micro\meter} at rates up to \SI[per-mode=symbol,group-separator={,}]{10000}{\count\per\second} and classifying them into 16 particle bins. It outputs the particle counts (\SI[per-mode=symbol]{}{\count\per\milli\litre} and \SI[per-mode=symbol]{}{\count\per\second}) of each particle bin and the mass concentrations (\SI[per-mode=symbol]{}{\micro\gram\per\cubic\meter}) of PM\textsubscript{1.0}, PM\textsubscript{2.5}, and PM\textsubscript{10} over a user-configurable sampling interval from \SIrange{1}{10}{\second} via a SPI interface. Note that it calculates the mass concentrations of certain PM following the definition in European Standard EN 481, which extends the particle size of PM\textsubscript{10} beyond the OPCN's upper size limit (\SI{17}{\micro\meter}), resulting in underestimation of PM\textsubscript{10} values. Superior to the other three models, the OPCN offers a user-configurable particle density (default \SI[per-mode=symbol]{1.65}{\gram\per\cubic\centi\meter}) and weighting factor of each particle bin for a more accurate determination of PM, such as accounting for undetection of extremely small particles, undercount at low particle bins, and underweight of urban carbon particles, when additional information of the aerosol's properties is available. This sensor model can operate for a very long period of time ($>\SI{1}{\year}$) without maintenance or cleaning.

The OPCN has been extensively evaluated in the literature \cite{crilley2018evaluation,badura2018evaluation,aqmd2019field,sousan2016evaluation,bulot2019long}. A wide range of coefficient of determination values ($R^2;$ \SIrange{0.38}{0.98}{}) as well as coefficient of variation values ($CV$; \SIrange{4.2}{32}{\percent}) have been reported. Different slopes and intercepts were observed as well when performing linear regressions on the mass concentration pairs of the OPCNs and the reference instruments; nonlinear relationships were reported in \cite{aqmd2019field}. These studies suggested that the OPCNs typically overestimated the mass concentrations of PM\textsubscript{2.5} and PM\textsubscript{10}, but underestimated the mass concentrations of PM\textsubscript{1.0} resulted from undercount of small particles. While the OPCNs in \cite{crilley2018evaluation,badura2018evaluation} significantly overestimated the mass concentrations of respective PM when RH was $>\SI{80}{\percent}$, it had limited influence on the OPCNs' outputs in \cite{bulot2019long}; all of them suggested that the TEMP had negligible effects on the sensor performance.

The manufacturer recently upgraded the OPCN to the OPC-N3 that provides wider particle size range, more particle bins, and lower standby current. A new model called OPC-R1 \cite{alphasense2019opcr1}, which is more compact, light-weight, low-cost, and power-efficient compared to the OPCN, was also launched recently.

\subsubsection{Honeywell HPMA115S0 (HPMA)}
\label{SubSubSect:HPMAIntro}

This OPC also uses a red LD to generate the incident light beam. Although the exact wavelength of the LD is unknown, it was estimated at its typical value, that is, \SI{650}{\nano\meter}. As illustrated in Fig. \ref{subfig:DiagramHPMA}, the angle between the laser beam and the normal to the PD face is \SI{90}{\degree}, and there is no mirror in the scattering chamber. The scattered light from individual particle drawn by a micro-fan strikes on the PD directly; the intensity is used to determine its particle size (spherical) based on the Mie scattering theory with assumptions on its particle RI (unknown). The HPMA is able to detect and count particles in spherical diameter from \SIrange{0.3}{5}{\micro\meter}. Inside its datasheet, the definitions of PM\textsubscript{2.5} and PM\textsubscript{10} are simplified as particulate matter with diameter $\leq\SI{2.5}{\micro\meter}$ and $\leq\SI{10}{\micro\meter}$, respectively; the counts of particles with various sizes (not available to users) are used to derive the PM\textsubscript{2.5} mass concentrations (\SI[per-mode=symbol]{}{\micro\gram\per\cubic\meter}) with assumptions on the particle density (unknown) while the mass concentrations of PM\textsubscript{10} are calculated from the corresponding PM\textsubscript{2.5} values using proprietary formula. The HPMA was calibrated against cigarette smoke before shipment and the nominal accuracy for PM\textsubscript{2.5} is $\pm\SI[per-mode=symbol]{15}{\micro\gram\per\cubic\meter}$ in range \SIrange[per-mode=symbol]{0}{100}{\micro\gram\per\cubic\meter} or $\pm\SI{15}{\percent}$ in range \SIrange[per-mode=symbol]{100}{1000}{\micro\gram\per\cubic\meter}. Both PM\textsubscript{2.5} and PM\textsubscript{10} mass concentration values are transmitted via a UART interface automatically or upon request, and it is expected to operate continuously for \SI[group-separator={,}]{20000}{\hour}.

Although the authors were aware of the HPMA since mid-2017, few studies have yet been conducted to rigorously and scientifically evaluate its performances. Study \cite{giusto2018particulate} simply assessed the correlations among four HPMAs ($r>0.95$ for the PM\textsubscript{2.5} or PM\textsubscript{10}). Study \cite{johnston2019city} only reported the correlation coefficient ($r=0.85$) and root-mean-square error ($RMSE=\SI[per-mode=symbol]{8.38}{\micro\gram\per\cubic\meter}$) of the PM\textsubscript{2.5} measurements between the HPMA and the reference station. Although the intent was to develop an evaluation chamber for the low-cost sensors, studies \cite{hapidin2019aerosol,omidvarborna2020envilution} demonstrated that the HPMAs' PM\textsubscript{2.5} outputs correlated well with each other ($CV=\SI{3.68}{\percent}$ or $R^2>0.98$) and with the reference measurements ($R^2>0.97$). The HPMA, a low-cost air quality monitor named Awair 2\textsuperscript{nd} Edition embedded with the HPMA to be specific \cite{awair2020awair}, was thoroughly evaluated in study \cite{wang2020performance}. Its PM\textsubscript{2.5} outputs correlated well with those from the reference ($R^2>0.83$ for all residential sources and $R^2>0.75$ for outdoor air), but the linear regression slopes and intercepts varied significantly depending on the particle sources. The HPMA had little or no response to ultra small particles in diameter $<\SI{0.25}{\micro\meter}$ and overweighted the smallest particles that it could detect. High consistency across units ($CV<5\%$ for PM\textsubscript{2.5}) was observed except for sources with much or all particles in diameter $<\SI{0.25}{\micro\meter}$ or $>\SI{1}{\micro\meter}$. All studies mentioned above did not evaluate the TEMP and RH impacts on the sensor performance.

Compared to the HPMA, a more compact model named HPMA115C0 \cite{honeywell2019hpma115c0} was launched recently. It derives the PM\textsubscript{2.5} mass concentration from detected particles (\SIrange{0.3}{5}{\micro\meter}) and calculates the mass concentrations of PM\textsubscript{1.0}, PM\textsubscript{4.0}, and PM\textsubscript{10} from corresponding PM\textsubscript{2.5} value. Another model named HPMD115S0\footnote{The HPMD115S0 is not listed on the official website (\url{https://sensing.honeywell.com/sensors/particulate-matter-sensors} accessed on 5 June 2020) but available from the manufacturer upon request.} was launched recently as well. It has geometry identical to the HPMA, but detects and counts particles in spherical diameter from \SIrange{0.3}{10}{\micro\meter}; both concentrations and counts of PM\textsubscript{1.0}, PM\textsubscript{2.5}, and PM\textsubscript{10} are accessible from it.

\subsubsection{PlanTower PMS-A003 (PMSA)}
\label{SubSubSect:PMSAIntro}

This OPC utilizes a red LD with unspecified wavelength to generate the incident light beam as well. The LD's wavelength was estimated at \SI{650}{\nano\meter} considering that it is the typical wavelength of a red LD and was reported from its sibling models (PlanTower PMS-1003 and PMS-3003) \cite{kelly2017ambient}. Scattered light from individual particle drawn by a micro-fan strikes directly on the PD face whose normal vector is perpendicular to the laser beam as shown in Fig. \ref{subfig:DiagramPMSA}; the intensity is used to determine the particle size (spherical) based on the Mie scattering theory with assumptions on the particle RI (unknown). It can detect and count particles in spherical diameter from \SIrange{0.3}{10}{\micro\meter} (\SI{50}{\percent} and \SI{98}{\percent} nominal counting efficiency for particles in diameter $=\SI{0.3}{\micro\meter}$ and $\geq\SI{0.5}{\micro\meter}$, respectively) and divide them into six size ranges: $>\SI{0.3}{\micro\meter}$, $>\SI{0.5}{\micro\meter}$, $>\SI{1}{\micro\meter}$, $>\SI{2.5}{\micro\meter}$, $>\SI{5}{\micro\meter}$, and $>\SI{10}{\micro\meter}$. Mass concentrations (\SI[per-mode=symbol]{}{\micro\gram\per\cubic\meter}) of respective PM\footnote{The datasheet did not specify the definition of PM\textsubscript{x} while it is generally referred to as particulate matter in spherical diameter no larger than \SI[parse-numbers=false]{x}{\micro\meter}.} are derived from the particle counts with assumptions on the particle density (unknown). Two types of mass concentrations, i.e., ``standard'' ($CF=1$; based on the density of industrial metal particles) and ``atmospheric'' ($CF=atmos$; based on the density of particles commonly found in the atmosphere), are available from the PMSA; both types of concentrations have been tuned to best represent the PM in industrial areas and ambient/residential air, respectively. Calculation of as well as conversion between ``standard'' and ``atmospheric'' readings are proprietary. Particle number concentrations (\SI[per-mode=symbol]{}{\count\per\SI{0.1}{\liter}}) of all the six size ranges and mass concentrations (\SI[per-mode=symbol]{}{\micro\gram\per\cubic\meter}) of PM\textsubscript{1.0}, PM\textsubscript{2.5}, and PM\textsubscript{10} are transmitted via a UART interface automatically (active) or upon request (passive). A dynamic data transmission interval, which is inversely proportional to the aerosol concentration ranging from \SIrange{800}{200}{\milli\second}, is applied in active mode; triplicate readings will be transmitted consecutively when the concentration is stable that results in an actual update interval of \SI{2.3}{\second}. The unit-wise consistency (i.e., precision) of each PMSA was tested by the manufacturer before shipment (unspecified testing particles and calibration procedures); the nominal precision for PM\textsubscript{2.5} ($CF=1$) is $\pm\SI[per-mode=symbol]{10}{\micro\gram\per\cubic\meter}$ in range \SIrange[per-mode=symbol]{0}{100}{\micro\gram\per\cubic\meter} or $\pm\SI{10}{\percent}$ in range \SIrange[per-mode=symbol]{100}{500}{\micro\gram\per\cubic\meter}. The expected mean time to failure of a PMSA is $>\SI{3}{\year}$.

Although only limited amount of studies had evaluated the PMSA \cite{tan2017laboratory,zamora2018field,zusman2020calibration,stampfer2020use}, they provided comprehensive investigations on its short- and long-term performances in the laboratory and field under different meteorological conditions, plus the responses to PM with various size distributions and chemical compositions. The particle number concentrations or particle mass concentrations from the PMSAs generally correlated well with each other ($r>0.95$ or $CV<\SI{15}{\percent}$) and with the reference measurements ($R^2>0.80$), but demonstrated significant dependence on the particle size and source. A wide variety of regression coefficients were reported when comparing the measurements from the PMSAs and reference instruments exposed in different aerosols. Study \cite{zamora2018field} further evidenced that a PMSA frequently categorized the particle sizes incorrectly for various particle sources and appeared to handle particles in diameter $<\SI{1}{\micro\meter}$ and $>\SI{1}{\micro\meter}$ differently; the PMSA exhibited a strong dependence on the RH when it was $>\SI{50}{\percent}$ but not the TEMP. Studies \cite{zusman2020calibration,tan2017laboratory} stated that better calibration performances could be achieved if the mass concentrations from the PMSAs were adjusted with calibration coefficients developed by fitting their number concentrations, instead of the raw mass concentrations, and the corresponding reference mass concentrations into certain regression models; regional dependence of calibration coefficients was observed. Nevertheless, low intra-unit variabilities were also reported in existing studies evaluating the previous versions of PlanTower PM sensors, i.e., PMS-1003 \cite{kelly2017ambient}, PMS-3003 \cite{scaqmd2017laseregg}, PMS-5003 \cite{scaqmd2017sainsmart}, and PMS-7003 \cite{badura2018evaluation}.

Two new models, namely PMS-9003M and PMS-X003N\footnote{The PMS-9003M and PMS-X003N are not listed on the official website (\url{http://www.plantower.com} accessed on 5 June 2020) but available from the manufacturer upon request.}, were launched in early and late 2019, respectively. The PMSA and the PMS-9003M have quite similar specifications, but the latter has slightly larger size (\SI{48x40x12}{\milli\meter}) and completely different structure. The PMS-X003N employs a dual-laser design with enhanced capability for detecting large particles and a dual-duct design such that the airflow could be generated externally either from a micro-fan or an air-pump. Compared with the PMSA, although the PMS-X003N offers a wider detection range (\SIrange[per-mode=symbol]{0}{1000}{\micro\gram\per\cubic\meter}) for PM\textsubscript{2.5} and PM\textsubscript{10}, it has larger size (\SI{60x45x40}{\milli\meter}) and power consumption ($\leq\SI{300}{\milli\ampere}$) and is more costly (\SI{75}[US\$]).

\subsubsection{Sharp DN7C3CA007 (DN7C)}
\label{SubSubSect:DN7CIntro}

The core component of this PNM is the Sharp GP2Y1012AU0F \cite{sharp2019gp2y} dust sensor, in which an infrared LED together with a focus lens are utilized to form the incident light beam. The authors conjectured that a Sharp GL480E00000F \cite{sharp2019irled} infrared LED with wavelength at \SI{950}{\nano\meter} is used. The angle between the incident light beam and the normal to the PD face is \SI{120}{\degree} as depicted in Fig. \ref{subfig:DiagramDN7C}. The dust sensor has no accommodation to move the particles through the scattering zone and scatter light from all particles inside is focused on the PD. In contrast, a DN7C has a virtual impactor at the inlet and a micro-fan at the outlet; the virtual impactor uses the air pressure generated by the micro-fan to separate the particles such that only the micro particles can enter the dust sensor's scattering zone\footnote{No performance evaluation on the virtual impactor has been aware of.}. The DN7C detects an ensemble of particles in aerodynamic diameter from \SIrange{0,5}{2.5}{\micro\meter} and outputs a voltage proportional to the concentration of the aerosol being sampled. A formula is available in the datasheet for converting a voltage output to the PM\textsubscript{2.5}\footnote{The definition of PM\textsubscript{2.5} was not listed in the datasheet while generally it is referred to as PM with aerodynamic diameter no larger than \SI[parse-numbers=false]{2.5}{\micro\meter}.} mass concentration (with unspecified testing particles and calibration procedures). However, this formula varies depending on the aerosol properties and hence its transferability is limited. The expected mean time to failure of a DN7C is \SI[group-separator={,}]{15000}{\hour}.

The DN7C has gone through multiple iterations, including the alpha version DN7C3JA001 with core GP2Y1010AU0F, the earlier mass production version DN7C3CA006 with core GP2Y1012AU0F, and the latest version DN7C3CD015 with core GP2Y1023AU0F that outputs a pulse width modulation (PWM) signal whose duty cycle is proportional to the aerosol concentration. Identified by the authors, the only difference between a DN7C3CA006 and a DN7C is the latter utilizes a more power-efficient micro-fan; therefore, the abbreviation DN7C is then referring to either the DN7C3CA006 or the DN7C3CA007 hereinafter.

The literature of the DN7C's performance evaluation is relatively thin. In addition, no performance evaluation on the virtual impactor has been aware of, particularly its aerodynamic cut-point, which decides the upper size limit of particles that can be driven into the scattering zone. Moderate correlations between the DN7C's voltage outputs and the reference PM\textsubscript{2.5} mass concentrations ($r=0.86$ in the laboratory and $r=0.67$ in the field) were reported in \cite{guo2016mobile}. Study \cite{sousan2016inter} showed that the DN7Cs' reproducibilities were heavily dependent on the particle sources ($CV=\SI{17}{\percent}$ to \SI{30}{\percent} on voltage outputs) and significantly improved after calibration ($CV=\SI{0.8}{\percent}$ to \SI{7.1}{\percent} on estimated PM\textsubscript{2.5}); excellent correlations ($R^2>\SI{0.98}{}$) and acceptable biases ($<\pm\SI{10}{\percent}$) were reported when comparing the calibrated DN7Cs' PM\textsubscript{2.5} estimations with the reference measurements. In study \cite{tan2017laboratory}, the DN7C's voltage outputs correlated well ($R^2>\SI{0.95}{}$) with the PM\textsubscript{2.5} mass and number concentrations from the reference instrument; good precision (quantified as relative standard deviation; $RSD=\SI{1.02}{\percent}$ on voltage outputs) was observed when measuring aerosol with constant concentration. In study \cite{cao2017portable}, the DN7C's voltage outputs and the reference PM\textsubscript{2.5} mass concentrations showed excellent correlation ($R^2>\SI{0.99}{}$; \SIrange[per-mode=symbol]{0}{500}{\micro\gram\per\cubic\meter}) in laboratory condition but poor correlation ($R^2<\SI{0.03}{}$) in the field, which may result from the relatively low concentration levels in the field (\SIrange[per-mode=symbol]{0}{30}{\micro\gram\per\cubic\meter}) in comparison to the DN7C's limit of detection ($LOD=\SI[per-mode=symbol]{9}{\micro\gram\per\cubic\meter}$). Taken the RH and TEMP into account, multiple machine learning based regression models were proposed in study \cite{loh2019calibration} to improve the calibration performances (best $R^2=\SI{0.78}{}$ and minimal $RMSE=\SI[per-mode=symbol]{5}{\micro\gram\per\cubic\meter}$).

\subsubsection{TSI DustTrak II Aerosol Monitor 8532 (DUST)}
\label{SubSubSect:DUSTIntro}

There are three models of DustTrak II aerosol monitors, namely the 8530 desktop model, the 8530EP desktop with external pump model, and the 8532 hand-held model (i.e., DUST). They are single-channel photometric instruments used to determine the aerosol's mass concentrations in real time \cite{tsi2012dusttrak}. An ensemble of particles are driven through the scattering chamber by the internal diaphragm pump and illuminated by a thin sheet of laser light generated from a LD with wavelength at \SI{780}{\nano\meter} and a set of beam shaping optics. A spherical mirror focus a significant fraction of the scattered light on the PD whose normal vector is perpendicular to the incident light. A sheath air system confines the particles to a narrow stream inside the scattering chamber to isolate the particles from the optics to keep them clean. The voltages across the PD are proportional to the aerosol's mass concentrations. They are estimated via multiplying the voltage with a calibration constant determined by the Arizona test dust (ISO 12103-1, A1 test dust) that is representative for a wide variety of ambient aerosols. Several TSI developed impactors (PM\textsubscript{1.0}, PM\textsubscript{2.5}, PM\textsubscript{4.0}, and PM\textsubscript{10}) operating at volume flow rate of \SI[per-mode=symbol]{3.0}{\liter\per\minute} can be attached to the inlet to perform size-selective sampling. The biggest difference between hand-held and desktop models is that the latter equips with a gravimetric filter for conducting collocated calibration against gravimetric reference method.

The DUST is battery operated with runtime up to six hours. It can detect particles with aerodynamic diameter from \SIrange{0.1}{10}{\micro\meter} at concentration range \SIrange[per-mode=symbol,group-separator={,}]{1}{150000}{\micro\gram\per\cubic\meter}. The resolution of the mass concentration readings is $\SI{0.1}{\percent}$ or \SI[per-mode=symbol]{1}{\micro\gram\per\cubic\meter}, whichever is greater. The flow rate and the logging interval are user adjustable from \SI[per-mode=symbol]{1.4}{\liter\per\minute} to \SI[per-mode=symbol]{3.0}{\liter\per\minute} and from \SI[per-mode=symbol]{1}{\second} to \SI[per-mode=symbol]{1}{\hour}, respectively. The nominal accuracy of flow rate is $\pm\SI{5}{\percent}$ at \SI[per-mode=symbol]{3.0}{\liter\per\minute}. The on-board memory can log data at 1-minute interval for 45 days.

The DustTrak II aerosol monitor series has been calibrated against the gravimetric reference method (FRM) \cite{egondi2016measuring,zhao2017characteristics,nguyen2019fine} or other more advanced instruments (FEM) \cite{holstius2014field,gao2015distributed} in the literature, and generally demonstrated good data quality ($R^2\geq0.91$ on 24-hour averaged PM\textsubscript{2.5}). The portability of a DUST further makes it a popular candidate in air pollution associated studies requiring real-time field PM concentration measurements \cite{chen2013beijing,zhao2017characteristics,kang2019characteristics}. Additionally, the monitors in this series are frequently adopted as reference instruments in research investigating the performances of the low-cost PM sensors \cite{holstius2014field,budde2014distributed,gao2015distributed,pillarisetti2017small,nyarku2018mobile}.




\subsubsection{Thermo Scientific Model 5030i SHARP Monitor (REFN)}
\label{SubSubSect:REFNIntro}

This SHARP (Synchronized Hybrid Ambient Real-time Particulate) monitor is a nephelometric plus radiometric hybrid monitor capable of providing real-time PM\textsubscript{1.0}, PM\textsubscript{2.5}, or PM\textsubscript{10} mass concentration reading with superior accuracy, precision, and sensitivity. This is also a U.S. Environmental Protection Agency (EPA) approved Class III FEM instrument for PM\textsubscript{2.5} monitoring. The particles are driven by a diaphragm pump to pass though a size selective inlet\footnote{The aerosol goes through a \SI{10}{\micro\meter} (aerodynamic diameter) size-selective inlet (\SI{50}{\percent} efficiency cut-off) and a \SI{2.5}{\micro\meter} (aerodynamic diameter) cyclone inlet successively.} to discriminate those with equivalent aerodynamic diameters above \SI{2.5}{\micro\meter}. A dynamic heating system (DHS) is employed to condition the sampled aerosol such that its RH condition agrees with that in gravimetric reference method. The DHS powers on only when the RH is above a adjustable threshold (default \SI{58}{\percent} \cite{thermo2018manual}) in order to minimize the internal TEMP rise while reducing the particle bound water which results in positive artefacts in the measurements. Downstream of the DHS is a light scattering nephelometer utilizing an infrared LED with wavelength at \SI{880}{\nano\meter} to form the incident light beam. The estimated angle between the incident light and the normal to the PD face is \SI{120}{\degree}. Its nephelometry responses are linear with the aerosol concentrations and independent of the sample flow rate. Then, the particles are deposited and accumulated onto a glass fiber filter tape having automatic replacement capability between a proportional detector and a beta source. The particle mass on the filter tape is determined by the attenuated beta intensity and the sample volume is integrated from the flow rate; from both of them the mass concentration is calculated. Calibration on the nephelometric signal is continuously performed using the mass estimated on the filter tape to ensure that the real-time mass concentration derived from nephelometry response remains independent of aerosol properties. The REFN hence outputs the hybrid, beta attenuation, and nephelometer derived mass concentration readings, respectively.

The REFN can detect/measure PM\textsubscript{2.5} at concentration up to \SI[per-mode=symbol,group-separator={,}]{10000}{\micro\gram\per\cubic\meter} with resolution as \SI[per-mode=symbol]{0.1}{\micro\gram\per\cubic\meter} and detection limit as low as $<\SI[per-mode=symbol]{0.2}{\micro\gram\per\cubic\meter}$. The nominal precision of mass concentration measurement is $\pm\SI[per-mode=symbol]{2}{\micro\gram\per\cubic\meter}$ and $\pm\SI[per-mode=symbol]{5}{\micro\gram\per\cubic\meter}$ for concentration $<\SI[per-mode=symbol]{80}{\micro\gram\per\cubic\meter}$ and $>\SI[per-mode=symbol]{80}{\micro\gram\per\cubic\meter}$, respectively; the nominal accuracy of mass measurement is $\pm\SI{5}{\percent}$ of a mass foil set that is traceable to the National Institute of Standards and Technology (NIST). For the sample flow rate, its nominal value is \SI[per-mode=symbol]{16.67}{\liter\per\minute} with nominal precision and accuracy as $\pm\SI{2}{\percent}$ and $<\SI{5}{\percent}$, respectively. The REFN generates a mass concentration reading every \SI{1}{\second} and offers a user configurable averaging interval from \SIrange{60}{3600}{\second}, and \SI{24}{\hour}. The REFN and its predecessor without a proprietary iPort (Model 5030 SHARP Monitor) are widely adopted as reference instruments in existing studies for evaluating and calibrating the research-grade instruments and the low-cost PM sensors \cite{kelly2017ambient,li2017characterization,zheng2018field,becnel2019distributed}.


\subsection{Experimental Setup}
\label{SubSect:ExperimentalSetup}

\begin{figure*}[!t]
\centering
	\subfloat[]{
		\includegraphics[width=0.37\linewidth]{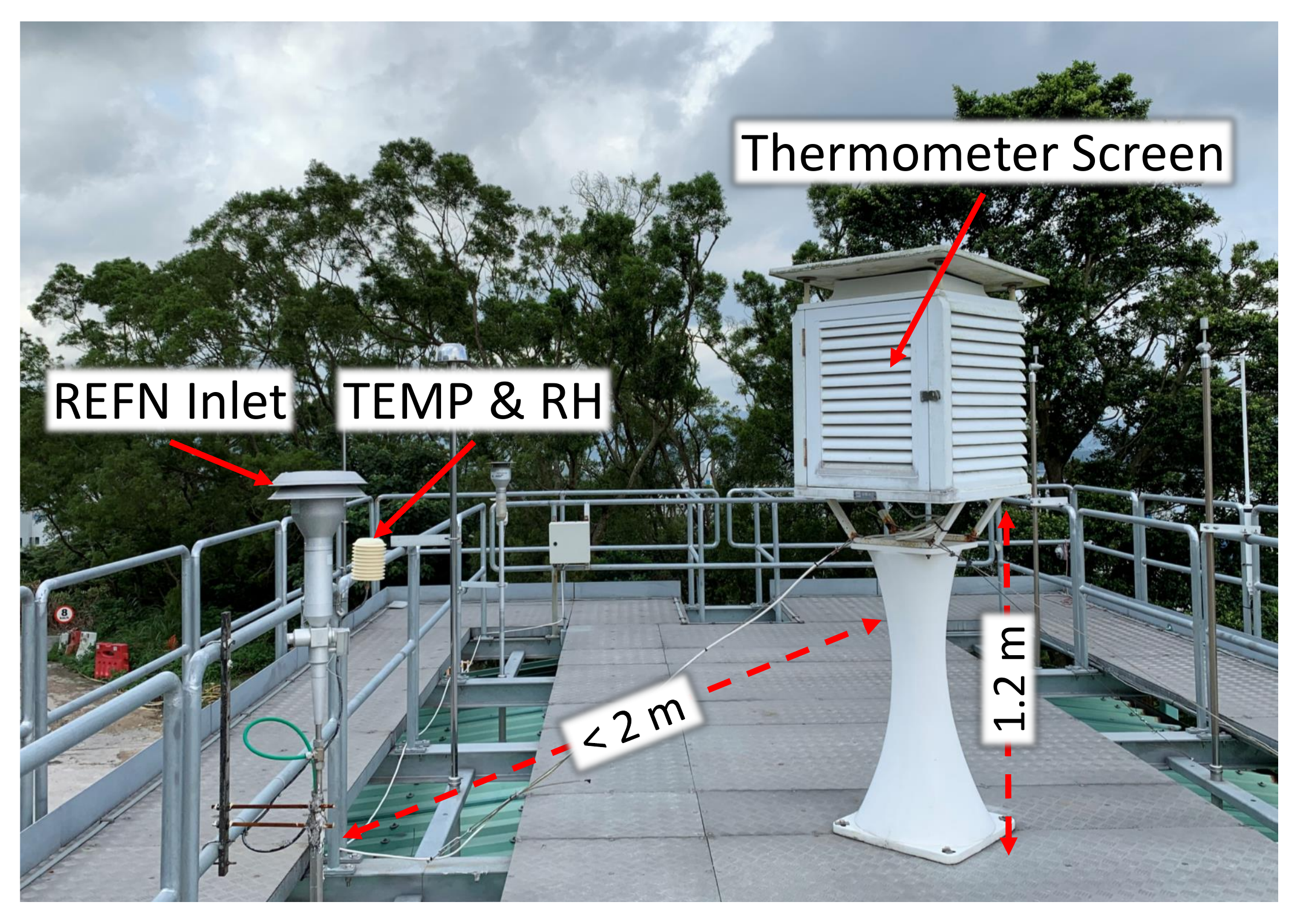}
		\label{subfig:RoofSupersite}}
	\subfloat[]{
		\includegraphics[width=0.2\linewidth]{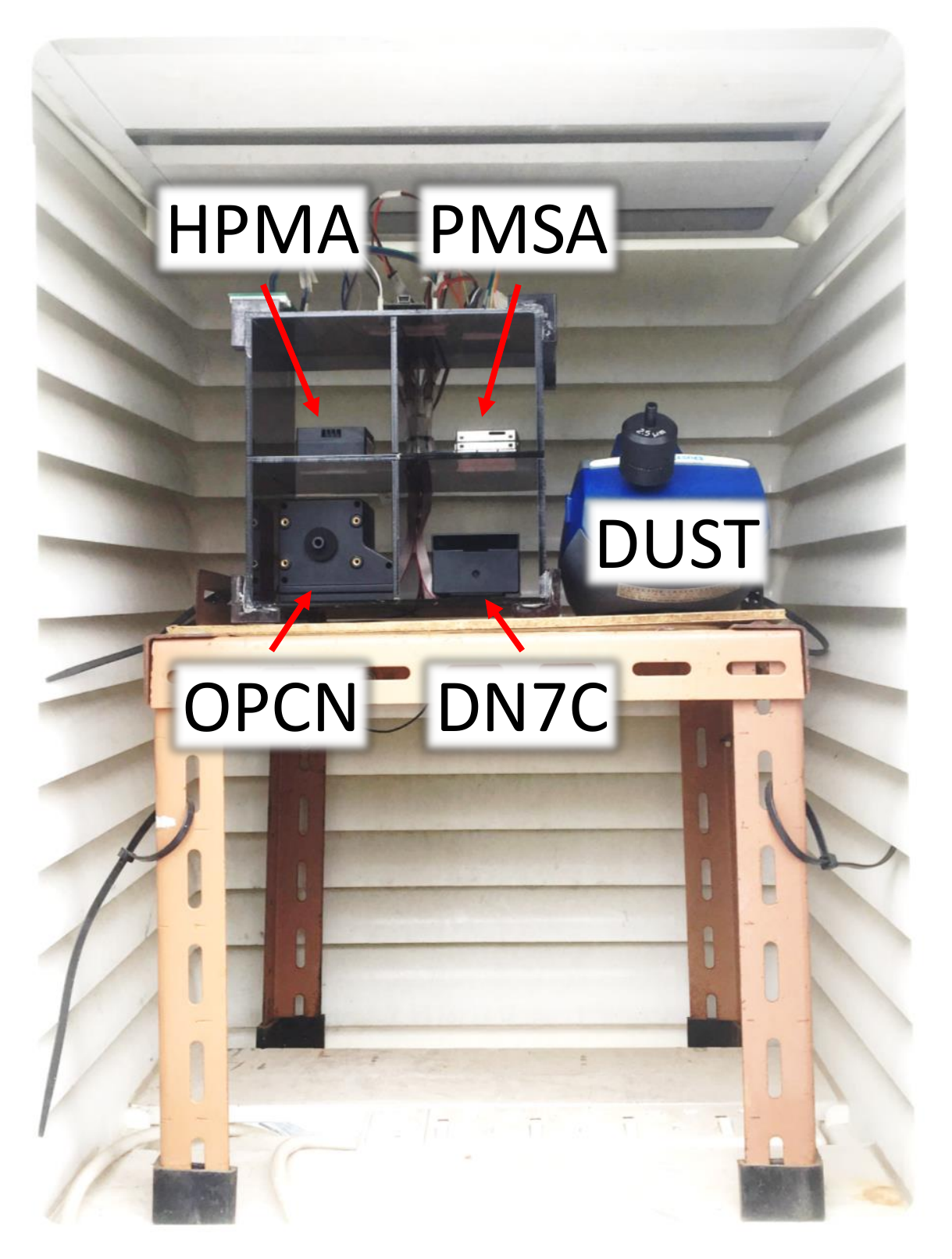}
		\label{subfig:PlacementPMSensors}}
	\subfloat[]{
		\includegraphics[width=0.2\linewidth]{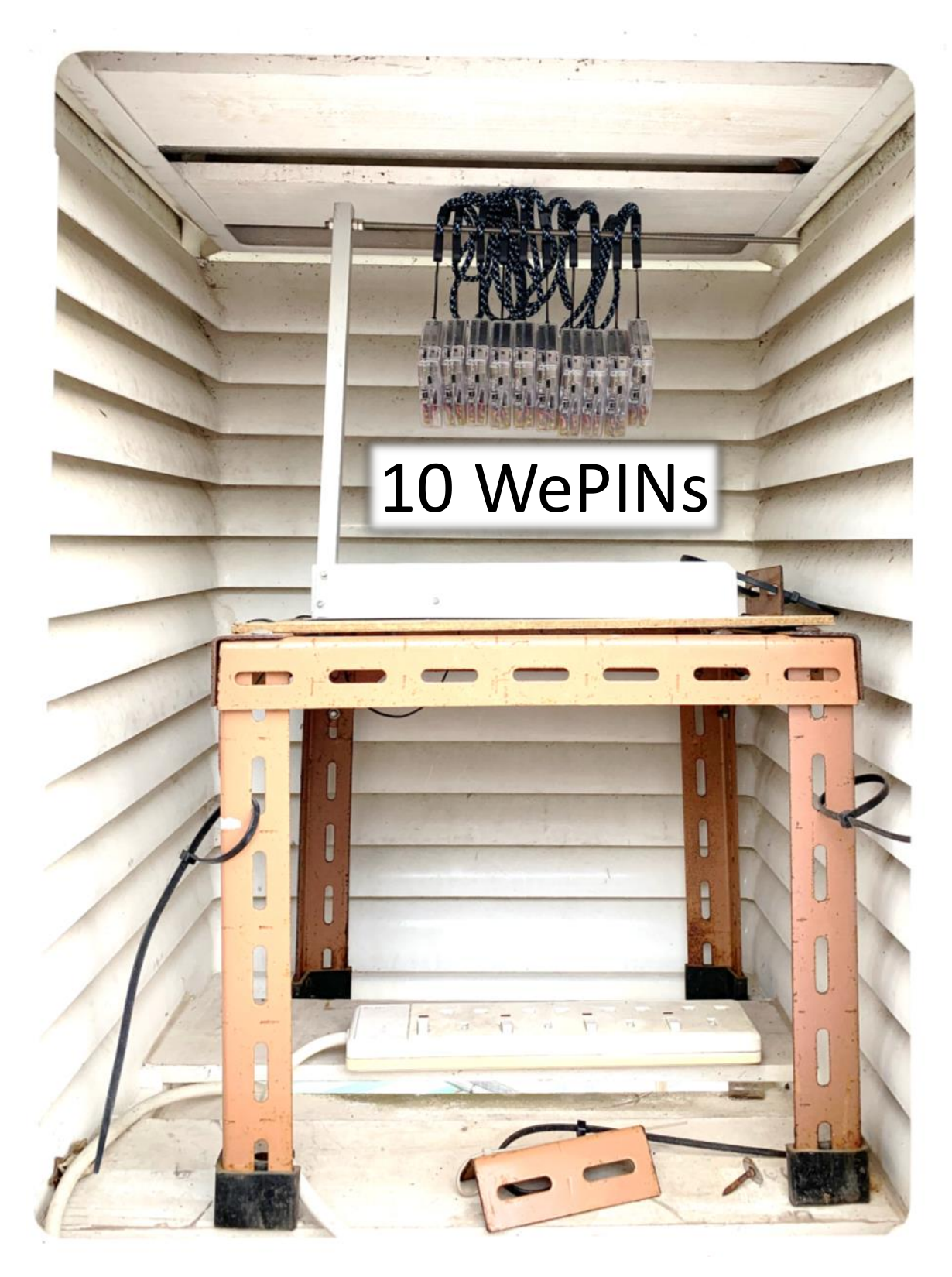}
		\label{subfig:PlacementPMDevices}}
\caption{Photos taken during the field experiments: (a) roof of the HKUST Air Quality Research Supersite; (b) placement of four PM sensors and the DUST inside the thermometer screen; and (c) placement of ten WePINs inside the thermometer screen.}
\label{fig:PhotoFieldExperimentSetup}
\end{figure*}

Unlike the laboratory evaluation and calibration of low-cost sensors detecting the gaseous pollutants including CO, SO\textsubscript{2}, O\textsubscript{3}, NO\textsubscript{x}, and volatile organic compounds (VOCs), which are typically conducted by challenging the sensors to pollutants with known and constant concentrations \cite{williams2014low}, it is especially difficult to maintain the concentration of aerosol at constant level primarily due to particle precipitation \cite{wang2015laboratory,rai2017end,badura2018evaluation}. Besides, earlier sections have highlighted the necessities and importances of evaluating and calibrating the PM sensors in environment close to the final deployment one because their responses are highly dependent on the aerosol properties and the atmospheric conditions that vary across locations. As a result, collocated evaluation and calibration of the PM sensor against more established instrument, which are considered as the default methods that could always be applied \cite{williams2014low}, are preferred in this manuscript. In accordance with the U.S. EPA, the term \textbf{collocated} is defined as:
\begin{quotation}
Collocated means two or more air samplers, analyzers, or other instruments that are operated simultaneously while located side by side, separated by a distance that is large enough to preclude the air sampled by any of the devices from being affected by any of the other devices, but small enough so that all devices obtain identical or uniform ambient air samples that are equally representative of the general area in which the group of devices is located \cite{ecfr2020part53subpartA}.
\end{quotation}
Further specified in the PM\textsubscript{2.5} performance evaluation program (PEP) procedures issued by the U.S. EPA, the horizontal and vertical distances between a primary (i.e., more established instrument) and a collocated sampler (e.g., PM sensor) should be within four meters and one meter, respectively \cite{ecfr2020part58appendixA}. In addition to the component level evaluation and calibration, the low-cost sensor utilized as part of a monitoring device should be evaluated and calibrated at system level instead of directly adopting the results from component level \cite{bychkovskiy2003collaborative} because the sensor measurements might subject to regulated air flow \cite{wang2015laboratory}, sensor housing \cite{kelly2017ambient}, system integration \cite{wang2020performance}, etc.

In this study, two field experiments were conducted in the spring of 2018 and the summer of 2019 in order to evaluate the performances of and calibrate the outputs from the four nominated PM sensors as well as the ten implemented wearable monitoring devices (i.e., the WePINs) against the collocated reference instrument(s), respectively. In each experiment, the PM sensors or the WePINs were placed inside a thermometer screen on the rooftop of the Hong Kong University of Science and Technology (HKUST) Air Quality Research Supersite as illustrated in Fig. \ref{subfig:RoofSupersite}, which is located on the hillside of Clear Water Bay on the east coast of Hong Kong (\SI{22.3379}{\degree\north} and \SI{114.2675}{\degree\east}). Placed inside the Supersite is the REFN FEM instrument that provides the hybrid PM\textsubscript{2.5} mass concentration measurements together with the TEMP and RH readings in 1-minute interval. The REFN is regularly maintained by the Hong Kong EPD and its size selective inlet is located next to ($<\SI{2}{\meter}$ horizontally and $<\SI{0.5}{\meter}$ vertically) the thermometer screen. The Supersite also provides the atmospheric pressure, wind direction, and wind speed readings in 1-minute interval.

\subsubsection{Evaluation and Calibration of PM Sensors}
\label{SubSubSect:EvalAndCaliOfPMSensors}

This campaign was conducted from 13 to 23 February 2018 in order to determine the most suitable PM sensor for constructing the wearable monitoring device (i.e., the WePIN) in the proposed personal PM exposure monitoring system. As illustrated in Fig. \ref{subfig:PlacementPMSensors}, the HPMA, PMSA, OPCN, and DN7C were secured in an enclosure without the front and back panels and placed inside the thermometer screen; their inlets and outlets were facing towards and away from the audiences, respectively. A dedicated embedded system\footnote{The dedicated data acquisition and logging system is a simplified version of the main control unit in \cite{yi2018modular} developed by the authors earlier.} was developed to acquire the latest readings\footnote{Mass concentrations of PM\textsubscript{1.0}, PM\textsubscript{2.5}, and PM\textsubscript{10} from the OPCN; mass concentrations of PM\textsubscript{2.5} and PM\textsubscript{10} from the HPMA; mass concentrations of PM\textsubscript{1.0}, PM\textsubscript{2.5}, and PM\textsubscript{10} from the PMSA ($CF=atmos$); and ADC (12-bit) measurements of the voltage across the PD from the DN7C.} from the four PM sensors and log them into a microSD card in 5-second interval. To minimize the power supply noise, the embedded system and the PM sensors were continuously powered by a Keithley Instruments Model 2400 SourceMeter with \SI{5}{\volt} DC output. The zero calibrated DUST with a PM\textsubscript{2.5} impactor attached and its battery charger plugged in was placed inside the thermometer screen as well to provide PM\textsubscript{2.5} mass concentration measurements in 1-minute interval for additional reference. Once the setup had been completed, the thermometer screen was closed, and no interruption was made during the entire campaign.

\subsubsection{Evaluation and Calibration of WePINs}
\label{SubSubSect:EvalAndCaliOfWePINs}

This campaign was conducted from 26 June to 17 July 2019 to evaluate and calibrate the ten firstly implemented WePINs in the field. As illustrated in Fig. \ref{subfig:PlacementPMDevices}, they were hanged in the middle of the thermometer screen with their inlets facing \SI{45}{\degree} downward, which is considered to be the most common orientation when they are carried by the users. Each WePIN is programmed to acquire and log the TEMP and RH readings from the on-board temperature-humidity sensor and the PM\textsubscript{1.0}, PM\textsubscript{2.5}, and PM\textsubscript{10} mass concentration measurements ($CF=atmos$) from the PMSA in 1-minute interval. Because a WePIN is battery powered with lifetime $\approx\SI{24}{\hour}$, a daily charging routine was established from 13:30 to 16:00. However, since access to the roof of the Supersite was denied on rainy days and repairs of malfunctioning\footnote{The electronic components on the circuit boards of these 10 WePINs were hand-soldered causing most of the malfunctions. No malfunction has been reported for the latest 120 WePINs with machine-soldered circuit boards.} WePINs had been performed, the charging routine was not strictly followed and the WePINs were down for a few days during the campaign.

\subsubsection{Assumptions in Collocated Evaluation and Calibration}
\label{SubSubSect:AssumptionInEvalCali}

The key idea of collocated evaluation and calibration of low-cost sensor/device is that the sensor/device and the reference instrument, which are collocated with each other, are probably observing the same phenomenon such that the time synchronized data pairs acquired by them are highly correlated \cite{bychkovskiy2003collaborative}. Assuming the reference instrument as the gold standard and its outputs as true values (i.e., the exact concentrations of the sampled aerosol), from the correlated data pairs, metrics can be calculated to assess the performances of the sensor/device under different atmospheric conditions (i.e., evaluation), and parametric and/or nonparametric adjustment model(s) can be developed to improve the quality of measurements from the sensor/device (i.e., calibration).

In the two field experiments conducted in this manuscript, the following assumptions were made:
\begin{itemize}
	\item the aerosol sampled by the reference instrument and the sensor/device simultaneously has negligible difference;

	\item the aerosol properties (i.e., the particle density, refractive index, size distribution, and chemical composition) have negligible change during the entire campaign;

	\item the aerosol concentration has negligible change within a sampling interval, or the measurement from a sampling interval is representative for that period;

	\item the aerosol concentrations from the reference instrument are considered as true values with zero uncertainty;

	\item the influence factors (i.e., RH and TEMP) in the ambient and that in the instruments (i.e., the reference instruments, PM sensors, and WePINs) have negligible difference\footnote{In real world, the sampled aerosol inside an instrument is likely to have slightly higher TEMP and lower RH than that in the ambient due to the heat generated by the internal electronic components \cite{crilley2018evaluation}.};
		
	\item the influence factors have negligible difference within a sampling intervals, or the measurements from a sampling interval are representative for that period;

	\item the measurements of the influence factors acquired are viewed as true values with zero uncertainty;

	\item the duration of the campaign is sufficient to characterize the micro-environment in deployment area for a sufficient period of succeeding time\footnote{The calibration parameters developed from this campaign are applicable for a sufficient period of time before recalibration is needed.}; and

	\item the relationship between the pairs of time synchronized measurements from the sensor/device and the reference instrument can be properly represented by the parametric models, the ordinary least-squares (OLS) regression and the multiple linear regression (MLR) to be specific.
\end{itemize}
With regard to the last assumption above, the formulas for the OLS and MLR models with single dependent variable and $n$ observations are:
\begin{equation}
\label{Eq:OLSFormula}
	y_i=\beta_0+\beta_1x_{i1}+\epsilon_i \textrm{,}\quad i=1,\cdots,n
\end{equation}
and
\begin{equation}
\label{Eq:MLRFormula}
	\begin{aligned}
		y_i=\beta_0+\beta_1x_{i1}+\cdots+\beta_kx_{ik}+\epsilon_i \textrm{,}\\
		i=1,\cdots,n \ \textrm{and} \ k\in\mathbb{Z}^+
	\end{aligned}
\end{equation}
respectively, where:
\begin{itemize}
	\item $y_i$ is the $i^{th}$ observation of the dependent variable;
	\item $x_{ik}$ is the $i^{th}$ observation of the $k^{th}$ independent variable;
	\item $\beta_0$ is the intercept value;
	\item $\beta_k$ is the slope value of the $k^{th}$ independent variable; and
	\item $\epsilon_i$ is the $i^{th}$ error term or residual of the model.
\end{itemize}
In the context of collocated calibration, the dependent variable $\vec{y}\,$ denotes the measurements from a reference instrument and the independent variable $\vec{x}_1$ is referring to the measurements from a sensor/device. For MLR model, the other independent variables $\vec{x}_k \ (k>1)$ denote the measurements of influence factors, namely TEMP, RH, etc. The following assumptions were made on the data set to which the OLS model and the MLR model are applicable \cite{seber2012linear9,gross2012linear2}:
\begin{itemize}
	\item $\vec{x}_k$ is not a random variable;
	\item no correlation between the independent variables;
	\item $\vec{\epsilon}\,$ has zero mean, i.e., $\bm{E}(\vec{\epsilon}\,)=0$;
	\item $\vec{\epsilon}\,$ has constant variance, i.e., $\textit{Var}(\vec{\epsilon}\,)=\sigma^2\bm{I}_n$ and $\sigma^2>0$;
	\item $\epsilon_i$ of $\vec{\epsilon}\,$ are independent and normally distributed; and
	\item no autocorrelation of $\vec{\epsilon}\,$, i.e., $\textit{Cov}(\epsilon_i,\epsilon_j)=0$ for $i\neq j$.
\end{itemize}

Preliminary results demonstrated that the wind speed, the wind direction, and the atmospheric pressure have negligible effects on the performances of these nominated PM sensors; study \cite{kelly2017ambient} also reported that no correlation was seen between the wind speed and the PM\textsubscript{2.5} mass concentrations from the PM sensors. Besides, there is hardly any literature studying the relation between any one of the above influence factors and the performances of PM sensors. In this regard, the discussion of the above influence factors was omitted in this manuscript. When the amount of water vapor remains constant in the air, the RH values are inversely related to the TEMP values. Such phenomenon was observed in the RH and the TEMP readings obtained from the first field experiment as shown in Fig. \ref{fig:PM2p5RhTempFull}. Therefore, either RH or TMEP, but not both, is considered to be the additional independent variable in the MLR model. It is conceivable that the PM sensors with laboratory calibration, in which case the RH and the TEMP are controlled separately and independently \cite{wang2015laboratory}, may likely have distinct responses to certain PM compared to that with collocated field calibration.

\subsection{Terminology and Evaluation Metrics}
\label{SubSect:TerminologyAndEvaluationMetrics}

As mentioned in section \ref{SubSect:WhyEvaluation}, because of the absence of a unifying evaluation protocol, the datasheet and the existing experimental results of a PM sensor or device have limited reference values; it is exceedingly difficult to draw generalized conclusions based on the evaluation outcomes from various sources. In response, the authors are appealing to the research community to develop a standardized evaluation protocol for low-cost PM sensors or devices to characterize the quality of their measurements, particularly under collocated conditions, including standard methodology, definitive terminology, and unified evaluation metrics. It is the key to ensuring that the results from different groups can be properly communicated, correctly interpreted, and conveniently compared.

An excellent example is the Air Quality Sensor Performance Evaluation Center (AQ-SPEC) program \cite{scaqmd2020air} established by the South Coast Air Quality Management District (AQMD), which aims at evaluating the commercially available and low-cost air quality sensors \cite{scaqmd2020gas,scaqmd2020pm} under field (ambient) and laboratory (controlled) conditions following a unified protocol \cite{scaqmd2020field,scaqmd2020lab}. Nevertheless, the AQ-SPEC program could be pushed further because most of the field experimental results only reported the coefficient of determination ($R^2$) values.

Approaching the development of a standardized protocol for evaluating low-cost PM sensors or devices, the first step is to understand the terminology pertaining to measurement quality. However, partly due to the subtle differences and partly due to the incorrect or inconsistent usage, the terminology can be a bit confusing \cite{ndt2014uncertainty}. Terms such as accuracy and trueness, error and uncertainty, and precision and standard deviation, are examples of the many terms that are frequently appeared in the literature with ambiguous definitions and synonymous usage \cite{amc2003terminology}. Besides, most of the terminology describing the measurement quality is itself a qualitative concept\footnote{One should not use the qualitative concept quantitatively, that is, associating number with it \cite{taylor1994guidelines}. For example, the statement ``the precision of the measurement results, expressed as the standard deviation obtained under repeatability conditions, is \SI[per-mode=symbol]{5}{\micro\gram\per\cubic\meter}'' is acceptable, but the statement ``the precision of the measurement results is \SI[per-mode=symbol]{5}{\micro\gram\per\cubic\meter}'' is not.} that can be quantified using various quantitative approaches\footnote{For example, both statement ``the precision of the measurement results, expressed as the standard deviation obtained under repeatability conditions, is \SI[per-mode=symbol]{5}{\micro\gram\per\cubic\meter}'' and statement ``the precision of the measurement results, expressed as the coefficient of variation obtained under reproducibility conditions, is \SI{5}{\percent}'' are describing the precision of the measurement results.} raising ubiquitous issues for researchers when interpreting existing results \cite{mcalinden2015precision}. Therefore, it is also vital to unify the formulas for quantification and the criteria for assessment in the development of the standardized evaluation protocol.

In reality, all measurements of physical quantities are subject to measurement uncertainties that determine the quality of measurements. For measurement uncertainty assessment, there are two general approaches with relevant metrological terminology that are internationally recognized, namely the Error Approach and the Uncertainty Approach. Though both approaches have been introduced for decades, the debate on their benefits over the other continues \cite{theodorsson2017uncertainty,oosterhuis2018use,farrance2018uncertainty}. In this manuscript, no preference is intended to be given to any one of these particular approaches; they are commonly used in association with one and the other.

To the best of the authors' knowledge, the \textit{International Vocabulary of Metrology -- Basic and General Concepts and Associated Terms} \cite{jcgm2012vim}, developed by the Joint Committee for Guides in Metrology (JCGM) \cite{jcgm2012jcgm} and usually known as the VIM, provides the most complete, standard, and international reference for the definitions of metrological characteristics. Although the definitive document on the Uncertainty Approach, that is, the \textit{Evaluation of Measurement Data -- Guide to the Expression of Uncertainty in Measurement} \cite{jcgm2008gum}, developed by the JCGM as well and often known as the GUM, modified the terminology taken from the VIM (second edition) slightly to reflect its view, the definitions are conceptually equivalent. In addition, the latest edition (third edition) of the VIM has adopted the definition of precision, which was not included in its earlier editions, from the ISO 5725-1 \cite{iso5725-1}. To enable mutual communication and interpretation, one should use the terminology consistent with the VIM and/or other international documents. Since they are rather lengthy documents, one may get a brief understanding of the metrological terminology and its correct usage from \cite{taylor1994guidelines,bell2001beginner,amc2003terminology,ndt2014uncertainty}.

In regard to quantifying the quality of measurements from or assessing the field and/or laboratory performances of a PM sensor or device, the evaluation metrics, including formulas and criteria, detailed in the \textit{Appendix A of 40 CFR Part 58} \cite{ecfr2020part58appendixA} and the \textit{Subpart C of 40 CFR Part 53} \cite{ecfr2020part53subpartC} (hereinafter referred to as the Part 58AA and the Part 53SC, respectively) are recognized by the authors as the best practices because they were introduced by the U.S. EPA for determining an instrument's compliance with the National Ambient Air Quality Standards (NAAQS) and testing an instrument's comparability to corresponding collocated reference methods, respectively. Part of them are frequently employed in existing studies for evaluating the performances of PM sensors or devices under different environmental conditions \cite{sousan2016inter,badura2018evaluation,zamora2018field,hapidin2019aerosol}. In accordance with \cite{ecfr2020part53subpartA}, all PM sensors nominated in this manuscript are categorized as the Class III candidate methods for PM\textsubscript{2.5}, and therefore, the evaluation metrics documented in \cite{ecfr2020part53subpartCpoint35}, i.e., a subsection of the Part 53SC, are more precise. Besides, a PM sensor or device is usually not intended to be used to ascertain if the ambient air quality standards are being met, and to reflect the assumptions listed in section \ref{SubSubSect:AssumptionInEvalCali}, the formulas provided in the Part 58AA are modified accordingly.

To avoid confusing the audiences and to provide a quick reference to other researchers conducting similar studies, the definitions of relevant meteorological terminology taken from the VIM (third edition) along with the comments from other documents or the authors, and the evaluation metrics used to quantify the measurement quality or assess the performances of the PM sensors/devices, are provided in the following. The VIM, the GUM, the ISO 5725 package, the Part 58AA, and the Part 53SC may be consulted for additional details.

\subsubsection{General Terms}
\label{SubSubSect:GeneralTerms}

\begin{itemize}
	\item The definition of the term \textbf{quantity} is given as ``property of a phenomenon, body, or substance, where the property has a magnitude that can be expressed as a number and a reference''. Here, the number is usually a scalar and the reference is usually the measurement unit.

	\item The term \textbf{measurand} denotes ``the quantity intended to be measured''. The specification of a measurand requires knowledge of the kind of quantity, and description of the state of the phenomenon, body, or substance carrying the quantity (e.g., statements about quantities such as time, temperature, and pressure). The latter is essential because it might change the phenomenon, body, or substance such that the quantity being measured may differ from the measurand as defined.

	\item The term \textbf{measuring system} stands for the ``set of one or more measuring instruments and often other devices, including any reagent and supply, assembled and adapted to given information used to generate measured quantity values within specified intervals for quantities of specified kinds''. A measuring system may comprise of only one \textbf{measuring instrument}, which is the ``device used for making measurements, alone or in conjunction with one or more supplementary devices''.


	\item The term \textbf{measurement result} or \textbf{result of measurement} is defined as the ``set of quantity values being attributed to a measurand together with any other available relevant information''. The general expression of a measurement result has two essential components:
	\begin{itemize}
		\item a single measured quantity value which gives the best estimate of the measurand\footnote{This best estimate of the measurand may be the quantity value obtained through a single measurement or the arithmetic mean of a series of quantity values obtained though replicate measurements.}; and
		\item a measurement uncertainty associated with this estimated quantity value.
	\end{itemize}
	The measurement result may be expressed as a single measured quantity value if the measurement uncertainty could be considered to be negligible.

	\item The term \textbf{true quantity value} or \textbf{true value} represents the ``quantity value consistent with the definition of a quantity''. In principle, a measurand cannot be completely described without an infinite amount of information, and therefore, the true quantity value of a measurand is an idealized concept that cannot be absolutely determined in reality. Thus, the quantity value termed \textbf{conventional quantity value} or \textbf{conventional value} is used in practice, which denotes ``quantity value attributed by agreement to a quantity for a given purpose''. It is generally accepted that the conventional quantity value is associated with a suitably small or even negligible measurement uncertainty. One should not confuse these two terms with the term \textbf{reference quantity value} or \textbf{reference value} that stands for ``quantity value used as a basis for comparison with values of quantities of the same kind''. A reference quantity value can be a true quantity value (unknown) or a conventional quantity value (known) of a measurand, and its measurement uncertainty is usually provided with reference to a device, a material, a reference measurement procedure, or a comparison of measurement standards.
\end{itemize}

\subsubsection{Error Approach}
\label{SubSubSect:ErrorApproach}

In Error Approach, the true quantity value of a measurand is considered unique and the objective of a measurement is to determine an estimation of the true quantity value as close to that true quantity value as possible. The deviation from the true quantity value is raised by both the systematic and the random effects that should be treated differently. Since a true quantity value cannot be determined by nature, in practice, the conventional quantity value is used. Compared to the Uncertainty Approach, the Error Approach offers simpler yet more intuitive and practical procedures for assessing the measurement uncertainty and is widely adopted in various research fields, performance evaluation on low-cost sensors for example; however, the Uncertainty Approach is an internationally standardized approach with increasingly used in all research fields and will be the norm for measurement uncertainty analyses.

\begin{itemize}
	\item The definition of the term \textbf{measurement error} or \textbf{error} is given as ``measured quantity value minus a reference quantity value''. Hence, it is a quantitative concept and one can associate number with it. Note that a reference quantity value can be a conventional quantity value, in which case the measurement error is known, or a true quantity value, in which case the measurement error is unknown. The measurement error does not include any mistakes in measurements. It is composed of \textbf{systematic measurement error} and \textbf{random measurement error}, but how these distinguishable errors combine to form the total measurement error of a given measurement result is incognizable. Usually, only the upper limit of the absolute value of the total measurement error is estimated.

	\item The term \textbf{systematic measurement error} or \textbf{systematic error} represents the ``component of measurement error that in replicate measurements remains constant or varies in a predictable manner''. The causes may be known or unknown but a \textbf{correction}, defined as ``compensation for an estimated systematic effect'', should always be applied to compensate for a known systematic measurement error. After correction, the expected value of the error arising from a systematic effect is generally assumed to be zero though such error can never be eliminated in reality.

	\item The term \textbf{random measurement error} or \textbf{random error} is defined as the ``component of measurement error that in replicate measurements varies in an unpredictable manner''. It can occur for a variety of reasons such as lack of equipment sensitivity, noise in measurement, and imperfect definition of measurand. Compensation for the error arising from a random effect is impossible, not to speak of elimination. However, it can usually be reduced by increasing the number of replicate measurements. The distribution of the random measurement errors of a set of replicate measurements can be characterized by its expectation (generally assumed to be zero) and variance.

	\item To characterize the systematic measurement error, the term \textbf{measurement trueness} or \textbf{trueness} is used. It is an idealized qualitative concept denoting the ``closeness of agreement between the average of an infinite number of replicate measured quantity values and a reference quantity value''. The measurement trueness is inversely related to the systematic measurement error and usually expressed in terms of \textbf{measurment bias} or \textbf{bias} that is a quantitative concept defined as the ``estimate of a systematic measurement error'', i.e., the difference between the average of a large series of replicate measured quantity values and the reference quantity value. A correction is the compensation for bias, which can take the form of an addend, a factor, or both.

	\item To characterize the random measurement error, the term \textbf{measurement precision} or \textbf{precision} is employed. It is a qualitative concept defined as ``closeness of agreement between indications or measured quantity values obtained by replicate measurements on the same or similar objects under specified conditions''. Indicating how well a measurement can be without a reference quantity value, the measurement precision depends only on the distribution of random measurement errors and is usually quantified in terms of standard deviation, variance, coefficient of variation, etc. According to the conditions under which the replicate measurements were made, the measurement precision is further distinguished into:
	\begin{itemize}
		\item (\textbf{measurement}) \textbf{repeatability},
		\item \textbf{intermediate} (\textbf{measurement}) \textbf{precision}, and
		\item (\textbf{measurement}) \textbf{reproducibility},
	\end{itemize}
	which represent the measurement precision under a set of repeatability, intermediate precision, and reproducibility conditions of measurement, respectively, where:
	\begin{itemize}
		\item \textbf{repeatability condition} (\textbf{of measurement}) denotes ``condition of measurement, out of a set of conditions that includes the same measurement procedure, same operators, same measuring system, same operating conditions and same location, and replicate measurements on the same or similar objects over a short period of time'';

		\item \textbf{intermediate precision condition} (\textbf{of measurement}) represents ``condition of measurement, out of a set of conditions that includes the same measurement procedure, same location, and replicate measurements on the same or similar objects over an extended period of time, but may include other conditions involving changes'' (changes such as new calibrations, calibrators, operators, and measuring systems); and
		
		\item \textbf{reproducibility condition} (\textbf{of measurement}) stands for ``condition of measurement, out of a set of conditions that includes different locations, operators, measuring systems, and replicate measurements on the same or similar objects''.
	\end{itemize}
	Conceivably, factors that may contribute to the variability of the replicate measured quantity values include:
	\begin{itemize}
		\item the measurement procedure;
		\item the measuring system;
		\item the calibration of the measuring system;
		\item the operator or observer;
		\item the operating conditions or environment;
		\item the location; and 
		\item the time elapsed between measurements.
	\end{itemize}
	Under repeatability conditions, all factors are considered constants and do not contribute to the variability, while under reproducibility conditions, all factors vary and do contribute to the variability. Thus, measurement repeatability and measurement reproducibility are two extremes of measurement precision describing the minimum and the maximum dispersion characteristics of the replicate measured quantity values, respectively; when one or more factors are allowed to vary, the replicate measurements are under intermediate precision conditions because the magnitude of the variability associated with the varying factors lies between these two extremes \cite{iso5725-3}. The GUM and the earlier editions of the VIM did not distinguish the concepts of intermediate measurement precision and measurement reproducibility; they were merged together and represented by the latter one.

	\item Finally, the term \textbf{measurement accuracy} or \textbf{accuracy}, defined as ``closeness of agreement between a measured quantity value and a true quantity value of a measurand'', is a qualitative concept encompassing the measurement trueness and the measurement precision. It is characterizing the deviation of a measurement result from the true quantity value arising from both systematic and random effects when applied to a set of replicate measurements. The term measurement accuracy is a source of confusion because it was at one time used as a synonymy of the term currently named measurement trueness; though this synonymous usage is still active in research, it should be avoided because it has been deprecated for years in many international documents.
\end{itemize}

\subsubsection{Uncertainty Approach}
\label{SubSubSect:UncertaintyApproach}

In Uncertainty Approach, due to the inherently imperfect definition of a measurand, there exists not a single true quantity value but a set of true quantity values that are consistent with the measurand's definition. Therefore, the objective of measurement in the Uncertainty Approach is not to determine the true quantity value as closely as possible as in the Error Approach; rather, it is to determine an interval within which the set of unknowable true quantity values will lie with a particular probability. The range of the interval may be reduced when additional relevant information is available, but it cannot be mitigated to a single quantity value because the relevant information is finite in reality. Although such an approach may be a bit complicated for simple purposes such as conducting performance evaluation on low-cost sensors by ordinary users, it is useful for manufactures or authorities to identify various sources of uncertainties and make improvements in a measuring system.

\begin{itemize}
	\item In a broadest sense, the term \textbf{measurement uncertainty} or \textbf{uncertainty} is a qualitative concept representing the doubt about the validity of a measurement result. Due to lacking different words for this general concept and its specified quantities, it is as well utilized as a quantitative concept to quantify itself, which is established as ``non-negative parameter characterizing the dispersion of the quantity values being attributed to a measurand, based on the information used''. The non-negative parameter is usually a standard deviation value. In general, there are many possible sources of uncertainty in a measurement; therefore, the measurement uncertainty comprises many components. Part of them are arising from the systematic effects (e.g., the uncertainties associated with corrections and reference standards, and the imperfect definition of measurand), whilst others are arising from the random effects (e.g., the variations of influence quantities). These components can be grouped into either Type A or Type B category depending upon whether they were evaluated by the \textbf{Type A evaluation} or the \textbf{Type B evaluation}, but not their sources, i.e., random or systematic effects. This categorization is not meant to indicate there exists any difference in the nature of the components resulting from the Type A and the Type B evaluations that are based on probability distributions obtained experimentally and hypothetically, respectively, but for discussion's sake only. Both types of evaluations are assuming corrections have been applied to compensate for all significant systematic effects that can be identified, which may not be practical in some cases due to limited resources; the uncertainty associated with each recognized systematic effect is then the uncertainty of the applied correction. According to the definition, the measurement uncertainty of a measurement result should have characterized all dispersions arising from both systematic and random effects; therefore, it is probably the most appropriate means for expressing the measurement accuracy. When all uncertainty components arising from systematic effects are considered negligible, the estimated measurement uncertainty is a reflection of the measurement precision under specified conditions.

	\item The \textbf{Type A evaluation} (\textbf{of measurement uncertainty}) is defined as ``evaluation of a component of measurement uncertainty by a statistical analysis of measured quantity values obtained under defined measurement conditions''. Hence, a Type A standard uncertainty, i.e., the estimated standard deviation from Type A evaluation, is obtained based on the probability density function derived from a series of measurements under repeatability, intermediate precision, or reproducibility conditions of measurement.

	\item The \textbf{Type B evaluation} (\textbf{of measurement uncertainty}) is defined as ``evaluation of a component of measurement uncertainty determined by means other than a Type A evaluation of measurement uncertainty''. Such uncertainty component can be characterized by a standard deviation as well, which is called a Type B standard uncertainty and evaluated from an assumed probability distribution based on other information such as limits deduced from personal experience and general knowledge, judgments obtained from calibration certificates, and quantity values published by authorities.

	\item The \textbf{standard measurement uncertainty} or \textbf{standard uncertainty} is the ``measurement uncertainty expressed as a standard deviation''. When a measurement has multiple identifiable sources of measurement uncertainties or the result of measurement is determined from the values of a number of other quantities, the final standard measurement uncertainty is then obtained by combining the individual standard measurement uncertainties, whether arising from the Type A evaluation or Type B evaluation, following the \textbf{law of propagation of uncertainty} that is summarized in the GUM and explained with examples in \cite{farrance2012uncertainty}. The resulted standard measurement uncertainty is termed \textbf{combined standard measurement uncertainty} or \textbf{combined standard uncertainty} denoting the ``standard measurement uncertainty that is obtained using the individual standard measurement uncertainties associated with the input quantities in a measurement model''. To express the measurement uncertainty, the term \textbf{expanded measurement uncertainty} or \textbf{expanded uncertainty}, defined as ``product of a combined standard measurement uncertainty and the factor larger than the number one'', is more commonly used. This factor is dependent upon the probability distribution of the output quantity of a measurement model and on the selected coverage probability.
\end{itemize}

\subsubsection{Evaluation Metrics}
\label{SubSubSect:EvaluationMetrics}

It is worth highlighting again that the purpose of evaluating the nominated PM sensors and the WePINs is to accomplish sufficient degree of knowledge about their performances under deployed environment with various circumstances and ultimately improve the confidence in their measurement results; it is neither critical nor possible for a low-cost sensor or device, whose data are not intended to be used to ascertain if the ambient air quality standards are being met, to achieve the same performances as the research-grade instruments or the conventional monitors. As a result, though most of the evaluation metrics below, including the formulas and the criteria, are adopted from the Part 58AA and the Part 53SC, due to limited resources and distinct objectives, they are modified accordingly as detailed in this section. Additionally, limited to what is practical, the procedures and requirements described in these materials are not rigorously followed when conducting the two collocated field experiments presented in section \ref{SubSect:ExperimentalSetup}. However, it is yet a decent attempt at unifying the evaluation metrics for approaching the development of a standardized evaluation protocol.

The Part 58AA specifies the minimum data quality requirements of a monitoring system whose data are intended to be used to determine compliance with the NAAQS, and focuses on assessing and controlling measurement uncertainty. It states the goal for acceptable measurement uncertainty of automated and manual PM\textsubscript{2.5} methods, which is defined as follows:
\begin{itemize}
	\item for measurement trueness as a \SI{90}{\percent} confidence interval for the relative bias of $\pm\SI{10}{\percent}$; and

	\item for measurement precision as an upper \SI{90}{\percent} confidence limit for the coefficient of variation ($CV$) of \SI{10}{\percent}.
\end{itemize}
Notice that the Error Approach and the Uncertainty Approach introduced in section \ref{SubSubSect:ErrorApproach} and section \ref{SubSubSect:UncertaintyApproach}, respectively, are utilized collectively to assess the measurement uncertainty. Both measurement trueness and measurement precision stated in the goal are assessed via collocated data pairs acquired by the PM\textsubscript{2.5} monitoring system under examination, designated as the primary monitor whose concentrations are used to report air quality for the regulatory monitoring site, and a collocated quality control monitor.

The measurement trueness, with respect to the collocated quality control monitor, of the primary monitor is assessed by the PM\textsubscript{2.5} performance evaluation program (PEP) introduced in the Part 58AA. Though the PEP is intended to be used to provide collocated quality control assessments on a regulatory PM\textsubscript{2.5} monitor, in this manuscript, its formulas\footnote{A quick tutorial \cite{epa2007guideline} and a specific tool \cite{epa2017dacs} for quantifying measurement trueness under the PEP procedures are provided by the U.S. EPA.} and criteria are adopted for assessing the measurement trueness of a low-cost PM\textsubscript{2.5} sensor or device with respect to a collocated PM\textsubscript{2.5} reference instrument (i.e., REFN) while the requirements and procedures presented in the Part 58AA are loosely followed due to limited resources and different objectives. The formulas and criteria adopted from the PEP are listed in the following.
\begin{itemize}
	\item For each collocated data pair, calculate the corresponding value of the relative error $D$ using the following equation:
	\begin{equation}
	\label{Eq:PEPRelativeError}
		\begin{aligned}
			d_i = \frac{x_i - y_i}{y_i} \cdot 100
		\end{aligned}
	\end{equation}
	where $x_i$ and $y_i$ are measurements from a sensor/device and a reference instrument at time instant $i$, respectively, and $d_i$ denotes the corresponding value of random variable $D$ at time instant $i$. In accordance with the PEP, the relative error may be extremely poor at low concentration levels, i.e., $d_i$ is exceedingly large if $y_i$ is significantly small, and therefore, a collocated data pair is considered valid only when both measurements are equal to or above \SI[per-mode=symbol]{3}{\micro\gram\per\cubic\meter} for PM\textsubscript{2.5}.

	\item The PEP measurement trueness estimator depends upon the sample mean $\bar{d}_n$ of relative errors $\{d_1,\cdots,d_n\}$ and the probability distribution of random variable $\bar{D}_n$. The corresponding value of random variable $\bar{D}_n$ is achieved by the equation below:
	\begin{equation}
	\label{Eq:PEPMeanRelativeError}
		\begin{aligned}
			\bar{d}_n = \frac{1}{n} \cdot \sum_{i=1}^{n}d_i
		\end{aligned}
	\end{equation}
	where $n$ is the number of valid collocated data pairs, and $d_1,\cdots,d_n$ are the corresponding values of $D_1,\cdots,D_n$ that are considered to be the independent and identically distributed random variables with population mean $\mu$ and population standard deviation $\sigma$.

	\item Statistically, according to the central limit theorem (CLT), the distribution of random variable $\bar{D}_n$ converges to the normal distribution $\mathcal{N}(\mu, \sigma^2/n)$ as $n\to\infty$ regardless of the distribution of random variable $D$. When $\sigma$ is known, the \textit{z}-score, i.e., $z=(\bar{d}_n-\mu)/(\sigma/\sqrt{n})$, can be computed such that the population mean $\mu$ of relative error $D$ can be estimated using the standard normal distribution.

	\item In reality, the number of valid collocated data pairs $n$ is finite and the population standard deviation $\sigma$ is usually unknown. Therefore, the population mean $\mu$ is estimated based on the \textit{t}-score, i.e., $t=(\bar{d}_n-\mu)/(s_n/\sqrt{n})$, whose distribution follows a \textit{t}-distribution with $n-1$ degrees of freedom. The symbol $s_n$ in the \textit{t}-score formula is the sample standard deviation of relative errors $\{d_1,\cdots,d_n\}$ calculated as follows:
	\begin{equation}
	\label{Eq:PEPSampleStandardDeviation}
		\begin{aligned}
			s_n & = \sqrt{\frac{\sum_{i=1}^{n}{(d_i-\bar{d}_n)}^2}{n-1}} \\ 
				& = \sqrt{\frac{n\cdot\sum_{i=1}^{n}{d_i}^2-{(\sum_{i=1}^{n}d_i)}^2}{n(n-1)}}
		\end{aligned}
	\end{equation}
	where the number of valid collocated data pairs $n$, i.e., the sample size, should be sufficiently large (e.g., $\geq 30$).

	\item Finally, the \SI{90}{\percent} confidence interval, that is, two-sided significance level $\alpha=0.1$, for the population mean $\mu$ of relative error $D$ can be determined by the following equation:
	\begin{equation}
	\label{Eq:PEPBiasUCLandLCL}
		\begin{aligned}
			Bias_{PEP} = \bar{d}_n \pm t_{0.95,n-1} \cdot \frac{s_n}{\sqrt{n}}
		\end{aligned}
	\end{equation}
	where $t_{0.95,n-1}$ is the 95th percentile of a \textit{t}-distribution with $n-1$ degrees of freedom.

	\item If the upper and lower limits of interval $Bias_{PEP}$ lie within $\pm\SI{10}{\percent}$, the measurement trueness of the low-cost PM\textsubscript{2.5} sensor or device with reference to the collocated PM\textsubscript{2.5} reference instrument meets the requirement stated in the goal.
\end{itemize}

The measurement precision, to be specific, the intermediate measurement precision under intermediate precision conditions with change in the measuring systems whose measuring principles may or may not be identical, of the primary monitor is assessed following the collocated quality control sampling procedures provided in the Part 58AA. The calculations\footnote{A quick tutorial \cite{epa2007guideline} and a specific tool \cite{epa2017dacs} for quantifying measurement precision under collocated quality control sampling procedures are given by the U.S. EPA.} for quantifying such an intermediate measurement precision are based on the collocated data pairs measured by the primary monitor and the quality control monitor, both of which are used for regulatory purposes. In this regard, the best estimate of the exact PM\textsubscript{2.5} concentration under measurement is the average of the collocated measurements from both monitors. However, in the case of assessing a low-cost PM\textsubscript{2.5} sensor's or device's intermediate measurement precision with respect to a collocated reference instrument whose data quality is far superior, the reference measurements are considered as true PM\textsubscript{2.5} concentrations (consistent with the fourth assumption presented in section \ref{SubSubSect:AssumptionInEvalCali}). Therefore, the calculations for quantifying such intermediate measurement precision adopted from the Part 58AA are modified accordingly; the collocated quality control sampling procedures detailed in the Part 58AA are loosely followed owing to limited resources and different objectives. The modified formulas and criteria are presented in the following.
\begin{itemize}
	\item For each collocated data pair, calculate the corresponding value of the relative error $D$ using Eq. \ref{Eq:PEPRelativeError}; similarly, a collocated data pair is considered valid only when both measurements are equal to or above \SI[per-mode=symbol]{3}{\micro\gram\per\cubic\meter} for PM\textsubscript{2.5}.

	\item The intermediate measurement precision estimator in the Part 58AA is based on the sample variance $s^2_n$ of relative errors $\{d_1,\cdots,d_n\}$ and the probability distribution of random variable $S^2_n$. Here, symbols $d_1,\cdots,d_n$ are the corresponding values of $D_1,\cdots,D_n$ that are assumed to be the independent and identically distributed random variables with normal distribution $\mathcal{N}(\mu,\sigma^2)$ where $\mu$ and $\sigma$ are the population mean and the population standard deviation of random variable $D$, respectively; then, each \textit{z}-score, that is, $z_i=(d_i-\mu)/\sigma$, can be computed and the sum of their squares is distributed according to the $\chi^2$-distribution with $n$ degrees of freedom. Because the population mean $\mu$ is usually unknown and estimated by the sample mean $\bar{d}_n$ in reality; the \textit{z}-score is modified and denoted by $z_i=(d_i-\bar{d}_n)/\sigma$. Then, the sum of their squares is distributed following the $\chi^2$-distribution with $n-1$ degrees of freedom expressed as follows:
	\begin{equation}
	\label{Eq:CQCChiSquareDistribution}
		\begin{aligned}
			\sum_{i=1}^{n}{(\frac{d_i-\bar{d}_n}{\sigma})^2} = \frac{(n-1) \cdot {s_n}^2}{\sigma^2} \sim \chi_{n-1}^2
		\end{aligned}
	\end{equation}
	based on which the population standard deviation $\sigma$ of relative error $D$ can be estimated.

	\item Statistically, there is \SI{90}{\percent} of chance that the value of $(n-1) \cdot {s_n}^2 / \sigma^2$ is above $\chi^2_{0.1,n-1}$, the 10th percentile of a $\chi^2$-distribution with $n-1$ degrees of freedom. In other words, there is \SI{90}{\percent} of chance that $\sigma^2$ does not exceed the value of $(n-1) \cdot {s_n}^2 / \chi^2_{0.1,n-1}$. Finally, the upper \SI{90}{\percent} confidence limit for the population standard deviation $\sigma$ of relative error $D$ is determined using the equation below:
	\begin{equation}
	\label{Eq:CQCPopulationStandardDeviation}
		\begin{aligned}
			\sigma_{UCL} = s_n \cdot \sqrt{\frac{n-1}{\chi_{0.1,n-1}^2}}
		\end{aligned}
	\end{equation}
	where $s_n$ is the sample standard deviation of relative errors $\{d_1,\cdots,d_n\}$ calculated using Eq. \ref{Eq:PEPSampleStandardDeviation}. Note that in the Part 58AA, the result of Eq. \ref{Eq:CQCPopulationStandardDeviation} is used to denote the coefficient of variation ($CV$) stated in the goal and symbol $CV_{UCL}$ is adopted instead of $\sigma_{UCL}$. From the authors' perspective, it is more appropriate to denote the result of Eq. \ref{Eq:CQCPopulationStandardDeviation} as standard deviation with symbol $\sigma_{UCL}$.

	\item If upper limit $\sigma_{UCL}$ is less than \SI{10}{\percent}, the intermediate measurement precision, which is under intermediate precision conditions involving change in measuring systems with distinct measuring principles, of the low-cost PM\textsubscript{2.5} sensor or device meets the requirement stated in the goal.
\end{itemize}

Except the evaluation metrics introduced in the Part 58AA that are adopted for assessing the measurement trueness and measurement precision of a low-cost PM\textsubscript{2.5} sensor or device with respect to a collocated reference monitor, multiple evaluation metrics provided in \cite{ecfr2020part53subpartCpoint35}, subsection of the Part 53SC, are adopted as well to test the comparability of the low-cost sensors or devices, categorized as Class III candidate method samplers for PM\textsubscript{2.5}, against the collocated reference monitor (i.e., REFN) in this manuscript. The comparability test needs at least three units of reference method samplers and also at least three units of candidate method samplers are deployed in collocation and operated concurrently at multiple monitoring sites during winter and summer seasons. Limited to what is practical, a portion of these requirements are not fulfilled; in particular, there is only one unit of reference method sampler for PM\textsubscript{2.5} in the field experiments presented in section \ref{SubSect:ExperimentalSetup}. Therefore, assumptions and compromises were made on the adopted formulas and criteria accordingly as presented in the following.
\begin{itemize}
	\item Test the measurement precision, to be specific, the intermediate measurement precision under intermediate precision conditions involving change in measuring systems with identical measuring principle, usually known as the unit-wise precision (or the intra-unit variability), of the candidate method according to the following procedures.
	\begin{itemize}
		\item For each valid PM\textsubscript{2.5} measurement set acquired by $m$ candidate method samplers (e.g., ten WePINs) at time instant $i$, denoted by $\{c_{i,1},\cdots,c_{i,m}\}$, its mean concentration $\bar{c}_i$ is calculated by the equation below:
		\begin{equation}
		\label{Eq:MeanConcentrationTimeInstant}
			\begin{aligned}
				\bar{c}_i = \frac{1}{m} \cdot \sum_{j=1}^{m}{c_{i,j}}
			\end{aligned}
		\end{equation}
		where $c_{i,j}$ is the measurement acquired by the candidate method sampler $j$ at time instant $i$. A PM\textsubscript{2.5} measurement set from candidate method samplers is valid whenever the reference mean concentration $\bar{r}_i$ from multiple reference method samplers is within the concentration range from \SIrange[per-mode=symbol]{3}{200}{\micro\gram\per\cubic\meter}. However, because there is only one unit of reference method sampler in the field experiment conducted in this manuscript, the reference concentration $r_i$ is considered representative, that is, the value of $\bar{r}_i$ in the criterion above is replaced by the value of $r_i$.

		\item Then, calculate the corresponding $CV$ value for each valid measurement set $\{c_{i,1},\cdots,c_{i,m}\}$ utilizing the equation as follows:
		\begin{equation}
		\label{Eq:CVForEachValidMeasurementSet}
			\begin{aligned}
				CV_i & = \frac{s_{c_i}}{\bar{c}_i} \cdot 100 \\
					 & = \frac{100}{\bar{c}_i} \cdot \sqrt{\frac{\sum_{j=1}^{m}{(c_{i,j}-\bar{c}_i)}^2}{m-1}}
			\end{aligned}
		\end{equation}
		where $s_{c_i}$ is the sample standard deviation of the measurement set $\{c_{i,1},\cdots,c_{i,m}\}$. Note that the $CV$ value should be computed only for quantities on ratio scale, that is, scale that has a meaningful zero (e.g., the mass concentration measurements of PM\textsubscript{2.5}).

		\item Eventually, the unit-wise precision of the candidate method, expressed as the root mean square over the $CV$ values of all valid measurement sets acquired, is determined by the following equation:
		\begin{equation}
		\label{Eq:UnitWisePrecision}
			\begin{aligned}
				CV_{RMS} = \sqrt{\frac{1}{n}\sum_{i=1}^{n}{CV_i}^2}
			\end{aligned}
		\end{equation}
		where $n$ is the total number of valid measurement sets acquired by the $m$ candidate method samplers during the test campaign in one test site.

		\item For a PM\textsubscript{2.5} Class III candidate method to pass the precision test, the unit-wise precision, expressed as $CV_{RMS}$, must not be greater than \SI{15}{\percent} for each test site\footnote{An other statement of the quantification of unit-wise precision ($CV={\sum_{i=1}^{n}{CV_i}}/n$) and its acceptable criterion ($\leq\SI{10}{\percent}$) was identified in several existing studies presented in section \ref{Sect:RelatedWorks} and section \ref{SubSect:Instrumentation}. It was originated from studies \cite{sousan2016evaluation,sousan2016inter} that might mistake the corresponding concepts introduced in the \textit{40 CFR Part 58} from the authors' perspective.}. Note that, owing to limited resources, there is only one test site in the field experiment carried out in this manuscript, and therefore, it is considered to be representative as well.
		\item This evaluation metric is not applicable to the nominated PM sensors because there is only one unit of each sensor model in the field experiment presented in section \ref{SubSubSect:EvalAndCaliOfPMSensors}, while multiple units of each model (normally no less than three) are required.
	\end{itemize}

	\item Test the comparative slope and intercept of the candidate method respecting to the reference method based on the procedures in the following.
	\begin{itemize}
		\item First calculate the mean concentration measured by the candidate method and the reference method in a test site using the equations below:
		\begin{equation}
		\label{Eq:MeanConcentrationCandidateAndReference}
			\begin{aligned}
				\bar{c} = \frac{1}{n} \cdot \sum_{i=1}^{n}{\bar{c}_i} \\
				\bar{r} = \frac{1}{n} \cdot \sum_{i=1}^{n}{\bar{r}_i}
			\end{aligned}
		\end{equation}
		where $\bar{c}_i$ and $\bar{r}_i$ are the mean concentrations of the candidate method and the reference method at time instant $i$, respectively. Note that the value of $r_i$ is assumed to be representative for $\bar{r}_i$ within the field experiment having single unit of reference method. Additionally, because there is only one unit of each candidate method in the field experiment presented in section \ref{SubSubSect:EvalAndCaliOfPMSensors}, the value of $c_i$ is as well assumed to be representative for $\bar{c}_i$.

		\item Then the slope and the intercept results of a linear regression between the candidate method measurements $\{\bar{c}_1,\cdots,\bar{c}_n\}$ and the reference method measurements $\{\bar{r}_1,\cdots,\bar{r}_n\}$ are determined by equation:
		\begin{equation}
		\label{Eq:LinearRegressionSlope}
			\begin{aligned}
				slope = \frac{\sum_{i=1}^{n}{(\bar{r}_i-\bar{r})(\bar{c}_i-\bar{c})}}{\sum_{i=1}^{n}{(\bar{r}_i - \bar{r})}^2}
			\end{aligned}
		\end{equation}
		and equation:
		\begin{equation}
		\label{Eq:LinearRegressionIntercept}
			\begin{aligned}
				intercept = \bar{c} - slope \cdot \bar{r}
			\end{aligned}
		\end{equation}
		respectively, where $n$ is the number of valid PM\textsubscript{2.5} measurement sets acquired by the candidate method samplers during the test campaign in one test site.

		\item For a PM\textsubscript{2.5} Class III candidate method to pass the slope test, the result of Eq. \ref{Eq:LinearRegressionSlope} must be in interval \SI[separate-uncertainty=true]{1\pm0.10}{}. In regard to passing the intercept test, the result of Eq. \ref{Eq:LinearRegressionIntercept} must be between \SI[per-mode=symbol]{-2.0}{\micro\gram\per\cubic\meter} or $(15.05 - {slope} \times 17.32) \, \SI[per-mode=symbol]{}{\micro\gram\per\cubic\meter}$, whichever is larger, and \SI[per-mode=symbol]{2.0}{\micro\gram\per\cubic\meter} or $(15.05 - {slope} \times 13.20) \, \SI[per-mode=symbol]{}{\micro\gram\per\cubic\meter}$, whichever is smaller. Similar to the unit-wise precision test on the field experiment carried out in this manuscript, the slope and the intercept results from single test site are considered representative.
	\end{itemize}

	\item Test the correlation between candidate method measurements $\{\bar{c}_1,\cdots,\bar{c}_n\}$ and reference method measurements $\{\bar{r}_1,\cdots,\bar{r}_n\}$ by calculating the Pearson correlation coefficient, denote by $r$, using the equation below:
	\begin{equation}
	\label{Eq:CorrelationCoefficient}
		\begin{aligned}
			r = \frac{\sum_{i=1}^{n}{(\bar{r}_i - \bar{r})(\bar{c}_i - \bar{c})}}{\sqrt{\sum_{i=1}^{n}{(\bar{r}_i - \bar{r})}^2 \cdot \sum_{i=1}^{n}{(\bar{c}_i - \bar{c})}^2}}
		\end{aligned}
	\end{equation}
	where the set $\{\bar{r}_1,\cdots,\bar{r}_n\}$ should be replaced by the set $\{r_1,\cdots,r_n\}$ if only one unit of reference method sampler is available in a specific test site while assuming its measurements are representative. For the field experiment presented in section \ref{SubSubSect:EvalAndCaliOfPMSensors}, because there is single unit of each candidate method sampler, their measurements are considered to be representative and the set $\{\bar{c}_1,\cdots,\bar{c}_n\}$ is replaced by the set $\{c_1,\cdots,c_n\}$. For a PM\textsubscript{2.5} Class III candidate method to pass such a correlation test, the correlation coefficient $r$ in every test site must not be lower than a value between $0.93$ and $0.95$ depending on the PM\textsubscript{2.5} concentration levels during the test campaign.
\end{itemize}

Besides the evaluation metrics introduced in the Part 58AA and the Part 53SC, the following metrics that can be frequently identified in existing studies are adopted as well for assessing the field performances of the nominated PM sensors and the implemented WePINs.
\begin{itemize}
	\item According to the VIM, the term \textbf{detection limit} or \textbf{limit of detection} ($LOD$) is defined as a ``measured quantity value, obtained by a given measurement procedure, for which the probability of falsely claiming the absence of a component in a material is $\beta$, given a probability $\alpha$ of falsely claiming its presence''. It is widely understood as a measure of the inherent detection capability, i.e., the minimum detectable true value of the quantity of interest \cite{currie1999nomenclature}. One should not confuse this term with the ``limit of detection'' (translation of ``Nachweisgrenze'') earlier introduced by H. Kaiser and H. Specker \cite{kaiser1956bewertung}, which is more widely employed in existing studies; the latter was originally intended to serve as a measure of the detection capability but then employed as the ``decision criterion'' to distinguish an estimated signal from the background noise, that is, the minimum significant estimated value of the quantity of interest \cite{currie1999nomenclature}. Conveniently, the $LOD$ is calculated by the formula below:
	\begin{equation}
	\label{Eq:LODFormula}
		\begin{aligned}
			LOD = k \cdot s_{blk}
		\end{aligned}
	\end{equation}
	where $s_{blk}$ is the sample standard deviation of the blank sample or the sample containing a low concentration of the analyte (e.g., PM\textsubscript{2.5}), and $k$ is a multiplying factor based on statistical reasoning. The International Union of Pure and Applied Chemistry (IUPAC) recommends $\alpha=\beta=0.05$ as default values. In this regard, under several assumptions and approximations that are listed in \cite{currie1999nomenclature}, assumption of normally distributed sample in particular, the factor $k=3.30$ when $n\to\infty$ where $n$ is the sample size of the blank or low concentration sample; documents \cite{barwick2011terminology} and \cite{currie1999nomenclature} may be consulted for additional details. To assess the $LOD$ of a low-cost PM sensor or device, the blank sample is usually achieved by challenging the calibrated sensor/device to air cleaned by high-efficiency particulate air (HEPA) filters in laboratory, while the low concentration sample acquired in the field is composed of those sensor/device measurements having corresponding collocated reference measurements below a certain value, e.g., $<\SI[per-mode=symbol]{1}{\micro\gram\per\cubic\meter}$ of reference PM\textsubscript{2.5} as presented in study \cite{kelly2017ambient}. In this manuscript, the low concentration sample of a PM sensor or device contains its calibrated PM\textsubscript{2.5} mass concentrations if the corresponding measurements from the collocated reference instrument, that is, the REFN, is $<\SI[per-mode=symbol]{3}{\micro\gram\per\cubic\meter}$ at low RH conditions ($\leq\SI{50}{\percent}$)\footnote{In order to have sufficient numbers of low concentration pairs, the RH condition was changed to $\leq\SI{80}{\percent}$ for the first field experiment. No low concentration pairs were available in the second field experiment.}.

	\item Data completeness, denoted by $\eta$, is the percentage of the amount of valid data obtained from a measuring system compared to the amount that was expected to be obtained under correct and normal conditions.
\end{itemize}



\section{System Implementation}
\label{Sect:SystemImplementation}

\begin{figure*}[!t]
\centering
	\subfloat[]{
		\includegraphics[width=0.54\linewidth]{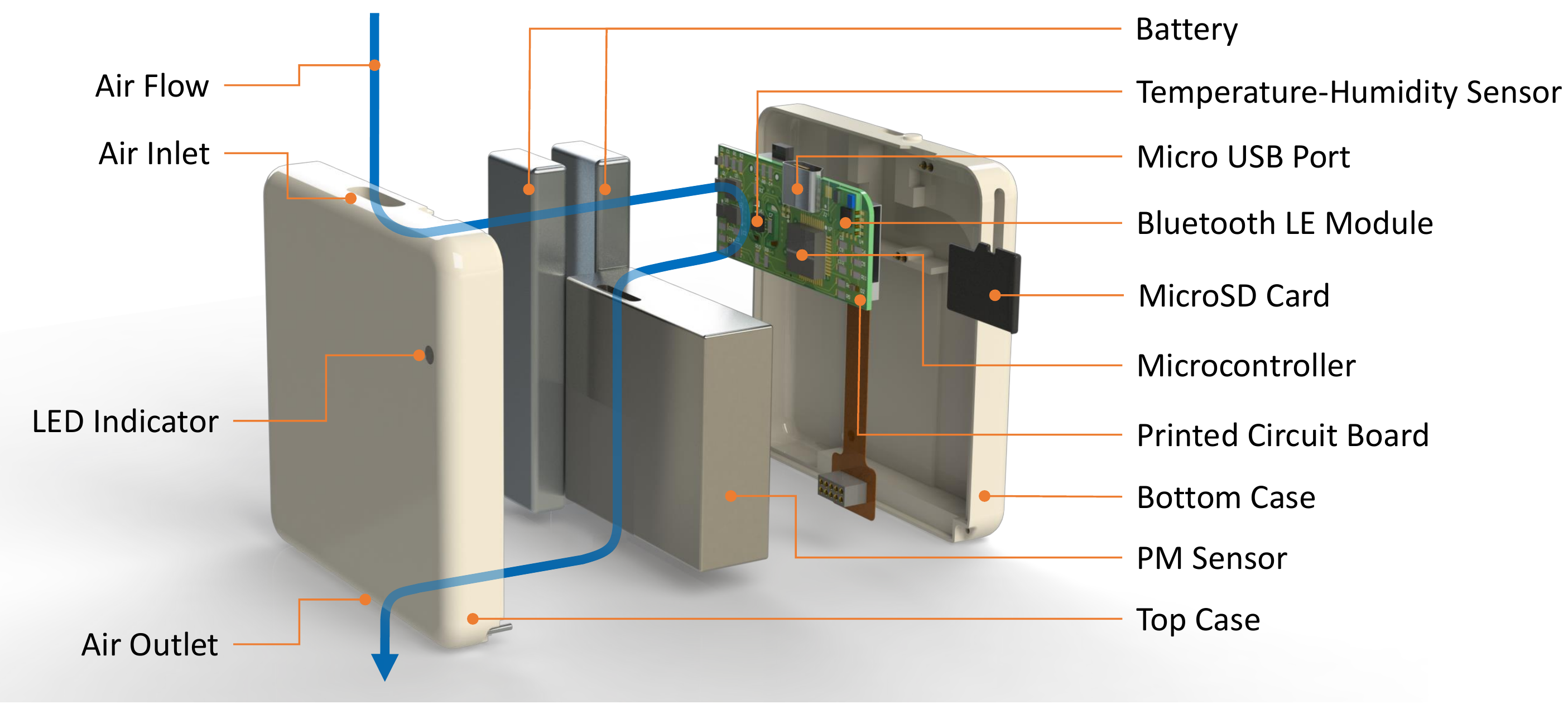}
		\label{subfig:PMDevice}}
	\subfloat[]{
		\includegraphics[width=0.23\linewidth]{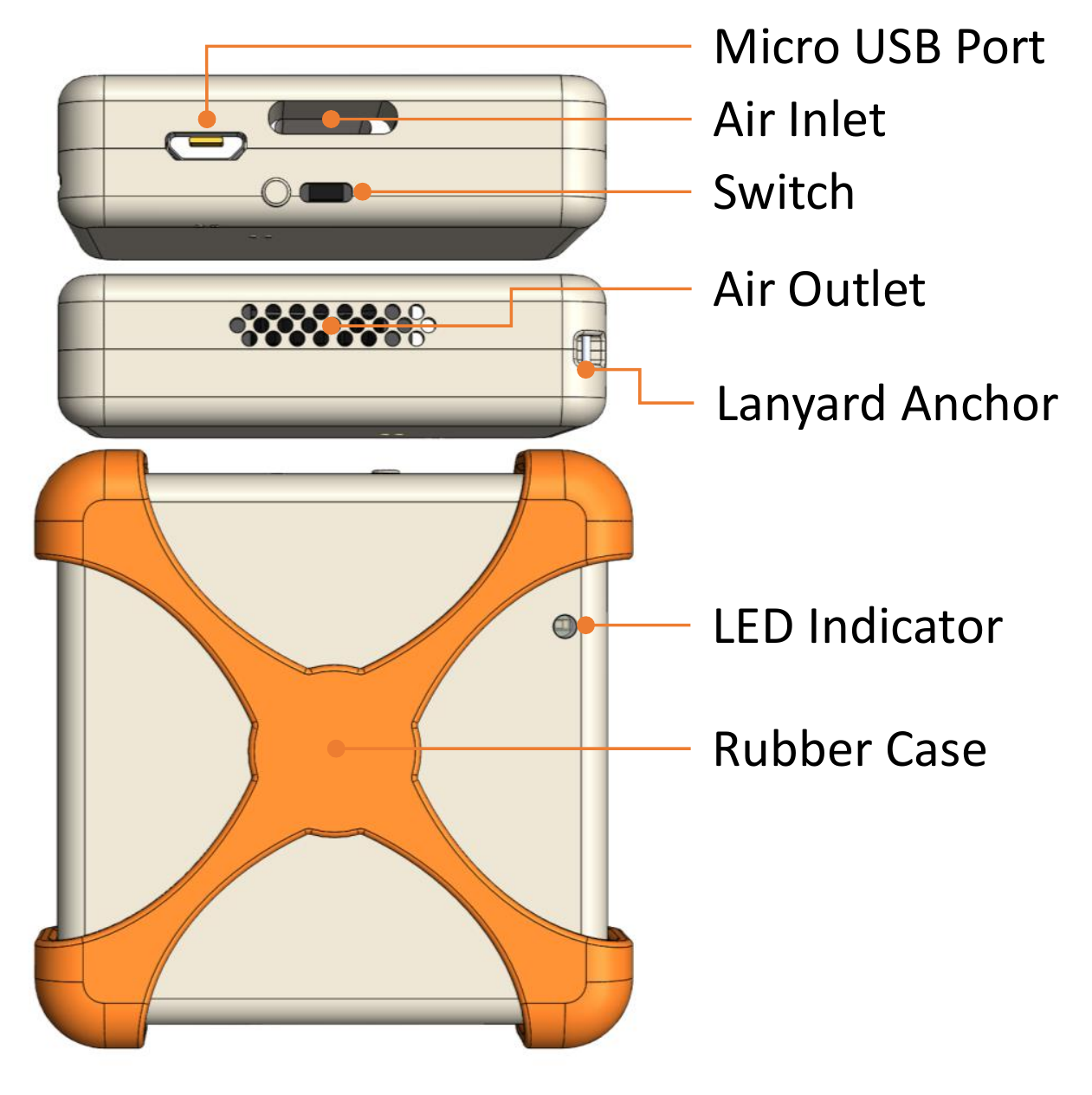}
		\label{subfig:PMDeviceViews}}
	\hfill
	\subfloat[]{
		\includegraphics[width=0.47\linewidth]{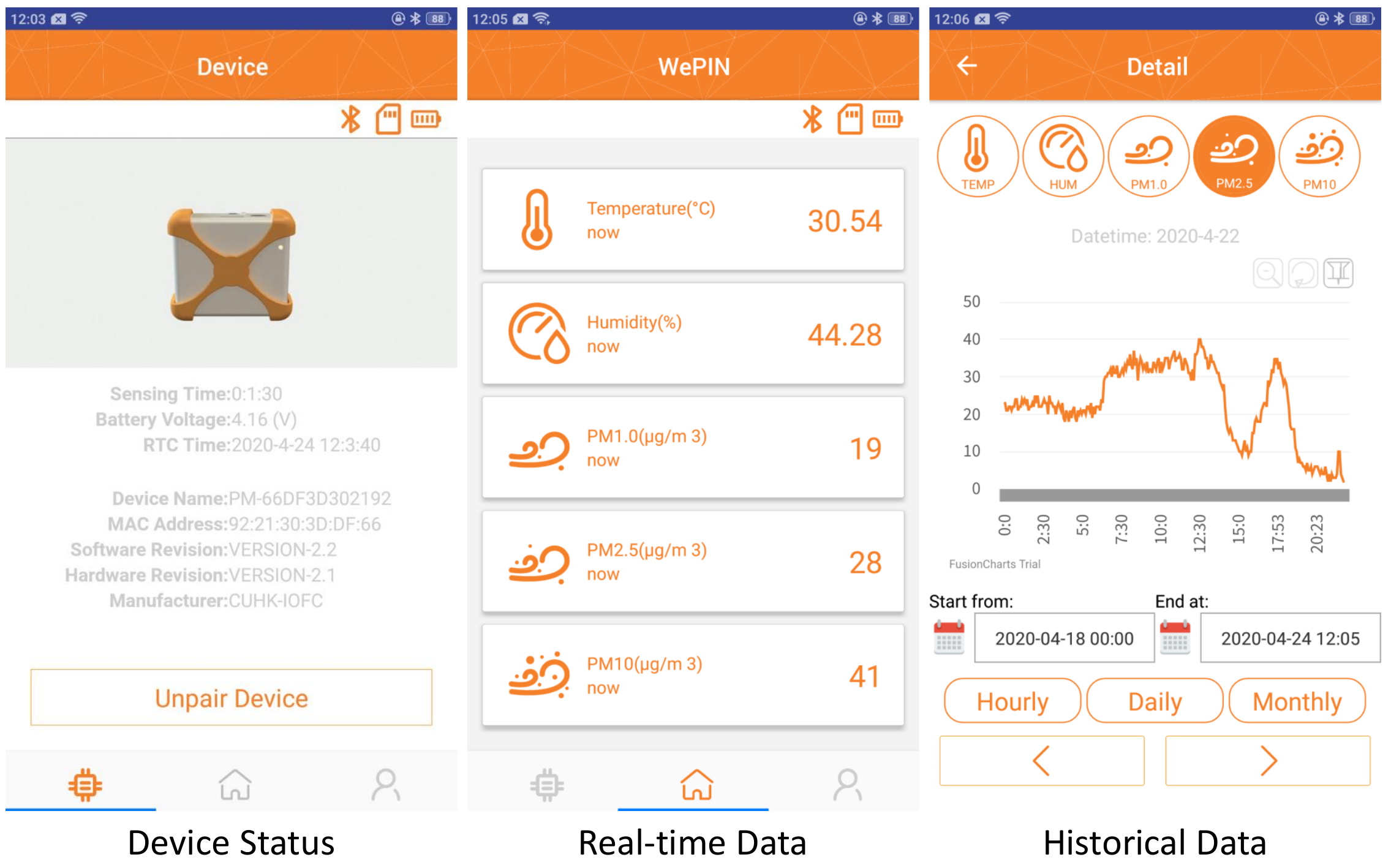}
		\label{subfig:PMApp}}
	\subfloat[]{
		\includegraphics[width=0.31\linewidth]{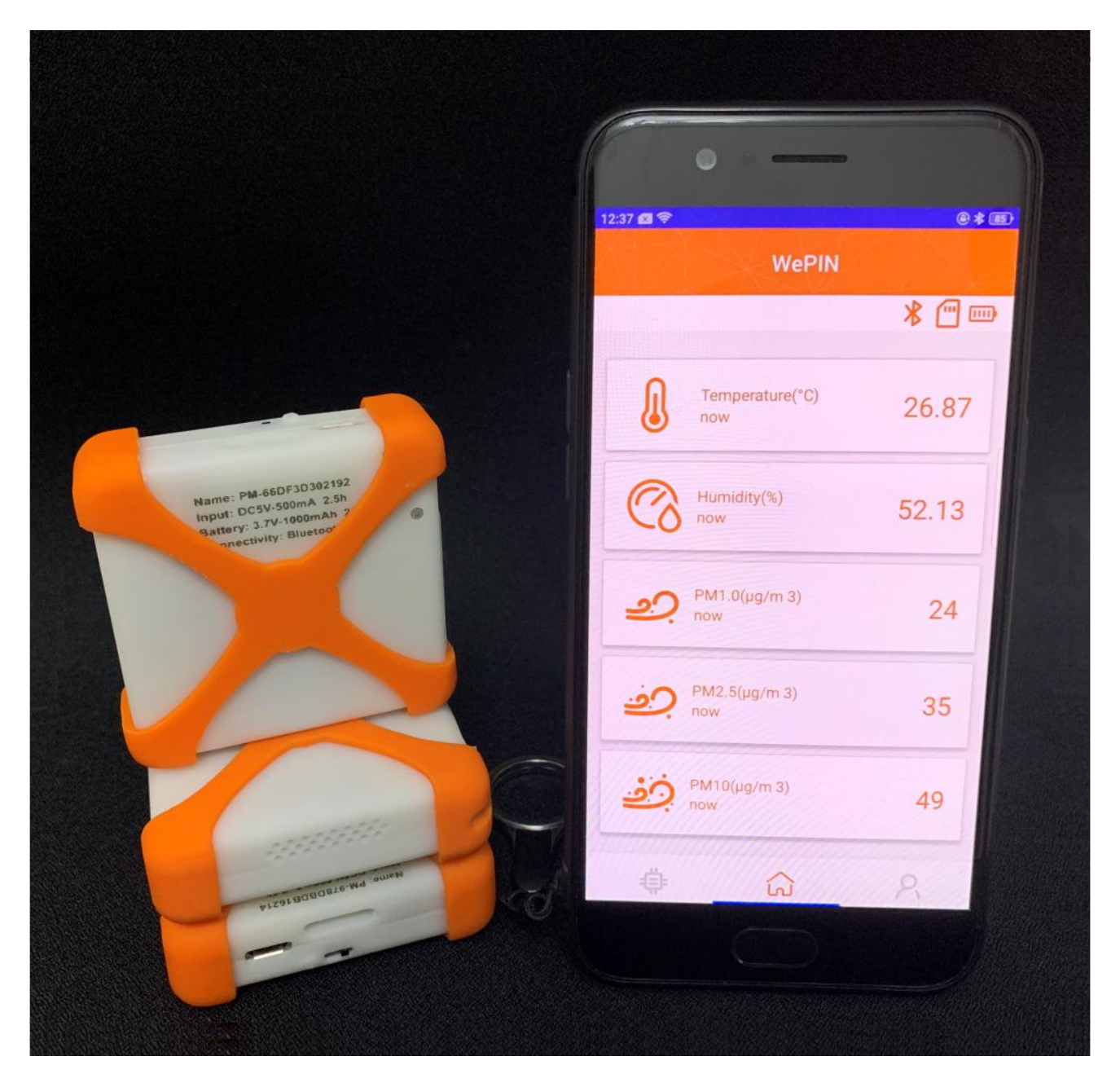}
		\label{subfig:PMDeviceApp}}
\caption{Illustration of the implemented WePIN and mobile application: (a) exploded drawing of a WePIN; (b) multiview (front, back, and top views) drawings of a WePIN; (c) screenshots of the mobile application; and (d) photo of three WePINs and the mobile application running on an Android phone.}
\label{fig:SystemImplementationDeviceApp}
\end{figure*}

Bearing the objectives (listed in section \ref{SubSect:WhyDevelopment}) of a personal pollution exposure monitoring system in mind, the PMSA was selected for constructing the wearable monitoring device, which is referred to as Wearable Particle Information Node (WePIN), of the proposed personal PM exposure monitoring system because it demonstrated a decent overall performance in accordance with the evaluation results detailed in section \ref{SubSect:EvaluationResultsOfPMSensors} and it has the minimum size and weight and the lowest cost compared to the other candidates regarding the technical specifications presented in Tab. \ref{tab:SensorSpec}. A mobile application for data visualization and relay, and a corresponding back-end server for data management and analytics were implemented as well. Initially, 10 WePINs were assembled and evaluated in the field; results demonstrated that the WePIN generally met the intended requirements listed in section \ref{SubSect:WhyDevelopment}. Subsequently, 120 WePINs with a few slight modifications that covered the earlier feedbacks were manufactured. The latest iteration of the proposed system as illustrated in Fig. \ref{fig:SystemImplementationDeviceApp} is introduced in the following. The system architecture of a WePIN as well as its hardware and software configurations are first introduced in section \ref{SubSect:SysImpWePIN}; comparisons between the WePIN and several recognized wearable devices were also performed. Then, the system architecture and detailed process flows of the mobile application and the back-end server are presented in section \ref{SubSect:SysImpMobileAppAndServer}. Finally, the limitation of the proposed system at current stage and the future work are discussed in section \ref{SubSect:SysImpFutureWork}.

\subsection{Wearable Particle Information Node (WePIN)}
\label{SubSect:SysImpWePIN}

\begin{figure}[!t]
\centering
	\includegraphics[width=0.84\columnwidth]{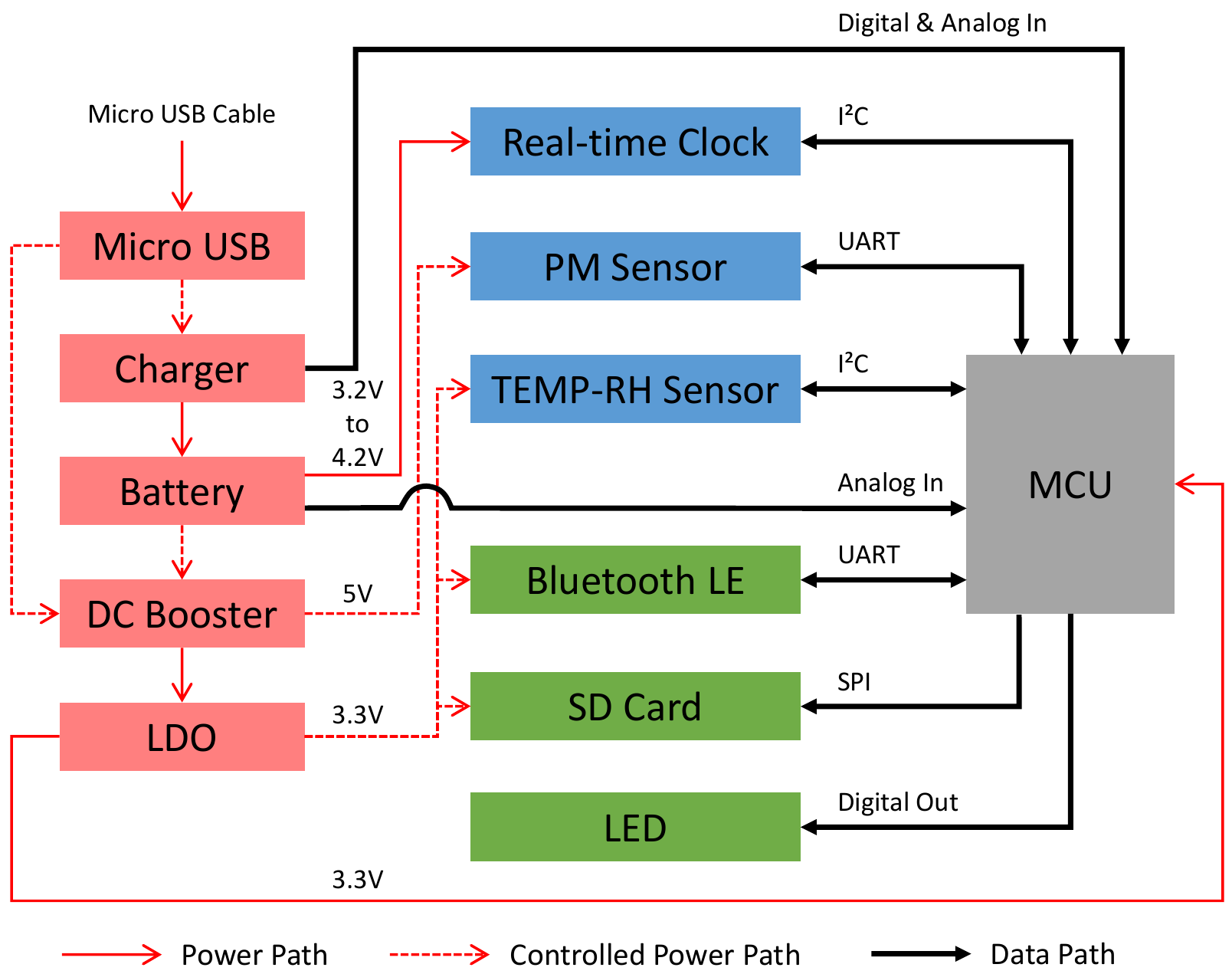}
\caption{System architecture diagram of the WePIN.}
\label{fig:WePINArchi}
\end{figure}

The system architecture of the WePIN is presented in Fig. \ref{fig:WePINArchi}. The core is an ultra-low-power Arm Cortex-M0+ microcontroller unit (MCU) running at \SI{32}{\mega\hertz} with \SI{64}{\kilo\byte} Flash and \SI{8}{\kilo\byte} RAM. The temperature-humidity (TEMP-RH) sensor measures the ambient TEMP and RH (nominal accuracies are $\pm\SI{0.1}{\celsius}$ and $\pm\SI{1.5}{\percent}$, respectively \cite{sensirion2021sht35}) for adjusting the PM sensor's measurements, whilst the real-time clock (RTC) is timestamping the data from both sensors and clocking the firmware. The MCU communicates with the RTC, the TEMP-RH sensor, and the PM sensor to acquire respective data via corresponding ports. Packed data are stored in the microSD card (up to \SI{32}{\giga\byte}) and passed to the Bluetooth low energy (BLE) module by the MCU through specific ports. The BLE module is responsible for transmitting data to and receiving commands from the mobile application. A LED is utilized for indication purpose. Battery (two paralleled with total capacity of \SI{1000}{\milli\ampere{}\hour}) information, charging status, and microSD card status are available to the MCU via various ports. Note that the MCU controls the power supplies of the PM sensor, the TEMP-RH sensor, the BLE module, and the microSD card, while the RTC is continuously powered by the batteries and the MCU is powered by a low-dropout (LDO) regulator after switching on. In addition, a DC-to-DC boost converter is used to step up the batteries' voltage output to \SI{5}{\volt} as required by the PM sensor.

The WePIN has three operating modes, namely the charge mode, the sleep mode, and the run mode. It enters the charge mode and charges the batteries once the micro USB cable is plugged in; all components except the RTC are then powered by the micro USB port. If it has been switched on and the BLE connection is established, the mobile application will first transmit a command to synchronize the date and time between the RTC and the smartphone. Then, information of battery voltage, charging current and status, and microSD card status are tagged with the current date and time acquired from the RTC and transmitted to the mobile application every second. The WePIN enters the sleep mode when battery is low (e.g., $\leq\SI{3.2}{\volt}$); power supplies of all peripherals except the RTC are shut off by the MCU which afterwards enters low-power mode resulting in a total current consumption $<\SI{1}{\milli\ampere}$ at \SI{3.7}{\volt} supply. It exits the sleep mode when the micro USB cable is plugged in and requires $\approx\SI{2.5}{\hour}$ to fully charge the batteries. Besides the charge and the sleep modes, the WePIN is operating in the run mode where data are acquired from the sensors, stored in the microSD card, and transmitted to the mobile application. At the beginning of each minute, the current date and time, the TEMP and RH measurements are acquired and temporally stored in the MCU's RAM; the PM sensor is switched on and set to passive configuration. The MCU is then running in low-power mode and will wake up after \SI{30}{\second} to acquire the ``atmospheric” ($CF=atmos$) PM\textsubscript{1.0}, PM\textsubscript{2.5}, and PM\textsubscript{10} mass concentrations from the PM sensor, which are further appended with the date and time, TEMP and RH measurements from the RAM before storing in the microSD card and transmitting to the mobile application. The MCU turns off the PM sensor and re-enters low-power mode afterwards until the start of next minute. Such a mechanism optimizes the overall power consumption (averaged current consumption $\approx\SI{35}{\milli\ampere}$ at \SI{3.7}{\volt} supply) and the performances of the PM sensor and the TEMP-RH sensor because the PM sensor's micro-fan needs $\approx\SI{30}{\second}$ to stabilize the airflow after turning on and powering the PM sensor heats up the circuit board that might bias the TEMP-RH sensor's measurements. In the run mode, the battery information and the microSD card status tagged with corresponding date and time are also transmitted to the mobile application every second with the MCU awake shortly. 

As illustrated in Fig. \ref{subfig:PMDevice} and Fig. \ref{subfig:PMDeviceViews}, the latest iteration of a WePIN accommodates all components in a 3D printed rigid case wrapped by a molded rubber case which protects it against impacts. The total cost of one WePIN is $\approx\SI{60}[US\$]{}$. Without the rubber case and lanyard, a WePIN weights \SI{60}{\gram} and has dimensions of \SI{58x58x16}{\milli\meter}. Estimated by performing five charge-discharge cycles on ten devices, a WePIN with 1-minute sampling interval has $\approx\SI{24}{\hour}$ battery lifetime\footnote{The mean and standard deviation values are \SI{24.45}{} and \SI{0.55}{}, respectively.} during which a data log file with size $<\SI{150}{\kilo\byte}$ is generated in the microSD card. Compared to the wearable PM monitoring devices recognized by the authors, i.e., the Flow2, the Atmotube PRO, and the MyPart presented in Tab. \ref{tab:CompareCommercialPMDevices} and Tab. \ref{tab:CompareDedicatedPMDevices}, the WePIN significantly outperforms the Flow2 and the Atmotube PRO in terms of affordability ($>\SI{62}{\percent}$ cheaper) and portability ($>\SI{43}{\percent}$ smaller and $>\SI{14}{\percent}$ lighter) while all of them have similar usability (automatic data recording and upload) and data accessibility (real-time and historical data visualization on mobile application). Although the wrist-worn MyPart is a bit cheaper ($\approx\SI{17}{\percent}$) and smaller ($\approx\SI{6}{\percent}$) compared to a WePIN, it is not reporting the mass concentrations but the counts of the small and the large sized particles; their exact size ranges require further investigations. Regarding the data reliability, comprehensive collocated evaluations on these devices should be conducted before drawing any conclusions. Nevertheless, according to the field evaluation results of the PMSA and the WePINs presented in section \ref{SubSect:EvaluationResultsOfPMSensors} and section \ref{SubSect:EvaluationResultsOfWePINs}, respectively, and the experimental results from existing studies evaluating the PMSA in the field and/or laboratory detailed in section \ref{SubSect:SensorEvaluation} and section \ref{SubSubSect:PMSAIntro}, PM measurements with adequate data quality are achievable given that the PM sensor or the PM monitoring device is properly calibrated and correctly RH adjusted. Recall that it is neither critical nor possible for a low-cost and compact PM sensor or monitoring device to meet the same performances as the research-grade instruments or the conventional monitors.

\subsection{Mobile Application and Back-end Server}
\label{SubSect:SysImpMobileAppAndServer}

The idea of improving the usability and data accessibility of the novel wearable monitoring device has been taken into sufficient consideration throughout the development process of the mobile application and the back-end server. The mobile application was built upon a cross-platform framework named React Native \cite{facebook2020react} which allows the same pieces of code to run on both Android and iOS devices. For the back-end server, all services were developed upon the Spring platform \cite{vmware2020spring} and deployed on a cloud cluster hosting two Tencent servers, making the development more convenient and improving the system robustness. For rapid prototyping, the back-end server and the mobile application are communicating via hypertext transfer protocol (HTTP), which will be switched to hypertext transfer protocol secure (HTTPS) for security reasons.

\begin{figure}[!t]
\centering
	\includegraphics[width=0.84\columnwidth]{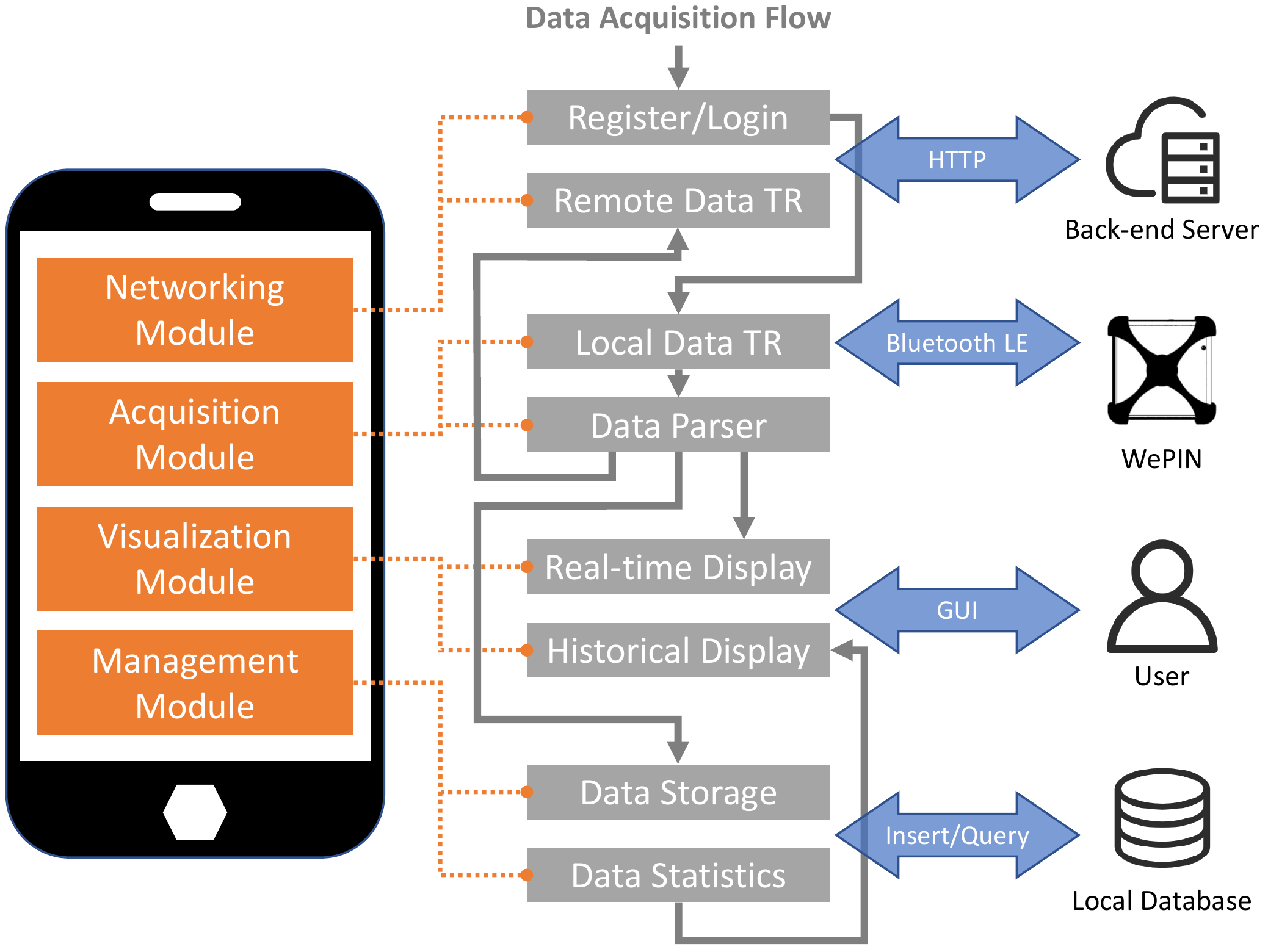}
\caption{System architecture diagram and data acquisition flowchart of the mobile application.}
\label{fig:AppArchi}
\end{figure}

The architecture of the mobile application is illustrated in Fig. \ref{fig:AppArchi}. In the networking module, the register/login submodule provides a register and login interface to the user. To register, user is asked to input a valid e-mail address, a password, and the verification code received from the back-end server. Login authentication is required for the user to proceed. Afterwards, the local data transmission and reception (TR) submodule is ready to establish BLE connection with a WePIN. Instead of manually selecting a specific WePIN, this submodule automatically connects to the device with maximum received signal strength indication (RSSI) above a predetermined threshold, that is, the closest device\footnote{The user only needs to place a WePIN as close to the smartphone as possible after clicking the ``Pair Device'' button on the mobile application. This button will be replaced by the ``Unpair Device'' button as shown in the first screenshot of Fig. \ref{subfig:PMApp} if connected. Connection or reconnection will be automatically issued whenever the paired WePIN is connectable.}. It records the BLE slave's media access control (MAC) address as well to automatically issue connection/reconnection. Whenever a WePIN is switched to the charge mode with BLE established, this submodule will transmit an encoded command to synchronize the date and time on the RTC with respect to that on the smartphone. Data packages received from a WePIN are decoded by the data parser submodule to retrieve respective information, including charging current and status, battery voltage, microSD card status, RTC date and time, TEMP and RH readings, and PM measurements. The real-time display submodule then renders the graphical user interface (GUI) for visualizing the retrieved information as illustrated in the first and second screenshots\footnote{The three icons on the upper right corner indicate the BLE status (gray if disconnected), the microSD card status (gray if not inserted), and the battery level, respectively.} of Fig. \ref{subfig:PMApp}. Concurrently, the RTC date and time, TEMP and RH readings, and PM measurements, which have been tagged with the MAC address of the BLE slave and the geographic location from the smartphone, are transmitted to the back-end server by the remote data TR submodule and inserted into the local Realm database \cite{realm2020database} by the data storage submodule. Clicking any taps in the second screenshot of Fig. \ref{subfig:PMApp} will activate the historical display submodule to render the GUI for visualizing the raw or processed data, which are generated by the data statistics module via querying and averaging (hourly, daily, or monthly) the data from local database, as indicated in the third screenshot of Fig. \ref{subfig:PMApp}. A photo of three WePINs and the mobile application running on an Android phone is illustrated in Fig. \ref{subfig:PMDeviceApp}.

\begin{figure}[!t]
\centering
	\includegraphics[width=0.72\columnwidth]{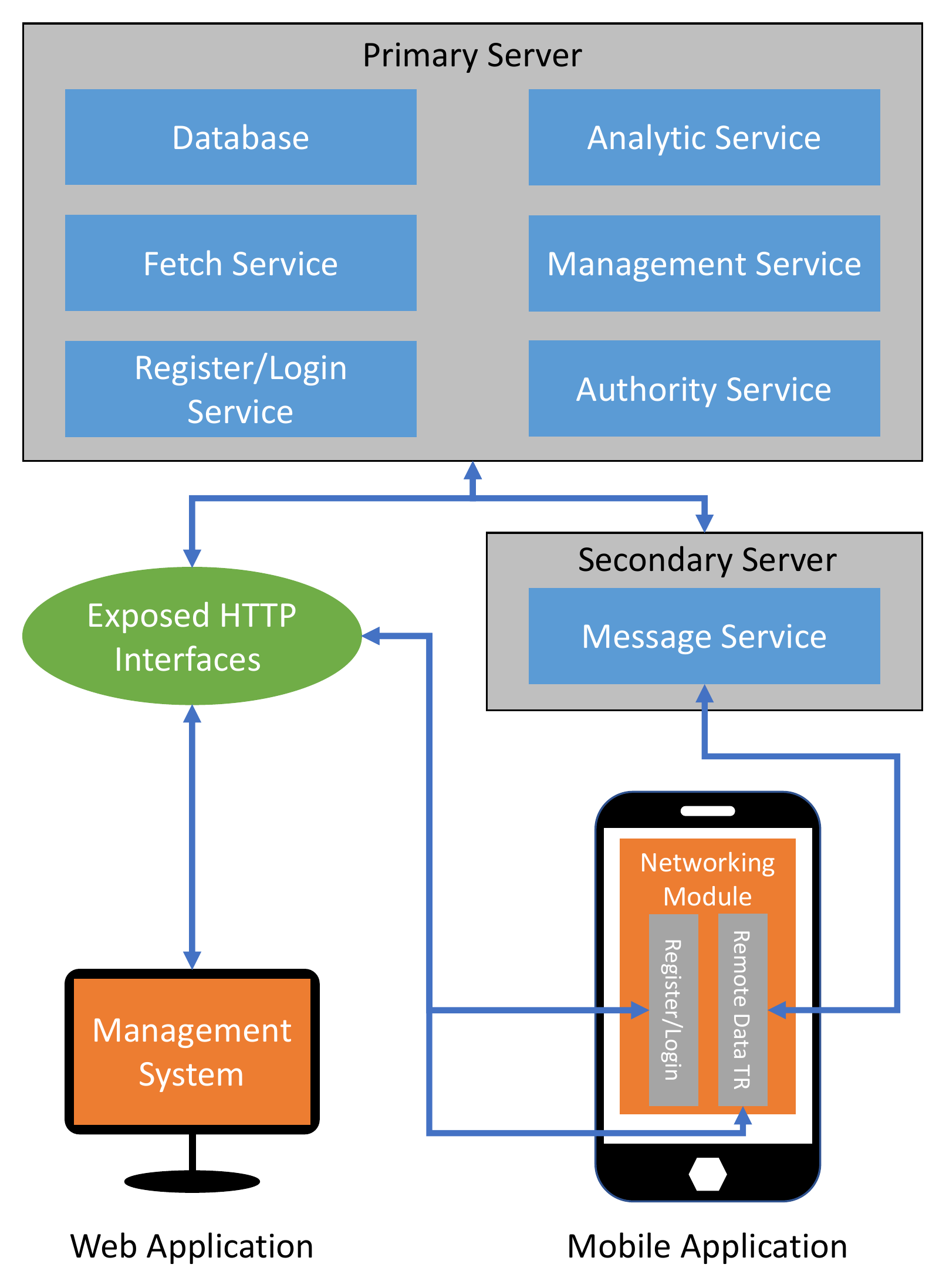}
\caption{System architecture diagram of the back-end server.}
\label{fig:ServerArchi}
\end{figure}

The architecture of the back-end server is shown in Fig. \ref{fig:ServerArchi}; most of the back-end services (accessible via exposed HTTP interfaces) and the database are deployed in the primary server while the message service is hosted in the secondary server. This mechanism significantly relaxes the network traffic on the primary server because the data streaming from the mobile applications (i.e., RTC dates and times, TEMP and RH readings, and PM measurements tagged with the MAC addresses of WePINs and the geographic locations from smartphones) are handled by the message service which was built upon the Apache Kafka \cite{apache2017kafka} high-performance distributed streaming platform. Thanks to the distributed and scalable nature of this platform, the back-end server can be easily scaled up to handle more streaming traffic by deploying more message services in additional servers. Enabled by the message service, data produced by the mobile application (``producer'') are first buffered in the secondary server and then consumed by the primary server (``consumer''), e.g., storing them into the MySQL database. The roles of a producer and a consumer are interchangeable enabling the primary server to deliver timely feedbacks, such as early health warnings automatically generated by the analytic service or manually issued via the management service, to the mobile application. In spite of the mobile application, a web-based management system (i.e., a web application) was also developed for administrators, the sole authority that can access the web application, to manage and visualize the data acquired from all users utilizing the management service; this service also forwards the feedbacks issued by an administrator to the message service to notify the users. To interact with the mobile or the web application, a successful login is required. The register/login service handles the corresponding requests from the mobile and web applications; it also guarantees that there is only one concurrent login per account. Although both mobile and web applications can use the fetch service for data download and the analytic service for data analysis via the exposed HTTP interfaces after login, different permissions are granted to a user and an administrator by the authority service; while users are restricted to the acquired data and analytical results of their own, an administrator has unlimited access.

\subsection{Future Work}
\label{SubSect:SysImpFutureWork}

Limited to what is practical, the analytic service in the back-end server is currently providing simple statistical results and its development is left for future work when sufficient amount of data such as PM exposures and subclinical atherosclerosis assessments are available. Even though the proposed personal PM exposure monitoring system aims at assessing the daily PM\textsubscript{2.5} exposures of users in personal level for investigation of specific pollution-health associations, it is also potentially a participatory or community sensing system in which users with personal mobile devices (i.e., WePINs and smartphones) collaboratively collect information (i.e., PM\textsubscript{2.5} concentrations) at different locations and contribute it for a joint cause (e.g., building pollution map with high spatio-temporal resolution). In order to further encourage public engagement, the idea of gameful environmental sensing introduced in \cite{budde2016sensified} is highly recognized by the authors and will be adopted in the future iterations. Additionally, all materials, both hardware and software, of the proposed monitoring system will be open source to improve the system cost-efficiency and sustainability. Open data APIs will also be provided such that other third parties can facilitate more innovations and applications based on the acquired data.

\section{Results and Discussion}
\label{Sect:ResultsAndDiscussion}

The collocated field evaluation and calibration results of the four nominated PM sensors and the ten implemented WePINs are presented in section \ref{SubSect:EvaluationResultsOfPMSensors} and section \ref{SubSect:EvaluationResultsOfWePINs}, respectively. In section \ref{SubSubSect:DataPreprocessingOfPMSensors}, preprocessing on the PM\textsubscript{2.5} measurements from the PM sensors and the reference instruments was carried out to prepare the collocated PM\textsubscript{2.5} data pairs. Data cleansing was then performed in section \ref{SubSubSect:DataCleansingOfPMSensors} in order to remove the PM\textsubscript{2.5} measurements that were considered to be affected by unusual occurrences. The pairwise correlations among the PM sensors and reference instruments as well as the collocated calibration of the PM sensors applying different parametric models were investigated in section \ref{SubSubSect:CorrelationAndCalibrationOfPMSensors}. After that, the residuals between the calibrated sensor measurements and the reference outputs were analyzed in section \ref{SubSubSect:ResidualAnalysisOfPMSensors}. Lastly, the evaluation metrics provided in section \ref{SubSubSect:EvaluationMetrics} were calculated from the calibrated PM\textsubscript{2.5} data pairs and discussed in section \ref{SubSubSect:EvaluationAndDiscussionOfPMSensors}; the limitations of the results or conclusions derived from the collocated field experiment were also discussed. Following a similar process, the field performances of the ten implemented WePINs were investigated from section \ref{SubSubSect:DataPreprocessingOfPWePINs} to section \ref{SubSubSect:EvaluationAndDiscussionOfWePINs}. Additionally, a simple yet effective approach that can automatically identify and eliminate the PM sensors' unusual measurements (that is, overestimation due to detecting the condensed tiny droplets as particles) in real time was proposed in section \ref{SubSubSect:AutoDataCleansingOfPMSensors}.

\subsection{Evaluation and Calibration of PM Sensors}
\label{SubSect:EvaluationResultsOfPMSensors}

The first evaluation campaign was conducted from 13 to 23 February 2018 in order to select the most suitable (in terms of affordability, portability, and performance) PM sensor for constructing the wearable monitoring device (i.e., the WePIN) in the proposed personal PM exposure monitoring system. To investigate the field performances of the four nominated PM sensors, namely the DN7C, PMSA, HPMA, and OPCN, they were compared against the collocated DUST and REFN; the DUST was also compared against the REFN. The details of the experimental setup are presented in section \ref{SubSubSect:EvalAndCaliOfPMSensors}. In section \ref{SubSubSect:DataPreprocessingOfPMSensors}, the PM\textsubscript{2.5} measurements from these nominated sensors were first averaged into 1-minute interval. Then, as presented in section \ref{SubSubSect:DataCleansingOfPMSensors}, three intervals of severe overestimation and fluctuations, which were speculated to be caused by detecting the condensed tiny droplets as particles, in the nominated PM sensors' measurements during substantially humid conditions were identified; they were manually eliminated for the sake of deriving effective RH correction factors and/or representative calibration parameters sets. Besides, dual linear relationships between the PM sensors plus the DUST's measurements and the REFN's concentrations, which were likely attributed to the change of aerosol properties since/on 17 February, were also identified; the collocated pairs before that date were manually eliminated as well when using the REFN as reference.

Afterwards, the pairwise correlations among the PM sensors and reference monitors as well as the collocated calibration of the PM sensors and DUST with different parametric models were examined in section \ref{SubSubSect:CorrelationAndCalibrationOfPMSensors}. The results suggested that the DN7C, the OPCN, the DUST, and the HPMA and PMSA had increasing correlations against the reference monitor(s). They also demonstrated an ascending trend in their insusceptibility to the hygroscopic growth of particles whilst the DN7C was as well susceptible to the TEMP; such atmospheric influences can be reduced by involving RH or TEMP in the MLR model. Their residuals from respective OLS and MLR models using the DUST and the REFN as reference were further analyzed in section \ref{SubSubSect:ResidualAnalysisOfPMSensors}. Those residuals associated with the DN7C, the OPCN, the DUST, and the HPMA and PMSA exhibited improved statistics than the former one; so did their model-wise residuals between the OLS and the MLR model and their reference-wise residuals between the REFN and the DUST. As evidenced by the analytical results, the hygroscopic growth of particles was also occurred in mildly humid conditions, and it had distinct influences on the respective PM\textsubscript{2.5} measurements compared to that in extremely humid conditions; involving RH in the MLR model typically improved the residuals' statistics but over adjusted the measurements at low concentrations and mildly humid conditions. Different techniques can be applied to further improve the residuals' statistical performance. The applicable evaluation metrics, presented in section \ref{SubSubSect:EvaluationMetrics}, for each PM sensor were calculated from the respective calibrated and reference concentration pairs in section \ref{SubSubSect:EvaluationAndDiscussionOfPMSensors}. The PMSA and the HPMA, having comparable or better evaluation results against the DUST, were considered to be the best candidates for constructing the wearable device; the PMSA was selected eventually since it typically outperformed the HPMA in terms of affordability, portability, detection capability, field performances, etc. The limitations of the evaluation and calibration results derived from this field experiment were also discussed in section \ref{SubSubSect:EvaluationAndDiscussionOfPMSensors}.

Finally, a simple yet effective approach based on the PM\textsubscript{10} and PM\textsubscript{1.0} mass concentration ratio is proposed and evaluated in section \ref{SubSubSect:AutoDataCleansingOfPMSensors} to automatically identify and eliminate the OPCN's and the PMSA's unusual measurements, which were probably attributed to detecting the condensed tiny droplets as particles during extremely humid conditions, in real time.

\subsubsection{Data Preprocessing}
\label{SubSubSect:DataPreprocessingOfPMSensors}

Considering that the reference instruments (i.e., DUST and REFN) have a minimum sampling interval of \SI{1}{\minute} and the PM sensors' PM\textsubscript{2.5} measurements (ADC readings from the DN7C and mass concentrations from the others) with \SI{5}{\second} sampling interval fluctuated significantly (the readings from the DN7C in particular), the PM sensors' readings retrieved from the microSD card were averaged into 1-minute interval before further analyses. For example, the 12 readings of a PM sensor sampled from 13:30:00 to 13:30:55 will be averaged into one single reading with timestamp as 13:30:00. In order to mitigate the interference from human interactions, the first and last hours of readings from the PM sensors and the DUST were trimmed resulting in the exact start and end times of this campaign as 2018-02-13 16:00:00 and 2018-02-23 13:59:00, respectively.

The data completenesses of the reference instruments were $\eta=\SI{100}{\percent}$, while only several corrupted data were identified from the PM sensors' readings. Although the corrupted data might be caused by the sensor itself and/or the dedicated data acquisition and logging system, it is acceptable to claim that the data completenesses of these PM sensors were $\eta>\SI{99.9}{\percent}$. Each corrupted sensor reading was replaced with the last valid one before averaging into 1-minute interval. The time series plots of these 1-minute interval PM\textsubscript{2.5} measurements from the PM sensors and the reference instruments together with the corresponding RH and TEMP readings are shown in Fig. \ref{fig:PM2p5RhTempFull}.

\subsubsection{Data Cleansing}
\label{SubSubSect:DataCleansingOfPMSensors}

As illustrated in Fig. \ref{fig:PM2p5RhTempFull}, in the nominated PM sensors' measurements, there were three intervals of significant overestimation and fluctuations (highlighted in salmon spans) accompanied by substantially high RH values ($>\SI{82}{\percent}$) and relatively low PM\textsubscript{2.5} concentrations from the REFN. In accordance with the zoomed views of these three intervals as depicted in Fig. \ref{fig:PM2p5RhZoom1}, Fig. \ref{fig:PM2p5RhZoom2}, and Fig. \ref{fig:PM2p5RhZoom3}, this kind of overestimation and fluctuation (hereinafter referred to as abnormal spike) was more pronounced in the measurements from the DN7C, the HPMA, and the OPCN compared with that from the PMSA and the DUST. Notice that a significant drop in the REFN's PM\textsubscript{2.5} mass concentrations can always be found along with the first abnormal spike in each interval.

Given the fact that neither the PM sensors nor the DUST equips with any heater or dryer, but the REFN uses a dynamic heating system to adjust the sampled aerosol's meteorological conditions (especially the RH), and there has been frequently reported in the literature that atmospheric RH can cause noticeable artefacts in a PM sensor's measurements, the authors believe that the abnormal spikes were likely attributed to the fairly high RH conditions ($>\SI{82}{\percent}$). However, the large RH values did not always imply the occurrence of the abnormal spikes; there were PM\textsubscript{2.5} measurements from the nominated sensors that were acquired under even higher RH conditions ($\approx\SI{90}{\percent}$) but demonstrated no abnormal spike. Referring to the causal mechanism underlying the RH influences on a PM sensor's measurements, which is presented in section \ref{SubSubSect:UncertaintyOfInstruments}, the overestimation can be mainly attributed to:
\begin{itemize}
	\item the hygroscopic growth of particles when the RH exceeds their deliquescent points; and
	\item the condensed tiny droplets (remained suspended in the air forming mist or fog) that are detected as particles.
\end{itemize}
The authors speculate that the abnormal spikes illustrated in Fig. \ref{fig:PM2p5RhTempFull} were more likely arose from the latter factor because the hygroscopic growth of particles does not change abruptly and manifests a hysteresis effect, while the presence or not of tiny liquid droplets can lead to significant differences in a PM sensor's measurements \cite{jayaratne2018influence}. Furthermore, during the morning of 17 February, fog and drizzle were reported at times \cite{timeanddate2018weather} though the RH did not reach \SI{100}{\percent}\footnote{The presence of fog at RH below \SI{100}{\percent} might due to the limitation of instruments \cite{jayaratne2018influence}, differences in the sensing area and the ambient atmosphere, etc.}.

Although it is of paramount importance to RH adjust the measurements from a PM sensor without any heater or dryer as described in the aforementioned sections, it is unlikely to determine the specific cause (hygroscopic growth of particles, counting condensed tiny droplets, or both) of the RH artefacts based on the corresponding RH levels only (e.g., extreme RH levels do not always result in the occurrence of the abnormal spikes). Besides, the extremely humid condition introduced by raining, during which the aerosol will be washed out, further complicates the determination of the cause of these artefacts. Since a PM sensor usually demonstrates distinct responses to RH among these circumstances, it is impossible to derive the effective RH correction factor(s) without identifying the exact circumstance in which the collocated data pairs are obtained, or the effectiveness of the RH correction factor(s) will likely be compromised when the collocated data pairs are acquired under two or more of these circumstances. Consequently, the PM\textsubscript{2.5} measurements within these three intervals of abnormal spikes were manually eliminated\footnote{The entire hour of measurements of the first abnormal spike and that of the last abnormal spike were also eliminated for the sake of convenience.} as illustrated in Fig. \ref{fig:PM2p5RhRmSpk}; excluding these measurements can also prevent the calibration models from being highly affected by the unusual occurrences (i.e., counting condensed droplets as particles). The remaining measurements are considered to be influenced by hygroscopic growth of particles only because the Supersite hardly recorded any precipitation data. An automated approach used to identify and remove the abnormal spikes in real time is proposed in section \ref{SubSubSect:AutoDataCleansingOfPMSensors}.

The scatter plots of the 1-minute interval PM\textsubscript{2.5} collocated data pairs between the DUST and the PM sensors as well as between the REFN and the PM instruments (PM sensors plus the DUST) are shown in Fig. \ref{fig:ScatterSensDust} and Fig. \ref{fig:ScatterSensDustRefn}, respectively. In accordance with the scatter plots in the first and the second columns (without and with elimination, respectively) of both figures, eliminating the abnormal spikes vastly improved the correlations (in terms of $R^2$ values) between the outputs from the DN7C ($>\SI{278}{\percent}$), the HPMA ($>\SI{355}{\percent}$), or the OPCN ($>\SI{41}{\percent}$) and the reference measurements from the DUST or the REFN, while the improvements for the PMSA ($<\SI{27}{\percent}$) and the DUST ($\approx\SI{38}{\percent}$) were marginal. Additionally, in the scatter plots of the first columns, the dispersion of data points associated with (horizontal axis) the OPCN, the HPMA, and the DN7C, but not the PMSA or the DUST, was increasingly widespread. It was suggested that the measurements from the DN7C, the HPMA, and the OPCN were more vulnerable than the next to circumstance in which the condensed tiny droplets remained suspended in the air might be detected as particles. These widespread data points also implied the unlikeliness of performing effective RH adjustments on the abnormal spikes of the DN7C, the HPMA, and the OPCN. With regard to the PMSA, its RH insusceptibility in comparison to the other PM sensors might be attributed to the usage of the tiniest micro-fan (see Fig. \ref{fig:SchematicDiagramOfSensorEvaluated}), which produces the smallest flow volume, and the laser source, which warms the sample flow, such that the sampled aerosol's water portion can be easily vaporized. It is recognized that the sheath flow in a research-grade instrument is generally warmed by the air-pump and proximal electronics, which reasons the RH insusceptibility of the DUST \cite{crilley2018evaluation}.

\begin{figure}[!t]
\centering
	\subfloat[]{
		\includegraphics[width=0.45\linewidth]{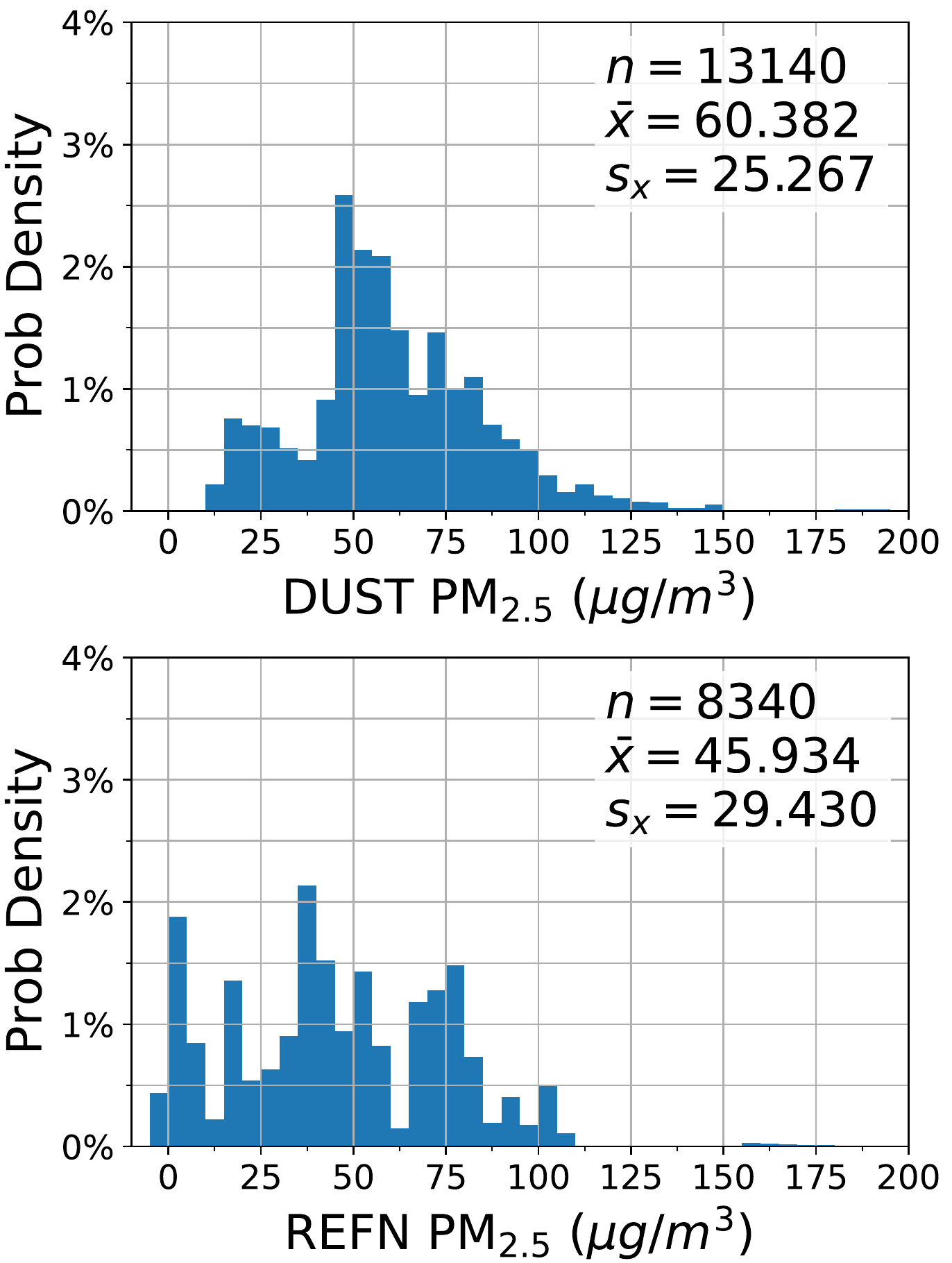}
		\label{subfig:StatDustRefnRhA}}
	\subfloat[]{
		\includegraphics[width=0.45\linewidth]{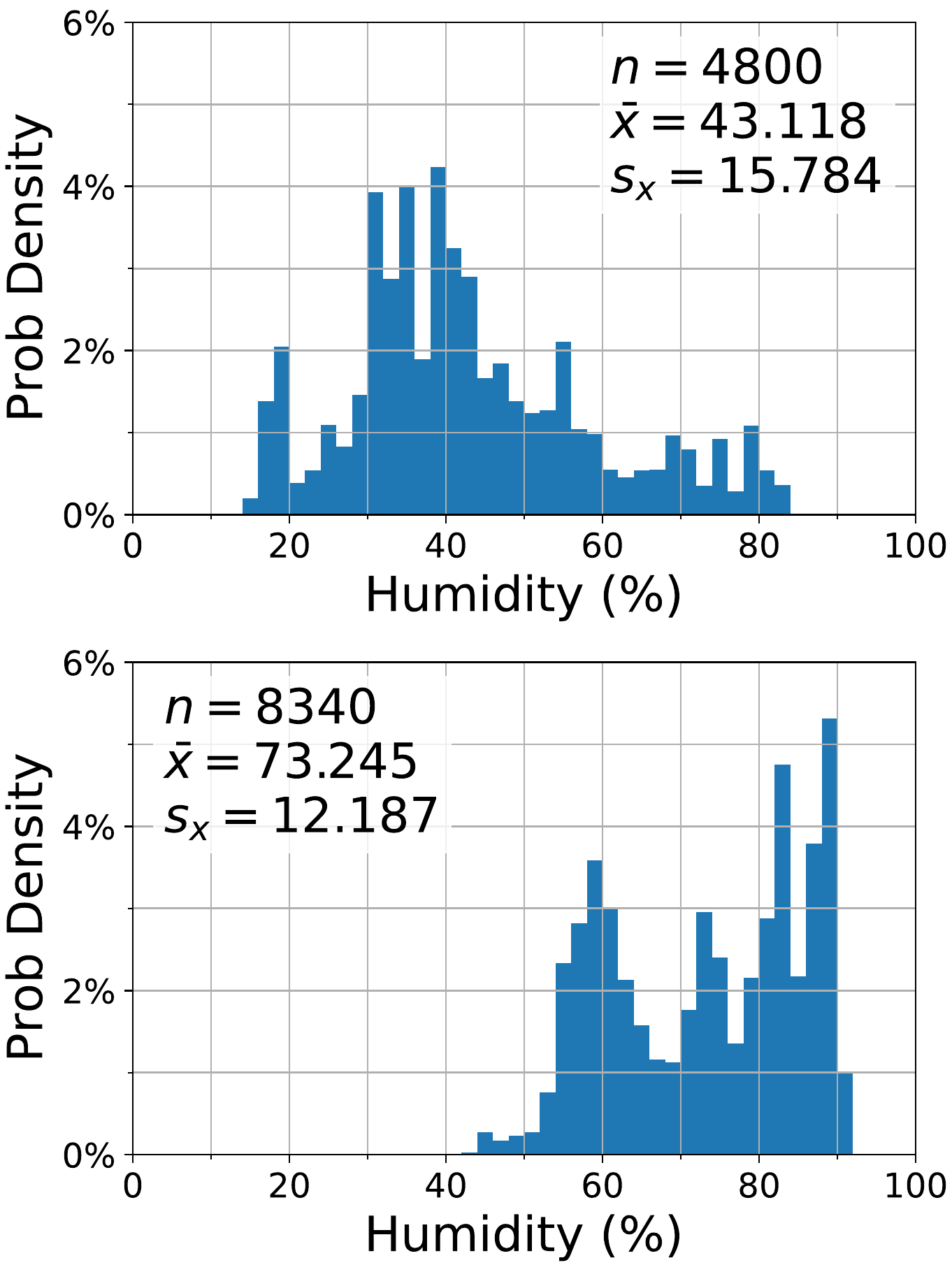}
		\label{subfig:StatDustRefnRhB}}
\caption{Histograms and statistics of 1-minute interval (a) PM\textsubscript{2.5} measurements from the DUST (top; from 13 to 23 February 2018) and the REFN (bottom; from 17 to 23 February 2018) with abnormal spikes eliminated (\SI[per-mode=symbol]{5}{\micro\gram\per\cubic\meter} bin width); and (b) RH measurements from the REFN before (top) and after (bottom) 17 February 2018 (\SI{2}{\percent} bin width). The symbols $n$, $\bar{x}$, and $s_{x}$ represent the number of valid data points, the sample mean, and the sample standard deviation, respectively.}
\label{fig:StatDustRefnRH}
\end{figure}

Apart from the abnormal spikes in the nominated sensors' measurements, the authors also noticed that there might exist dual linear relationships between the PM instruments' outputs and the REFN's measurements as demonstrated by the scatter plots in the second column of Fig. \ref{fig:ScatterSensDustRefn}. To exclude the impact from the dynamic heating system employed in the REFN, the correlations of the collocated data pairs having corresponding RH below \SI{58}{\percent} (REFN's heater was off) were investigated. As evidenced by the scatter plots in the third columns of Fig. \ref{fig:ScatterSensDust} and Fig. \ref{fig:ScatterSensDustRefn}, the RH thresholded data pairs between the PM instruments (particularly the PMSA, the HPMA, and the DUST) and the REFN, but not between the PM sensors and the DUST, clearly manifested two distinct linear relationships, where the data points in blue and in orange denoted the data pairs acquired before and after 17 February, respectively. The exact cause is inconclusive, but a plausible hypothesis would be the change of aerosol properties (probably the composition) since/on 17 February because the measurements from the PM instruments, but not the REFN, are heavily dependent on the aerosol properties as detailed in section \ref{SubSect:Instrumentation}. The change of aerosol properties might be attributed to the Lunar New Year holidays from 16 to 22 February, the RH differences before and after 17 February ($\overline{RH}\approx\SI{43}{\percent}$ vs. $\overline{RH}\approx\SI{73}{\percent}$ as listed in Fig. \ref{subfig:StatDustRefnRhB}), and/or the windy (generally $>\SI[per-mode=symbol]{30}{\kilo\meter\per\hour}$; east) and humid weather on 17 February \cite{timeanddate2018weather}. Substantial day-to-day variation in the aerosol composition was also conjectured in studies \cite{day2001aerosol,zheng2018field}. The importance of initially and frequently calibrating a PM sensor in the final deployment environment is again emphasized. Consequently, when utilizing the REFN as reference, all the collocated data pairs acquired before 17 February were manually eliminated as well (see Fig. \ref{fig:PM2p5RhRmSpkThld}).

To sum up, the valid data pairs in this field experiment were the collocated measurements, after elimination of the abnormal spikes, from 13 to 23 February when the DUST ($n=13140$) was selected as reference or from 17 to 23 February when the REFN ($n=8340$) was. From the presence of the abnormal spikes, it can be reasonably deduced that the PMSA and the DUST demonstrated better insusceptibility to condensed tiny droplets when compared with the OPCN, the HPMA, and the DN7C. The histograms along with the basic statistics of the preprocessed and cleansed data pairs are illustrated in Fig. \ref{subfig:StatDustRefnRhA}. For the sake of convenience, the collocated data pairs in the following analyses and discussion are referring to the valid data pairs after preprocessing and cleansing.

\subsubsection{Correlation and Calibration Analyses}
\label{SubSubSect:CorrelationAndCalibrationOfPMSensors}

\begin{figure}[!t]
\centering
	\includegraphics[width=0.95\columnwidth]{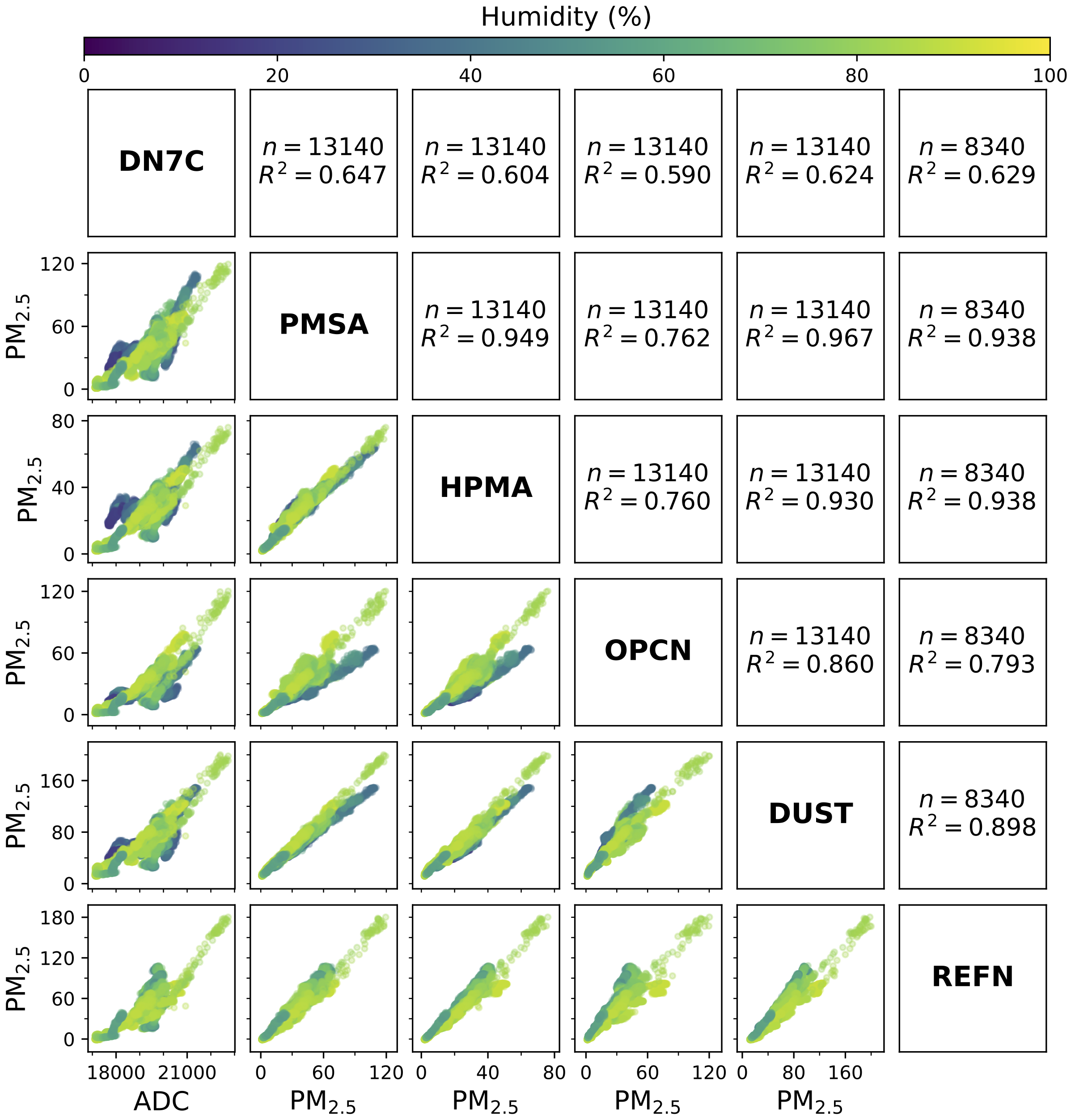}
\caption{Pairwise correlations (in terms of $R^2$) of the collocated data pairs (acquired from 13 to 23 February 2018 having abnormal spikes eliminated) among the PM sensors (1-minute averaged; the DN7C, PMSA, HPMA, and OPCN) and reference monitors (1-minute interval; DUST and REFN). The diagonal panels list the names of instruments. For the scatter plot to the left of each name panel, its vertical axis represents the values measured by that instrument; for that at the bottom of each name panel, its horizontal axis represents the values measured by that instrument. The upper right panels provide the numbers of collocated data pairs ($n=8340$ and $n=13140$ are the numbers of data pairs with and without elimination of measurements before 17 February, respectively) and the $R^2$ values of the corresponding scatter plots in the diagonal symmetry panels. The color of each data point indicates the corresponding RH value.}
\label{fig:ScatterPlotsSensDustRefn}
\end{figure}

For each pair of instruments among the PM sensors and reference monitors, the respective correlations of the collocated PM\textsubscript{2.5} measurements were investigated and the results (scatter plots and $R^2$ values) are presented in Fig. \ref{fig:ScatterPlotsSensDustRefn}. Apparently, the DN7C's ADC outputs were poorly correlated with the mass concentrations from the others because the five lowest $R^2$ values ($R^2<0.65$; see the panels in the first row) were all associated with the DN7C; it might, as will be discussed later in this section, stem from the susceptibility to both TEMP and RH of the DN7C while the TEMP had negligible effects on the others. Excluding the instrument pairs involving the DN7C, moderate correlations ($0.76<R^2<0.86$; see those panels above and to the right of the OPCN panel) were observed between the PM\textsubscript{2.5} mass concentrations from the OPCN and that from the others. The dispersion pattern in the HPMA-OPCN and the PMSA-OPCN scatter plots (moist data points were mostly on the upper left of the dry ones) implied that the overestimation attributed to hygroscopic growth of particles was more acute for the OPCN when compared to the HPMA and the PMSA. Such a strong RH susceptibility of the OPCN might be owing to its largest dimensions (distances the laser source from the sensing zone most greatly) and its biggest micro-fan (generates the largest flow volume) among the PM sensors (see Tab. \ref{tab:SensorSpec} and Fig. \ref{fig:SchematicDiagramOfSensorEvaluated}), which result in warming the sampled aerosol and vaporizing the water portion less effectively; it might as well explain the pronounced abnormal spikes in the OPCN's measurements as discussed in section \ref{SubSubSect:DataCleansingOfPMSensors}. As for the HPMA and the PMSA, their measurements were well correlated with each other and with that from the reference monitors ($0.93<R^2<0.97$; see those panels to the right of the HPMA and the PMSA panels). The strong correlations might partly benefit from their small dimensions and the adoption of laser sources and tiny micro-fans (see Tab. \ref{tab:SensorSpec} and Fig. \ref{fig:SchematicDiagramOfSensorEvaluated}), which likely contribute to their insusceptibility to hygroscopic growth of particles by quickly warming the sample flow. Additionally, in the PMSA-HPMA scatter plot, the low dispersion between the moist and dry data points implied that the PMSA's and the HPMA's responses to hygroscopic growth were similar, though they showed distinct responses to condensed tiny droplets in section \ref{SubSubSect:DataCleansingOfPMSensors}; it might be ascribed to their similar measuring principles and internal structures but different sized micro-fans. In the OPCN-DUST, the HPMA-DUST, and the PMSA-DUST scatter plots, their respective dispersion patterns of the moist and dry data points implied that the insusceptibility to hygroscopic growth of the DUST was between that of the OPCN and the HPMA or the PMSA; it might also partly justify the $R^2$ ranking of the PM instruments with reference to the REFN (see the panels in the last column of Fig. \ref{fig:ScatterPlotsSensDustRefn}).

Afterwards, the four nominated PM sensors were calibrated against the DUST and the REFN, respectively, and the sets of calibration parameters were achieved by fitting the respective collocated data pairs to the OLS or the MLR model, in which the reference (DUST or REFN) measurements represented the observations of the dependent variable. The calibrated PM\textsubscript{2.5} mass concentrations of the PM sensors were then obtained by feeding their measurements into the respective fitted models. Comparisons between the calibrated and the reference PM\textsubscript{2.5} mass concentrations were performed in order to evaluate the sensors' and the calibration models' performances. The DUST was also calibrated and evaluated against the REFN, and the results were served as baselines for justifying whether or not the calibrated sensors' data quality was adequate for research purposes. Note that multiple assumptions about the collocated calibration and evaluation were made in section \ref{SubSubSect:AssumptionInEvalCali}.

The collocated calibration results (scatter plots, $R^2$ values, regression parameters, etc.) of the PM sensors with reference to the DUST and the REFN are shown in Fig. \ref{fig:ScatterCaliSensDust} and Fig. \ref{fig:ScatterCaliSensRefn}, respectively. As demonstrated in the scatter plots of the first columns, nearly every data point associated with the PM sensors, except the DN7C\footnote{The DN7C was excluded because it outputs ADC readings instead of PM\textsubscript{2.5} mass concentrations.}, was located to the upper left of the identity lines, indicating that the PMSA, the HPMA, and the OPCN typically underestimated the mass concentrations of PM\textsubscript{2.5}. Moreover, the angles between the OLS regression lines and the identity lines (i.e., the slope values of respective OLS regression lines) varied from sensor to sensor, which is expected because they were initially calibrated with different types of aerosols by their manufacturers (presented in section \ref{SubSect:Instrumentation}). In contrast, the DUST typically overestimated the mass concentrations of PM\textsubscript{2.5} with the OLS regression line slightly shifted to the right of the identity line, though it had been zero calibrated few months ahead this evaluation campaign. Thus, it was suggested that frequent calibration is essential for any light-scattering based instruments, regardless of research-grade or low-cost, that derive the mass concentrations of certain PM solely from the scattered light. Although other studies tended to show markedly better correlations between PM sensors and optical references (e.g., the DUST), compared to gravimetric or gravimetric-related references (e.g., the REFN) \cite{kelly2017ambient}, the $R^2$ values listed in the first columns argued against this primarily because eliminating the REFN's data pairs acquired before 17 February significantly lowered the variations of RH conditions and aerosol properties such that the corresponding correlations were generally improved (see those scatter plots in the second columns of Fig. \ref{fig:ScatterSensDust} and Fig. \ref{fig:ScatterSensDustRefn}).

Within the second to fourth columns of Fig. \ref{fig:ScatterCaliSensDust} and Fig. \ref{fig:ScatterCaliSensRefn}, the scatter plots of the calibrated versus reference mass concentration pairs of PM\textsubscript{2.5} were established. The calibrated concentrations in these columns were obtained from the OLS, the RH involved MLR (MLH), and the TEMP involved MLR (MLT) models, respectively. The set of calibration parameters (i.e., $\{\beta_0,\cdots,\beta_k|k\in\mathbb{Z}^+\}$ with a \SI{95}{\percent} confidence interval) and the $R^2$ value (calibrated vs. reference pairs) are presented in each of the scatter plots. These scatter plots, visually depicting the concentration pairs' correlations and their dispersions attributed to the RH levels, provide an intuitive sense of the degrees of correlation improvement from different calibration models and the levels of RH or TEMP insusceptibility of the PM sensors that will be discussed in the paragraphs below.

\begin{table}[!t]
\centering
\caption{$R^2$ values of the 1-minute interval calibrated and reference PM\textsubscript{2.5} mass concentration pairs.}
\label{tab:RSquareValuesSensDustRefn}
\renewcommand{\arraystretch}{1.15}
\begin{threeparttable}
\begin{tabular}{|c|c|r|r|r|r|r|}
\hline
Reference & Model & \multicolumn{1}{c|}{DN7C} & \multicolumn{1}{c|}{PMSA} & \multicolumn{1}{c|}{HPMA} & \multicolumn{1}{c|}{OPCN} & \multicolumn{1}{c|}{DUST} \\ 
\hline
\multirow{4}{*}{DUST\tnote{a}} 
& OLS 			& $0.624$ & $0.967$ & $0.930$ & $0.860$ & n/a \\ \cline{2-7} 
& MLH\tnote{c} 	& $0.624$ & $0.977$ & $0.963$ & $0.894$ & n/a \\ \cline{2-7} 
& MLT\tnote{d} 	& $0.825$ & $0.973$ & $0.932$ & $0.870$ & n/a \\ \cline{2-7} 
& MLHT\tnote{e} & $0.915$ & $0.980$ & $0.963$ & $0.894$ & n/a \\
\hline
\multirow{4}{*}{REFN\tnote{b}} 
& OLS 			& $0.629$ & $0.938$ & $0.938$ & $0.793$ & $0.898$ \\ \cline{2-7} 
& MLH\tnote{c} 	& $0.631$ & $0.949$ & $0.955$ & $0.873$ & $0.943$ \\ \cline{2-7} 
& MLT\tnote{d} 	& $0.772$ & $0.938$ & $0.938$ & $0.801$ & $0.907$ \\ \cline{2-7} 
& MLHT\tnote{e} & $0.926$ & $0.956$ & $0.968$ & $0.897$ & $0.948$ \\
\hline
\end{tabular}
\begin{tablenotes}
	\item [a] The collocated data pairs were acquired from 13 to 23 February 2018 with abnormal spikes eliminated ($n=13140$).
	\item [b] The collocated data pairs were acquired from 17 to 23 February 2018 with abnormal spikes eliminated ($n=8340$).
	\item [c] The instruments were calibrated using MLR model with RH.
	\item [d] The instruments were calibrated using MLR model with TEMP.
	\item [e] The instruments were calibrated using MLR model with RH and TEMP.
\end{tablenotes}
\end{threeparttable}
\end{table}

The $R^2$ values, including those associated with the RH and TEMP involved MLR model (MLHT), were also summarized in Tab. \ref{tab:RSquareValuesSensDustRefn}, from which it could be deduced that the DN7C's MLT calibrated outputs, in comparison to its OLS calibrated ones, correlated much better with the reference concentrations ($\approx\SI{32}{\percent}$ or \SI{23}{\percent} improvement of $R^2$), that is, the DN7C's outputs were severely influenced by the TEMP. Additionally, the DN7C's MLH calibrated outputs showed no improvement over its OLS calibrated ones in the correlations against the reference concentrations ($<\SI{0.4}{\percent}$ of $R^2$), suggesting that the RH effects on its outputs were negligible. However, it did not necessarily imply that the outputs from the DN7C were not influenced by the RH because they might be obscured by the relatively significant TEMP effects. Further investigating the noticeable improvements in correlations against the reference concentrations for the DN7C's MLHT calibrated outputs compared to its MLT calibrated ones ($\approx\SI{11}{\percent}$ or \SI{20}{\percent} of $R^2$), it is reasonable to claim that the DN7C was susceptible to both TEMP and RH (i.e., hygroscopic growth of particles), though involving both TMEP and RH in the MLR model may likely violate its fundamental assumptions. The dispersion pattern of those DN7C associated scatter plots (the MLT calibrated dry data points were mostly on the upper left of the moist ones) in the fourth columns also substantiated this claim.

In accordance with the same philosophy, it could as well be inferred that the other PM instruments (the PMSA, HPMA, OPCN, and DUST) were insusceptible to the TEMP in view of their respective negligible improvements in the MLT vs. OLS correlations ($<\SI{0.7}{\percent}$, $<\SI{0.3}{\percent}$, $<\SI{1.2}{\percent}$, and $\approx\SI{1.0}{\percent}$ of $R^2$, respectively) and the MLHT vs. MLH correlations ($<\SI{0.8}{\percent}$, $<\SI{1.4}{\percent}$, $<\SI{2.8}{\percent}$, and $\approx\SI{0.5}{\percent}$ of $R^2$, respectively). Given the fact that no TEMP nor RH sensor has been identified in the DN7C, the PMSA, and the HPMA (but the OPCN has a TEMP-RH sensor installed), it is unlikely for the firmware of the PMSA or the HPMA to employ any TEMP adjustments or corrections on the measurements. Thus, a probable cause for the distinct TEMP insusceptibility between the DN7C and the other PM instruments might be that the heat produced by the laser sources used in the latter is capable of maintaining their sensing zones' temperature, while the infrared LED employed in the DN7C generally emits far less heat. Further considering the large micro-fan adopted in the DN7C (see Fig. \ref{fig:SchematicDiagramOfSensorEvaluated}), its poor insusceptibility to condensed tiny droplets (see section \ref{SubSubSect:DataCleansingOfPMSensors}) and hygroscopic growth of particles is no surprise.

Similarly, the improvements in the other PM instruments' (the PMSA's, HPMA's, OPCN's, and DUST's) MLH vs. OLS correlations, ranged from negligible to moderate ($\approx\SI{1.0}{\percent}$ or \SI{1.2}{\percent}, $\approx\SI{3.5}{\percent}$ or \SI{1.8}{\percent}, $\approx\SI{4.0}{\percent}$ or \SI{10}{\percent}, and $\approx\SI{5.0}{\percent}$ of $R^2$, respectively), further justified the above-mentioned ranking of their insusceptibility to hygroscopic growth of particles; so do the signs and the absolute values of their MLH models' RH associated parameters (i.e., $\beta_2$ in the scatter plots of the third columns). According to the scatter plots of the OLS calibrated and reference concentration pairs as illustrated in the second columns, the dry and the moist data points relevant to these PM instruments were commonly distributed on opposite sides of the identity lines and the dispersions increased along with the reference concentrations, indicating that the atmospheric RH typically affected the slopes of respective OLS regression lines. Although involving RH in the OLS model (i.e., MLH model) can slightly to moderately improve their correlations against the references (in terms of $R^2$ values), the dispersions between the dry and the moist data points were still noticeable as shown in the scatter plots, particularly in the PMSA-DUST and the OPCN-DUST ones, of the third columns. Additionally, the MLH model seemed to over adjust these PM instruments' outputs that were acquired under low concentrations and dry conditions, for instance, the set of dry data points (dark blue) appeared on the lower left of the PMSA-DUST scatter plot or the set of dry data points (dark green) formed a curved shape on the lower left of the OPCN-REFN scatter plot. Both issues can be considerably eased by applying the thresholded MLH model, that is, OLS model for the dry data pairs and MLH model for the moist ones. However, the RH threshold value for identifying the dry and the moist data points depends on the sampled aerosol's deliquescent point (determined by the chemical composition and varying on occasion), limiting the calibration parameters' transferability furthermore compared to the simple MLH or OLS. Other more advanced calibration models can largely address the aforementioned issues as well, but they are typically specific to the data sets and determining the best performing calibration model is out of scope of this study. Served as a quick example, the PMSA and OPCN were calibrated against the DUST and the REFN, respectively, with an empirical model (ADV) considering the product of sensor's output and corresponding RH to be an additional independent variable in the MLH model (see the third column of Fig. \ref{fig:ScatterCaliAdvExample}).

According to the evaluation results in other studies investigating the sensor models that were identical or similar to the nominated ones as reported in section \ref{SubSect:SensorEvaluation} and section \ref{SubSect:Instrumentation}, although their reference values may be limited primarily due to inconsistencies in the experimental setups and variations in the sampled aerosols' properties across studies and locations, respectively, a group of similar degrees of correlations (against reference) and atmospheric influences compared to this study were observed. For instance, the OPCN manifested moderate correlations with the reference monitors ($R^2=0.63$ to $0.76$ \cite{crilley2018evaluation} or $R^2=0.49$ to $0.60$ \cite{badura2018evaluation}) and showed severe artefacts at extreme RH ($RH>\SI{85}{\percent}$ \cite{crilley2018evaluation} or $RH>\SI{80}{\percent}$ \cite{badura2018evaluation}), while the PMSA and/or its sibling models correlated well with the reference monitors ($R^2=0.81$ to $0.88$ \cite{badura2018evaluation}, $R^2\geq0.89$ \cite{zamora2018field}, or $R^2=0.93$ \cite{zheng2018field}) and were slightly influenced by extreme RH ($RH>\SI{80}{\percent}$ \cite{badura2018evaluation} or $RH>78\%$ \cite{jayaratne2018influence}) but not the TEMP \cite{zheng2018field,zamora2018field}.

Briefly, the TEMP and RH induced significant impacts on the DN7C's ADC outputs; the mass concentrations from the PMSA, the HPMA, the DUST, and the OPCN were negligibly affected by the TEMP, but slightly to moderately influenced by the RH, which reflected their descending insusceptibility to hygroscopic growth of particles. Such atmospheric influences on the PM\textsubscript{2.5} measurements from the PM instruments can be considerably mitigated via proper adjustment and calibration. The calibrated and reference concentration pairs between the reference instruments (DUST and REFN) and the DN7C, the OPCN, and the HPMA or PMSA exhibited a set of ascending $R^2$ values.

\subsubsection{Residual Analysis}
\label{SubSubSect:ResidualAnalysisOfPMSensors}

According to the analyses above, to compensate the atmospheric impacts on the PM instruments' measurements, the DN7C was calibrated with TEMP involved MLR model, and the PMSA, the HPMA, the OPCN, and the DUST were calibrated with RH involved MLR model. These instruments were also calibrated with the OLS model, which is the baseline parametric model for collocated calibration. In this section, the relative residuals and the (absolute) residuals of the respective fitted parametric models were examined. For convenience's sake, the relative residual(s) and the residual(s) are hereinafter referred to as Rel Resid and Resid, respectively.

In accordance with section \ref{SubSubSect:EvaluationMetrics}, to avoid the significantly large values, a Rel Resid was considered valid only when the calibrated and reference concentration pair were both no less than $\SI[per-mode=symbol]{3}{\micro\gram\per\cubic\meter}$. For the OLS models fitted with the collocated data pairs between the PM sensors and the DUST, the scatter plots of RH versus Rel Resid and RH versus Resid together with the histograms of Rel Resid and Resid are presented in the first, the third, the second, and the fourth column of Fig. \ref{subfig:CaliErrorOlsDust}, respectively. Likewise, the corresponding scatter plots and histograms of the PM sensors' MLR models (the DUST as reference) and the PM instruments' OLS and MLR models (the REFN as reference) are shown in Fig. \ref{subfig:CaliErrorMlrDust}, Fig. \ref{subfig:CaliErrorOlsRefn}, and Fig. \ref{subfig:CaliErrorMlrRefn}, respectively. The data points in each scatter plot were colored (from purple to yellow) depending on their respective reference concentrations (from low to high).

From the shapes of the scatter plots and the histograms in Fig. \ref{subfig:CaliErrorOlsDust} and Fig. \ref{subfig:CaliErrorOlsRefn}, it can be intuitively concluded that the performance ranks, in terms of Rel Resid and Resid from OLS models, of the PMSA and the DN7C were the best and the worst, respectively, while the HPMA, the DUST, and the OPCN were ranked in between. The ranking principle behind is that a scatter plot with data points more congregated along the horizontal zero line or a histogram in narrower bell shape generally indicates a lower sample standard deviation, thus a narrower confidence interval for the Rel Resid or Resid that is assumed to be normally distributed with zero mean. According to the Error Approach detailed in section \ref{SubSubSect:ErrorApproach} as well as the evaluation metrics of Part 58AA listed in section \ref{SubSubSect:EvaluationMetrics}, the sample means and the sample standard deviations of the Resid or Rel Resid can be utilized to characterize the measurement trueness and the intermediate measurement precision (under intermediate precision conditions having change in measuring systems with distinct measuring principles) of the PM instruments, respectively.

Exploring deeper into the first columns of Fig. \ref{subfig:CaliErrorOlsDust} and Fig. \ref{subfig:CaliErrorOlsRefn}, on the right sides of most of the RH vs. Rel Resid scatter plots, the data points inclining to purple were broadly dispersed (mostly above the horizontal zero lines\footnote{The purple points associated with the PMSA and the HPMA in scatter plots using the DUST as reference were dispersed below the horizontal zero lines because they have better insusceptibility to hygroscopic growth compared to the DUST as discussed previously.}), implying that hygroscopic growth of particles at low concentrations ($\leq\SI[per-mode=symbol]{20}{\micro\gram\per\cubic\meter}$ and $RH>\SI{82}{\percent}$) introduced considerable relative differences (overestimation) between the OLS calibrated and the reference measurements. Excluding the DN7C, the degrees of such purple dispersion associated with the PMSA and the HPMA, the DUST, and the OPCN were in an ascending order, which reflected their increasing susceptibility to hygroscopic growth as elaborated earlier. It is noteworthy that noticeable dispersions of purple points associated with the DN7C were observed as well at RH conditions as low as \SI{50}{\percent}, suggesting that the DN7C's outputs were affected by factor(s) other than hygroscopic growth, that is, the TEMP as discussed previously. As shown in the first columns of Fig. \ref{subfig:CaliErrorMlrDust} and Fig. \ref{subfig:CaliErrorMlrRefn}, involving TEMP in the DN7C's MLR model can considerably mitigate the purple dispersions at low RH conditions but raise those roughly or over twice at high RH conditions ($>\SI{82}{\percent}$). Conceivably, the Rel Resid of the DN7C's calibrated outputs against the reference measurements can be further diminished when employing more advanced calibration models with both TEMP and RH involved. For the remaining PM instruments, the OPCN and the DUST in particular, involving RH in their MLR models can dramatically lower the purple dispersions at high RH condition ($>\SI{82}{\percent}$). However, within their Rel Resid scatter plots utilizing the REFN as reference (see first column of Fig. \ref{subfig:CaliErrorMlrRefn}), clumps of purple points were appeared on top of the horizontal zero lines at low RH conditions ($\approx\SI{60}{\percent}$), signifying that the low concentration measurements under dry conditions from these PM instruments were over adjusted by the MLH model as evidenced earlier. Although the cause of such over adjustment is inconclusive, it might be attributed to the REFN's dynamic heating system which turns on when the RH is above \SI{58}{\percent}. In general, the Rel Resid from the MLR models, compared to that from the OLS models, formed a set of better bell-shaped histograms with reduced sample means ($\bar{d}\,$) and sample standard deviations ($s_d$).

In the third columns of Fig. \ref{subfig:CaliErrorOlsDust} and Fig. \ref{subfig:CaliErrorOlsRefn}, dispersions of data points inclining to purple were observed as well on the right sides ($RH>\SI{82}{\percent}$) of most of the RH vs. Resid scatter plots. However, the degrees of such purple dispersions were marginal compared to that of data points colored from cyan to yellow (mostly on top of the horizontal zero lines\footnote{The PMSA and the HPMA associated data points colored from cyan to yellow in scatter plots using the DUST as reference were dispersed below the horizontal zero lines because they have better insusceptibility to hygroscopic growth compared to the DUST as discussed previously.}), which suggested that the differences (overestimation) between the OLS calibrated and the reference measurements, attributed to extreme RH conditions, were proportional to the (reference) mass concentrations of PM\textsubscript{2.5}. Such proportional characteristic satisfied the influencing mechanism of hygroscopic growth of particles on these PM instruments' measurements (see section \ref{SubSect:LightScattering}), and thus in turn, justified the aforesaid assumption that the high RH attributed overestimation in their cleansed measurements were subject to hygroscopic growth only. Excluding those associated with the DN7C, the purple points on the left side ($RH\leq\SI{82}{\percent}$) of a particular RH vs. Resid scatter plot were commonly positioned on the horizontal zero line, while the data points colored from cyan to yellow were extensively dispersed. This cyan to yellow dispersion contributed largely to the sample standard deviation of the Resid ($s_{\epsilon}$) from the OLS model, and typically, the higher the concentration level, the larger the dispersion (from zero) was. Notice that a portion of the corresponding purple points associated with the DN7C formed a sloped line and the other portion was observed far above the horizontal zero line. It was likely attributed to the foregoing claim that the DN7C's outputs were influenced by both TEMP and RH because the MLT model can well flatten the sloped lines and mitigate the dispersions on the left sides (but raise those on the right sides) as demonstrated in the third columns of Fig. \ref{subfig:CaliErrorMlrDust} and Fig. \ref{subfig:CaliErrorMlrRefn}. The Resid from MLT model associated with the DN7C exhibited similar dispersion patterns as those from OLS models associated with the other PM instruments; the DN7C's MLT calibrated measurements were considered not to be susceptible to the TEMP, but RH.

For the MLT Resid of the DN7C and the OLS Resid of the others, the cyan to yellow data points on the left sides of their scatter plots were commonly below the horizontal zero lines, reflecting that their calibrated measurements at comparatively high concentration levels ($\geq\SI[per-mode=symbol]{50}{\micro\gram\per\cubic\meter}$) under mildly humid conditions ($RH\leq\SI{82}{\percent}$) typically underestimated the mass concentrations of PM\textsubscript{2.5} (the degrees were proportional to the concentration levels; see the corresponding scatter plots of the calibrated and reference concentration pairs given in Fig. \ref{fig:ScatterCaliSensDust} and Fig. \ref{fig:ScatterCaliSensRefn}). Additionally, these cyan or yellow data points (close color indicated similar reference concentration levels) within the corresponding Resid scatter plots using the REFN as reference established a number of sloped lines approaching the horizontal zero ones, suggesting that, at certain (reference) concentration level, the PM instruments' measurements after calibration increased along with the growth of RH conditions. A reasonable inference would be that hygroscopic growth was occurred as well at RH from \SIrange{42}{82}{\percent}, in which case the consequent overestimation was much less severe compared to that at RH conditions over \SI{82}{\percent} (most acute at $RH\approx\SI{90}{\percent}$), due to the MLH model (excluding the DN7C) can well flatten the sloped lines in cyan or yellow and considerably decrease the dispersions at extreme RH conditions. However, the purple points that were horizontally distributed in the first place were slanted (see the third columns of Fig. \ref{subfig:CaliErrorOlsRefn} and Fig. \ref{subfig:CaliErrorMlrRefn}), especially for those associated with the OPCN and the DUST, reflecting the prior deduction that these PM instruments' low concentration measurements at mildly humid conditions were over adjusted by the MLH model. The empty spaces between the purple and the yellow points on the left sides were likely resulted from the absence of PM\textsubscript{2.5} measurements at medium concentrations under low RH ($<\SI{60}{\percent}$). In the corresponding Resid scatter plots utilizing the DUST as reference, the event of cyan or yellow points forming obliquely oriented lines was inconspicuous; it might be ascribed to the fact that the DUST and the nominated sensors were all heater- or dryer-less light-scattering based instruments that manifested little fundamental heterogeneity in their responses to hygroscopic growth at RH below \SI{82}{\percent}. Compared to OLS models, the Resid from MLR models fitted with collocated measurements between the PM sensors and the DUST established a set of better bell-shaped histograms with zero sample means ($\bar{\epsilon}\,$) and decreased sample standard deviations ($s_{\epsilon}$) (see the fourth columns of Fig. \ref{subfig:CaliErrorOlsDust} and Fig. \ref{subfig:CaliErrorMlrDust}). When using the REFN as reference, although the PM instruments' MLR Resid, compared to the OLS ones, also had improved statistics, they did not necessarily result in better bell-shaped histograms due to over adjustments (see the fourth columns of Fig. \ref{subfig:CaliErrorOlsRefn} and Fig. \ref{subfig:CaliErrorMlrRefn}).

Overall, as evidenced by the dispersion patterns of the data points in the OLS Rel Resid and the OLS Resid scatter plots as well as the over adjusted data points in the MLR Rel Resid and the MLR Resid scatter plots, the responses to hygroscopic growth of particles from these PM instruments, excluding the DN7C owing to its profound TEMP effects, varied according to the RH levels and manifested distinct behaviors at mildly ($\leq\SI{82}{\percent}$) and extremely ($>\SI{82}{\percent}$) humid conditions (gradually increased with the rise of RH and dramatically overestimated once over damp), respectively. Regarding the DN7C, its measurements were primarily suffered from the TEMP influences and the hygroscopic influences at mildly humid and extremely humid conditions, respectively. Note that the splitting value of \SI{82}{\percent} was mainly depended on the sampled aerosol's chemical composition that varies on occasion (see section \ref{SubSubSect:UncertaintyOfInstruments}). More advanced calibration models such as TEMP and RH involved parametric model (for the DN7C) or thresholded MLH model (for the others) are considered to have improved performance in terms of Rel Resid and Resid.

Although the collocated data pairs between the PM sensors and the optical references (e.g., the DUST), compared to the gravimetric or gravimetric-related references (e.g., the REFN), did not tend to manifest markedly stronger correlations (\SIrange{-1}{9}{\percent} increase of $R^2$ values; see Tab. \ref{tab:RSquareValuesSensDustRefn}) as in other studies (see section \ref{SubSubSect:CorrelationAndCalibrationOfPMSensors}), the Rel Resid and Resid of the calibrated PM sensors against the DUST, in contrast to that against the REFN, exhibited obviously improved statistics (\SIrange{64}{99}{\percent} drop of $|\bar{d}\,|$, \SIrange{17}{67}{\percent} drop of $s_{d}$, and \SIrange{9}{43}{\percent} drop of $s_{\epsilon}$) as summarized in Tab. \ref{tab:RelativeErrorAnalysisSensDustRefn} and Tab. \ref{tab:ErrorAnalysisSensDustRefn} (with the MLHT model being excluded\footnote{Involving both TMEP and RH in the MLR model may likely violate its fundamental assumptions (see section \ref{SubSubSect:AssumptionInEvalCali}). Therefore, the MLHT model should be excluded in most cases except for discussing both the TEMP and RH effects on the DN7C.}). For the PM sensors excluding the DN7C, the MLHT model, compared to the MLH one, slightly improved the correlations (\SIrange{0.0}{0.3}{\percent} and \SIrange{0.7}{2.7}{\percent} rise of $R^2$ values using the DUST and the REFN as reference, respectively) of their calibrated and reference collocated pairs and moderately diminished the sample standard deviations of Resid (\SIrange{0.2}{4.8}{\percent} and \SIrange{7.6}{16.2}{\percent} reduction of $s_{\epsilon}$, respectively), but slightly or considerably boosted the sample standard deviations of Rel Resid (\SIrange{1.6}{2.3}{\percent} and \SIrange{32.8}{44.5}{\percent} growth of $s_{d}$, respectively)\footnote{The sample means of Resid ($\bar{\epsilon}\,$) were normally close to zero; the sample means of Rel Resid ($\bar{d}\,$) were negligibly diminished, if not boosted, compared to the boosted sample standard deviations of Rel Resid ($s_d$). Therefore, the differences between the MLHT's and the MLH's $\bar{\epsilon}\,$ and $\bar{d}\,$ were omitted in the discussion.}. It might result from that the MLHT model further over adjusted these PM sensors' measurements at low concentrations, in which case Rel Resid with relatively large values would be developed more likely, but better adjusted those at high concentrations, in which case the values of Resid would be typically diminished. Regarding the DN7C, the corresponding aspects plus the sample mean of Rel Resid from MLHT model, against that from MLT model, were usually improved substantially ($\approx\SI{11}{\percent}$ or \SI{20}{\percent} rise of $R^2$, $\approx\SI{30}{\percent}$ or \SI{43}{\percent} drop of $s_{\epsilon}$, $\approx\SI{93}{\percent}$ or \SI{27}{\percent} drop of $|\bar{d}\,|$, and $\approx\SI{35}{\percent}$ drop or $\approx\SI{27}{\percent}$ rise of $s_{d}$) because its outputs were considered to be susceptible to both TEMP and RH.

In accordance with the earlier discussion on the scatter plots in Fig. \ref{fig:ErrorAnalysisSensDust} and Fig. \ref{fig:ErrorAnalysisSensDustRefn}, variations of the Rel Resid and the Resid were primarily contributed by the PM\textsubscript{2.5} measurements at low and high concentration levels (especially under extreme humid conditions), respectively. Therefore, it is intuitive that removing the components at low and high concentration levels will result in numerically better statistical performance of Rel Resid and Resid, respectively. As shown in Tab. \ref{tab:RelativeErrorAnalysisSensDustRefn} and Tab. \ref{tab:ErrorAnalysisSensDustRefn}, the splitting value of \SI[per-mode=symbol]{50}{\micro\gram\per\cubic\meter} of reference concentration was empirically selected to separate the components into two groups. Regarding those Rel Resid from the OLS and MLR models with reference concentrations $\geq\SI[per-mode=symbol]{50}{\micro\gram\per\cubic\meter}$, compared to that having full range, the sample standard deviation values were usually moderately reduced when utilizing the DUST as reference (\SI{43}{\percent} or \SI{33}{\percent}, \SI{15}{\percent} or \SI{22}{\percent}, \SI{23}{\percent} or \SI{21}{\percent}, and \SI{19}{\percent} or \SI{7}{\percent} drop of $s_{d}$ for the DN7C, the PMSA, the HPMA, and the OPCN, respectively) and often significantly lowered when utilizing the REFN as reference (\SI{67}{\percent} or \SI{44}{\percent}, \SI{37}{\percent} or \SI{36}{\percent}, \SI{20}{\percent} or \SI{24}{\percent}, \SI{50}{\percent} or \SI{48}{\percent}, and \SI{59}{\percent} or \SI{49}{\percent} drop of $s_{d}$ for the PM sensors and the DUST, respectively). Although the absolute values of the sample means of corresponding Rel Resid failed to demonstrate reductions consistently (\SIrange{34}{77}{\percent} drop of $|\bar{d}\,|$ in most cases), the absolute values of those increased $\bar{d}$ were negligible ($|\bar{d}\,|<\SI{1.3}{\percent}$ for the PMSA and HPMA with reference to the DUST). As for those Resid from the corresponding models, eliminating the components having reference concentrations $\geq\SI[per-mode=symbol]{50}{\micro\gram\per\cubic\meter}$ resulted in reductions of the sample standard deviation values but rises of the absolute sample mean values. The respective $s_{\epsilon}$ values were moderately to significantly diminished (\SIrange{7}{38}{\percent}, \SIrange{22}{30}{\percent}, \SIrange{14}{42}{\percent}, \SIrange{44}{50}{\percent}, and \SIrange{33}{35}{\percent} drop, respectively), while the new $|\bar{\epsilon}\,|$ values became negligible to significant for different instruments (\SIrange[per-mode=symbol]{0.1}{1.3}{\micro\gram\per\cubic\meter}, \SIrange[per-mode=symbol]{0.3}{1.4}{\micro\gram\per\cubic\meter}, \SIrange[per-mode=symbol]{1.5}{2.0}{\micro\gram\per\cubic\meter}, \SIrange[per-mode=symbol]{2.1}{3.7}{\micro\gram\per\cubic\meter}, and \SIrange[per-mode=symbol]{4.6}{9.3}{\micro\gram\per\cubic\meter} for the HPMA, the PMSA, the DUST, the OPCN, and the DN7C, respectively). The increases of the new $|\bar{\epsilon}\,|$ values were anticipated, particularly for those associated with the OPCN and the DN7C, because the Resid from a simple calibration model would rarely follow a perfect normal distribution (see the fourth columns of Fig. \ref{fig:ErrorAnalysisSensDust} and Fig. \ref{fig:ErrorAnalysisSensDustRefn}). They might be mitigated to a negligible amount if their respective splitting values were determined carefully and individually. 

Overall, compared to using the REFN as reference, the OLS or the MLR Rel Resid and Resid of the PM sensors with the DUST as reference exhibited significantly improved statistics. The MLHT Rel Resid and Resid of the DN7C, but not those of the other three PM sensors, showed much better statistical performance than the MLR ones primarily because its outputs were considered to be susceptible to both the TEMP and RH. Rel Resid and Resid manifesting numerically better statistical performance can be further achieved by properly removing the components at low and high concentration levels, respectively.

\subsubsection{Evaluation Results and Discussion}
\label{SubSubSect:EvaluationAndDiscussionOfPMSensors}

After calibrating the nominated PM sensors and the DUST against the REFN with OLS and MLR models\footnote{The PM sensors were not calibrated against the DUST because some of them had demonstrated comparable or better performances when compared to the DUST according to the previous analyses.}, all the evaluation metrics introduced in section \ref{SubSubSect:EvaluationMetrics}, except the unit-wise precision $CV_{RMS}$ and the data completeness $\eta$, were calculated from their respective calibrated and reference concentration pairs; the results were presented in Tab. \ref{tab:EvaluationMetricsSensDustRefn}. Note that a concentration pair is valid when both the calibrated and the reference concentrations are $\geq\SI[per-mode=symbol]{3}{\micro\gram\per\cubic\meter}$. Moreover, in order to have sufficient numbers of low concentration pairs ($<\SI[per-mode=symbol]{3}{\micro\gram\per\cubic\meter}$) for calculating the limit of detection $LOD$, the RH condition is set to $\leq\SI{80}{\percent}$ in this field experiment instead of $\leq\SI{50}{\percent}$ as defined earlier.

In accordance with the acceptance criteria or requirements of these evaluation metrics, the cells in Tab. \ref{tab:EvaluationMetricsSensDustRefn} containing the acceptable values were highlighted in gray. It is apparent that none of the four nominated PM sensors (i.e., the DN7C, the PMSA, the HPMA, and the OPCN) nor the research-grade PM instrument (i.e., the DUST) met all the criteria, which is expected because the evaluation metrics are mainly originated from the data quality assessments for regulatory/conventional monitors. As aforementioned in section \ref{SubSubSect:EvaluationMetrics}, the purpose of performing field evaluation and calibration on a PM sensor is to get a better understanding of and improve the confidence in its measurement results; its data quality is not intended to be comparable with the conventional monitor's. All the evaluation metrics introduced in section \ref{SubSubSect:EvaluationMetrics} or presented in Tab. \ref{tab:EvaluationMetricsSensDustRefn} are the products of a decent attempt at unifying the collocated field evaluation for low-cost PM sensors or devices.

Although none of the PM instruments had fulfilled all the acceptance criteria, the PMSA and the HPMA usually yielded better results (i.e., lower $|Bias_{PEP}|$, $\sigma_{UCL}$, $|intercept|$, and $LOD$; $slope$ and $r$ closer to one) in comparison to the other sensors, even outperformed the DUST, a research-grade PM instrument, in most evaluation aspects. Additionally, the MLR calibrated measurements, compared to the OLS ones, of these instruments generally manifested better performances in most aspects, with the HPMA's $Bias_{PEP}$, the PMSA and HPMA's $slope$, the former two sensors and OPCN's $intercept$, and the former three sensors and DUST's $LOD$ as exceptions. Notice that the calculation results of $Bias_{PEP}$ and $\sigma_{UCL}$ are based on the values of both $\bar{d}$ and $s_{d}$ and that of $s_{d}$, respectively, which have been discussed in section \ref{SubSubSect:ResidualAnalysisOfPMSensors}. Therefore, the rise of the HPMA's MLR $|Bias_{PEP}|$ was attributed to the RH over adjusted measurements at low concentrations and mildly humid conditions (see those corresponding RH vs. Rel Resid scatter plots in Fig. \ref{fig:ErrorAnalysisSensDustRefn}). The rises of the $LOD$ might result from the same cause; and the dramatic boosts of the OPCN's and the DUST's $LOD$ were anticipated regarding the clumps of purple points above the horizontal zero lines at RH around \SI{60}{\percent} (over adjusted low concentration measurements; see the corresponding RH vs. Rel Resid scatter plots in Fig. \ref{subfig:CaliErrorMlrRefn}). The drops of the PMSA's and the HPMA's $slope$ might stem from the joint effects of their slight susceptibility to RH (refer to the hygroscopic growth) and the lack of high concentration measurements, but the variations of $slope$ (between OLS and MLH calibrated measurements) were relatively small ($\pm\SI{3}{\percent}$). For the OPCN's $|intercept|$, the rise was comparatively small ($\approx\SI{1}{\percent}$); though the relative increases of the PMSA's and the HPMA's $|intercept|$ were significant ($\approx\SI{14}{\percent}$ and $\approx\SI{31}{\percent}$, respectively), the absolute increases were minor ($<\SI[per-mode=symbol]{1.7}{\micro\gram\per\cubic\meter}$) considering their measurement resolutions (i.e., \SI[per-mode=symbol]{1}{\micro\gram\per\cubic\meter}).

Specifically comparing the evaluation results of the PMSA and the HPMA, calibrated with the MLH model, against that of the same calibrated DUST, of whom the evaluation results were considered as research-grade baselines, the PMSA's and the HPMA's $|Bias_{PEP}|$ were \SIrange{59}{71}{\percent} and \SIrange{69}{90}{\percent} of the DUST's, respectively. Further, their $\sigma_{UCL}$, $|intercept|$, and $LOD$ values were \SI{72}{\percent} and \SI{60}{\percent}, \SI{94}{\percent} and \SI{133}{\percent}, and \SI{55}{\percent} and \SI{71}{\percent} of the baselines, respectively. As for the $slope$ or $r$ values, they were $0.007$ and $-0.022$, or $0.003$ and $0.008$ closer to one, respectively, than the baseline.

Overall, the PMSA and the HPMA are the best candidates for constructing the wearable PM\textsubscript{2.5} monitor in the proposed system considering their collocated calibration and evaluation results in the field that are presented in section \ref{SubSubSect:CorrelationAndCalibrationOfPMSensors}, section \ref{SubSubSect:ResidualAnalysisOfPMSensors}, and section \ref{SubSubSect:EvaluationAndDiscussionOfPMSensors}. The properly calibrated and correctly RH adjusted PMSA and HPMA showed comparable or better field performances against the DUST. Further referring to the insusceptibility to condensed tiny droplets (detailed in section \ref{SubSubSect:DataCleansingOfPMSensors}) and the technical specifications including affordability, portability, and detection capability (illustrated in Tab. \ref{tab:SensorSpec}), the PMSA typically outperformed the HPMA. In spite of that, the PMSA and its sibling models are more extensively evaluated in environmental research than the HPMA (see section \ref{SubSubSect:HPMAIntro} and section \ref{SubSubSect:PMSAIntro}). Eventually, the PMSA (i.e., the PlanTower PMS-A003 sensor) was selected as the PM\textsubscript{2.5} monitoring core of the wearable device WePIN presented in section \ref{SubSect:SysImpWePIN}. The field performances of ten implemented WePIN with reference to the collocated REFN were investigated in section \ref{SubSect:EvaluationResultsOfWePINs}.

As discussed in section \ref{SubSubSect:CorrelationAndCalibrationOfPMSensors} and section \ref{SubSubSect:ResidualAnalysisOfPMSensors}, employing more advanced calibration models can improve the nominated sensors' data quality at the cost of limited transferability of the calibration parameter sets, and therefore, higher recalibration frequency is conceivable. In the context of developing a unified evaluation protocol for low-cost PM sensors/devices, a simple yet comparatively effective calibration model, e.g., an OLS or MLR model, is more appreciated owing to limited resources, varieties of models, convenience of access, etc. Given that the RH influences on a PM sensor's measurements were referred to as artefacts throughout this manuscript or in other existing studies, based on the authors' understanding, it is an improper usage because the PM sensors measure/detect what is actually present in the air, including moisten particles and condensed droplets; it is a real observation instead of an artefact \cite{jayaratne2018influence}. In this regard, evaluating and calibrating the PM sensors against collocated high-end instruments equipped with heater or dryer might compromise the performance results; the definition and regulation of certain PM specially for the PM sensors should be established.

Although multiple meaningful conclusions have been drawn from this collocated field experiment as stated in the previous sections, there exist limitations in the calibration and evaluation results. One of them is that only single unit of each nominated sensor and reference instrument was available in this experiment; there is a rare possibility that their measurements might not be representative of their kind or involve unusual occurrence(s). Additionally, this experiment was conducted on the coast side in Hong Kong during spring, in which case the marine climate introduces a significant portion of hygroscopic particles in the sampled aerosol, such that the results derived might not be applicable in locations with completely distinct climates. Moreover, the duration of this campaign might not be sufficient to characterize the micro-environment in the final deployment area(s), e.g., the nominated PM sensors were not confronted to the full range of aerosol concentrations and/or atmospheric conditions during this campaign. As presented in section \ref{SubSubSect:EvaluationMetrics}, the procedures and requirements of an adopted evaluation metric were not rigorously followed in some cases, which might bias its calculated values. Besides, as proposed in section \ref{SubSubSect:AutoDataCleansingOfPMSensors}, the automated approach for data cleansing and its user-configurable parameters were determined empirically based on the data pairs acquired in this experiment; therefore, they are not necessarily optimal and their transferability might be limited as well. Although the wearable devices integrating with the selected PM sensors are supposed to be carried along by the users, none of them have been evaluated for biases due to movement in this manuscript; however, study \cite{zamora2018field} suggested that the PMSA was not affected by movement.

\subsubsection{Automated Data Cleansing}
\label{SubSubSect:AutoDataCleansingOfPMSensors}

In this section, a real-time automated approach is proposed to identify and eliminate the PM sensors' unusual measurements, detecting the condensed tiny droplets as particles during extremely humid conditions to be specific. The foundation of this automated approach is the ratio between the PM\textsubscript{10} and PM\textsubscript{1.0} mass concentrations from a PM sensor measured at a certain moment. According to the statistics of a previous window of PM\textsubscript{10} and PM\textsubscript{1.0} ratios, in which no unusual measurement has occurred, the occurrence likelihood of the current ratio can be determined under proper assumptions; a threshold value can be empirically established to decide whether or not to accept the current ratio. A rejected ratio is considered to be indicating the occurrence of unusual measurement, that is, the PM sensor is probably counting the condensed tiny droplets as particles. Consequently, the current PM\textsubscript{2.5} mass concentration measurement should be eliminated such that an effective RH correction is achievable in the case of collocated calibration and the data integrity is improved in the case of real-world deployment. Because the DN7C outputs PM\textsubscript{2.5} measurements only and the HPMA's PM\textsubscript{10} outputs are calculated from the corresponding PM\textsubscript{2.5} mass concentrations showing differences around \SIrange[per-mode=symbol]{1}{2}{\micro\gram\per\cubic\meter} for most of the time\footnote{This phenomenon was observed in this field experiment as well as in study \cite{tiele2018design}.}, this automated approach is only applicable to the PMSA and the OPCN. The underlying philosophy along with the detailed implementation of this approach are presented in the following paragraphs.


Referring to section \ref{SubSubSect:DataCleansingOfPMSensors}, a PM sensor generally overestimates the mass concentrations of certain PM during extremely humid conditions compared to a reference instrument equipped with a heater or a dry at its inlet. Such overestimation can be primarily attributed to the hygroscopic growth of particles or misdetecting the condensed tiny droplets as particles, and the impacts on the sensor measurements from the latter factor are much more substantial. As discussed previously, the unusual measurements that have been highlighted in the salmon spans manually as depicted in Fig. \ref{fig:PM2p5RhTempFull} to Fig. \ref{fig:PM2p5RhZoom3} were considered to be caused by the latter factor\footnote{The influences from the hygroscopic growth of particles may also occur concurrently during these highlighted intervals but they were considered to be comparatively negligible.}. Upon further investigating the scatter plots in Fig. \ref{fig:ScatterSensDust} and Fig. \ref{fig:ScatterSensDustRefn}, the authors believe that it is impossible to perform any effective RH correction on these unusual measurements with the information available in this field experiment. Conceivably, the unusual measurements that involved misdetecting the condensed droplets as particles should be recognized and eliminated such that the remaining measurements were influenced by hygroscopic growth only, in which case the overestimated measurements could be properly adjusted via MLH model during the calibration phase. During the deployment phase, the ability of identifying such unusual occurrences, in which dramatic differences between the sensor measurements and the actual concentrations that are unlikely to be adjusted will be introduced, enhances the data integrity by improving the confidence in a PM sensor's measurements. However, manually recognizing and eliminating these unusual occurrences via post-processing is neither feasible in deployment phase nor practical for large-scale deployment. Overall, a real-time automated approach that exclusively employs the information available from a PM sensor or device is essential.

Given that the unusual occurrences can not be determined by the corresponding RH readings only as analyzed in section \ref{SubSubSect:DataCleansingOfPMSensors}, attention was drawn to the analytic of sensor responses during these events. In accordance with studies \cite{price2011radiation,atmospheric2018clouds,liu2020fog}, the diameters of those condensed tiny droplets forming fog or mist have means ranged from \SIrange{5}{15}{\micro\meter}, and can be widely distributed in range of \SIrange{1}{100}{\micro\meter}. During the events of misdetecting the condensed droplets as particles by PM sensors, the authors anticipate that the mass concentration measurements of the large sized particles, but not small sized ones, will rise dramatically and cause huge ratios between the PM\textsubscript{10} and the PM\textsubscript{2.5} or PM\textsubscript{1.0}. The particle number counts of the large and the small sized particles are as well anticipated to manifest a similar phenomenon. In fact, both of the above phenomena were observed during the presence of fog in study \cite{jayaratne2018influence}. Although the particle number count information was not recorded in this field experiment, it is believed that analyzing the ratios between different particle size bins, if available, may probably lead to better recognition of the misdetecting events due to typically higher resolution of as well as lower limit on particle diameter (6 and 16 particle bins with lowest particle diameters $=\SI{0.3}{\micro\meter}$ and $=\SI{0.38}{\micro\meter}$ for the PMSA and OPCN, respectively).

Considering that the aerosols' properties, including the particle size distribution, may vary over time and across locations as presented in section \ref{SubSubSect:UncertaintyOfInstruments}, instead of employing a constant threshold of ratio between the large and the small sized PM to determine a misdetecting event, a more lenient threshold is appreciated. Therefore, in this approach, the ratio threshold at a certain moment is decided by a moving window containing a set of recent ratios that are carefully selected and considered to be involved no misdetecting event. The ratios in the moving window can be initialized with typical values and values that are calculated from recent PM measurements obtained under dry or slightly humid conditions, during which a misdetecting event is not anticipated to occur (see Fig. \ref{fig:PM2p5RhTempFull} to Fig. \ref{fig:PM2p5RhZoom3}). To avoid generating exceptionally large ratios due to significantly small denominators, any ratios calculated from fairly low PM concentrations are excluded. Instead, the corresponding PM\textsubscript{2.5} measurements are directly accepted as valid ones because no extreme outliers that indicated the occurrences of misdetecting were observed in the OPCN and the PMSA associated scatter plots when their reported PM\textsubscript{2.5} were fairly low (see the first columns of Fig. \ref{fig:ScatterSensDust} and Fig. \ref{fig:ScatterSensDustRefn}). 

\begin{algorithm}[H]
	\caption{PM ratio based real-time data cleansing method}
	\label{algo:AutoDataCleansing}

	\begin{algorithmic}[1]
		\REQUIRE
		$\beta \in \mathbb{R}_{0}^{+}$: multiplying factor; \\
		$c_{l} \in \mathbb{R}_{0}^{+}$: threshold for low PM\textsubscript{2.5} mass concentration;\\
		$h_{l} \in \{h \in \mathbb{R}_{0}^{+} \mid h \leq 100\}$: threshold for low RH level; \\
		$w_{s} \in \{w \in \mathbb{Z}^{+} \mid w \geq3\}$: size of moving window; \\

		$\vec{p}_{t}=(o_{t}, c_{t}, e_{t}) \in (\mathbb{R}_{0}^{+})^{3}$: mass concentration measurements from a PM sensor at current time instant $t \in \mathbb{N}_{0}$, where $o_{t}$, $c_{t}$, and $e_{t}$ are the mass concentrations of PM\textsubscript{1.0}, PM\textsubscript{2.5}, and PM\textsubscript{10} at current time instant $t$ ($o_{t} \leq c_{t} \leq e_{t}$), respectively; \\

		$h_{t} \in \{h \in \mathbb{R}_{0}^{+} \mid h \leq 100\}$: RH condition at current time instant $t \in \mathbb{N}_{0}$; \\

		$\vec{w}=(r_{i},r_{j}, \cdots, r_{k}) \in (\mathbb{R}^{+})^{w_{s}}$: moving window containing previous PM\textsubscript{10} and PM\textsubscript{1.0} ratios (without ratios from unusual measurements), where $i,j,\cdots,k \in \mathbb{N}_{0}$ are time instants ($i < j < \cdots < k < t$), and $r_{i}, r_{j}, \cdots, r_{k}$ are the respective ratio values; \\

		$mean(\vec{w})$ and $std(\vec{w})$: functions for calculating the sample mean and the sample standard deviation of ratios in the moving window $\vec{w}$, respectively.\\

		\ENSURE
		$|{\vec{w}}'| = w_{s}$. \\

		\STATE $\beta \leftarrow 2.5$, $c_{l} \leftarrow 20.0$, $h_{l} \leftarrow 80.0$, $w_{s} \leftarrow 30$ \COMMENT{empirically determined user configurable parameters}
		\STATE $r_{t} \leftarrow e_{t} / o_{t}$ \COMMENT{calculate the current ratio value}
		\STATE ${\vec{w}}' \leftarrow \vec{w}$

		\IF {$c_{t} < c_{l}$}
		\RETURN ${\vec{w}}'$, $c_{t}$ \COMMENT{no update on the moving window}
		\ELSIF {$h_{t} < h_{l}$ \OR $r_{t} < mean({\vec{w}}') + \beta \cdot std({\vec{w}}')$}
		\STATE ${\vec{w}}' \leftarrow (r_{j}, \cdots, r_{k}, r_{t})$ \COMMENT{updated moving window}
		\RETURN ${\vec{w}}'$, $c_{t}$
		\ELSE
		\RETURN ${\vec{w}}'$ \COMMENT{unusual measurement detected; no update on the moving window}
		\ENDIF
	\end{algorithmic}
\end{algorithm}

The moving window will be updated from time to time by replacing the most previous ratios with the most recent ones if no misdetecting events are considered to be occurring. Such a mechanism enables the variations in particle size distribution, which are unlikely attributed to any misdetecting events, to be captured promptly such that the timeliness of a ratio threshold together with the transferability of this approach are improved. Given the current concentrations of small sized PM are fairly high, an update will be initiated if the current RH condition is non-extreme or the calculated ratio is not statistically unusual (assuming the ratio distribution in a reasonable period of time with misdetecting events excluded is Gaussian); otherwise, an unusual ratio under extreme humid condition is considered to be indicating a misdetecting event, and the corresponding PM measurements should be eliminated. The pseudocode for this update procedure is presented in Algo. \ref{algo:AutoDataCleansing}. Note that the values of the user configurable parameters, namely $\beta$, $c_{l}$, $h_{l}$, and $w_{s}$ that determine the upper limits of acceptable ratios, fairly low PM concentrations, and non-extreme RH conditions, and the size of moving window, respectively, were empirically chosen according to the performance on the collocated pairs acquired in this field experiment. In the pseudocode, the PM\textsubscript{2.5} outputs from a PM sensor are used to determine the PM concentration levels. Also, the RH outputs from the REFN are employed to quantify the humidity conditions as they are often unavailable from a PM sensor; the applicability of this approach will not be compromised because a RH sensor can be easily integrated into a PM device such as the WePIN. The mass concentration ratios between PM\textsubscript{10} and PM\textsubscript{1.0}, instead of that between PM\textsubscript{10} and PM\textsubscript{2.5}, are calculated considering the distribution range of condensed droplets (i.e., \SIrange{1}{100}{\micro\meter}); preliminary results from experiments using the former ratios generally manifested better performance as well.

The PMSA's and the OPCN's PM ratios along with their corresponding acceptable thresholds at each 1-minute interval are depicted in Fig. \ref{fig:AutoCleansingRatio}, in which the salmon spans represent the three intervals of unusual sensor measurements that were manually identified in Fig. \ref{fig:PM2p5RhTempFull}. For both PM sensors, most of the ratios within the first interval were recognized as unusual by the proposed approach, while a handful of the ratios right after the first interval were as well recognized as unusual for the PMSA. Specifically, within the second and third intervals, few of the ratios were recognized as unusual for the PMSA, while in contrast, besides most of the ratios in both of these intervals, additional ratios were recognized as unusual for the OPCN. Such behavior difference may probably be attributed to the PMSA's superior insusceptibility to condensed droplets as discussed in section \ref{SubSubSect:DataCleansingOfPMSensors}. Note that the sets of remaining ratios with comparatively large values and fluctuations on the right sides of Fig. \ref{subfig:AutoCleansingPMSARatio} and Fig. \ref{subfig:AutoCleansingOPCNRatio} (started around 14:00 on 22 February) were resulted from low concentration levels of PM\textsubscript{1.0} reported by these PM sensors. The scatter plots and pairwise correlations (in terms of $R^2$) between these sensors' processed PM measurements (labeled as either remaining or eliminated PM\textsubscript{2.5} by the proposed approach) and the reference PM\textsubscript{2.5} mass concentrations are provided in the first column of Fig. \ref{fig:AutoCleansingScatter}. The calibration results including scatter plots, $R^2$ values, and regression parameters of the OLS, the MLH, and the advanced empirical (ADV) models on the remaining PM\textsubscript{2.5} pairs determined by this approach are shown in the second, the third, and the fourth columns of Fig. \ref{fig:AutoCleansingScatter}, respectively. Note that, due to the issue of dual linear relationships as presented in section \ref{SubSubSect:DataCleansingOfPMSensors}, the collocated data pairs acquired before 17 February were excluded when using the REFN as reference.

Given the difficulty in accurately labeling all the misdetecting events in this field experiment, the proposed approach was empirically evaluated by investigating the scatter plots and the correlations of the three sets of collocated data pairs that were not-cleansed, automatically-cleansed, and manually-cleansed. Referring to the scatter plots in the first column of Fig. \ref{fig:AutoCleansingScatter}, the data points associated with unusual sensor measurements that should be eliminated according to the proposed approach were marked in orange, while those points associated with the remaining sensor measurements were highlighted in blue. The scatter plots within Fig. \ref{fig:ScatterSensDust} to Fig. \ref{fig:ScatterCaliSensRefn} should be referenced to from time to time in the discussion below. Also, instead of evaluating the proposed approach against those RH correction methods introduced in existing studies \cite{zhang1994mie,zheng2018field,crilley2018evaluation} (every approach is highly specific to the study site owing to its great reliance on the aerosol's properties), it can be adopted as their preprocessing stage to remove sensor measurements that were unlikely to be corrected.

In the PMSA-DUST scatter plot, it is apparent that only a small cluster of data points (the orange ones at PMSA PM\textsubscript{2.5} $\approx\SI[per-mode=symbol]{100}{\micro\gram\per\cubic\meter}$) on the identify line can be certainly identified as outliers by visual inspection; the majority of orange points were overlaying on top of the blue ones. Apart from incorrect labeling by the proposed approach (referring to those unusual ratios right after the first salmon span as shown Fig. \ref{subfig:AutoCleansingPMSARatio}), most of these overlaid orange points were likely attributed to the PMSA's excellent RH insusceptibility as discussed earlier such that the PMSA and the DUST yielded similar responses to (not or slightly influenced by) condensed droplets for most of the time. Hence, it is expected that the correlations of the automatically- and the manually-cleansed PMSA-DUST data pairs manifested marginal improvements over that of the not-cleansed pairs ($\approx\SI{4}{\percent}$ and $\approx\SI{6}{\percent}$ of $R^2$, respectively). The calibration results differences between the automatically- and the manually-cleansed data pairs are also expected to be small (e.g., $\approx\SI{-2}{\percent}$ in $R^2$).

As for the PMSA-REFN scatter plot, to the lower right of the identify line, a noticeable cluster (the blue and the orange points at PMSA PM\textsubscript{2.5} $\approx\SI[per-mode=symbol]{50}{\micro\gram\per\cubic\meter}$) and a small cluster (the orange ones at PMSA PM\textsubscript{2.5} $\approx\SI[per-mode=symbol]{100}{\micro\gram\per\cubic\meter}$) of data points can be visually identified as outliers. It should be note that these clusters represented the data pairs acquired on the first half of 17 February (that is, within the first salmon span of Fig. \ref{fig:PM2p5RhTempFull}), since when the aerosol's composition (chemical composition and/or size distribution) was changed causing the dual-linear-relationships issue (see section \ref{SubSubSect:DataCleansingOfPMSensors}). As the variation in the remaining ratios before and after 17 February is rather small, and the corresponding remaining ratios within the first salmon span are comparatively low (see Fig. \ref{subfig:AutoCleansingPMSARatio}), the blue outliers within the noticeable cluster were probably resulted from the change of chemical composition rather than the misdetecting events. In this regard, these blue outliers did not signify that the proposed approach had failed to recognize a subset of the unusual sensor measurements but suggest that it would not be influenced by the variation in aerosol's chemical composition. No blue outliers were visually identified in the PMSA-DUST scatter plot primarily due to their similar detection principles. To the upper left of the identity line, those orange data points overlaying on top of the blue ones were corresponding to the incorrectly labeled ratios. The significant differences between the correlations or the calibration results of the automatically- and the manually-cleansed PMSA-REFN data pairs should be primarily contributed by the blue outliers ($\approx\SI{-12}{\percent}$ in $R^2$).

In the OPCN-DUST and the OPCN-REFN scatter plots, no apparent blue outliers, which were attributed to the change of chemical composition, were discovered since almost all of the ratios in the first salmon span of Fig. \ref{subfig:AutoCleansingOPCNRatio} were recognized as unusual. It also partly justified the previous conjecture that the OPCN has higher RH susceptibility than the PMSA. The group of blue points to the lower right of the identity line (at OPCN PM\textsubscript{2.5} $\approx\SI[per-mode=symbol]{75}{\micro\gram\per\cubic\meter}$ to \SI[per-mode=symbol]{140}{\micro\gram\per\cubic\meter}) in the latter scatter plot, which appeared to be outliers, should not be counted as unusual measurements caused by misdetecting events as they could be properly adjusted by the advanced empirical model. Due to incorrect labeling by the proposed approach (referring to those unusual ratios right ahead the second salmon span as shown in Fig. \ref{subfig:AutoCleansingOPCNRatio}), the orange points overlaying on top of the blue ones in both scatter plots are anticipated. Additional overlaid orange points were noticed within the former scatter plot; it might be ascribed to the similar detection principle but different RH susceptibility between the OPCN and the DUST as discussed earlier. It is noteworthy that, within both scatter plots, all the outliers that can be certainly identified by visual inspection have been recognized by this automated approach; therefore, the correlations of the automatically-cleansed pairs exhibited dramatic improvements over that of the not-cleansed ones ($\approx\SI{41}{\percent}$ and $\approx\SI{505}{\percent}$ of $R^2$, respectively). Whilst the correlations together with the calibration results between the automatically- and the manually-cleansed OPCN-DUST pairs were demonstrating negligible differences ($\approx\SI{-0.7}{\percent}$ in $R^2$), the corresponding differences of the OPCN-REFN pairs were noticeable ($\approx\SI{-10}{\percent}$ in $R^2$; mainly contributed by that group of blue points aforementioned in this paragraph).

To sum up, this real-time automated approach based on the PM sensor's PM\textsubscript{10} and PM\textsubscript{1.0} mass concentration ratio is able to identify and eliminate all the unusual sensor measurements, which were considered to be caused by miscounting the tiny condensed droplets as particles (significantly lower the sensor-reference correlation and can not be properly adjusted), at the cost of incorrectly labeling a few of the sensor measurements. The incorrectly labeled sensor measurements exert negligible influences on the collocated calibration and evaluation results given that the number of sensor-reference measurement pairs is sufficient.

\subsection{Evaluation and Calibration of WePINs}
\label{SubSect:EvaluationResultsOfWePINs}

The second field evaluation campaign was carried out from 26 June to 17 July 2019 in order to evaluate and calibrate the ten implemented WePINs in the field. To investigate the field performances of these wearable PM devices that were named from WP00 to WP09, they were first compared against each other; then, their unit-wise average measurements, denoted by WPUW, were compared against the corresponding collocated measurements from the REFN. The experimental setup details are provided in section \ref{SubSubSect:EvalAndCaliOfWePINs}. Firstly, the availability of these devices' trimmed readings and the calculation of the WPUW readings are presented in section \ref{SubSubSect:DataPreprocessingOfPWePINs}. Then in section \ref{SubSubSect:DataCleansingOfWePINs}, three intervals of unusual reference PM\textsubscript{2.5} measurements with significant and rapid fluctuations or showing distinct trends to the WPUW PM\textsubscript{2.5} were identified; collocated data pairs within these intervals were excluded when evaluating and calibrating the WePINs against the REFN.

In section \ref{SubSubSect:CorrelationAndCalibrationOfWePINs}, the pairwise correlations of PM\textsubscript{2.5} among the PM devices and the collocated calibration of the WPUW-REFN PM\textsubscript{2.5} data pairs with different parametric models were investigated. The results indicated that these PM devices were moderately correlated with each other owing to the significant random fluctuations in their PM\textsubscript{2.5} measurements and the low concentration level in this evaluation campaign. Compared to those PMSA-REFN data pairs, the WPUW-REFN ones (both raw and calibrated) yielded diverse or contrary interpretations of the field performances; subsequent experiments are needed to get better understanding of the WePIN' field performances. The residuals between the calibrated WPUW PM\textsubscript{2.5} measurements and the REFN outputs were analyzed in section \ref{SubSubSect:ResidualAnalysisOfWePINs}. Compared to the OLS model, involving RH in the MLR model typically improved the statistical performance of residuals but over adjusted those measurements at low concentration levels and mildly humid conditions. Such over adjustment was fairly mitigated by the ADV model whose residuals yielded the best statistical performance. Lastly, the evaluation metrics detailed in section \ref{SubSubSect:EvaluationMetrics} were calculated from the calibrated WPUW-REFN PM\textsubscript{2.5} data pairs (with the OLS and MLH models) and discussed in section \ref{SubSubSect:EvaluationAndDiscussionOfWePINs}. Compared with the OLS calibrated WPUW-REFN data pairs, the MLH calibrated ones produced better results in most evaluation aspects. The limitations that were identified in this field experiment were also discussed in section \ref{SubSubSect:EvaluationAndDiscussionOfWePINs}.

\subsubsection{Data Preprocessing}
\label{SubSubSect:DataPreprocessingOfPWePINs}

The availability of device readings (TEMP, RH, PM\textsubscript{1.0}, PM\textsubscript{2.5}, and PM\textsubscript{10}) with \SI{1}{\minute} sampling interval retrieved from the WePINs' microSD cards is shown in Fig. \ref{fig:WePINsDataAvali}. All of these PM devices operated normally in the first two days, and the WP00, WP05, and WP08 managed to operate stably through the entire campaign. However, the rest of them encountered various malfunctions, which are believed to be resulting from the defects introduced during the manual manufacturing process, that dramatically diminished their data availability. Note that a daily charging routine was established from around 13:30 to 16:00, during which no readings would be recorded. In order to mitigate the interference from human interactions, the first and last hours of readings were trimmed resulting in the exact start and end times of this campaign as 2019-06-26 16:30:00 and 2019-07-17 12:59:00, respectively. Besides, the half hours of readings acquired before and after the charging routine were trimmed as well. Since access to the roof of the Supersite to charge the PM devices was denied on rainy days, they were not operational from 29 June to 4 July, on 7 July, and from 11 to 14 July.

The data completeness of the REFN was $\eta\approx\SI{99.9}{\percent}$ with 24 missing readings identified from 12:00 to 16:00 on 16 July. For the ten WePINs, the data completenesses of most of them were $\eta>\SI{99.8}{\percent}$, while the $\eta$ values for the WP02 and WP07 were $\approx\SI{97.9}{\percent}$ and $\approx\SI{91.8}{\percent}$, respectively; the WP07 had the minimum data availability as well. Given that there existed a consecutive number of corrupted readings from these devices, they were manually eliminated instead of being replaced with the last valid readings.

The time series plots illustrating the 1-minute interval readings retrieved from the WePINs and the REFN are provided in Fig. \ref{fig:WePINsPMDataPlot} and Fig. \ref{fig:WePINsRefnAllDataPlot}; the ratios between the PM\textsubscript{10} and the PM\textsubscript{1.0} mass concentrations measured by each WePIN are also depicted in Fig. \ref{fig:WePINsPMDataPlot}. The unit-wise averages (WPUW) of the five types of measurements acquired by the PM devices were calculated by Eq. \ref{Eq:MeanConcentrationTimeInstant} and highlighted in respective time series plots (black); so did that of the PM ratios. As stated in section \ref{SubSubSect:EvaluationMetrics}, at any time instant, measurements from at least three units were required to calculate the WPUW reading. They are considered to be representative measurements for the ten PM devices when evaluating and calibrating against the REFN.

\subsubsection{Data Cleansing}
\label{SubSubSect:DataCleansingOfWePINs}

As shown in the top and middle plots of Fig. \ref{fig:WePINsRefnAllDataPlot}, abrupt variations in the TEMP and RH readings of the WP09 were observed at around 20:00 on 5 July, which were probably resulted from the glitches on the hand-soldered circuit board. Although the corresponding PM readings were normal, all types of measurements retrieved from this device from 19:30 to 20:29 on 5 July were removed for the sake of convenience. In the bottom plot, three time intervals that were considered to contain abnormal reference measurements were highlighted in salmon, within which the WPUW and reference PM\textsubscript{2.5} collocated pairs should be eliminated when evaluating and calibrating the WePINs against the REFN. The rationale is provided in the following.

Through visual inspection, significant and rapid fluctuations with extreme values in the REFN's PM\textsubscript{2.5} measurements were identified on the left and right sides of the bottom plot in Fig. \ref{fig:WePINsRefnAllDataPlot}. Because the Supersite is located on the hillside next to the bay without any apparent PM sources in the surrounding areas, it is monitoring the PM\textsubscript{2.5} concentration in the ambient background. Therefore, these significant and rapid fluctuations are considered as unusual occurrences. For simplicity reasons, the entire hours of REFN's PM\textsubscript{2.5} measurements experiencing such fluctuations were marked as unusual and highlighted in salmon (the left salmon span ended at 20:59 on 26 June and the right one started at 12:00 on 17 July). The exact cause of such fluctuations is inconclusive.

Besides, apparently distinct trends between the WPUW and the reference PM\textsubscript{2.5} measurements were observed from 18:00 on 15 July to 23:59 on 16 July (i.e., the middle salmon span), during which the reference PM\textsubscript{2.5} values were markedly larger than the WPUW ones. Specifically at around 8:00 to 12:00 on 16 July, the WPUW and the reference PM\textsubscript{2.5} values exhibited two completely opposite trends and respectively reached their local minimum and local maximum at about 10:00. Given that the WePINs' PM\textsubscript{2.5} readings within the corresponding interval showed comparable trends (see Fig. \ref{fig:WePINsPMDataPlot}), the above distinct trending behavior between the WPUW and the REFN should not be ascribed to these PM devices' operating improperness. It was also not resulted from these devices' misdetecting issue considering the fairly constant PM ratios (see the bottom plot of Fig. \ref{fig:WePINsPMDataPlot}) and the non-extreme humidity readings ($RH\leq\SI{80}{\percent}$; see the middle plot of Fig. \ref{fig:WePINsRefnAllDataPlot}) in the corresponding interval. Note that the comparatively large ratio values from 5 to 6 July and at about 15:00 on 10 July were attributed to the low PM\textsubscript{1.0} concentrations ($<\SI[per-mode=symbol]{5}{\micro\gram\per\cubic\meter}$) instead of misdetecting of condensed tiny droplets.

\begin{figure}[!t]
\centering
	\includegraphics[width=0.67\linewidth]{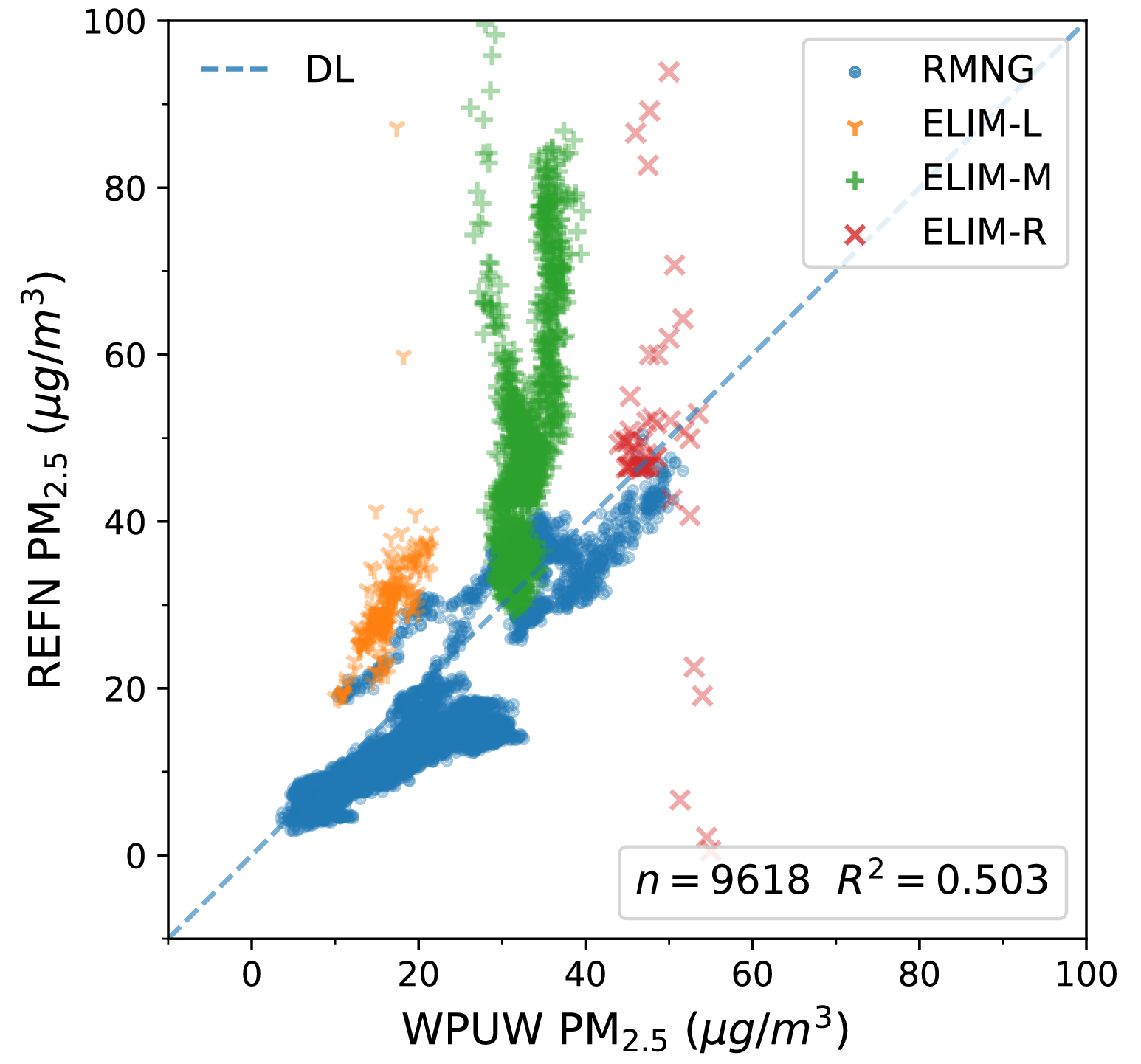}
\caption{Scatter plot and pairwise correlation (in term of $R^2$) of the collocated data pairs (second field experiment conducted from 26 June to 17 July 2019; $n=9618$) between the REFN (vertical axis; 1-minute interval PM\textsubscript{2.5} mass concentrations) and the WePINs (horizontal axis; 1-minute interval unit-wise average PM\textsubscript{2.5} mass concentrations). The blue dashed line (DL) is the identity line. The ELIM-L (orange), ELIM-M (green), and ELIM-R (red) data points represent the WPUW-REFN data pairs in the left, middle, and right salmon spans of Fig. \ref{fig:WePINsRefnAllDataPlot}, respectively, while the RMNG (blue; $n=7698$) points represent the remaining data pairs.}
\label{fig:ScatterElimWePINsRefn}
\end{figure}

Although the exact cause of this distinct trending behavior is also inconclusive, a plausible hypothesis has been conceived empirically. That is, it might be associated with the occasional construction work conducted near the Supersite, during which higher density particles that were difficult to be drawn into the scattering zone of a WePIN's PM sensor were introduced. As illustrated in Fig. \ref{fig:ScatterElimWePINsRefn}, the data points associated with the three highlighted intervals (the orange, green, and red points) were significantly differed from the majority (the blue ones), which could be considered as outliers and should be excluded when evaluating and calibrating the WePINs against the REFN. The colored points overlaying on the blue ones were the data pairs used to pad the three highlighted intervals such that each one of them was started and ended on the hour. The correlation of the cleansed WPUW-REFN PM\textsubscript{2.5} data pairs was significantly improved ($\approx\SI{49}{\percent}$ of $R^2$).

\subsubsection{Correlation and Calibration Analyses}
\label{SubSubSect:CorrelationAndCalibrationOfWePINs}

\begin{figure}[!t]
\centering
	\includegraphics[width=0.98\linewidth]{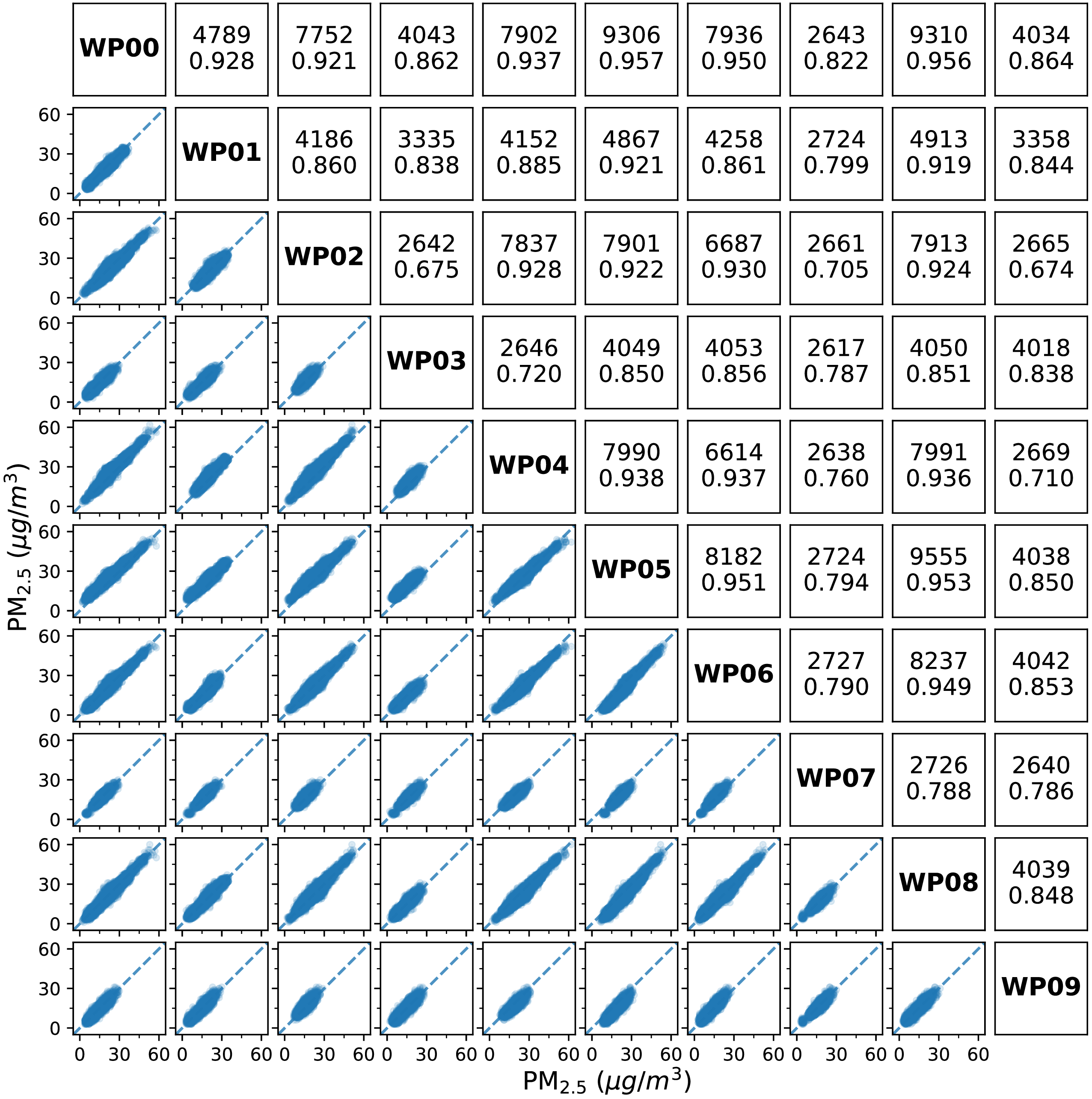}
\caption{Pairwise correlations (in terms of $R^2$) of the collocated data pairs (second field experiment from 26 June to 17 July 2019) among the ten WePINs (1-minute interval). The diagonal panels list the names assigned to these PM devices. For the scatter plot to the left of each name panel, its vertical axis represents the values measured by that device; for that at the bottom of each name panel, its horizontal axis represents the values measured by that device. The upper right panels provide the numbers of collocated data pairs (upper values) and the $R^2$ values (lower values) of the corresponding scatter plots in the diagonal symmetry panels. The blue dashed lines are the identity lines.}
\label{fig:ScatterPlotsWePINs}
\end{figure}

Among these ten implemented WePINs, the pairwise correlations of their PM\textsubscript{2.5} mass concentrations were investigated and the results (scatter plots and $R^2$ values) are provided in Fig. \ref{fig:ScatterPlotsWePINs}. The numbers of collocated data pairs varied from $n=2617$ to $n=9555$ due to different data availability in these devices. The $R^2$ values, ranged from $0.674$ to $0.957$, were typically proportional to the numbers of data pairs given that the mass concentration levels of the small sets of data pairs were fairly low compared to the measurement resolution (i.e., concentration level $\leq\SI[per-mode=symbol]{30}{\micro\gram\per\cubic\meter}$ with \SI[per-mode=symbol]{1}{\micro\gram\per\cubic\meter} resolution). The pairwise correlations of PM\textsubscript{2.5} among these devices were considered moderate (sample mean $\overline{R^2}=0.859$ and sample standard deviation $s_{R^2}=0.081$).

Referring to section \ref{SubSect:SysImpWePIN}, a WePIN takes a single measurement at each sampling interval, in which case no measure has been adopted to mitigate the errors/uncertainties arising from random effects, such as noise, in the measurement results. As a result, the measurement results might be heavily subject to the random effects. An example would be these devices' PM measurements that exhibited significant random fluctuations, particularly for the large sized PM, as illustrated in Fig. \ref{fig:WePINsPMDataPlot} (negligible random fluctuations were observed in the devices' TEMP and RH readings in Fig. \ref{fig:WePINsRefnAllDataPlot}). Apparently, the PM\textsubscript{10} mass concentrations from the WP05 (in brown) were typically fluctuating under those from the WP09 (in cyan), whereas the contrary was the case for their PM\textsubscript{2.5} measurements. It might result from that the utilized PM sensors were undersizing (for WP05) or oversizing (for WP09) those particles with marginal diameters (i.e., around \SI{2.5}{\micro\meter}).

Except the fairly low PM\textsubscript{2.5} concentrations in this field experiment, the moderate pairwise correlations of PM\textsubscript{2.5} among these devices were attributed to the random fluctuations in the PM measurements as well. However, the random fluctuations can be greatly mitigated by enabling a WePIN to take multiple measurements and output the average value at each sampling interval at the cost of slightly greater power consumption. Or simply average their 1-minute interval PM measurements over a longer time period via post-processing at the cost of lower temporal resolution. For instant, averaging these devices' PM measurements into 5-minute interval would result in $\approx\SI{9}{\percent}$ to $\approx\SI{31}{\percent}$ improvements in the mean $R^2$ (the $\overline{R^2}$ of the PM\textsubscript{1.0}, PM\textsubscript{2.5}, and PM\textsubscript{10} are increased from $0.889$ to $0.971$, $0.859$ to $0.964$, and $0.694$ to $0.908$, respectively\footnote{The mean pairwise correlations among the PM devices' RH and TEMP readings, both 1-minute interval and 5-minute averaged, are $\overline{R^2}=0.926$ and $\overline{R^2}=0.939$, respectively.}. Conceivably, it also would bring about improved unit-wise precision of these PM devices (e.g., $CV_{RMS}\approx\SI{15}{\percent}$ and $\approx\SI{12}{\percent}$ for the 1-minute interval and the 5-minute averaged PM\textsubscript{2.5} measurements). As the unit-wise precision is quantified by the $CV_{RMS}$ calculated from Eq. \ref{Eq:MeanConcentrationTimeInstant} to Eq. \ref{Eq:UnitWisePrecision}, the evaluation results are also heavily dependent on the PM\textsubscript{2.5} concentration levels in the evaluation campaign. Therefore, it is specified in the Part 53SC that the collocated reference PM\textsubscript{2.5} concentrations should be between \SI[per-mode=symbol]{3}{\micro\gram\per\cubic\meter} and \SI[per-mode=symbol]{200}{\micro\gram\per\cubic\meter} when calculating the $CV_{RMS}$\footnote{Without limitation on the reference concentrations, the $CV_{RMS}$ of the 1-minute interval and the 5-minute averaged measurements are $\approx\SI{11}{\percent}$ and $\approx\SI{6}{\percent}$ for the PM\textsubscript{1.0}, and $\approx\SI{20}{\percent}$ and $\approx\SI{18}{\percent}$ for the PM\textsubscript{10}.}. As aforementioned, a WePIN can be enabled to acquire multiple measurements and take average within each sampling interval (firmware upgrade needed) to reduce the random fluctuations in its PM measurements, the results (e.g., pairwise correlation and unit-wise precision) from the 5-minute averaged readings would be more representative for its typical performances in the field, whereas those from the 1-minute interval ones could be considered as lower bounds.

In this field experiment, rather than calibrating every single WePIN against the REFN, their WPUW PM\textsubscript{2.5} measurements are considered representative and the calibration results were achieved by fitting the WPUW-REFN PM\textsubscript{2.5} data pairs into a set of parametric models, namely the OLS, the MLH, and the ADV model\footnote{The MLT and MLHT models were tested but not shown. Similar to the PMSA sensor, the TEMP (\SIrange{28}{38}{\celsius} from WPUW or \SIrange{25}{36}{\celsius} from REFN) had negligible effects on the WePINs' PM\textsubscript{2.5} measurements.}\textsuperscript{,}\footnote{The ADV model served as evidence of the existence of better calibration models only because it violated the key assumptions of a MLR model detailed in section \ref{SubSubSect:AssumptionInEvalCali}; the calibrated WPUW PM\textsubscript{2.5} measurements from it should be omitted when evaluating against the REFN in section \ref{SubSubSect:EvaluationAndDiscussionOfWePINs}.}, in which case the results, such as calibration parameters, would not be significantly biased owing to narrow concentration range (for devices with limited data availability) and/or random fluctuations (partially mitigated by unit-wisely averaging their PM measurements). The fundamental rationale behind the representativeness of the WPUW measurements is detailed in the following paragraph.

Considering the strong pairwise correlations among the ten PM devices, the excellent correlation of each WePIN-WPUW pair is anticipated ($\overline{R^{2}}=0.990$ and $s_{R^2}=0.006$ on 5-minute averaged PM\textsubscript{2.5}). Hence, the PM\textsubscript{2.5} measurements from every WePIN, denoted by $\vec{x}_{w}$, can be represented as:
\begin{equation}
\label{Eq:WpuwOLSFormula}
	\vec{x}_{w} = \beta^w_0 + \beta^w_1\vec{x}_{u} + \vec{\epsilon}_{w}
\end{equation}
where $\vec{x}_{u}$ are the WPUW PM\textsubscript{2.5} measurements. Applying the OLS model's calibration parameters $\beta^u_0$ and $\beta^u_1$ derived from the WPUW-REFN PM\textsubscript{2.5} pairs on those measurements from a WePIN then results in:
\begin{equation}
\label{Eq:WpuwRefnOLStoWepin}
		\beta^u_0 + \beta^u_1\vec{x}_{w} = \beta^u_0 + \beta^u_1\beta^w_1\vec{x}_{u} + \beta^u_1\beta^w_0 + \beta^u_1\vec{\epsilon}_{w}.
\end{equation}
In view of that the WePIN-WPUW PM\textsubscript{2.5} data pairs (5-minute averaged) typically demonstrated unity slopes ($\overline{slope}=0.997$ and $s_{slope}=0.045$) but varying intercepts ($\overline{intercept}=0.107$ and $s_{intercept}=1.920$), the Eq. \ref{Eq:WpuwRefnOLStoWepin} can be approximated as:
\begin{equation}
\label{Eq:WpuwRefnOLStoWepinSimplfy1}
	\beta^u_0 + \beta^u_1\vec{x}_{w} \approx \beta^u_0 + \beta^u_1\vec{x}_{u} + \beta^u_1\beta^w_0 + \beta^u_1\vec{\epsilon}_{w} \textrm{,}
\end{equation}
that is:
\begin{equation}
\label{Eq:WpuwRefnOLStoWepinSimplfy2}
	\begin{aligned}
		\vec{y}_{r} & \approx \beta^u_0 - \beta^u_1\beta^w_0 + \beta^u_1\vec{x}_{w} - \beta^u_1\vec{\epsilon}_{w} - \vec{\epsilon}_{u} \\
					& \approx \beta_0 + \beta^u_1\vec{x}_{w} + \vec{\epsilon}
	\end{aligned}
\end{equation}
where $\vec{y}_{r}$ are the REFN PM\textsubscript{2.5} measurements that are usually considered to be true concentrations. At certain time instant $i$, the OLS model adjusted value, denoted by $\hat{y}_{i}$, of a WePIN's PM\textsubscript{2.5} reading $x_i$ is calculated using the following equation:
\begin{equation}
\label{Eq:OLSCalibratedWepin}
	\hat{y}_{i} = \beta_0 + \beta^u_1x_i
\end{equation}
where the value of $\beta_0$ depends upon individual device and $\beta^u_1$ is one of the calibration parameters derived from the WPUW-REFN PM\textsubscript{2.5} pairs with OLS model. Similarly, parameters $\beta^u_1$ and $\beta^u_2$ of the MLH model and parameters $\beta^u_1$ and $\beta^u_3$ of the ADV model can be used to adjust a WePIN's PM\textsubscript{2.5} readings, while the remaining parameter(s) is/are device dependent. The values of these parameters could be individually estimated by challenging a WePIN to the clean air, i.e., via zero calibration. As a reference instrument is not necessity for zero calibration, it can be conveniently carried out, even at user level and in an automated manner, with simple accessories (e.g., employing a HEPA filter at the WePIN's inlet). Enabling the WePINs with auto-zero calibration capability is left for future development. Notice that, it is often stated in the existing studies that sensor or device model with high unit-wise precision (implying unity slopes and negligible intercepts among units) could relax the needs for individual calibration.

For simplicity reasons, the WPUW TEMP and RH readings are considered representative with unity slopes and negligible intercepts among units. Given that the TEMP-RH sensors used in these WePINs had been well calibrated by the manufacture before shipment (nominal accuracies are $\pm\SI{0.1}{\celsius}$ and $\pm\SI{1.5}{\percent}$, respectively), plus they maintain good stability over time and across locations, the substantial differences in the TEMP and RH readings between the WPUW and the REFN as illustrated in Fig. \ref{fig:WePINsRefnAllDataPlot} were most likely ascribed to their distinct micro-environments. For example, the WPUW TEMP readings were always higher than the REFN ones because the circuit boards were warmed up by the on-board components after switching on and considerable heat was yet dissipated in the TEMP-RH sensors although the thermal path had been reduced (see Fig. \ref{subfig:PMDevice}). Therefore, the WPUW TEMP and RH readings would not be calibrated against that from the REFN, and they were used to indicate the atmospheric conditions of the aerosol sampled by the WePINs when performing calibration.

The collocated calibration results (scatter plots, $R^2$ values, regression parameters, etc.) of the WPUW-REFN PM\textsubscript{2.5} data pairs are presented in Fig. \ref{fig:ScatterCaliWePINsRefn}. As demonstrated in the first scatter plot, most of the data points were located to the lower right of the identity line, suggesting that the WePINs typically overestimated the mass concentrations of PM\textsubscript{2.5}. This was in contrast to the observations in the first filed experiment (refer to section \ref{SubSubSect:CorrelationAndCalibrationOfPMSensors}). It again emphasizes the necessity of initially and frequently calibrating the PM sensor or device in the final deployment environment. For those data points having REFN PM\textsubscript{2.5} $\leq\SI[per-mode=symbol]{20}{\micro\gram\per\cubic\meter}$, the moist ones were mostly dispersed on the lower right of the dry ones, whereas the contrary was the case for those with REFN PM\textsubscript{2.5} $>\SI[per-mode=symbol]{20}{\micro\gram\per\cubic\meter}$. As presented in section \ref{SubSubSect:DataCleansingOfPMSensors} and section \ref{SubSubSect:CorrelationAndCalibrationOfPMSensors}, the former dispersion pattern resulted from the hygroscopic growth of particles that caused overestimation in these devices' PM measurements, while the cause of the latter pattern is inconclusive. Additionally, these two sets of data points appeared to exhibit two distinct linear relationships, or their combination seemed to have a nonlinear relationship. Notice that the latter set of points represented the PM\textsubscript{2.5} pairs acquired in those days having abnormal reference measurements (i.e., salmon spans in Fig. \ref{fig:WePINsRefnAllDataPlot}), in which case the linear relationship may be altered due to variations in the aerosol composition. Subsequent experiments are necessary to draw any solid conclusions. The analyses below are based on the limited data/information available in this field experiment.

Comparing with the PMSA-REFN PM\textsubscript{2.5} data pairs ($R^2=0.938$; see Fig. \ref{fig:ScatterCaliSensRefn}), the WPUW-REFN ones ($R^2=0.748$) yielded a correlation much lower even than those with similar concentration range, that is, REFN PM\textsubscript{2.5} $\leq\SI[per-mode=symbol]{50}{\micro\gram\per\cubic\meter}$ ($R^2=0.912$; see Fig. \ref{fig:ScatterCaliPMSARefn}). The fairly low $R^2$ was probably owing to the WePIN's single-shot sampling mechanism summarized in section \ref{SubSect:SysImpWePIN}. While the REFN takes a measurement every second and generates an average output per minute, in which case the rapid concentration variations could be captured, the WePIN outputs an instantaneous measurement per minute, in which case the PM\textsubscript{2.5} value may not be representative enough for that interval and results in weaker correlation against the REFN. A WePIN could be configured to operate continuously (with minimum sampling interval $\leq\SI{2.3}{\second}$) at the cost of half the battery lifetime, i.e., $\approx\SI{12}{\hour}$ at best. Although it rained a lot through this field experiment, no apparent multiple linear relationships, such as that presented in section \ref{SubSubSect:DataCleansingOfPMSensors} and Fig. \ref{fig:ScatterSensDustRefn}, between the WPUW PM\textsubscript{2.5} readings and the REFN ones were observed. Thus, the sufficient conditions for determining whether recalibration is needed require further investigations.

The scatter plots between the calibrated WPUW PM\textsubscript{2.5} and the reference measurements were established in the second to the fourth columns of Fig. \ref{fig:ScatterCaliWePINsRefn}. The calibrated WPUW PM\textsubscript{2.5} mass concentrations in these scatter plots were obtained from the OLS, the MLH, and the ADV models, respectively. From visual inspection, the calibrated WPUW-REFN data pairs had distinct dispersion compared to those calibrated PMSA-REFN pairs with REFN PM\textsubscript{2.5} $\leq\SI[per-mode=symbol]{50}{\micro\gram\per\cubic\meter}$ as depicted in Fig. \ref{fig:ScatterCaliPMSARefn}, which might result in diverse or even contrary interpretations of their field performance. For instant, in contrast to the linear trend between the PMSA's OLS calibrated concentrations and the reference outputs, the curved shape of the OLS calibrated WPUW-REFN data points implied that the WePINs exhibited nonlinear responses when the PM\textsubscript{2.5} concentrations exceeded \SI[per-mode=symbol]{20}{\micro\gram\per\cubic\meter}. Besides, the dispersion patterns of their moist and dry data points in respective scatter plots were suggesting that the WePINs, but not the PMSA, had better RH insusceptibility than the REFN at low RH conditions (purple points associated with those OLS calibrated WPUW readings were on the lower right of the cyan ones). However, the MLH calibrated WPUW-REFN data pairs yielded a stronger correlation than the OLS calibrated ones ($\approx\SI{12}{\percent}$ of $R^2$), which indicated the WePINs were susceptible to RH (i.e., hygroscopic growth of particles), whereas the corresponding improvement of the PMSA-REFN data pairs was negligible ($\approx\SI{1.2}{\percent}$ of $R^2$). Nevertheless, the MLH model seemed to over adjust the WPUW PM\textsubscript{2.5} readings at low concentrations and humid conditions ($\leq\SI[per-mode=symbol]{10}{\micro\gram\per\cubic\meter}$ and $RH\approx\SI{65}{\percent}$); the ADV model can considerably mitigate such over adjustment with additional \SI{1.6}{\percent} increase of $R^2$. Similar occurrence was not observed for the PMSA-REFN data pairs. Such diversity in the field performances of the PMSA and the WePIN might partly caused by the fact that the measurements from a PM sensor, used as part of a monitoring device, could be affected by the housing, system integration, etc., as stated in existing studies. For instant, the sibling models of the PMSA sensor with a housing were less responsive to change in PM\textsubscript{2.5} concentrations and exhibited varied slope \cite{kelly2017ambient}.

In summary, the pairwise correlations of PM\textsubscript{2.5} among the ten implemented WePINs were moderate due to the fairly low concentration level in this field experiment and the significant random fluctuations in their PM measurements. The latter can be greatly mitigated through averaging the PM measurements of each WePIN and resulted in improved pairwise correlation and unit-wise precision. The correlation and calibration results of the WPUW-REFN PM\textsubscript{2.5} data pairs yielded diverse or even contrary interpretations of the field performances compared to that of the PMSA-REFN ones, which illustrated the necessity of initial and frequent calibration of a PM sensor or device in deployment environment as well as the essential of calibrating the PM sensor at system level. Owing to the limited ranges of PM\textsubscript{2.5} concentrations and atmospheric conditions in this field experiment, subsequent investigations are needed to facilitate better understanding of the field performances of a WePIN.

\subsubsection{Residual Analysis}
\label{SubSubSect:ResidualAnalysisOfWePINs}

The residuals between the calibrated WPUW PM\textsubscript{2.5} and the REFN ones are depicted in Fig. \ref{fig:ErrorAnalysisWePINsRefn}, where the first, the second, and the third rows are illustrating the residuals from OLS model, MLH model, and ADV model, respectively. Within each row, the scatter plots of RH vs. Rel Resid and RH vs. Resid plus the histograms of Rel Resid and Resid are presented in the first, the third, the second, and the fourth columns, respectively. Identical to the residual analyses of the PM sensors provided in section \ref{SubSubSect:ResidualAnalysisOfPMSensors}, a Rel Resid was considered valid only if the pair of PM\textsubscript{2.5} mass concentrations from the calibrated WPUW and the REFN were both equal to or above \SI[per-mode=symbol]{3}{\micro\gram\per\cubic\meter}. Following the same ranking principle, it is intuitively that the calibration performances (in terms of Rel Resid and Resid) of the OLS, the MLH, and the ADV models were in ascending order according to the shapes of the scatter plots and the histograms. Same conclusion can be drawn from the statistics of Rel Resid and Resid from respective models.

As depicted in the OLS model's RH vs. Rel Resid and RH vs. Resid scatter plots, upward tends of data points (especially the purple ones) were observed. It suggested that the WePINs' PM\textsubscript{2.5} measurements were probably susceptible to hygroscopic growth of particles that caused overestimation. Upward trends of those cyan to yellow data points ($>\SI[per-mode=symbol]{20}{\micro\gram\per\cubic\meter}$) were not as apparent as that of the purple ones primarily due to the fairly low RH conditions and limited data availability ($RH<\SI{74}{\percent}$ and $n\approx1000$). Applying the MLH model can well flatten the upward trends in both scatter plots and resulted in improved sample standard deviations of Rel Resid ($\approx\SI{18}{\percent}$ drop of $s_{d}$) and Resid ($\approx\SI{19}{\percent}$ drop of $s_{\epsilon}$) but worsened sample mean of Rel Resid ($\approx\SI{22}{\percent}$ rise of $|\bar{d}\,|$). Although the upward trend in the Rel Resid scatter plot was fairly flattened, the purple data points on the left side were mirrored along the horizontal zero line. The extremum of these purple points had been increased from $\approx\SI{-60}{\percent}$ to $\approx\SI{80}{\percent}$, indicating that the WPUW PM\textsubscript{2.5} measurements at low concentration levels and slightly humid conditions ($\leq\SI[per-mode=symbol]{10}{\micro\gram\per\cubic\meter}$ and $RH<\SI{68}{\percent}$) were over adjusted by the MLH model. As shown in the ADV model's Rel Resid scatter plot, the data points were uniformly distributed on both sides of the horizontal zero line, suggesting that the MLH over adjustment can be considerably mitigated by the ADV model. The downward trend of the cyan to yellow data points in the Resid scatter plot of the ADV model was probably attributed to the limited data points at high concentrations ($>\SI[per-mode=symbol]{20}{\micro\gram\per\cubic\meter}$) because they exhibited dispersion pattern as expected, that is, the dispersion (from zero) of a data point was proportional to the reference concentration level. Compared to the OLS or the MLH model, employing the ADV model would significantly improve the statistics of Rel Resid and Resid ($\approx\SI{30}{\percent}$ or \SI{43}{\percent} drop of $s_{d}$, $\approx\SI{23}{\percent}$ or \SI{3.9}{\percent} drop of $s_{\epsilon}$, and $\approx\SI{29}{\percent}$ or \SI{13}{\percent} drop of $|\bar{d}\,|$).

\subsubsection{Evaluation Results and Discussion}
\label{SubSubSect:EvaluationAndDiscussionOfWePINs}

From the OLS and MLH calibrated WPUW-REFN PM\textsubscript{2.5} pairs, all the evaluation metrics presented in section \ref{SubSubSect:EvaluationMetrics} except the data completeness $\eta$ were calculated and their results were summarized in Tab. \ref{tab:EvaluationMetricsSensDustRefn}. The cells with acceptable values were highlighted in gray. A calibrated WPUW-REFN PM\textsubscript{2.5} pair is considered valid when both PM\textsubscript{2.5} concentrations are $\geq\SI[per-mode=symbol]{3}{\micro\gram\per\cubic\meter}$. Notice that the limit of detection $LOD$ was not calculated given that no low concentration pairs (reference PM\textsubscript{2.5} $<\SI[per-mode=symbol]{3}{\micro\gram\per\cubic\meter}$) were available in this field experiment, and the unit-wise precision $CV_{RMS}$ had been discussed in section \ref{SubSubSect:CorrelationAndCalibrationOfWePINs}.

It is noteworthy that, the $CV_{RMS}$ value listed in Tab. \ref{tab:EvaluationMetricsSensDustRefn} was calculated form the WePINs' not-calibrated 1-minute interval PM\textsubscript{2.5} measurements. Using the averaged measurements from each device typically resulted in improved $CV_{RMS}$; for instant, about \SI{20}{\percent} drop ($CV_{RMS}=\SI{12.271}{\percent}$) was achieved with 5-minute averaged measurements since the random fluctuations in devices' measurements were greatly mitigated. The $CV_{RMS}$ values calculated from their calibrated measurements (a WePIN was calibrated using the parameter sets achieved by fitting the WPUW-REFN PM\textsubscript{2.5} pairs into the OLS, MLH, and ADV models) yielded worsened results ($\approx\SI{21}{\percent}$, $\approx\SI{99}{\percent}$, and $\approx\SI{18}{\percent}$ for respective models using their calibrated 1-minute interval measurements). The $CV_{RMS}$ values were moderately to substantially increased primarily due to the lower values of the calibrated measurements, particularly those from the MLH model. Restricting the concentration range of these calibrated measurements (i.e., from \SIrange[per-mode=symbol]{3}{200}{\micro\gram\per\cubic\meter}) would bring about improved results ($CV_{RMS}\approx\SI{19}{\percent}$, \SI{17}{\percent}, and \SI{17}{\percent} for respective models). Further improvements were achievable by averaging these devices' calibrated measurements ($CV_{RMS}\approx\SI{16}{\percent}$, \SI{15}{\percent}, and \SI{15}{\percent} with 5-minute averaged measurements, respectively). Differed from the Part 53SC, the authors prefer calculating the $CV_{RMS}$ values from the calibrated sensors' or devices' measurements given that the calibrated PM\textsubscript{2.5} values along with the collocated reference outputs were both within range from \SIrange[per-mode=symbol]{3}{200}{\micro\gram\per\cubic\meter}.

For the remaining evaluation metrics, in comparison to the OLS calibrated WPUW-REFN data pairs, the MLH calibrated ones usually yielded better results in most evaluation aspects (i.e., lower $\sigma_{UCL}$ and $|intercept|$; $slope$ and $r$ closer to one) except the $Bias_{PEP}$. The growth of the MLH $|Bias_{PEP}|$ was ascribed to the RH over adjustment at low concentrations and mildly humid conditions as stated in sections \ref{SubSubSect:CorrelationAndCalibrationOfWePINs} and \ref{SubSubSect:ResidualAnalysisOfWePINs}. Despite such over adjustment, the MLH $\sigma_{UCL}$ and $|intercept|$ were \SI{18}{\percent} and \SI{22}{\percent} smaller than the OLS ones, respectively. Notice that the absolute decrease of $|intercept|$ was marginal ($\approx\SI[per-mode=symbol]{0.8}{\micro\gram\per\cubic\meter}$) considering the resolution of a WePIN's PM\textsubscript{2.5} readings (i.e., \SI[per-mode=symbol]{1}{\micro\gram\per\cubic\meter}). As for the MLH $slope$ and $r$ values, they were \SI{0.068}{} and \SI{0.051}{} closer to unity than the OLS ones, respectively.

By contrast, the evaluation results of the PMSA-REFN data pairs, which were calibrated with the OLS and MLH models, were usually better than those of the same calibrated WPUW-REFN pairs. However, it did not necessarily indicate that the data quality and field performances of the PMSA sensor were more superior than those of the WePINs since the differences in the evaluation results might be attributed to various factors. For instant, field experiment with lower PM\textsubscript{2.5} concentrations will generally produce larger $|Bias_{PEP}|$ and $\sigma_{UCL}$ according to the calculations (smaller denominator in Eq. \ref{Eq:PEPRelativeError}). Likewise, a narrower PM\textsubscript{2.5} concentration range will cause lower $r$ and varied $slope$ and $intercept$. Other factors such as the single-shot sampling mechanism, the housing and system integration, and the aerosol properties and meteorological conditions during these two field experiments might as well contribute to the differences. Yet the comparison evidenced that no major flaws existed in the design and implementation of the WePIN as the differences in their calibration and evaluation results were not tremendously huge.

Similar to the first field experiment, there exist limitations in the calibration and evaluation results derived from this field experiment as well. In addition to the limitations presented in section \ref{SubSubSect:EvaluationAndDiscussionOfPMSensors}, the construction works carried out occasionally near the Supersite might greatly alter the correlations of PM\textsubscript{2.5} between the WePINs and the reference (see section \ref{SubSubSect:DataCleansingOfWePINs}), so did the single-shot sampling mechanism of a WePIN (refer to section \ref{SubSubSect:CorrelationAndCalibrationOfWePINs}), in which case the results would get distorted. Also, rather than individually calibrating and evaluating every WePIN against the REFN, their WPUW PM\textsubscript{2.5} measurements were considered representative; applying the set of calibration parameters derived from the WPUW-REFN PM\textsubscript{2.5} pairs to the WePINs might result in varied data quality (e.g., the values of $|Bias_{PEP}|$). Considerable heat from other components on a WePIN could still dissipate in the TEMP-RH sensor through the reduced thermal path such that its TEMP and RH readings might not reflect the actual atmospheric conditions of aerosol being sampled (better design is needed). Moreover, these PM devices' measurements have not been evaluated for biases due to the housing and system integration. Nevertheless, both the four nominated PM sensors and the ten implemented WePINs were calibrated and evaluated against an ambient background station, while the proposed system is intended for monitoring the daily PM exposures of an individual who spends most of the time indoor. Therefore, evaluating the data quality and the performances of a WePIN in the indoor air (residential, office, etc.) is needed.

\section{Conclusion}
\label{Sect:Conclusion}

Given the aerosol distributions in urban areas may present considerable spatio-temporal inhomogeneity, the conventional PM monitoring paradigm based upon a few fixed stations that equipped with well established instruments hinders the access to effective personalized PM exposure assessments, which are critical for recognizing and understanding the PM-associated health impacts. Such deficiency has fostered the development of personal PM exposure monitoring systems, which comprise portable/wearable devices utilizing sensors that are compact, low-cost, power-efficient, and yet providing instantaneous data with sufficient quality. Enable by these emerging systems, the participants can promptly be aware of the PM concentrations in their vicinity. In order to address the limitations in existing devices or systems and enable the authors with more controls, a novel personal PM exposure monitoring system comprising a wearable PM device (i.e., the WePIN), a mobile application, and a back-end server is proposed. This system is intended to collect subjects' daily PM\textsubscript{2.5} exposure levels for investigating specific PM-health association. However, owing to the absence of a standardized evaluation protocol for low-cost PM sensors, the evaluation results in existing studies and the performance specifications from the datasheets are of limited values when attempting to determine the best candidate PM sensor for the proposed system. Besides, the low-cost PM sensors should be evaluated in environment close to the deployment one due to their responses are highly dependent on the sampled aerosol's properties. Thus, two experiments were respectively conducted at the rooftop of the HKUST Supersite (on 1-minute interval PM\textsubscript{2.5}) to select the best PM sensor for a WePIN and evaluate the field performances of the implemented WePINs.

The first field evaluation campaign was conducted from 13 to 23 February 2018. Four models of PM sensors, denoted by DN7C, PMSA, HPMA, and OPCN, were empirically selected and compared against two collocated reference monitors, denoted by DUST (research-grade) and REFN (Class III FEM). Results illustrated that the condensed water droplets remained suspended in the air might be detected as particles by the PM sensors and caused significant overestimation and fluctuations in their PM\textsubscript{2.5} measurements that were unable to be corrected, particularly for the DN7C, HPMA, and OPCN. An automated approach based on the PM\textsubscript{10} and PM\textsubscript{1.0} ratios was developed to remove the condensation related measurements. Moreover, the sensors and DUST's responses were highly dependent on the aerosol's properties; the alteration of aerosol composition since/on 17 February caused two different linear relationships between their responses and the REFN's readings. With these unusual measurements eliminated, the DN7C, OPCN, HPMA, and PMSA exhibited increasing correlations against the reference monitors; these sensors also demonstrated an ascending trend in their RH insusceptibility (i.e., hygroscopic growth of particles) while the DN7C was as well affected by the TEMP. In comparison with the research-grade DUST, the HPMA and PMSA even exhibited stronger correlations against the REFN ($R^2=0.898$ vs. $R^2=0.938$) and higher RH insusceptibility. The atmospheric influences on the sensors' measurements can be mitigated by involving TEMP (for the DN7C) or RH (for the others) in respective MLR models and resulted in slightly to greatly improved correlations (\SIrange{1}{32}{\percent} of $R^2$) against the reference monitors. The results also demonstrated that the sensors' responses to hygroscopic growth varied according to the RH levels and had different behaviors at mildly ($\leq\SI{82}{\percent}$) and extremely ($>\SI{82}{\percent}$) humid conditions; involving RH in a MLR model would over adjust measurements at mildly humid and low concentration conditions. Evaluation metrics adopted for developing a unifying evaluation protocol were calculated using the calibrated sensor-reference PM\textsubscript{2.5} pairs; the HPMA and PMSA generally showed better results than the others and even outperformed the DUST in most aspects. The PMSA was selected eventually considering that it had better affordability, portability, detection capability, and insusceptibility to RH and condensation than the HPMA. The measurement trueness as well as the measurement precision of the PMSA's calibrated PM\textsubscript{2.5} measurements, expressed as $Bias_{PEP}$ and $\sigma_{UCL}$, were \SI[separate-uncertainty=true]{3.696\pm0.328}{\percent} and \SI{17.461}{\percent}, respectively.

Utilizing the selected PM sensor, ten WePINs were initially implemented. They provide better affordability and portability than the existing devices or systems recognized by the authors while maintaining similar usability and data accessibility. The WePINs were compared against each other and the collocated REFN in the second field evaluation campaign from 26 June to 17 July 2019. The pairwise PM\textsubscript{2.5} correlations (on 1-minute interval) were moderate ($\overline{R^2}=0.859$) primarily owing to the random fluctuations in their readings, which could be reduced by averaging each device's readings over a longer time period and resulted in noticeably improved correlations ($\overline{R^2}=0.964$ on 5-minute averaged PM\textsubscript{2.5}). The unit-wise precision of their 1-minute interval and 5-minute averaged readings, expressed as $CV_{RMS}$, were \SI{14.719}{\percent} and \SI{12.271}{\percent}, respectively. Rather than comparing each WePIN against the REFN, the unit-wise average PM\textsubscript{2.5} readings, denoted by WPUW, were considered representative. Compared to the PMSA-REFN data pairs with similar concentration range, the WPUW-REFN ones (1-minute interval) yielded a considerably lower correlation ($R^2=0.912$ vs. $R^2=0.748$), which was probably ascribed to the single-shot sampling mechanism of these devices such that the rapid variations in concentration may not be captured. Besides, the results (raw and calibrated) suggested diverse or even contrary interpretations of field performances between the PMSA and the WePINs, emphasizing the necessity of evaluating the PM sensors at system level. Overall, owing to the limited ranges of PM\textsubscript{2.5} concentrations and meteorological conditions in this campaign, subsequent experiments are required to facilitate a better understanding of these devices' field performances. The measurement trueness and the measurement precision of the calibrated WPUW PM\textsubscript{2.5} readings, expressed as $Bias_{PEP}$ and $\sigma_{UCL}$, were \SI[separate-uncertainty=true]{3.882\pm0.407}{\percent} and \SI{21.501}{\percent}, respectively.

Another 120 WePIN units of the latest iteration have been manufactured and they are being delivered to the subjects to collect their daily PM\textsubscript{2.5} exposure levels for in investigating the association with subclinical atherosclerosis. The limitations of this study were discussed in the last paragraphs of section \ref{SubSubSect:EvaluationAndDiscussionOfPMSensors} and section \ref{SubSubSect:EvaluationAndDiscussionOfWePINs}.

\clearpage 
\onecolumn

\appendix
\counterwithin{figure}{section}
\renewcommand\thefigure{\thesection.\arabic{figure}}
\setcounter{figure}{0}

\counterwithin{table}{section}
\renewcommand\thetable{\thesection.\Roman{table}}
\setcounter{table}{0}

\begin{figure*}[hbt!]
\centering
	\includegraphics[width=0.98\linewidth]{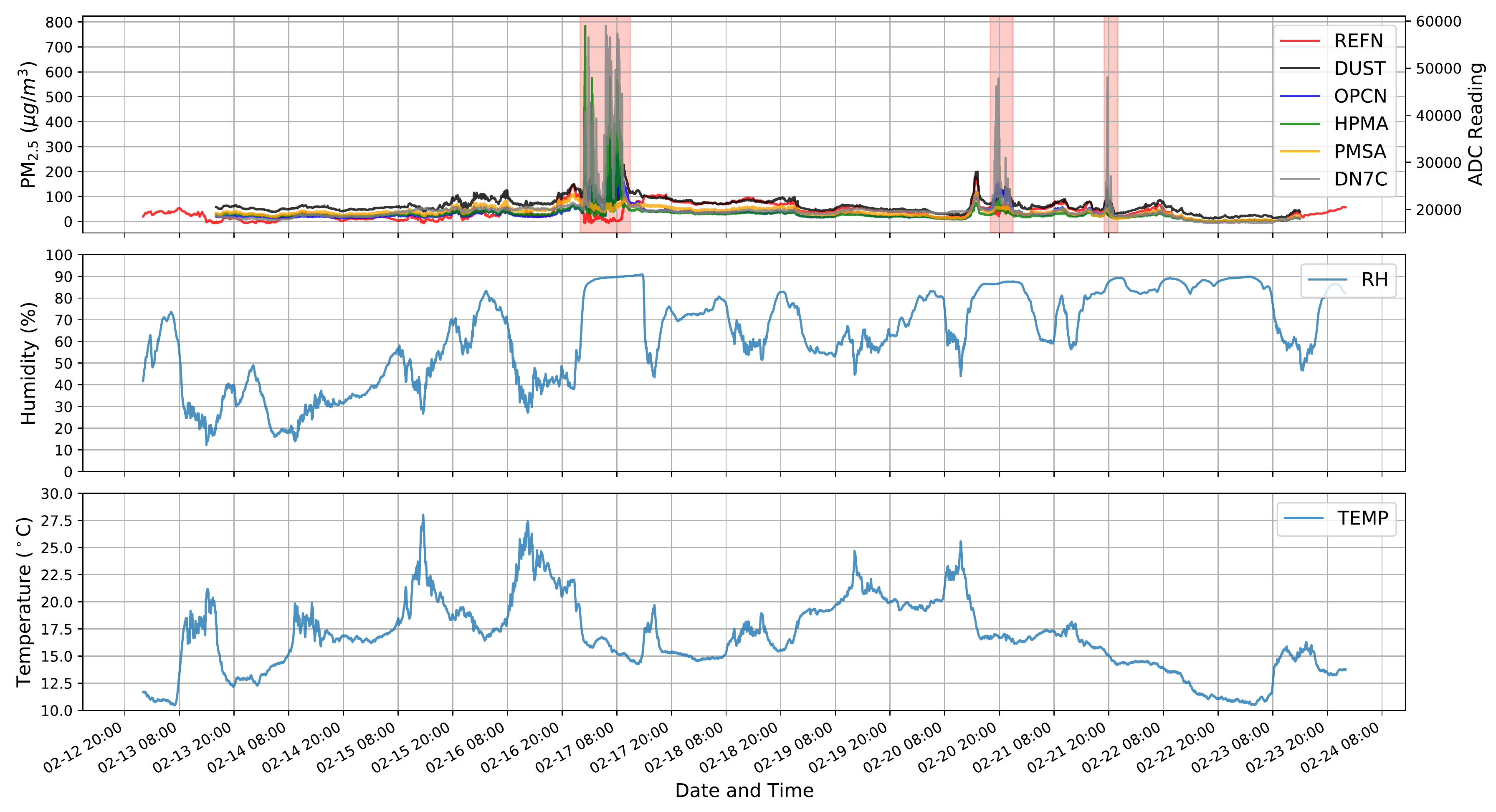}
\caption{PM\textsubscript{2.5} mass concentrations (1-minute interval) measured by the REFN (red) and DUST (black) reference instruments, PM\textsubscript{2.5} mass concentrations (1-minute averaged) measured by the OPCN (blue), HPMA (green), and PMSA (orange) PM sensors, and ADC readings (1-minute averaged) acquired from the DN7C (gray) PM sensor, as well as the RH and TEMP readings (1-minute interval) measured by the REFN during the first field experiment conducted from 13 to 23 February 2018. Three intervals with abnormal spikes (significant overestimation and fluctuations in the measurements from PM sensors, particularly the DN7C, HPMA, and OPCN) were identified and highlighted (salmon span) in the top time series plot.}
\label{fig:PM2p5RhTempFull}
\end{figure*}

\begin{figure*}[hbt!]
\centering
	\includegraphics[width=0.98\linewidth]{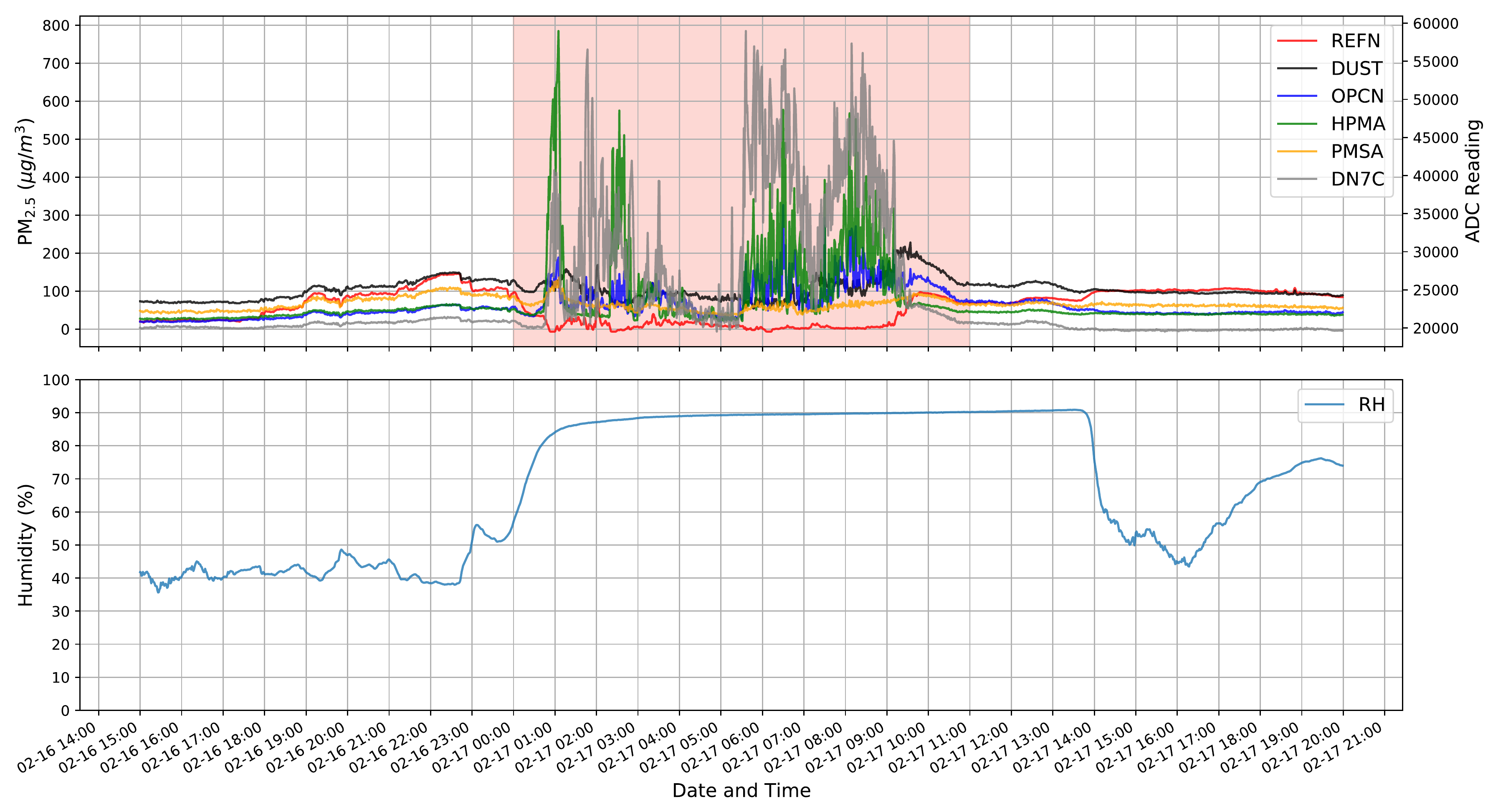}
\caption{Zoomed view of the first interval with abnormal spikes (salmon span) in PM sensors' measurements together with the corresponding RH readings (the entire hour of measurements of the first spike and that of the last spike were also included).}
\label{fig:PM2p5RhZoom1}
\end{figure*}

\begin{figure*}[hbt!]
\centering
	\includegraphics[width=0.98\linewidth]{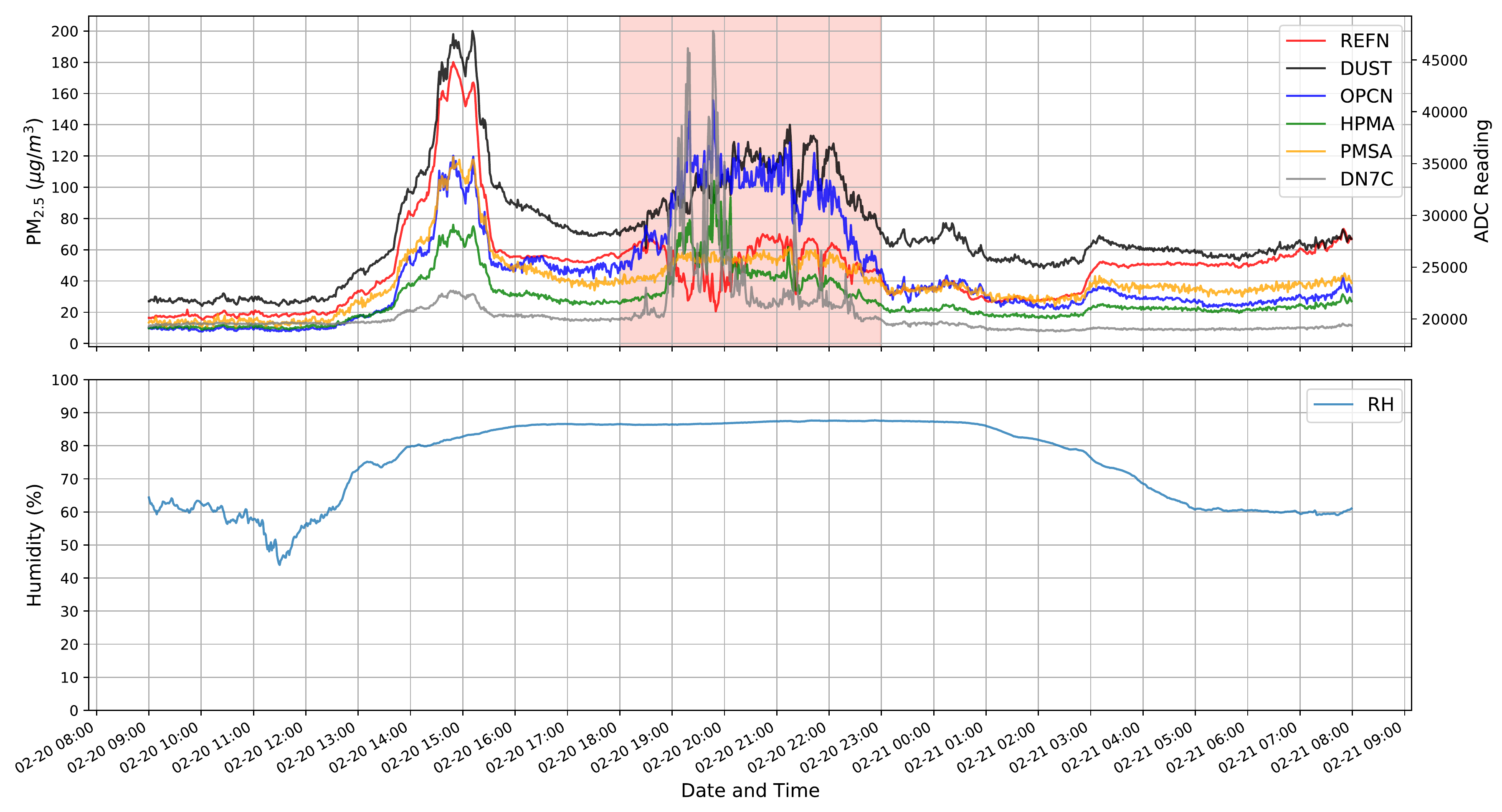}
\caption{Zoomed view of the second interval with abnormal spikes (salmon span) in PM sensors' measurements together with the corresponding RH readings (the entire hour of measurements of the first spike and that of the last spike were also included).}
\label{fig:PM2p5RhZoom2}
\end{figure*}

\begin{figure*}[hbt!]
\centering
	\includegraphics[width=0.98\linewidth]{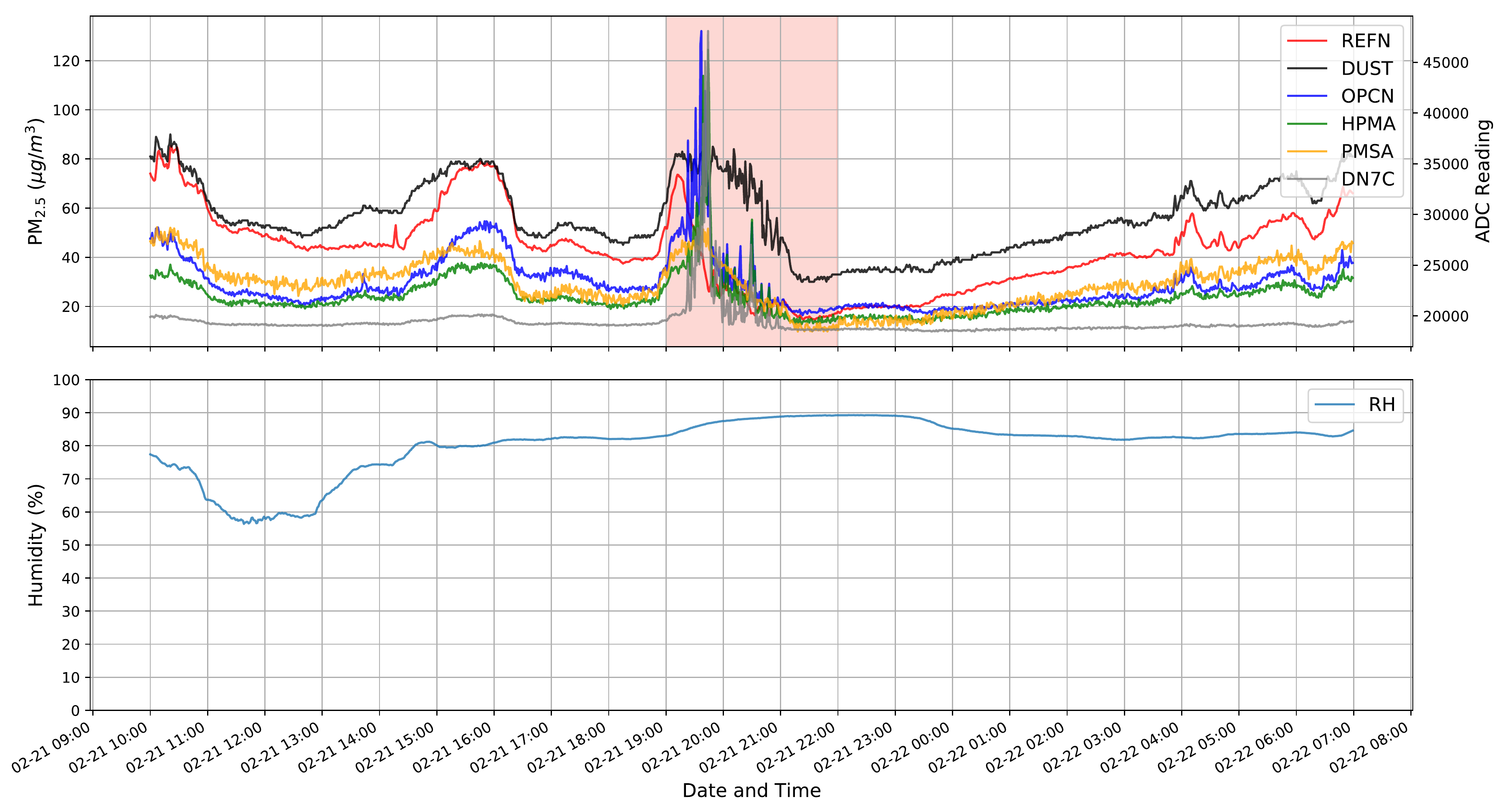}
\caption{Zoomed view of the third interval with abnormal spikes (salmon span) in PM sensors' measurements together with the corresponding RH readings (the entire hour of measurements of the first spike and that of the last spike were also included).}
\label{fig:PM2p5RhZoom3}
\end{figure*}

\begin{figure*}[hbt!]
\centering
	\includegraphics[width=0.98\linewidth]{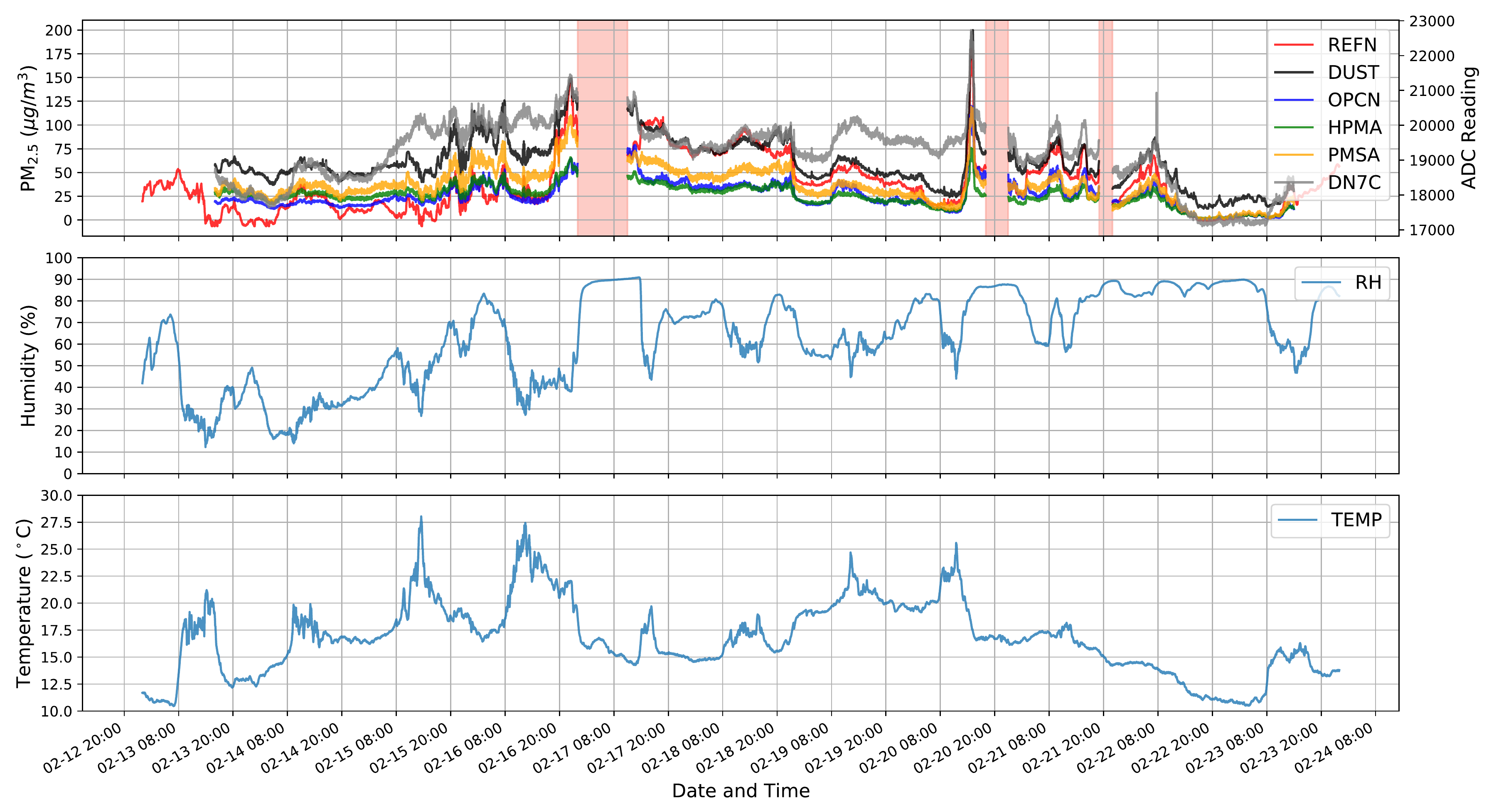}
\caption{PM\textsubscript{2.5} mass concentrations (1-minute interval) measured by the REFN (red) and DUST (black) reference instruments, PM\textsubscript{2.5} mass concentrations (1-minute averaged) measured by the OPCN (blue), HPMA (green), and PMSA (orange) PM sensors, and ADC readings (1-minute averaged) acquired from the DN7C (gray) PM sensor, as well as the RH and TEMP readings (1-minute interval) measured by the REFN during the first field experiment conducted from 13 to 23 February 2018. The three intervals with abnormal spikes (salmon spans) in PM sensors' measurements have been manually eliminated.}
\label{fig:PM2p5RhRmSpk}
\end{figure*}

\begin{figure*}[hbt!]
\centering
	\includegraphics[width=0.98\linewidth]{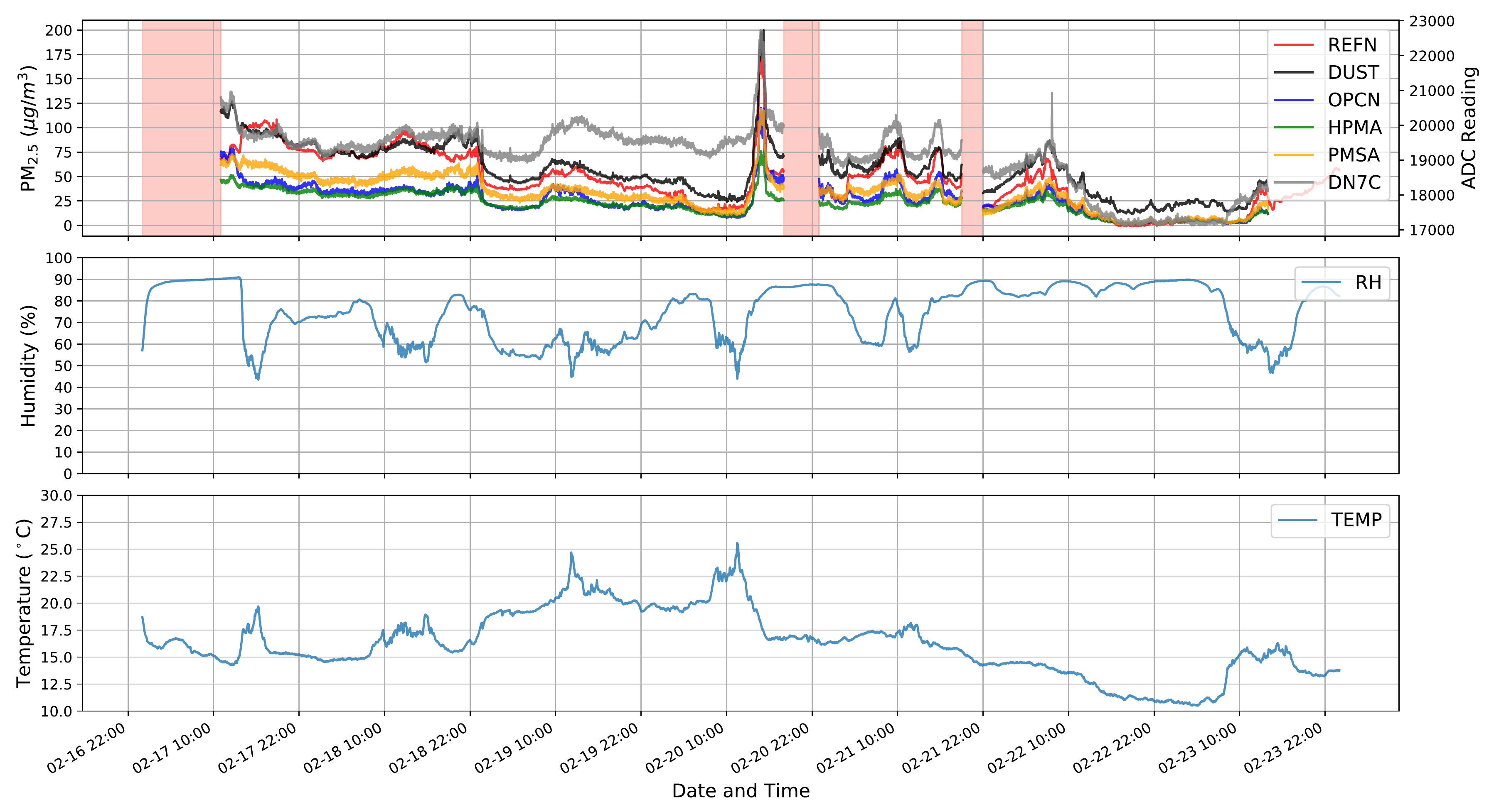}
\caption{PM\textsubscript{2.5} mass concentrations (1-minute interval) measured by the REFN (red) and DUST (black) reference instruments, PM\textsubscript{2.5} mass concentrations (1-minute averaged) measured by the OPCN (blue), HPMA (green), and PMSA (orange) PM sensors, and ADC readings (1-minute averaged) acquired from the DN7C (gray) PM sensor, as well as the RH and TEMP readings (1-minute interval) measured by the REFN during the first field experiment conducted from 13 to 23 February 2018. The three intervals with abnormal spikes (salmon spans) in PM sensors' measurements together with their measurements before 17 February have been manually eliminated.}
\label{fig:PM2p5RhRmSpkThld}
\end{figure*}

\begin{figure*}[hbt!]
\centering
	\includegraphics[width=0.82\linewidth]{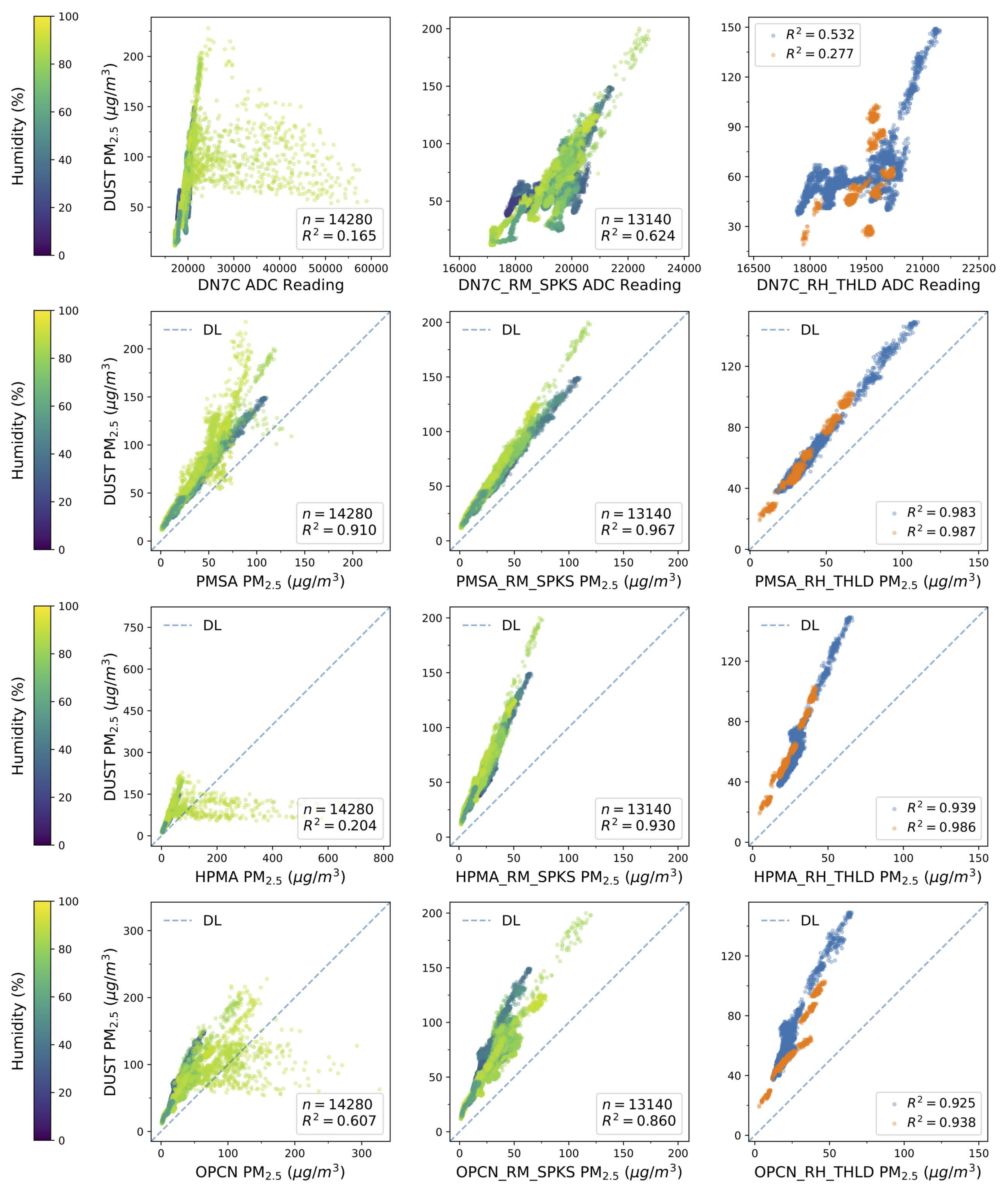}
\caption{Scatter plots and pairwise correlations (in terms of $R^2$) of the collocated data pairs (first field experiment conducted from 13 to 23 February 2018) between the DUST (vertical axis; 1-minute interval PM\textsubscript{2.5} mass concentrations) and the corresponding PM sensors, namely DN7C, PMSA, HPMA, and OPCN (horizontal axis; 1-minute averaged ADC readings or PM\textsubscript{2.5} mass concentrations). The diagonal dashed lines in blue (DL) are the identity lines. Scatter plots in the first column and the second column contain the collocated data pairs of the entire campaign ($n=14280$) and the remaining collocated data pairs with abnormal spikes eliminated ($n=13140$), respectively; the color of each data point indicates the corresponding RH value. Each scatter plot in the third column contains the collocated data pairs with abnormal spikes eliminated and under low RH conditions ($RH<\SI{58}{\percent}$ at which the REFN's dynamic heating system is off) that are separated into two groups as before (blue) and after (orange) 17 February.}
\label{fig:ScatterSensDust}
\end{figure*}

\begin{figure*}[hbt!]
\centering
	\includegraphics[width=0.82\linewidth]{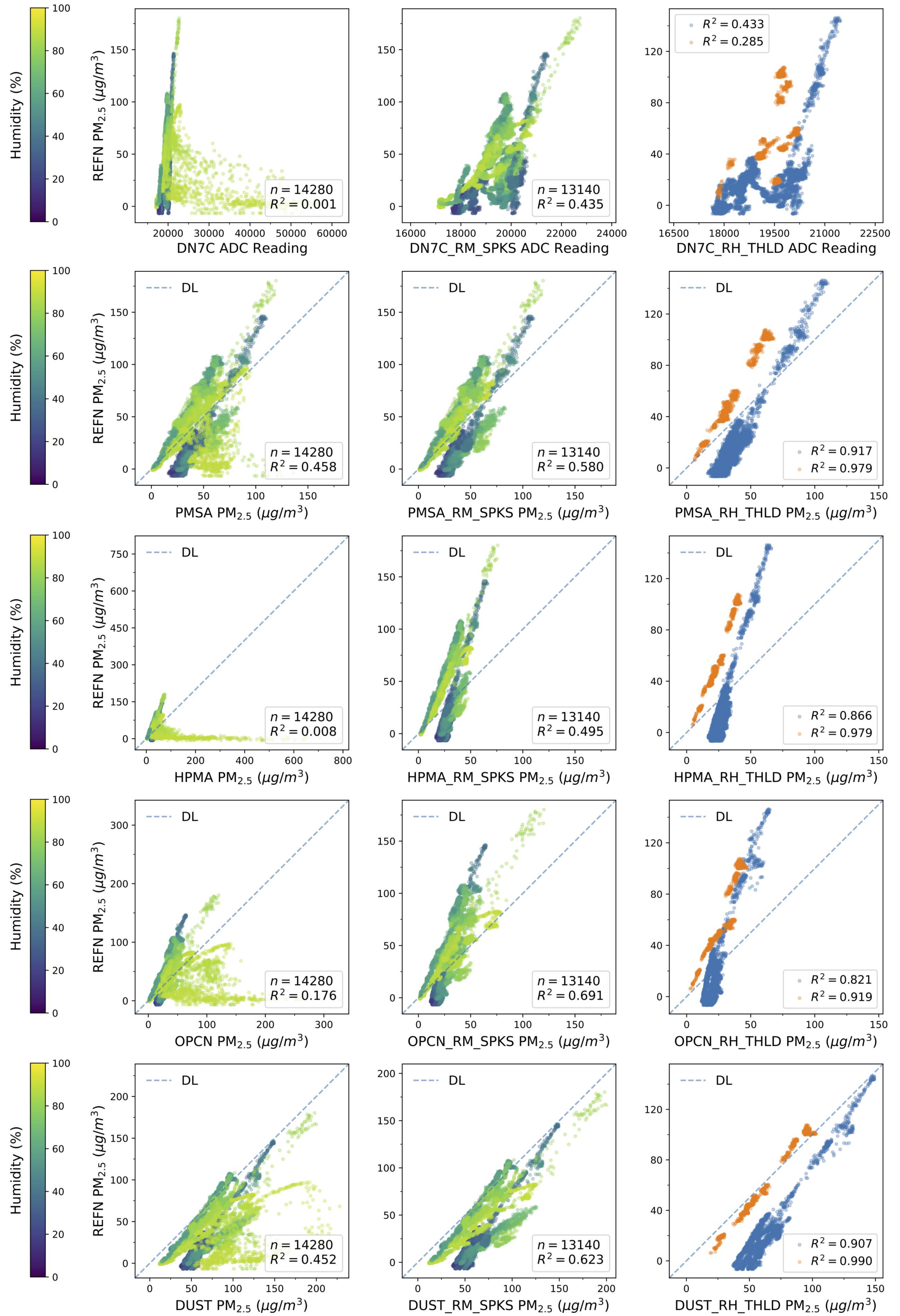}
\caption{Scatter plots and pairwise correlations (in terms of $R^2$) of the collocated data pairs (first field experiment conducted from 13 to 23 February 2018) between the REFN (vertical axis; 1-minute interval PM\textsubscript{2.5} mass concentrations) and the corresponding instruments, namely DN7C, PMSA, HPMA, OPCN, and DUST (horizontal axis; 1-minute averaged/interval ADC readings or PM\textsubscript{2.5} mass concentrations). The explanations of the scatter plots and the notation are detailed in Fig. \ref{fig:ScatterSensDust}.}
\label{fig:ScatterSensDustRefn}
\end{figure*}

\begin{figure*}[hbt!]
\centering
	\includegraphics[width=0.87\linewidth]{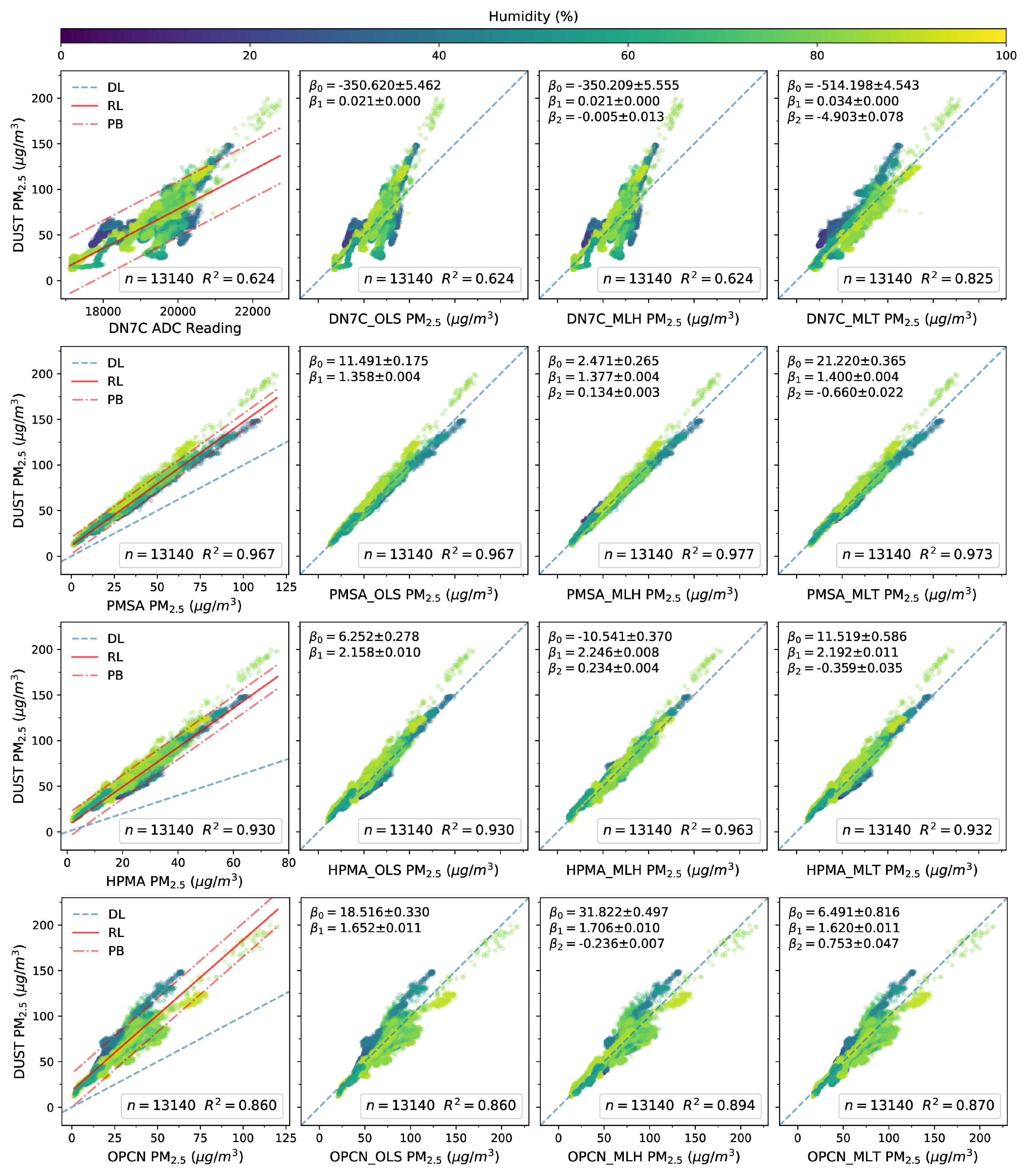}
\caption{Scatter plots and pairwise correlations (in terms of $R^2$) of the collocated data pairs (first field experiment conducted from 13 to 23 February 2018 with abnormal spikes eliminated; $n=13140$) between the DUST (vertical axis; 1-minute interval PM\textsubscript{2.5} mass concentrations) and the corresponding PM sensors, namely DN7C, PMSA, HPMA, and OPCN (horizontal axis; 1-minute averaged ADC readings or PM\textsubscript{2.5} mass concentrations in the first column; 1-minute interval calibrated PM\textsubscript{2.5} mass concentrations from the OLS, the MLH, and the MLT models in the second, the third, and the fourth columns, respectively). The color of each data point indicates the corresponding RH value. The blue dashed lines (DL) are the identity lines. The red solid lines (RL) and the paired red dot-dashed lines (PB) in the first column are the OLS regression lines and the upper and lower prediction bounds (\SI{95}{\percent} confidence level), respectively. The parameters (\SI{95}{\percent} confidence interval) of the OLS models, the MLH models, and the MLT models are listed on the upper left corners of the scatter plots in the second to fourth columns.}
\label{fig:ScatterCaliSensDust}
\end{figure*}

\begin{figure*}[hbt!]
\centering
	\includegraphics[width=0.87\linewidth]{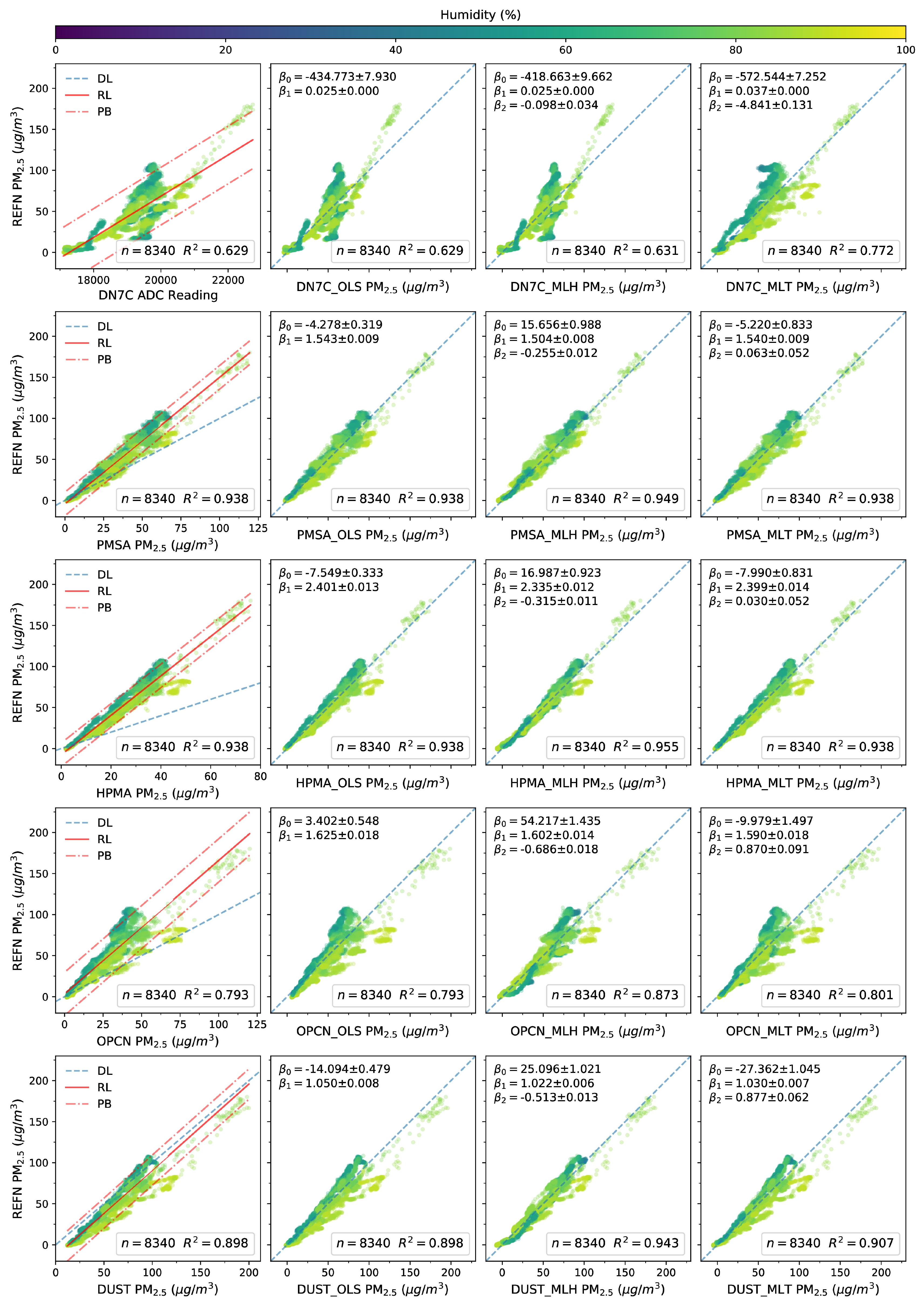}
\caption{Scatter plots and pairwise correlations (in terms of $R^2$) of the collocated data pairs (first field experiment conducted from 13 to 23 February 2018 with abnormal spikes and measurements before 17 February eliminated; $n=8340$) between the REFN (vertical axis; 1-minute interval PM\textsubscript{2.5} mass concentrations) and the corresponding instruments, namely DN7C, PMSA, HPMA, OPCN, and DUST (horizontal axis; 1-minute averaged/interval ADC readings or PM\textsubscript{2.5} mass concentrations in the first column; 1-minute interval calibrated PM\textsubscript{2.5} mass concentrations from the OLS, the MLH, and the MLT models in the second, the third, and the fourth columns, respectively). The notation definitions are detailed in Fig. \ref{fig:ScatterCaliSensDust}.}
\label{fig:ScatterCaliSensRefn}
\end{figure*}

\begin{figure*}[hbt!]
\centering
	\subfloat[]{
		\includegraphics[width=0.98\linewidth]{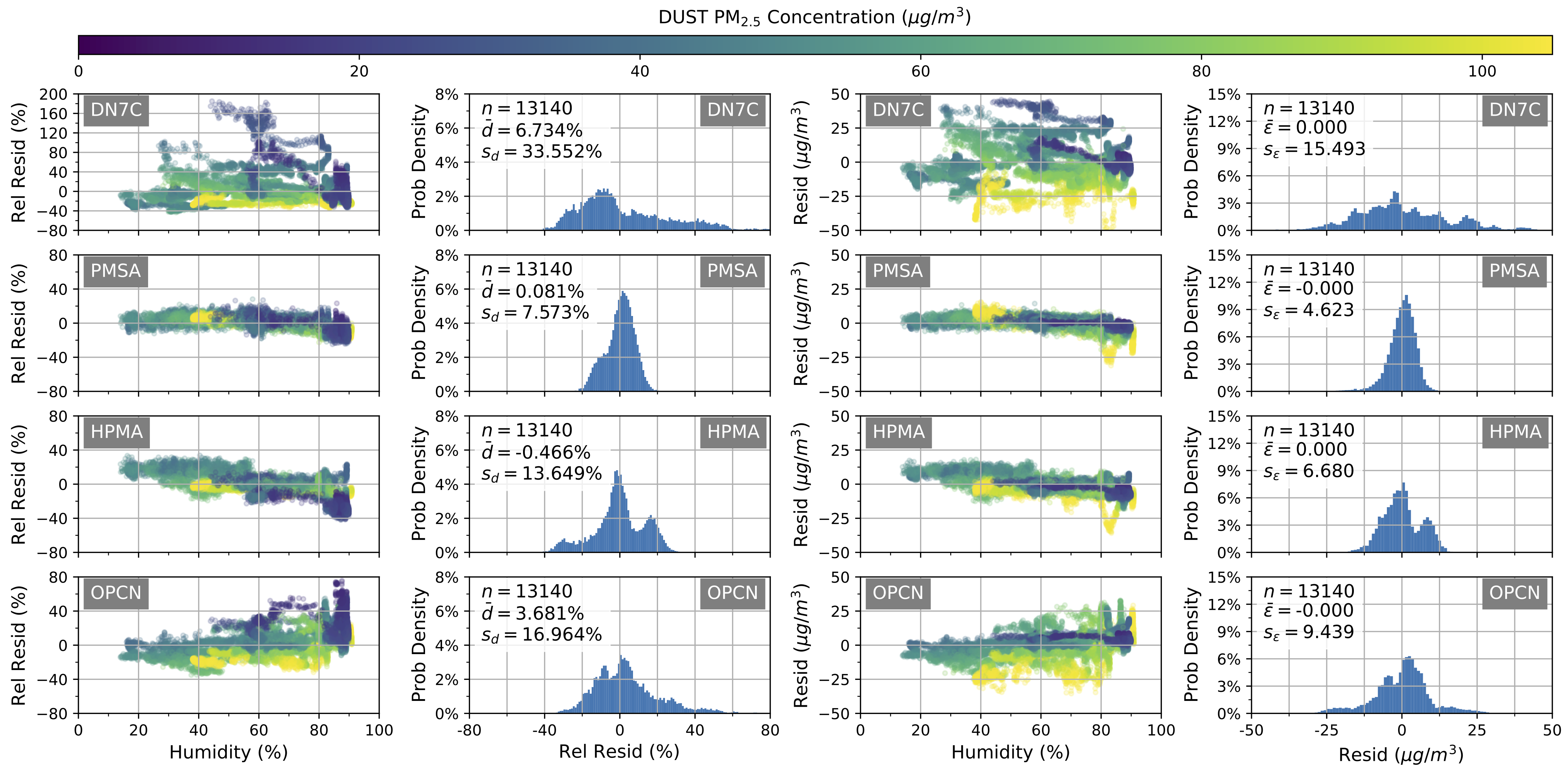}
		\label{subfig:CaliErrorOlsDust}}
	\\
	\subfloat[]{
		\includegraphics[width=0.98\linewidth]{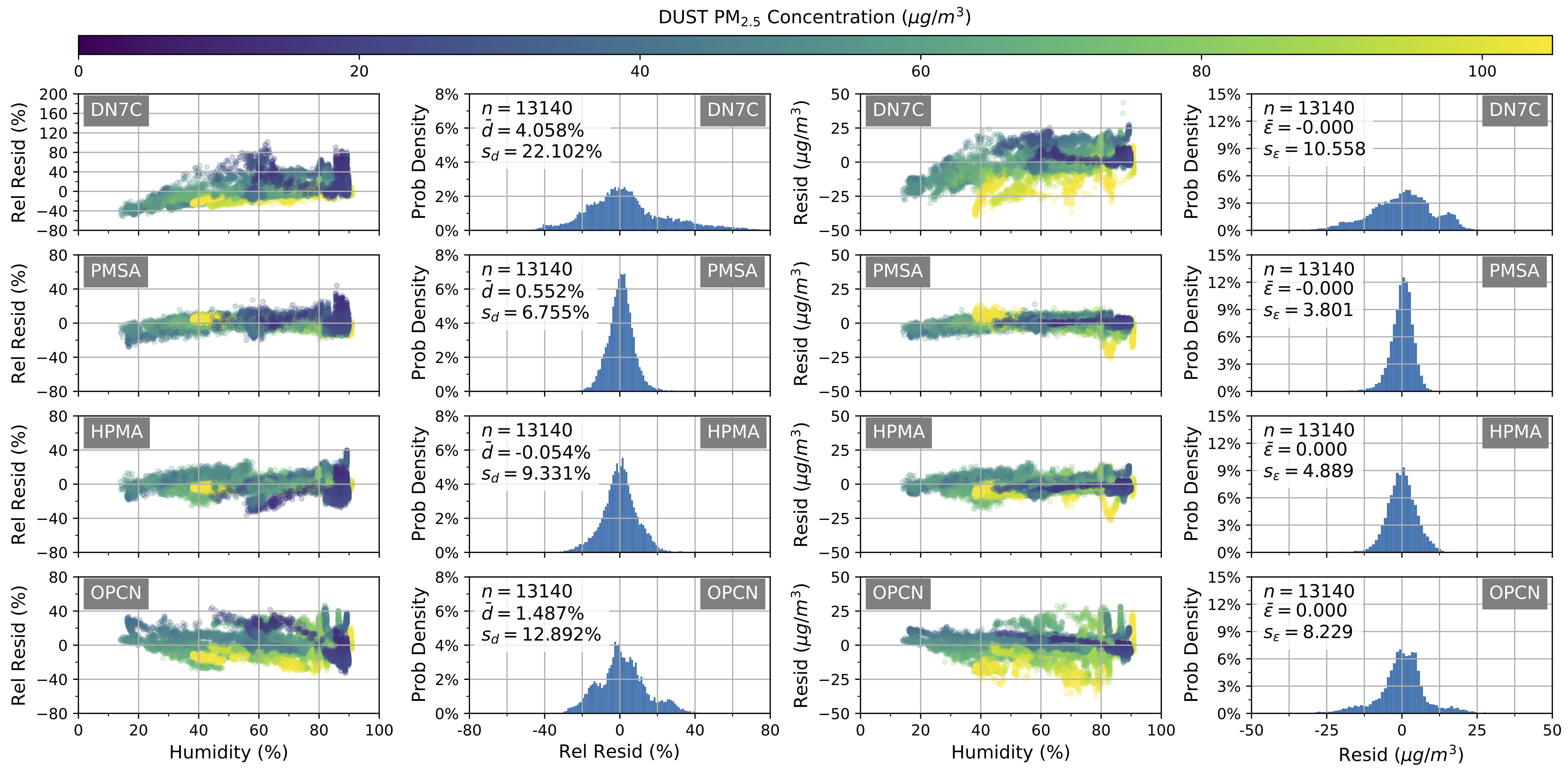}
		\label{subfig:CaliErrorMlrDust}}
\caption{Residual analysis of the PM sensors that have been calibrated against the DUST using (a) OLS model and (b) MLR model (MLT for the DN7C and MLH for the others) on the 1-minute interval collocated data pairs (first field experiment conducted from 13 to 23 February 2018 with abnormal spikes eliminated; $n=13140$). The Rel Resid and Resid represent the relative residuals and the (absolute) residuals of PM\textsubscript{2.5} mass concentration between the calibrated PM sensors and the DUST, respectively; the Rel Resid value is considered valid when both concentrations are $\geq\SI[per-mode=symbol]{3}{\micro\gram\per\cubic\meter}$. Scatter plots in the first and the third columns are illustrating the relationships between the RH and the Rel Resid and the Resid, respectively; the color of each data point indicates the corresponding PM\textsubscript{2.5} mass concentration measured by the reference instrument. Histograms in the second and the fourth columns are illustrating the probability distributions of Rel Resid (\SI{1}{\percent} bin width) and Resid (\SI[per-mode=symbol]{1}{\micro\gram\per\cubic\meter} bin width), respectively; the number of valid data pairs ($n$), as well as the sample mean and the sample standard deviation of Rel Resid ($\bar{d}$ and $s_{d}$) or Resid ($\bar{\epsilon}$ and $s_{\epsilon}$) are listed on the upper left corner of each histogram.}
\label{fig:ErrorAnalysisSensDust}
\end{figure*}

\begin{figure*}[hbt!]
\centering
	\subfloat[]{
		\includegraphics[width=0.98\linewidth]{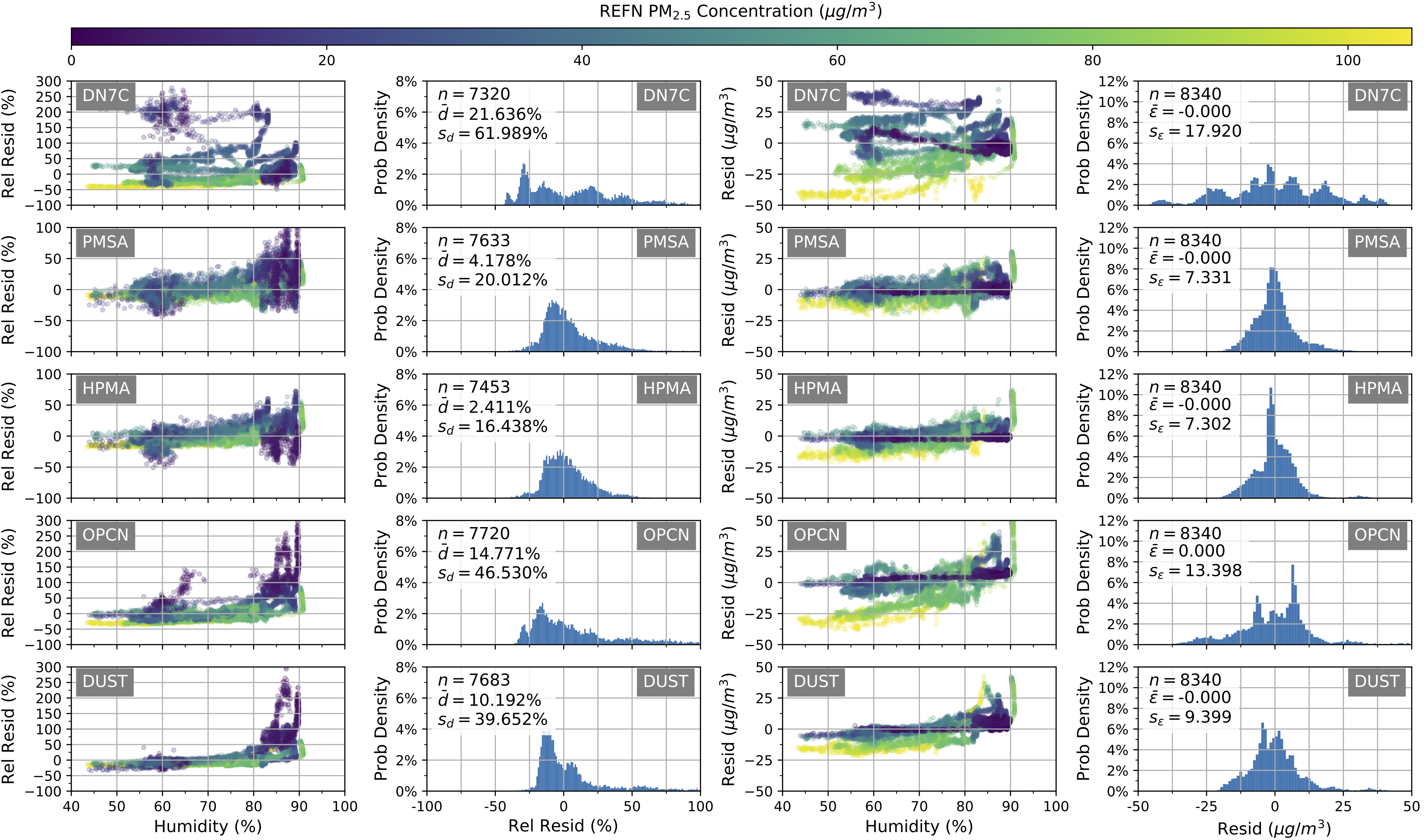}
		\label{subfig:CaliErrorOlsRefn}}
	\\
	\subfloat[]{
		\includegraphics[width=0.98\linewidth]{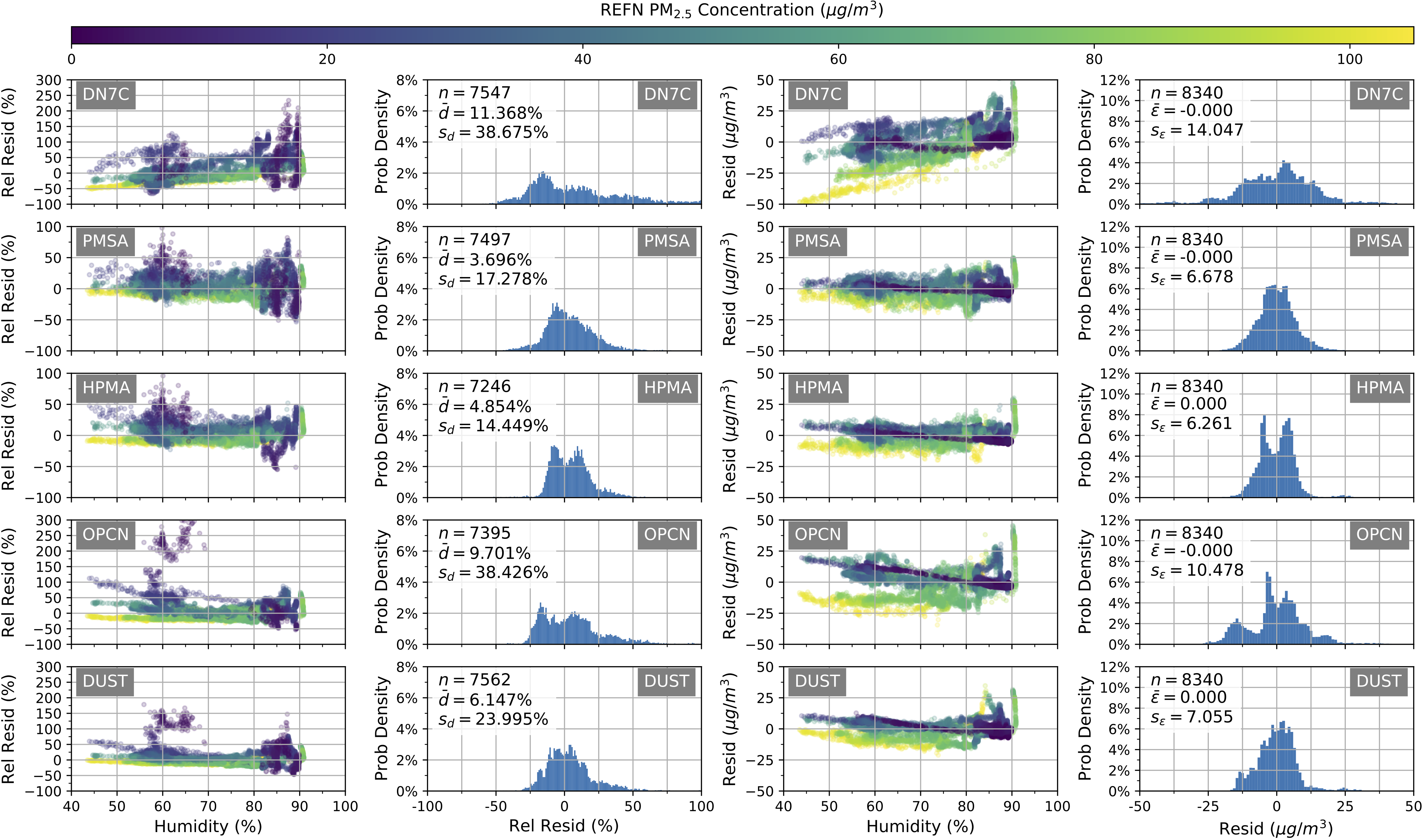}
		\label{subfig:CaliErrorMlrRefn}}
\caption{Residual analysis of the PM sensors and the DUST that have been calibrated against the REFN using (a) OLS model and (b) MLR model (MLT for the DN7C and MLH for the others) on the 1-minute interval collocated data pairs (first field experiment conducted from 13 to 23 February 2018 with abnormal spikes and measurements before 17 February eliminated; $n=8340$). The Rel Resid and Resid represent the relative residuals and the (absolute) residuals of PM\textsubscript{2.5} mass concentration between the calibrated instruments and the REFN, respectively; the Rel Resid value is considered valid when both concentrations are $\geq\SI[per-mode=symbol]{3}{\micro\gram\per\cubic\meter}$. The explanations of the scatter plots, the histograms, and the notation are detailed in Fig. \ref{fig:ErrorAnalysisSensDust}.}
\label{fig:ErrorAnalysisSensDustRefn}
\end{figure*}

\begin{table*}[ht]
\centering
\caption{Relative residuals of the PM instruments calibrated with difference parametric models at different concentration ranges.}
\label{tab:RelativeErrorAnalysisSensDustRefn}
\renewcommand{\arraystretch}{1.3}
\begin{threeparttable}
\begin{tabular}{|c|c|c|c|r|r|r|r|r|}
\hline
\textbf{Reference} & \textbf{Error Type} & \textbf{Model} & \textbf{Characteristics}\tnote{g} & \multicolumn{1}{c|}{\textbf{DN7C}} & \multicolumn{1}{c|}{\textbf{PMSA}} & \multicolumn{1}{c|}{\textbf{HPMA}} & \multicolumn{1}{c|}{\textbf{OPCN}} & \multicolumn{1}{c|}{\textbf{DUST}} \\
\hline
\multirow{18}{*}{DUST\tnote{a}} & \multirow{9}{*}{Relative Residual ($\%$)\tnote{c}} & \multirow{3}{*}{OLS}
& $n$ & \multicolumn{4}{c|}{$13140$} & n/a \\
\cline{4-9} & & & $\bar{d}$ & $6.734$ & $0.081$ & $-0.466$ & $3.681$ & n/a \\
\cline{4-9} & & & $s_{d}$ & $33.552$ & $7.573$ & $13.649$ & $16.964$ & n/a \\
\cline{3-9} & & \multirow{3}{*}{MLR\tnote{e}} 
& $n$ & \multicolumn{4}{c|}{$13140$} & n/a \\
\cline{4-9} & & & $\bar{d}$ & $4.058$ & $0.552$ & $-0.054$ & $1.487$ & n/a \\
\cline{4-9} & & & $s_{d}$ & $22.102$ & $6.755$ & $9.331$ & $12.892$ & n/a \\
\cline{3-9} & & \multirow{3}{*}{MLHT\tnote{f}} 
& $n$ & \multicolumn{4}{c|}{$13140$} & n/a \\
\cline{4-9} & & & $\bar{d}$ & $0.295$ & $0.690$ & $-0.143$ & $1.405$ & n/a \\
\cline{4-9} & & & $s_{d}$ & $14.457$ & $6.861$ & $9.529$ & $13.188$ & n/a \\
\cline{2-9} & \multirow{9}{*}{\begin{tabular}[c]{@{}c@{}}Relative Residual ($\%$)\\ (DUST $\geq\SI[per-mode=symbol]{50}{\micro\gram\per\cubic\meter}$)\tnote{d}\end{tabular}} & \multirow{3}{*}{OLS}
& $n$ & \multicolumn{4}{c|}{$8683$} & n/a \\
\cline{4-9} & & & $\bar{d}$ & $-3.639$ & $0.214$ & $1.235$ & $-1.808$ & n/a \\
\cline{4-9} & & & $s_{d}$ & $19.238$ & $6.473$ & $10.452$ & $13.796$ & n/a \\
\cline{3-9} & & \multirow{3}{*}{MLR\tnote{e}} 
& $n$ & \multicolumn{4}{c|}{$8683$} & n/a \\
\cline{4-9} & & & $\bar{d}$ & $-2.512$ & $-0.257$ & $0.639$ & $-0.336$ & n/a \\
\cline{4-9} & & & $s_{d}$ & $14.724$ & $5.272$ & $7.349$ & $11.938$ & n/a \\
\cline{3-9} & & \multirow{3}{*}{MLHT\tnote{f}} 
& $n$ & \multicolumn{4}{c|}{$8683$} & n/a \\
\cline{4-9} & & & $\bar{d}$ & $0.068$ & $-0.257$ & $0.649$ & $-0.326$ & n/a \\
\cline{4-9} & & & $s_{d}$ & $10.463$ & $5.062$ & $7.148$ & $11.890$ & n/a \\
\hline
\multirow{18}{*}{REFN\tnote{b}} & \multirow{9}{*}{Relative Residual ($\%$)\tnote{c}} & \multirow{3}{*}{OLS}
& $n$ & $7320$ & $7633$ & $7453$ & $7720$ & $7683$ \\
\cline{4-9} & & & $\bar{d}$ & $21.636$ & $4.178$ & $2.411$ & $14.771$ & $10.192$ \\
\cline{4-9} & & & $s_{d}$ & $61.989$ & $20.012$ & $16.438$ & $46.530$ & $39.652$ \\ \cline{3-9} & & \multirow{3}{*}{MLR\tnote{e}}
& $n$ & $7547$ & $7497$ & $7246$ & $7395$ & $7562$ \\
\cline{4-9} & & & $\bar{d}$ & $11.368$ & $3.696$ & $4.854$ & $9.701$ & $6.147$ \\
\cline{4-9} & & & $s_{d}$ & $38.675$ & $17.278$ & $14.449$ & $38.426$ & $23.995$ \\
\cline{3-9} & & \multirow{3}{*}{MLHT\tnote{f}} 
& $n$ & $7407$ & $7701$ & $7553$ & $7716$ & $7688$ \\
\cline{4-9} & & & $\bar{d}$ & $8.317$ & $4.613$ & $2.525$ & $9.676$ & $7.282$ \\
\cline{4-9} & & & $s_{d}$ & $49.113$ & $22.945$ & $20.874$ & $52.039$ & $32.833$ \\
\cline{2-9} & \multirow{9}{*}{\begin{tabular}[c]{@{}c@{}}Relative Residual ($\%$)\\ (REFN $\geq\SI[per-mode=symbol]{50}{\micro\gram\per\cubic\meter}$)\tnote{d}\end{tabular}} & \multirow{3}{*}{OLS}
& $n$ & \multicolumn{5}{c|}{$3579$} \\
\cline{4-9} & & & $\bar{d}$ & $-13.965$ & $-1.722$ & $-1.389$ & $-4.574$ & $-2.792$ \\
\cline{4-9} & & & $s_{d}$ & $20.536$ & $12.658$ & $13.113$ & $23.270$ & $16.185$ \\
\cline{3-9} & & \multirow{3}{*}{MLR\tnote{e}} 
& $n$ & \multicolumn{5}{c|}{$3579$} \\
\cline{4-9} & & & $\bar{d}$ & $-7.485$ & $-1.702$ & $-1.259$ & $-2.736$ & $-1.949$ \\ 
\cline{4-9} & & & $s_{d}$ & $21.797$ & $11.078$ & $10.918$ & $19.898$ & $12.206$ \\ 
\cline{3-9} & & \multirow{3}{*}{MLHT\tnote{f}} 
& $n$ & \multicolumn{5}{c|}{$3579$} \\ 
\cline{4-9} & & & $\bar{d}$ & $-2.220$ & $-1.200$ & $-0.461$ & $-1.386$ & $-1.543$ \\ 
\cline{4-9} & & & $s_{d}$ & $12.441$ & $10.305$ & $9.500$ & $17.366$ & $11.234$ \\ 
\hline
\end{tabular}
\begin{tablenotes}
	\item [a] The PM sensors (1-minute averaged ADC readings or PM\textsubscript{2.5} mass concentrations) are calibrated against the DUST (1-minute interval PM\textsubscript{2.5} mass concentrations) using the collocated data pairs acquired from 13 to 23 February 2018 with abnormal spikes eliminated ($n=13140$).
	\item [b] The PM sensors (1-minute averaged ADC readings or PM\textsubscript{2.5} mass concentrations) and the DUST (1-minute interval PM\textsubscript{2.5} mass concentrations) are calibrated against the REFN (1-minute interval PM\textsubscript{2.5} mass concentrations) using the collocated data pairs acquired from 17 to 23 February 2018 with abnormal spikes eliminated ($n=8340$).
	\item [c] The relative residual is considered valid when the calibrated and the reference PM\textsubscript{2.5} mass concentrations are both $\geq\SI[per-mode=symbol]{3}{\micro\gram\per\cubic\meter}$.
	\item [d] The relative residual will be included in the analysis when its corresponding reference PM\textsubscript{2.5} mass concentration is $\geq\SI[per-mode=symbol]{50}{\micro\gram\per\cubic\meter}$.
 	\item [e] The DN7C is calibrated using MLT model while the others are calibrated using MLH model.
	\item [f] The symbol MLHT represents the MLR parametric model involving both RH and TEMP.
	\item [g] The symbols $n$, $\bar{d}$, and $s_{d}$ represent the number of valid data pairs, the sample mean of relative residuals, and the sample standard deviation of relative residuals, respectively.
\end{tablenotes}
\end{threeparttable}
\end{table*}

\begin{table*}[ht]
\centering
\caption{Residuals of the PM instruments calibrated with difference parametric models at different concentration ranges.}
\label{tab:ErrorAnalysisSensDustRefn}
\renewcommand{\arraystretch}{1.3}
\begin{threeparttable}
\begin{tabular}{|c|c|c|c|r|r|r|r|r|}
\hline
\textbf{Reference} & \textbf{Error Type} & \textbf{Model} & \textbf{Characteristics}\tnote{g} & \multicolumn{1}{c|}{\textbf{DN7C}} & \multicolumn{1}{c|}{\textbf{PMSA}} & \multicolumn{1}{c|}{\textbf{HPMA}} & \multicolumn{1}{c|}{\textbf{OPCN}} & \multicolumn{1}{c|}{\textbf{DUST}} \\
\hline
\multirow{18}{*}{DUST\tnote{a}} & \multirow{9}{*}{Residual ($\SI[per-mode=symbol]{}{\micro\gram\per\cubic\meter}$)\tnote{c}} & \multirow{3}{*}{OLS}
& $n$ & \multicolumn{4}{c|}{$13140$} & n/a \\
\cline{4-9} & & & $\bar{\epsilon}$ & $0.000$ & $-0.000$ & $0.000$ & $-0.000$ & n/a \\
\cline{4-9} & & & $s_{\epsilon}$ & $15.493$ & $4.623$ & $6.680$ & $9.439$ & n/a \\
\cline{3-9} & & \multirow{3}{*}{MLR\tnote{e}} 
& $n$ & \multicolumn{4}{c|}{$13140$} & n/a \\
\cline{4-9} & & & $\bar{\epsilon}$ & $-0.000$ & $-0.000$ & $0.000$ & $0.000$ & n/a \\
\cline{4-9} & & & $s_{\epsilon}$ & $10.558$ & $3.801$ & $4.889$ & $8.229$ & n/a \\
\cline{3-9} & & \multirow{3}{*}{MLHT\tnote{f}} 
& $n$ & \multicolumn{4}{c|}{$13140$} & n/a \\
\cline{4-9} & & & $\bar{\epsilon}$ & $0.000$ & $0.000$ & $-0.000$ & $-0.000$ & n/a \\
\cline{4-9} & & & $s_{\epsilon}$ & $7.358$ & $3.617$ & $4.861$ & $8.214$ & n/a \\
\cline{2-9} & \multirow{9}{*}{\begin{tabular}[c]{@{}c@{}}Residual ($\SI[per-mode=symbol]{}{\micro\gram\per\cubic\meter}$)\\ (DUST $<\SI[per-mode=symbol]{50}{\micro\gram\per\cubic\meter}$)\tnote{d}\end{tabular}} & \multirow{3}{*}{OLS}
& $n$ & \multicolumn{4}{c|}{$4457$} & n/a \\
\cline{4-9} & & & $\bar{\epsilon}$ & $8.834$ & $0.349$ & $0.160$ & $3.603$ & n/a \\
\cline{4-9} & & & $s_{\epsilon}$ & $14.482$ & $3.305$ & $5.748$ & $4.724$ & n/a \\
\cline{3-9} & & \multirow{3}{*}{MLR\tnote{e}} 
& $n$ & \multicolumn{4}{c|}{$4457$} & n/a \\
\cline{4-9} & & & $\bar{\epsilon}$ & $4.559$ & $0.450$ & $-0.046$ & $2.120$ & n/a \\
\cline{4-9} & & & $s_{\epsilon}$ & $9.663$ & $2.967$ & $4.125$ & $4.461$ & n/a \\
\cline{3-9} & & \multirow{3}{*}{MLHT\tnote{f}} 
& $n$ & \multicolumn{4}{c|}{$4457$} & n/a \\
\cline{4-9} & & & $\bar{\epsilon}$ & $1.374$ & $0.399$ & $-0.053$ & $2.112$ & n/a \\
\cline{4-9} & & & $s_{\epsilon}$ & $6.278$ & $2.557$ & $4.315$ & $4.638$ & n/a \\
\hline
\multirow{18}{*}{REFN\tnote{b}} & \multirow{9}{*}{Residual ($\SI[per-mode=symbol]{}{\micro\gram\per\cubic\meter}$)\tnote{c}} & \multirow{3}{*}{OLS}
& $n$ & \multicolumn{5}{c|}{$8340$} \\
\cline{4-9} & & & $\bar{\epsilon}$ & $-0.000$ & $-0.000$ & $-0.000$ & $0.000$ & $-0.000$ \\
\cline{4-9} & & & $s_{\epsilon}$ & $17.920$ & $7.331$ & $7.302$ & $13.398$ & $9.399$ \\ \cline{3-9} & & \multirow{3}{*}{MLR\tnote{e}}
& $n$ & \multicolumn{5}{c|}{$8340$} \\
\cline{4-9} & & & $\bar{\epsilon}$ & $-0.000$ & $-0.000$ & $0.000$ & $-0.000$ & $0.000$ \\
\cline{4-9} & & & $s_{\epsilon}$ & $14.047$ & $6.678$ & $6.261$ & $10.478$ & $7.055$ \\
\cline{3-9} & & \multirow{3}{*}{MLHT\tnote{f}} 
& $n$ & \multicolumn{5}{c|}{$8340$} \\
\cline{4-9} & & & $\bar{\epsilon}$ & $-0.000$ & $-0.000$ & $0.000$ & $-0.000$ & $-0.000$ \\
\cline{4-9} & & & $s_{\epsilon}$ & $8.022$ & $6.168$ & $5.245$ & $9.435$ & $6.693$ \\
\cline{2-9} & \multirow{9}{*}{\begin{tabular}[c]{@{}c@{}}Residual ($\SI[per-mode=symbol]{}{\micro\gram\per\cubic\meter}$)\\ (REFN $<\SI[per-mode=symbol]{50}{\micro\gram\per\cubic\meter}$)\tnote{d}\end{tabular}} & \multirow{3}{*}{OLS}
& $n$ & \multicolumn{5}{c|}{$4761$} \\
\cline{4-9} & & & $\bar{\epsilon}$ & $9.348$ & $1.330$ & $1.282$ & $3.675$ & $1.976$ \\
\cline{4-9} & & & $s_{\epsilon}$ & $12.974$ & $5.114$ & $4.224$ & $7.517$ & $6.065$ \\
\cline{3-9} & & \multirow{3}{*}{MLR\tnote{e}} 
& $n$ & \multicolumn{5}{c|}{$4761$} \\
\cline{4-9} & & & $\bar{\epsilon}$ & $5.186$ & $1.350$ & $1.240$ & $2.656$ & $1.538$ \\ 
\cline{4-9} & & & $s_{\epsilon}$ & $8.758$ & $5.113$ & $4.135$ & $5.550$ & $4.735$ \\ 
\cline{3-9} & & \multirow{3}{*}{MLHT\tnote{f}} 
& $n$ & \multicolumn{5}{c|}{$4761$} \\ 
\cline{4-9} & & & $\bar{\epsilon}$ & $1.756$ & $0.929$ & $0.604$ & $1.660$ & $1.190$ \\ 
\cline{4-9} & & & $s_{\epsilon}$ & $6.430$ & $4.862$ & $3.184$ & $6.269$ & $5.028$ \\ 
\hline
\end{tabular}
\begin{tablenotes}
	\item [a] The PM sensors (1-minute averaged ADC readings or PM\textsubscript{2.5} mass concentrations) are calibrated against the DUST (1-minute interval PM\textsubscript{2.5} mass concentrations) using the collocated data pairs acquired from 13 to 23 February 2018 with abnormal spikes eliminated ($n=13140$).
	\item [b] The PM sensors (1-minute averaged ADC readings or PM\textsubscript{2.5} mass concentrations) and the DUST (1-minute interval PM\textsubscript{2.5} mass concentrations) are calibrated against the REFN (1-minute interval PM\textsubscript{2.5} mass concentrations) using the collocated data pairs acquired from 17 to 23 February 2018 with abnormal spikes eliminated ($n=8340$).
	\item [c] All residuals are considered valid for the entire PM\textsubscript{2.5} mass concentration range.
	\item [d] The residual will be included in the analysis when its corresponding reference PM\textsubscript{2.5} mass concentration is $<\SI[per-mode=symbol]{50}{\micro\gram\per\cubic\meter}$.
	\item [e] The DN7C is calibrated using MLT model while the others are calibrated using MLH model.
	\item [f] The symbol MLHT represents the MLR parametric model involving both RH and TEMP.
	\item [g] The symbols $n$, $\bar{\epsilon}$, and $s_{\epsilon}$ represent the number of valid data pairs, the sample mean of residuals, and the sample standard deviation of residuals, respectively.
\end{tablenotes}
\end{threeparttable}
\end{table*}

\begin{table*}[ht]
\centering
\caption{Performances of the nominated PM sensors and the implemented WePINs after calibrating with difference parametric models.}
\label{tab:EvaluationMetricsSensDustRefn}
\renewcommand{\arraystretch}{1.3}
\begin{threeparttable}
\begin{tabular}{|c|c|c|r|r|r|r|r|r|}
\hline
\textbf{Reference} & \textbf{Model} & \textbf{Metrics}\tnote{c} & \multicolumn{1}{c|}{\textbf{DN7C}} & \multicolumn{1}{c|}{\textbf{PMSA}} & \multicolumn{1}{c|}{\textbf{HPMA}} & \multicolumn{1}{c|}{\textbf{OPCN}} & \multicolumn{1}{c|}{\textbf{DUST}} & \multicolumn{1}{c|}{\textbf{WePIN}\tnote{f}} \\
\hline
\multirow{17}{*}{REFN\tnote{a}} & \multirow{8}{*}{OLS} & $n$ & $7320$ & $7633$ & $7453$ & $7720$ & $7683$ & $7610$ \\ 
\hhline{*{2}{|~}*{7}{|-}|} & 
& $Bias_{PEP}$ & $21.636\pm1.192$ & \cellcolor{gray!50}$4.178\pm0.377$ & \cellcolor{gray!50}$2.411\pm0.313$ & $14.771\pm0.871$ & $10.192\pm0.744$ & \cellcolor{gray!50}$3.186\pm0.491$ \\
\hhline{*{2}{|~}*{7}{|-}|} & 
& $\sigma_{UCL}$ & $62.655$ & $20.222$ & $16.613$ & $47.017$ & $40.068$ & $26.308$ \\ 
\hhline{*{2}{|~}*{7}{|-}|} & 
& $CV_{UCL}$\tnote{d} & $44.304$ & $14.299$ & $11.747$ & $33.246$ & $28.332$ & $18.602$ \\ 
\hhline{*{2}{|~}*{7}{|-}|} & 
& $slope$ & $0.435$ & \cellcolor{gray!50}$0.916$ & \cellcolor{gray!50}$0.901$ & $0.779$ & $0.885$ & $0.740$ \\ 
\hhline{*{2}{|~}*{7}{|-}|} & 
& $intercept$ & $29.851$ & $4.272$ & $5.291$ & $10.411$ & $5.546$ & $3.953$ \\ 
\hhline{*{2}{|~}*{7}{|-}|} & 
& $r$ & $0.689$ & \cellcolor{gray!50}$0.960$ & \cellcolor{gray!50}$0.958$ & $0.867$ & \cellcolor{gray!50}$0.935$ & $0.864$ \\ 
\hhline{*{2}{|~}*{7}{|-}|} & 
& $LOD$\tnote{e} & $12.470$ & $2.304$ & $1.909$ & $1.016$ & $1.689$ & n/a \\ 

\hhline{*{1}{|~}*{8}{|-}|} & \multirow{7}{*}{MLR\tnote{b}} & $n$ & $7547$ & $7497$ & $7246$ & $7395$ & $7562$ & $7382$ \\ 
\hhline{*{2}{|~}*{7}{|-}|} & 
& $Bias_{PEP}$ & $11.368\pm0.732$ & \cellcolor{gray!50}$3.696\pm0.328$ & \cellcolor{gray!50}$4.854\pm0.279$ & $9.701\pm0.735$ & \cellcolor{gray!50}$6.147\pm0.454$ & \cellcolor{gray!50}$3.882\pm0.407$ \\ 
\hhline{*{2}{|~}*{7}{|-}|} & 
& $\sigma_{UCL}$ & $39.084$ & $17.461$ & $14.605$ & $38.837$ & $24.248$ & $21.501$ \\ 
\hhline{*{2}{|~}*{7}{|-}|} & 
& $CV_{UCL}$\tnote{d} & $27.637$ & $12.347$ & $10.327$ & $27.8462$ & $17.146$ & $15.204$ \\ 
\hhline{*{2}{|~}*{7}{|-}|} & 
& $slope$ & $0.708$ & \cellcolor{gray!50}$0.910$ & $0.881$ & $0.802$ & \cellcolor{gray!50}$0.903$ & $0.808$ \\ 
\hhline{*{2}{|~}*{7}{|-}|} & 
& $intercept$ & $14.704$ & $4.889$ & $6.919$ & $10.533$ & $5.206$ & $3.114$ \\ 
\hhline{*{2}{|~}*{7}{|-}|} & 
& $r$ & $0.839$ & \cellcolor{gray!50}$0.966$ & \cellcolor{gray!50}$0.971$ & $0.909$ & \cellcolor{gray!50}$0.963$ & $0.915$ \\ 
\hhline{*{2}{|~}*{7}{|-}|} & 
& $LOD$\tnote{e} & 6.783 & 4.136 & 5.326 & 10.129 & 7.535 & n/a \\ 

\hhline{*{1}{|~}*{8}{|-}|} & \multirow{2}{*}{n/a} & $n$ & n/a & n/a & n/a & n/a & n/a & $9618$ \\ 
\hhline{*{2}{|~}*{7}{|-}|} &
& $CV_{RMS}$ & n/a & n/a & n/a & n/a & n/a & \cellcolor{gray!50}$14.719$ \\ 
\hline
\end{tabular}
\begin{tablenotes}
	\item [a] The nominated PM sensors (1-minute averaged ADC readings or PM\textsubscript{2.5} mass concentrations) and the DUST (1-minute interval PM\textsubscript{2.5} mass concentrations) are calibrated against the REFN (1-minute interval PM\textsubscript{2.5} mass concentrations) using the collocated data pairs acquired from 17 to 23 February 2018 with abnormal spikes eliminated ($n=8340$); the implemented WePINs (1-minute interval unit-wise average PM\textsubscript{2.5} mass concentrations) are calibrated against the REFN (1-minute interval PM\textsubscript{2.5} mass concentrations) using the collocated data pairs acquired from 26 June to 17 July 2019 with abnormal reference measurements eliminated ($n=7698$).
	\item [b] The DN7C is calibrated using MLT model while the others are calibrated using MLH model.
	\item [c] The symbol $n$ represents the number of valid data pairs; the definitions of the remaining symbols are detailed in section \ref{SubSubSect:EvaluationMetrics}.
	\item [d] The $CV_{UCL}$ value is calculated with the original formula in the Part 58AA to quantify the intermediate measurement precision ($CV_{UCL}=\sqrt{1/2}\cdot\sigma_{UCL}$).
	\item [e] In order to have sufficient numbers of low concentration pairs (reference PM\textsubscript{2.5} mass concentrations $<\SI[per-mode=symbol]{3}{\micro\gram\per\cubic\meter})$, the RH condition is set to $\leq\SI{80}{\percent}$ for the first field experiment ($n=64$). No low concentration pairs were available in the second field experiment.
	\item [f] For the WePINs, all the metrics except the $CV_{RMS}$ are calculated from the calibrated (1-minute interval unit-wise average PM\textsubscript{2.5}) and reference (1-minute interval PM\textsubscript{2.5}) concentration pairs, whereas the $CV_{RMS}$ are calculated using the PM\textsubscript{2.5} measurements (1-minute interval) from all WePINs.
\end{tablenotes}
\end{threeparttable}
\end{table*}

\begin{figure*}[!t]
\centering
	\includegraphics[width=0.65\linewidth]{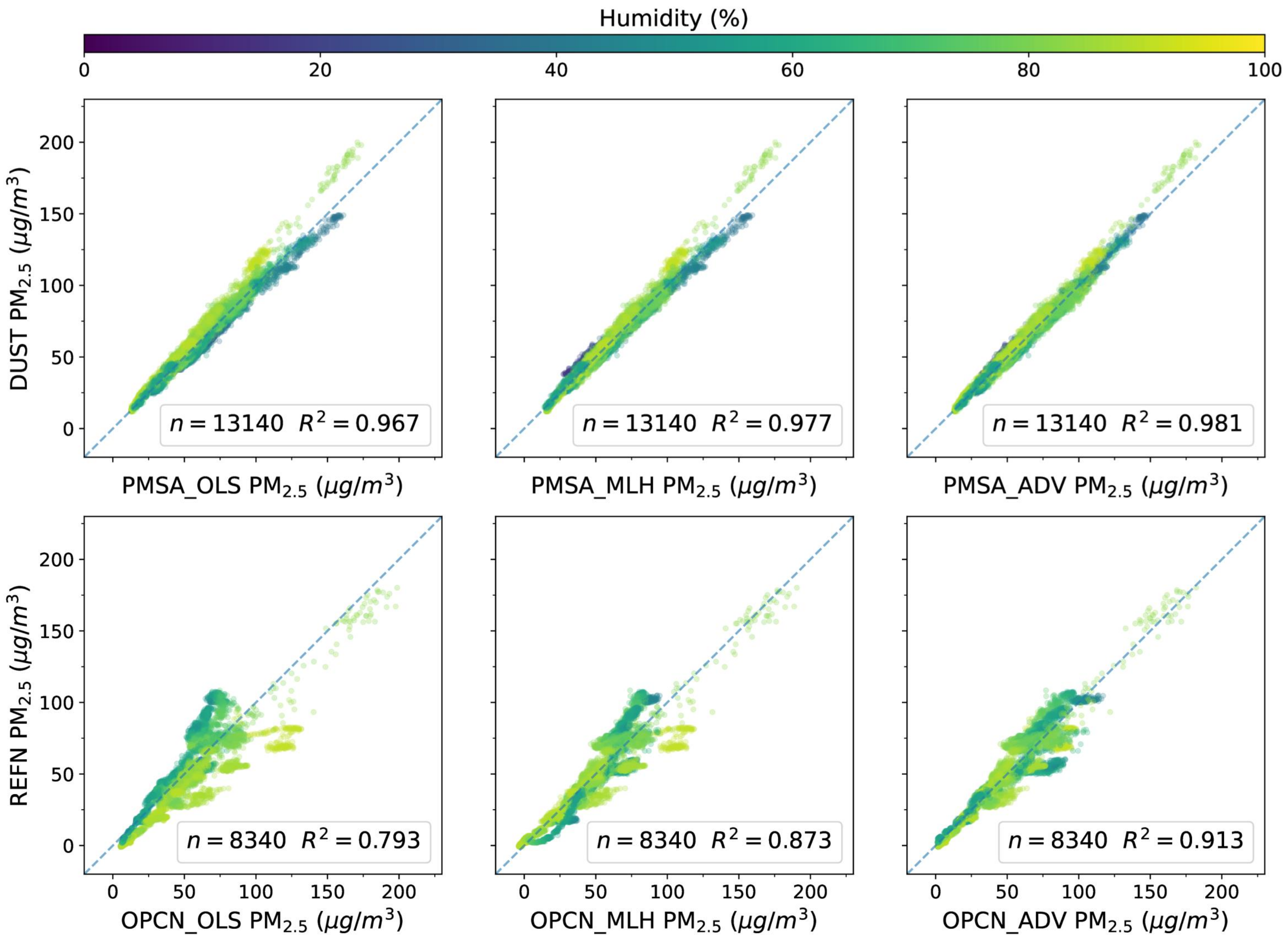}
\caption{Scatter plots and pairwise correlations (in terms of $R^2$) of the collocated data pairs between the DUST and the PMSA (acquired from 13 to 23 February 2018 with abnormal spikes eliminated; $n=13140$) and between the REFN and the OPCN (acquired from 17 to 23 February 2018 with abnormal spikes eliminated; $n=8340$). The vertical axes of the first and the second rows represent the 1-minute interval PM\textsubscript{2.5} mass concentrations from the DUST and the REFN, respectively. The horizontal axes of the first, the second, and the third columns represent the PMSA's or OPCN's 1-minute interval calibrated PM\textsubscript{2.5} mass concentrations from the OLS, the MLH, and the ADV models, respectively. The color of each data point indicates the corresponding RH value. The blue dashed lines are the identity lines.}
\label{fig:ScatterCaliAdvExample}         
\end{figure*}

\begin{figure*}[!hbt]
\centering
	\subfloat[]{
		\includegraphics[width=0.936\linewidth]{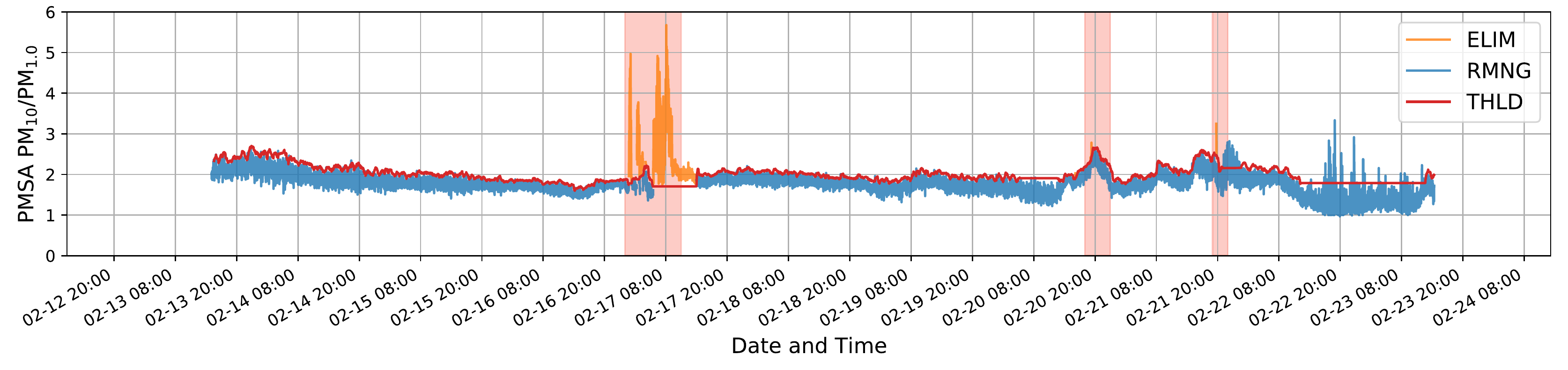}
		\label{subfig:AutoCleansingPMSARatio}}
	\\
	\subfloat[]{
		\includegraphics[width=0.95\linewidth]{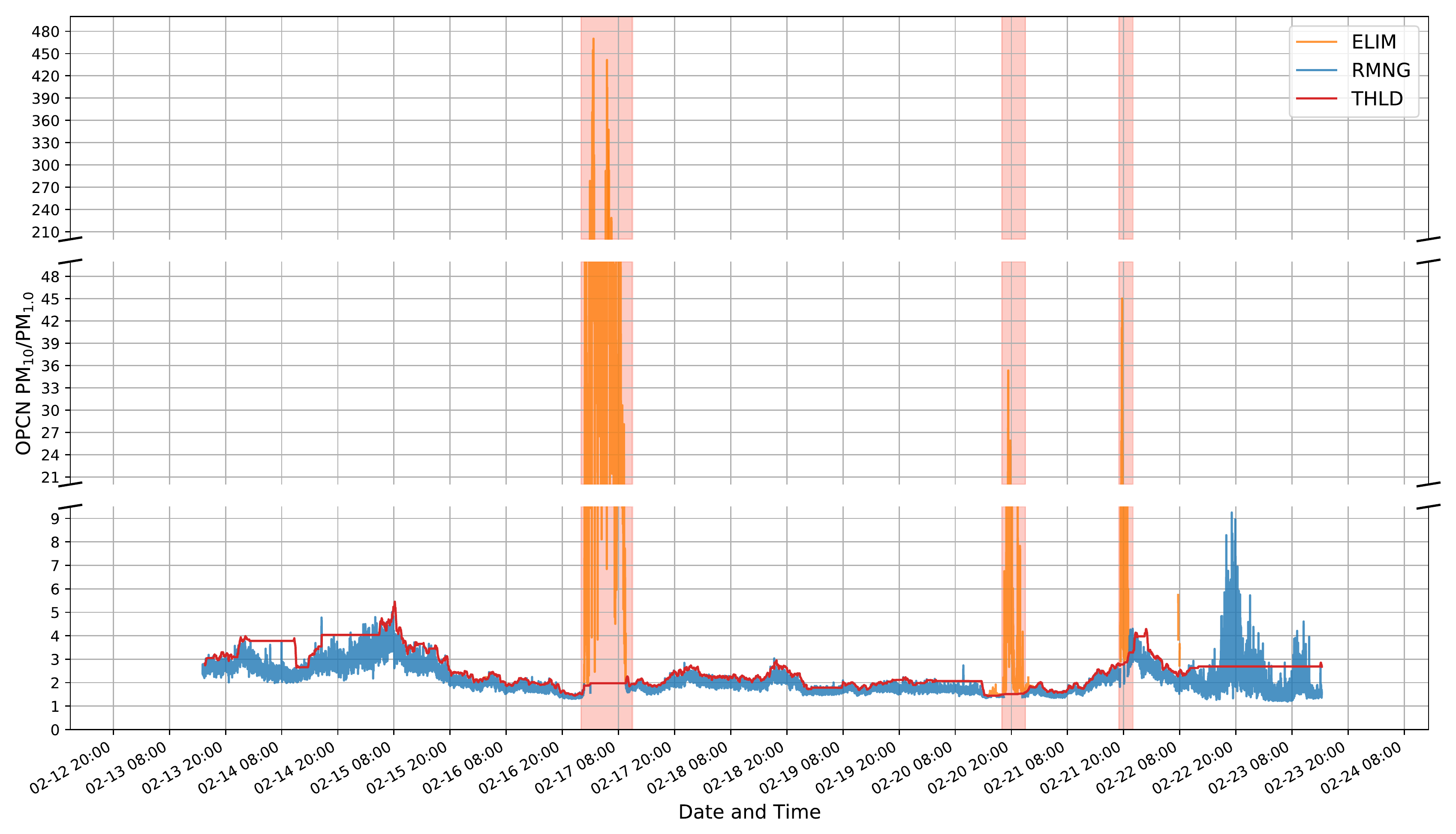}
		\label{subfig:AutoCleansingOPCNRatio}}
\caption{Ratios between the PM\textsubscript{10} and PM\textsubscript{1.0} mass concentrations (1-minute averaged measurements in the first field experiment conducted from 13 to 23 February 2018 with abnormal spikes included; $n=14280$) measured by (a) the PMSA and (b) the OPCN. The three intervals of abnormal spikes that were manually identified and eliminated in Fig. \ref{fig:PM2p5RhTempFull} are highlighted (salmon span) accordingly. The ELIM (orange) and the RMNG (blue) solid lines represent the eliminated and the remaining ratios that are determined by the proposed automated data cleansing approach, respectively; an eliminated ratio value is considered to be indicating the occurrence of an unusual measurement and the corresponding PM\textsubscript{2.5} mass concentration will be eliminated. The THLD (red) solid line represents the ratio thresholds determined by the moving window (containing a number of selected previous ratios) of the proposed approach; any ratio over the threshold value at a certain time will be eliminated if the corresponding PM\textsubscript{2.5} mass concentration is reasonably large and measured under extremely humid condition.}
\label{fig:AutoCleansingRatio}         
\end{figure*}

\begin{figure*}[!hbt]
\centering
	\subfloat[]{
		\includegraphics[width=0.95\linewidth]{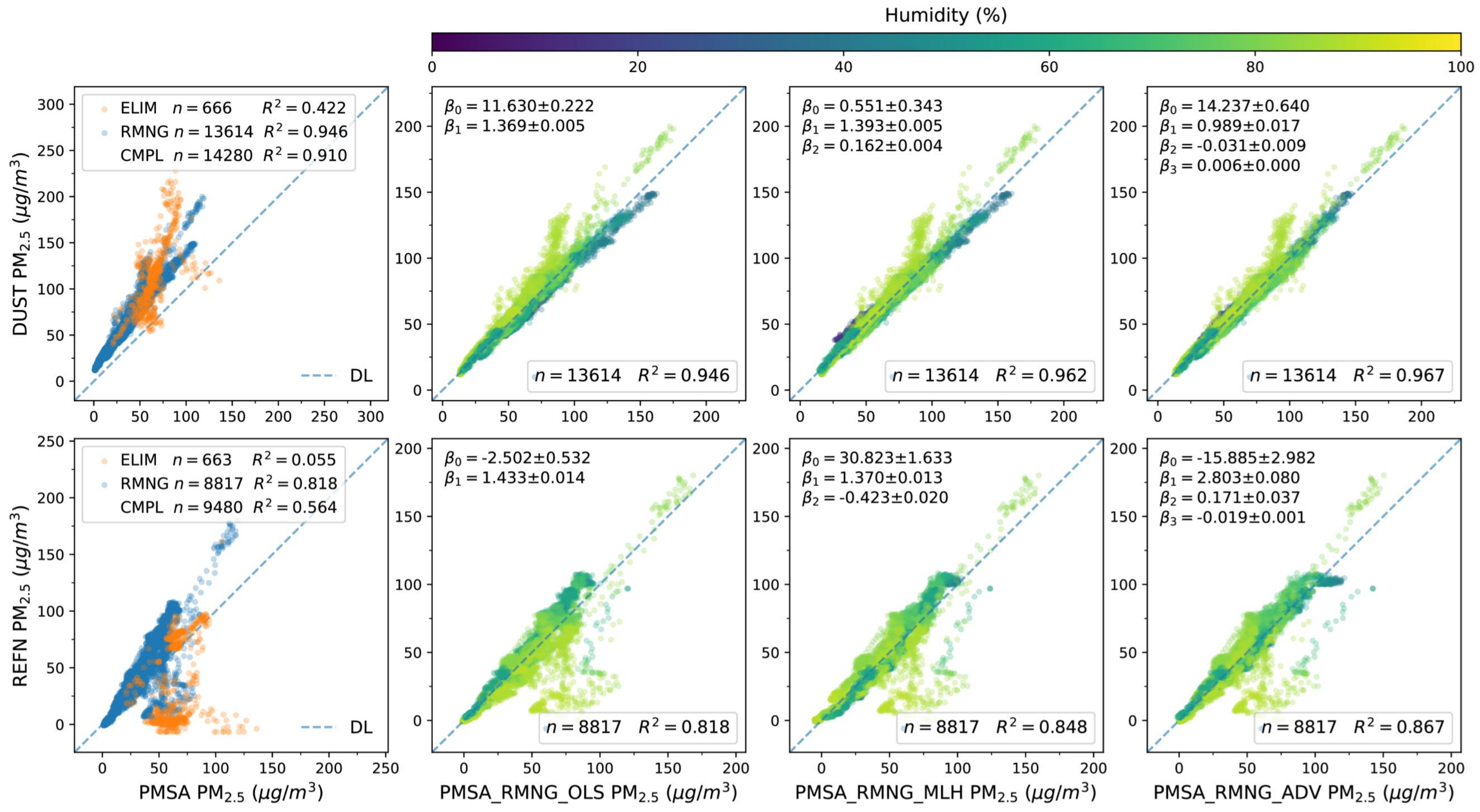}
		\label{subfig:AutoCleansingPMSAScatter}}
	\\
	\subfloat[]{
		\includegraphics[width=0.95\linewidth]{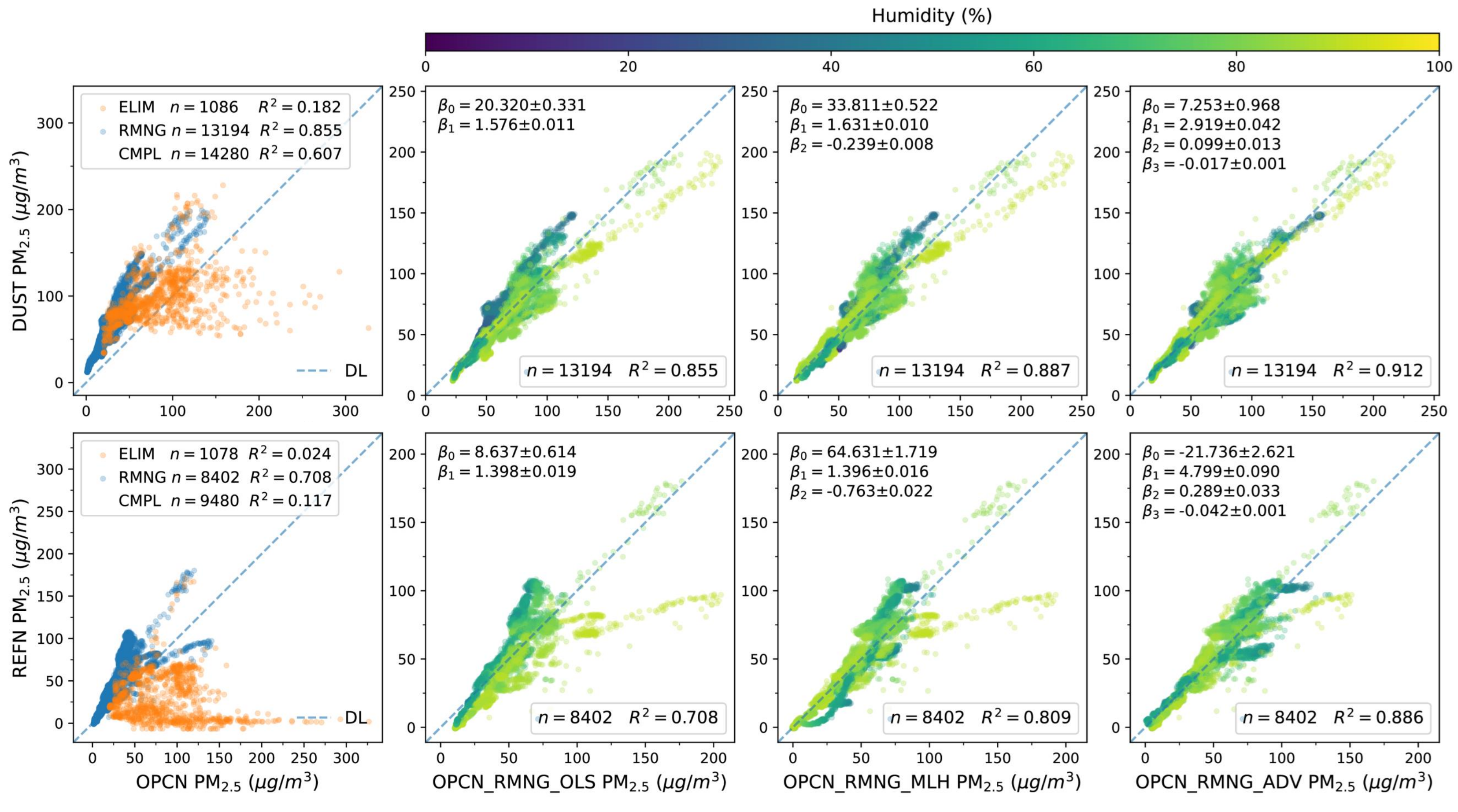}
		\label{subfig:AutoCleansingOPCNScatter}}
\caption{Scatter plots and pairwise correlations (in terms of $R^2$) of the collocated data pairs (first field experiment conducted from 13 to 23 February 2018) between the reference instruments (the DUST or the REFN; vertical axis; 1-minute interval PM\textsubscript{2.5} mass concentrations) and (a) the PMSA or (b) the OPCN (horizontal axis; 1-minute averaged PM\textsubscript{2.5} mass concentrations). The diagonal dashed lines (DL) in blue are the identity lines. Scatter plots in the first column contain the collocated data pairs of the entire campaign (CMPL; $n=14280$) and that after 17 February (CMPL; $n=9480$) when using the DUST and the REFN as reference, respectively; the ELIM (orange) and RMNG (blue) data points represent the eliminated and the remaining data pairs determined by the proposed automated data cleansing approach, respectively. The scatter plots in the second to fourth columns contain the remaining collocated data pairs; the color of each data point indicates the corresponding RH value. The horizontal axes of the second, the third, and the fourth columns represent the PMSA's or OPCN's calibrated PM\textsubscript{2.5} mass concentrations from the OLS, the MLH, and the ADV models, respectively; the parameters ($\SI{95}{\percent}$ confidence interval) of these calibration models are listed on the upper left corners of respective scatter plots.}
\label{fig:AutoCleansingScatter}
\end{figure*}

\begin{figure*}[!hbt]
\centering
	\includegraphics[width=0.95\linewidth]{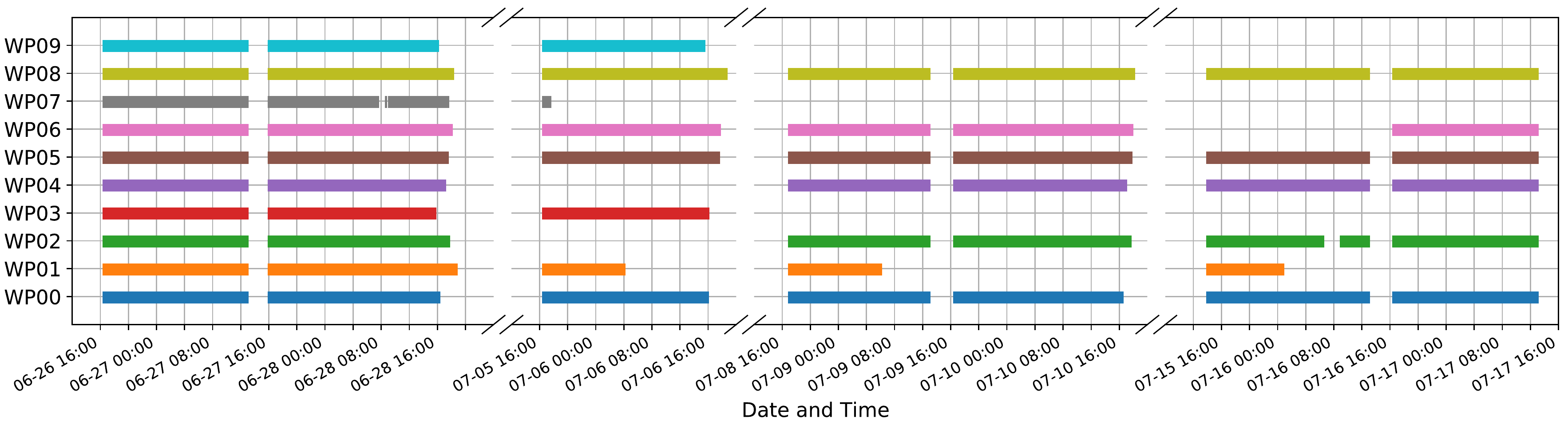}
\caption{Data availability of the ten WePINs. These devices were not operational from 29 June to 4 July, on 7 July, and from 11 to 14 July because they could not be charged on rainy days. The missing data of a device were primarily caused by malfunctions. Note that a daily charging routine was established from around 13:30 to 16:00, during which no measurements were recorded.}
\label{fig:WePINsDataAvali}
\end{figure*}

\begin{figure*}[!hbt]
\centering
	\includegraphics[width=0.95\linewidth]{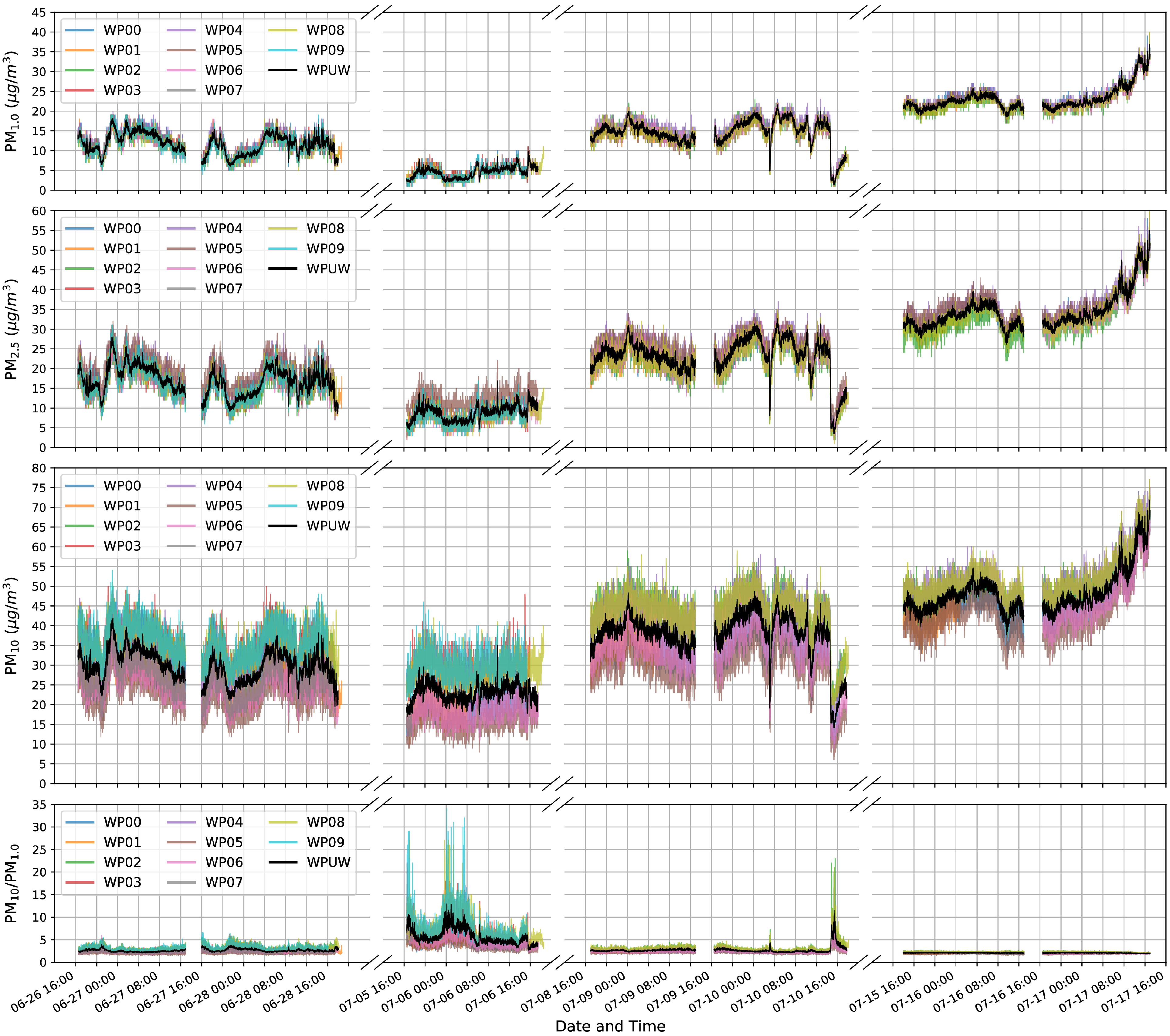}
\caption{Mass concentrations of PM\textsubscript{1.0}, PM\textsubscript{2.5}, and PM\textsubscript{10} (1-minute interval) measured by the ten WePINs plus the corresponding ratios between the PM\textsubscript{10} and PM\textsubscript{1.0} during the second field experiment conducted from 26 June to 17 July 2019. The unit-wise average PM mass concentrations and PM ratios (black; averaged from at least three WePIN units at a certain time; $n=9618$) are also shown in respective time series plots.} 
\label{fig:WePINsPMDataPlot}
\end{figure*}

\begin{figure*}[!hbt]
\centering
	\includegraphics[width=0.95\linewidth]{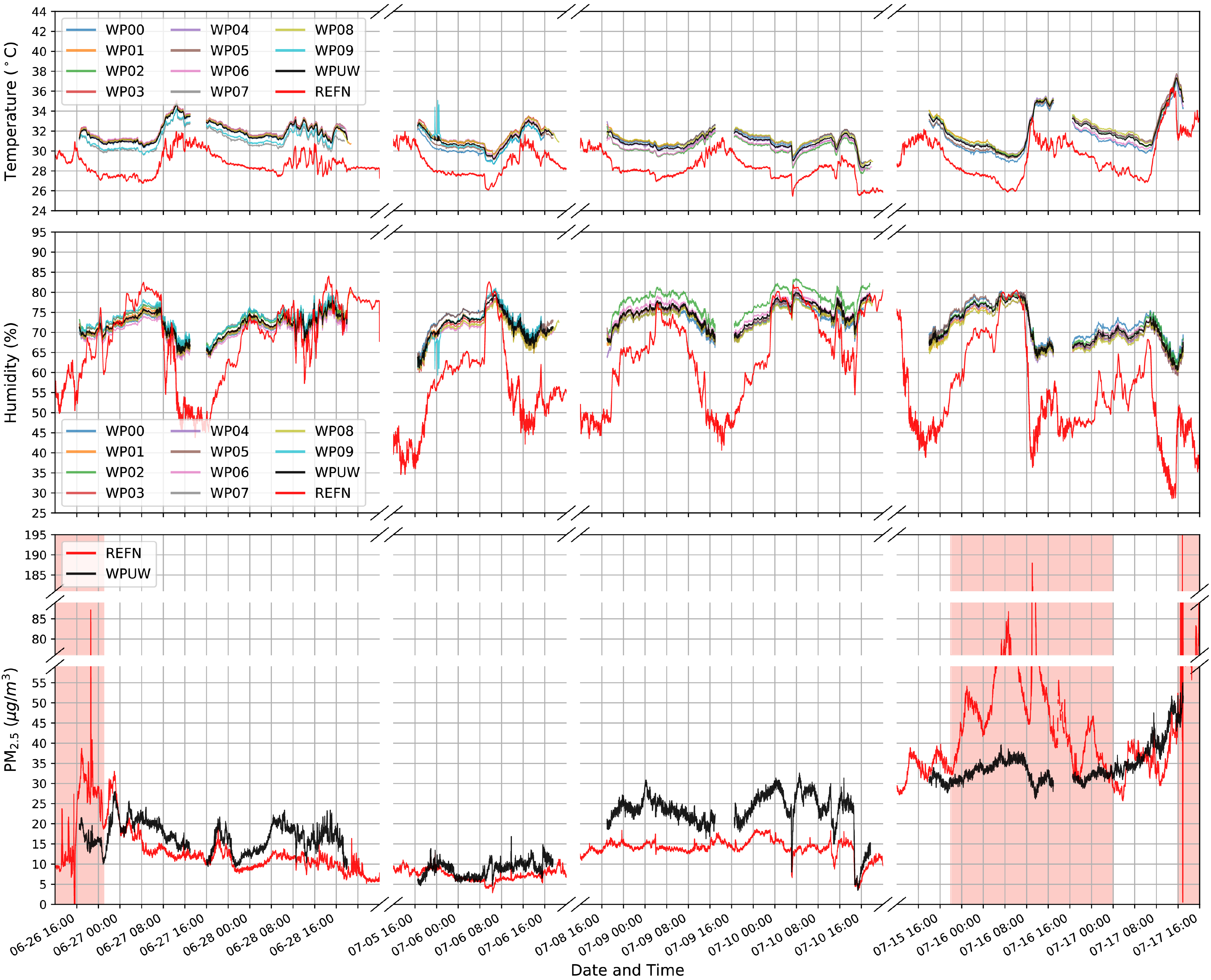}
\caption{TEMP and RH readings and PM\textsubscript{2.5} mass concentrations (1-minute interval) measured by the ten WePINs and the REFN (red) reference instrument during the second field experiment conducted from 26 June to 17 July 2019. The unit-wise average TEMP and RH readings (black; averaged from at least three WePIN units at a certain time; $n=9618$) are also shown in the top and the middle time series plots, respectively. In the bottom time series plot, only the unit-wise average (black) and the reference (red) PM\textsubscript{2.5} mass concentrations are presented; three intervals with abnormal reference measurements were identified and highlighted (salmon span).}
\label{fig:WePINsRefnAllDataPlot}
\end{figure*}

\begin{figure*}[!hbt]
\centering
	\includegraphics[width=0.95\linewidth]{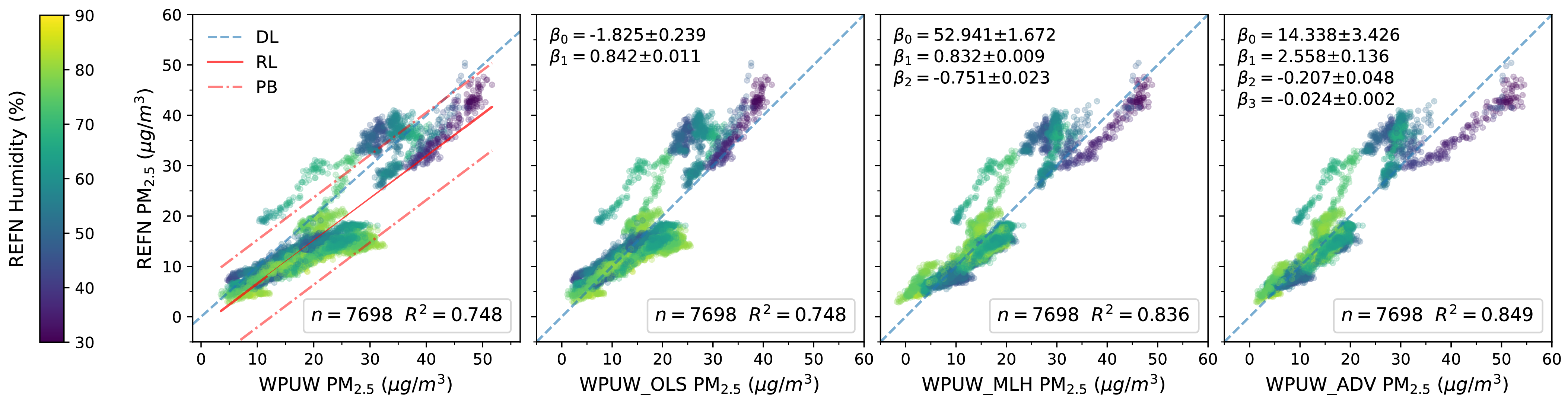}
\caption{Scatter plots and pairwise correlations (in terms of $R^2$) of the collocated data pairs (second field experiment conducted from 26 June to 17 July 2019 with abnormal reference measurements eliminated; $n=7698$) between the REFN (vertical axis; 1-minute interval PM\textsubscript{2.5} mass concentrations) and the WePINs (horizontal axis; 1-minute interval unit-wise average PM\textsubscript{2.5} mass concentrations in the first scatter plot; 1-minute interval calibrated unit-wise average PM\textsubscript{2.5} mass concentrations from the OLS, the MLH, and the ADV models in the second, the third, and the fourth scatter plots, respectively; the unit-wise average RH readings from the WePINs were used for calibration). The color of each data point indicates the corresponding RH value (1-minute interval RH readings from the REFN). The blue dashed lines (DL) are the identity lines. The red solid line (RL) and the paired red dot-dashed lines (PB) in the first scatter plot are the OLS regression line and the upper and lower prediction bounds (\SI{95}{\percent} confidence level), respectively. The parameters (\SI{95}{\percent} confidence interval) of the OLS, the MLH, and the ADV models are listed on the upper left corners of the second, the third, and the fourth scatter plots, respectively.}
\label{fig:ScatterCaliWePINsRefn}
\end{figure*}

\begin{figure*}[!hbt]
\centering
	\includegraphics[width=0.95\linewidth]{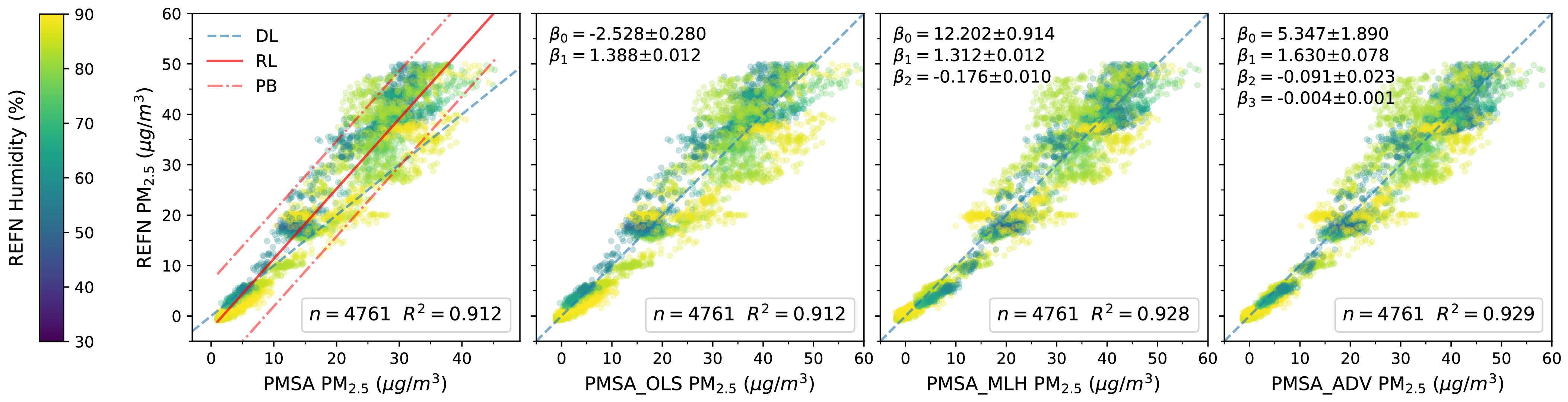}
\caption{Scatter plots and pairwise correlations (in terms of $R^2$) of the collocated data pairs (first field experiment conducted from 13 to 23 February 2018 with abnormal spikes and measurements before 17 February and above \SI[per-mode=symbol]{50}{\micro\gram\per\cubic\meter} eliminated; $n=4761$) between the REFN (vertical axis; 1-minute interval PM\textsubscript{2.5} mass concentrations) and the PMSA (horizontal axis; 1-minute averaged PM\textsubscript{2.5} mass concentrations in the first scatter plot; 1-minute interval calibrated PM\textsubscript{2.5} mass concentrations from the OLS, the MLH, and the ADV models in the second, the third, and the fourth scatter plots, respectively). The color of each data point indicates the corresponding RH value (1-minute interval RH readings from the REFN). The blue dashed lines (DL) are the identity lines. The red solid line (RL) and the paired red dot-dashed lines (PB) in the first scatter plot are the OLS regression line and the upper and lower prediction bounds (\SI{95}{\percent} confidence level), respectively. The parameters (\SI{95}{\percent} confidence interval) of the OLS, the MLH, and the ADV models are listed on the upper left corners of the second, the third, and the fourth scatter plots, respectively.}
\label{fig:ScatterCaliPMSARefn}
\end{figure*}

\begin{figure*}[!hbt]
\centering
	\includegraphics[width=0.98\linewidth]{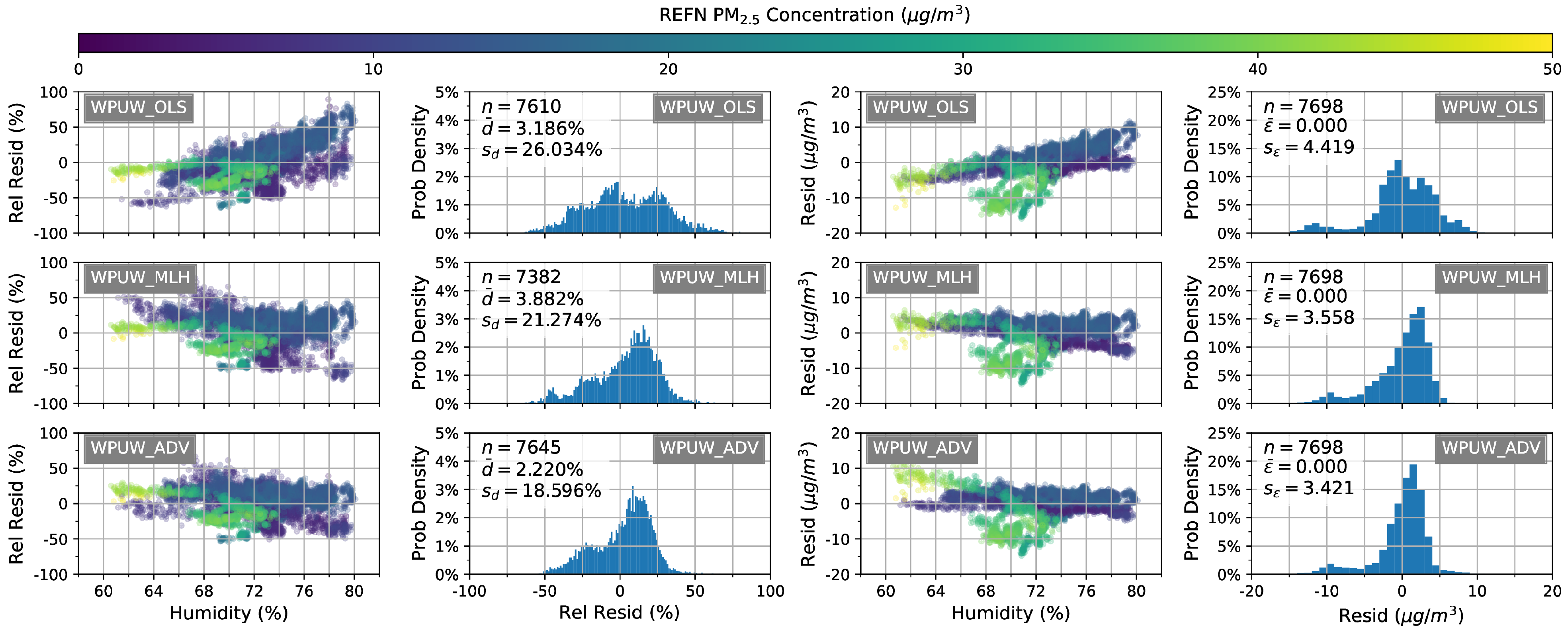}
\caption{Residual analysis of the ten WePINs' unit-wise average PM\textsubscript{2.5} mass concentrations (1-minute interval; averaged from at least three WePIN units at a certain time) that have been calibrated against the REFN's reference PM\textsubscript{2.5} mass concentrations (1-minute interval) using the OLS, the MLH, and the ADV models (collocated data pairs acquired in the second field experiment conducted from 26 June to 17 July 2019 with abnormal reference measurements eliminated; $n=7698$). The Rel Resid and Resid represent the relative residuals and the (absolute) residuals between the calibrated unit-wise average PM\textsubscript{2.5} mass concentrations and the corresponding reference ones, respectively; the Rel Resid value is considered valid when both concentrations are $\geq\SI[per-mode=symbol]{3}{\micro\gram\per\cubic\meter}$. Scatter plots in the first and the third columns are illustrating the relationships between the RH (unit-wise average RH readings from the WePINs) and the Rel Resid and the Resid, respectively; the color of each data point indicates the corresponding PM\textsubscript{2.5} mass concentration measured by the REFN. Histograms in the second and the fourth columns are illustrating the probability distributions of Rel Resid (\SI{1}{\percent} bin width) and Resid (\SI[per-mode=symbol]{1}{\micro\gram\per\cubic\meter} bin width), respectively; the number of valid data pairs ($n$), as well as the sample mean and the sample standard deviation of Rel Resid ($\bar{d}$ and $s_{d}$) or Resid ($\bar{\epsilon}$ and $s_{\epsilon}$) are listed on the upper left corner of each histogram.}
\label{fig:ErrorAnalysisWePINsRefn}
\end{figure*}

\clearpage
\twocolumn

\section*{Acknowledgment}

The authors would like to thank... 


\bibliographystyle{IEEEtran}
\bibliography{./citation/citation.bib}



\end{document}